\documentclass[
	fontsize=10pt, 
	twoside=true, 
	numbers=noenddot, 
]{kaobook}

\newcommand{\future}[1]{}
\usepackage{fix-cm}
\usepackage{styles/environments}
\usepackage{styles/mdftheorems}
\usepackage{blindtext}
\usepackage{mathtools}
\usepackage{xcolor}

\graphicspath{{images/}{./}}

\makeindex[columns=3, title=Alphabetical Index, intoc] 
\makeglossaries 
\makenomenclature 

\begin{document}

\titlehead{}

\title[Path Integral Methods in Atomistic Modelling]{Path Integral Methods in Atomistic Modelling: An Introduction}

\author{
Michele Ceriotti \\ 
\normalsize{Institut des Matériaux, École Polytechnique Fédérale de Lausanne} \\
\normalsize{Lausanne, Switzerland} 
\and
David E. Manolopoulos \\
\normalsize{Department of Chemistry, University of Oxford} \\
\normalsize{Oxford, United Kingdom}
\and
Thomas E. Markland \\
\normalsize{Department of Chemistry, Stanford University} \\
\normalsize{Stanford, California, USA}
\and
Mariana Rossi \\
\normalsize{MPI for the Structure and Dynamics of Matter} \\
\normalsize{Hamburg, Germany} \vspace{0.2cm} \\
\normalsize{Yusuf Hamied Chemistry Department, University of Cambridge} \\
\normalsize{Cambridge, UK}
}

\date{2021}

\frontmatter
\KOMAoptions{twoside=semi}
\maketitle
\KOMAoptions{twoside=true}

\begingroup 
\setlength{\textheight}{23cm} 
\etocstandarddisplaystyle 
\etocstandardlines

\tableofcontents 
\endgroup

\mainmatter

\cleardoublepage

\newcommand{\TODO}[1]{{\color{blue}\bfseries #1 } }
\newcommand{\mbf}[1]{\mathbf{#1}}
\newcommand{\mrm}[1]{\mathrm{#1}}
\newcommand{\mcal}[1]{\mathcal{#1}}
\newcommand{\bq}{\ensuremath{\mbf{q}}}
\newcommand{\bp}{\ensuremath{\mbf{p}}}
\newcommand{\bx}{\ensuremath{\mbf{x}}}
\newcommand{\bv}{\ensuremath{\mbf{v}}}
\newcommand{\bk}{\ensuremath{\mbf{k}}}

\newcommand{\Prob}[0]{\ensuremath{\mcal{P}}}

\newcommand{\kB}{\ensuremath{k_\mrm{B}}}
\newcommand{\rG}{\ensuremath{r_\mrm{G}}}
\newcommand{\TCV}{\ensuremath{\mcal{T}_\mrm{CV}}}

\newcommand{\CV}{\ensuremath{\mcal{V}}}
\newcommand{\CJ}{\ensuremath{\mcal{J}}}
\newcommand{\D}[2][]{\operatorname{d}^{#1}{#2}\,}
\newcommand{\pder}[2][]{\frac{\partial#1}{\partial#2}}
\newcommand{\avg}[1]{\ensuremath{\left\langle{#1}\right\rangle}}
\newcommand{\I}{\mathrm{i}}
\renewcommand{\Re}{\ensuremath{\operatorname{Re}}}
\renewcommand{\Im}{\ensuremath{\operatorname{Im}}}
\newcommand{\Arg}{\ensuremath{\operatorname{Arg}}}
\newcommand{\sign}{\ensuremath{\operatorname{sgn}}}

\setchapterpreamble[u]{\margintoc}
\chapter{Molecular dynamics and sampling}
\labch{sampling}

This chapter provides a brief overview of the problem of computing the properties of an ensemble of $N$ atoms or molecules, which can be fully characterized, at the classical level, by specifying their positions $\bq$ and momenta $\bp$. 
\nomenclature{$N$}{Number of particles in a simulation}
\nomenclature{$\bq$}{$3N$-dimensional vector containing the positions of all particles}
\nomenclature{$\bp$}{$3N$-dimensional vector containing the momenta of all particles}
It provides a summary of molecular dynamics, including a discussion of integrators, that will play a central role in many of the techniques discussed in subsequent chapters. 
It also discusses the problem of sampling, the link between autocorrelation functions and  statistical error, as well as the use of Langevin dynamics to generate an ensemble of configurations consistent with the constant-temperature, Boltzmann distribution. 

\nomenclature{$V(\bq)$}{Potential-energy surface}
\nomenclature{$H\left(\bp,\bq\right)$}{Classical Hamiltonian for a configuration of atoms with positions $\bq$ and momenta $\bp$} 
Throughout this chapter we assume that the dynamics of the electronic degrees of freedom is completely decoupled from that of the nuclei, and that the electrons occupy the ground state at for each configuration $\bq$ of the nuclei (Born-Oppenheimer approximation); we indicate with $V(\bq)$ the potential energy associated with such configuration.
Furthermore, we assume that the nuclei behave as classical, distinguishable particles, subject to the Hamiltonian
\[
H\left(\bp,\bq\right)=\sum_{i}\frac{\bp_{i}^{2}}{2m_{i}}+V\left(\bq\right),
\]
$m_{i}$ and $\bp_{i}$ being the mass and the momentum of each nucleus, respectively\footnote{At times we will just use expressions such as $\bp/m$ to mean the vector whose elements are $p_{i\alpha}/m_{i}$. }.

\section{Thermodynamics and phase space sampling}

Different thermodynamic ensembles are defined by considering three macroscopic observables (energy, pressure, temperature, volume, composition \ldots) as fixed, and by using their value to define the state of the system.
We focus in particular on the canonical ($N\CV T$) ensemble, in which temperature $T$, volume $\CV$ and number of atoms $N$ are assumed constant. This ensemble often corresponds to experimental conditions, and allows us to discuss most of sampling issues and the techniques to solve them. 
We do not discuss the derivation of the canonical ensemble, but just state that it implies that the probability of observing a configuration $\left(\bp,\bq\right)$ corresponds to
\begin{equation}
\Prob\left(\bp,\bq\right)=e^{-\beta H\left(\bp,\bq\right)}/Z,\qquad Z=\int\D{\bp}\D{\bq}e^{-\beta H\left(\bp,\bq\right)},\label{eq:p-canonical}
\end{equation}
where we have introduced the inverse temperature $\beta=1/k_{B}T$, and the canonical partition function $Z$.
\nomenclature{$Z$}{The canonical partition function}
\nomenclature{$\Prob(\bp,\bq)$}{A probability distribution in phase space}

An important feature of the (classical) canonical ensemble -- one that simplifies considerably analytical and numerical treatment -- is that position and momentum are not correlated so that the $\bp$ and $\bq$ parts of the partition function and of the probability distribution can be factored exactly, and treated separately
\begin{equation}
\Prob\left(\bp,\bq\right)=\Prob\left(\bp\right)\cdot \Prob\left(\bq\right)=\frac{e^{-\beta\sum_{i}\frac{\bp_{i}^{2}}{2m_{i}}}}{\int\D{\bp} e^{-\beta\sum_{i}\frac{\bp_{i}^{2}}{2m_{i}}}}\cdot\frac{e^{-\beta V\left(\bq\right)}}{\int\D{\bq}e^{-\beta V\left(\bq\right)}}.\label{eq:pp-pq-factoring}
\end{equation}
Note that $\Prob\left(\bp\right)$ is just a multi-variate Gaussian, so the distribution of momenta is trivial and the normalization can be computed analytically. 
The difficulty is in determining the \emph{configurational} part $\Prob\left(\bq\right)=e^{-\beta V\left(\bq\right)}/\int\D{\bq} e^{-\beta V\left(\bq\right)}$, that depends on the potential and which is typically a very complicated function of the atomic coordinates. 

\nomenclature{\avg{\cdot}}{Expectation value for the quantity $\cdot$}
Knowing $\Prob\left(\bq\right)$ is important because the expectation value of any configuration-dependent property $A\left(\bq\right)$ (structure factors, average bond lengths, \ldots ) can be computed as an integral over the configurational probability distribution:
\begin{equation}
\avg{A} =\int\D{\bq}\,A\left(\bq\right)\,\Prob\left(\bq\right)=\frac{\int\D{\bq}\,A\left(\bq\right)\,e^{-\beta V\left(\bq\right)}}{\int\D{\bq}e^{-\beta V\left(\bq\right)}}.\label{eq:canonical-average}
\end{equation}
For simple, low-dimensional problems, computing an integral of the form~(\ref{eq:canonical-average}) by some kind of quadrature (e.g. computing the value of the integrand on a grid of $\bq$ points) is a sensible proposition, but it becomes completely impractical as the number of atoms increases: even with just two grid points per degree of freedom, the evaluation of the integrand on a grid requires $2^{3N}$ points.

The exponential increase of the computational cost involved in an integration by quadrature can be in principle circumvented by using \emph{importance sampling}, i.e. by generating a sequence of $M$ configurations $\bq(i)$ that are distributed according to the target canonical probability $\Prob(\bq)$. 
Few or no sample points are present where the probability distribution has a negligible value, and instead samples naturally concentrate in regions with a high value of $\Prob\left(\bq\right)$  -- which typically correspond to a tiny fraction of the phase space. 
Expectation values can then be obtained from this sequence of points simply as $\avg{A} \approx\frac{1}{M}\sum_{i}A\left(\bq(i)\right)$, since the exponential Boltzmann factor is implicitly accounted for by the uneven distribution of samples. The problem is then how to generate a set of atomic configuration consistent with the canonical distribution. 

In general, techniques to generate this canonically-distributed set of points proceed iteratively, by taking one point $\bq(i)$ and providing a rule to generate a new point $\bq(i+1)$.
In many cases, the rule that generates a new configuration is not deterministic, but is characterized by a transition probability distribution $p\left(\bq\rightarrow\bq'\right)=p\left(\bq({i+1})=\bq'|\bq({i})=\bq\right)$.
A necessary condition for $p\left(\bq\rightarrow\bq'\right)$ to generate the correct distribution is that the canonical distribution itself is left invariant under the action of the operation that generates the sequence of points, i.e. that 
\begin{equation}
\int\D{\bq}\Prob\left(\bq\right)p\left(\bq\rightarrow\bq'\right)=\Prob\left(\bq'\right).\label{eq:mc-necessary}
\end{equation}
A more stringent, sufficient condition that is however easier to prove in most cases is that of \emph{detailed balance}, that relates the probabilities of performing a move $\bq\rightarrow\bq'$ and that of the reverse move $\bq'\rightarrow\bq$ to the relative probability of the initial and final configurations:
\begin{equation}
\Prob\left(\bq\right)p\left(\bq\rightarrow\bq'\right)=\Prob\left(\bq'\right)p\left(\bq'\rightarrow\bq\right).\label{eq:mc-detailedbalance}
\end{equation}
It is easy to show that Eq.~(\ref{eq:mc-detailedbalance}) implies Eq.~(\ref{eq:mc-necessary}), by integrating both sides over $\bq$ and realizing that the probability of going \emph{anywhere }starting from $\bq'$ has to integrate to one.

\begin{exercise}[label={ex:canonical-ho},title={Canonical harmonic oscillator}]

Consider the case of a one-dimensional harmonic oscillator, with mass $m$ and frequency $\omega$, described by the Hamiltonian
\[
H(p,q) = \frac{p^2}{2m} + \frac{1}{2}m\omega^2 q^2
\]
The thermodynamic conditions correspond to canonical sampling at inverse temperature $\beta=1/k_B T$. 

\Question Write the expression for the canonical partition function for the oscillator, and compute its value, remembering the Gaussian integral $\int_{\infty}^{\infty}\D{x} e^{-a x^2} = \sqrt{\pi/a}$.

\Question Compute the expectation value of the kinetic energy $\avg{p^2/2m}$ and of the position fluctuations $\avg{q^2}$

\end{exercise}

\section{MD and integrators}
\label{sec:molecular-dynamics}

Molecular dynamics (MD) provides a deterministic strategy to generate a continuous sequence of configurations that are consistent with the canonical ensemble. Let us consider what happens if we choose a configuration of position and momentum $\left(\bp,\bq\right)$ that is consistent with Boltzmann statistics, and evolve it in time based on Hamilton's equations
\begin{equation}
\dot{\bq}=\frac{\partial H}{\partial\bp}=\frac{\bp}{m},\qquad\dot{\bp}=-\frac{\partial H}{\partial\bq}=-\frac{\partial V}{\partial\bq}.\label{eq:hamilton}
\end{equation}
One can prove~\cite{gard03book} that MD satisfies a simple generalization of the detailed balance condition, in which variables can change sign upon a time reversal operation. It is however more instructive to show that a MD time step fulfills the more general necessary condition~(\ref{eq:mc-necessary}). 

\begin{figure}[tbph]
\begin{centering}
\includegraphics[width=0.7\textwidth]{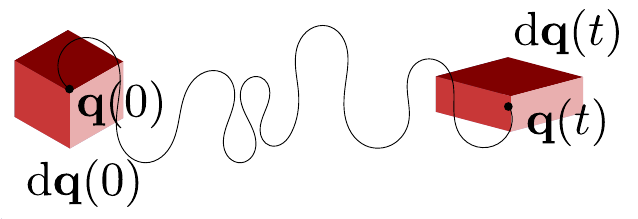}
\par\end{centering}
\caption{\label{fig:traj-dens}Evolution of a configuration in phase space
and of the corresponding volume element along a molecular dynamics
simulation.}
\end{figure}

\subsection{Conservation of density and phase-space volume}

First, let us define the position of an atomistic system in phase space as the $6N$-dimensional vector $\bx=\left(\bp,\bq\right)$ that combines position and momentum of all the $N$ atoms. 
Then, it is easy to see that the MD trajectory conserves the probability distribution, as
\begin{equation}
\frac{\D{\Prob}}{\D{t}}\propto e^{-\beta H}\frac{\D{H}}{\D{t}},
\qquad\frac{\D{H}}{\D{t}}=\frac{\partial H}{\partial\bp}\cdot\dot{\bp}+\frac{\partial H}{\partial\bq}\cdot\dot{\bq}=-\frac{\partial H}{\partial\bp}\cdot\frac{\partial H}{\partial\bq}+\frac{\partial H}{\partial\bq}\cdot\frac{\partial H}{\partial\bp}=0.\label{eq:h-conservation}
\end{equation}
In order to be able to perform the integral~(\ref{eq:mc-necessary}) one also needs to work out how the volume element $\D{\bx\left(0\right)}$ is transformed to $\D{\bx\left(t\right)}$. 
Here it is useful to imagine the evolution of the volume element as that of a swarm of trajectories starting off around $\bx\left(0\right)$ -- a picture that is very useful as it naturally links a description of dynamics in terms of trajectories in phase space to one that deals with the time evolution of a probability density. 
The change of variables $\bx\left(0\right)\rightarrow\bx\left(t\right)$ is associated with the Jacobian determinant~\cite{tuck08book} 
\[
\CJ\left(t\right)=\det\mbf{J},\qquad J_{ij}=\frac{\partial x_{i}\left(t\right)}{\partial x_{j}\left(0\right)}.
\label{eq:jacobian}
\]
It is possible to show (see Exercise \ref{ex:jacobian}) that  $\CJ\left(t\right)=1$, implying that the volume element is conserved, and so the probability distribution is left invariant by Hamiltonian evolution of the classical trajectory.
The conservation of the Hamiltonian and of the phase space differential
means that the probability distribution is conserved by the MD propagation,
i.e. that the necessary condition for canonical sampling is satisfied.

\begin{exercise}[label={ex:jacobian},title={Phase-space volume in MD}]
Show that the Jacobian determinant 
\[\CJ\left(t\right)=\det\mbf{J},\qquad J_{ij}=\frac{\partial x_{i}\left(t\right)}{\partial x_{j}\left(0\right)}\]
is equal to $1$ for Hamiltonian dynamics. You can use Jacobi's formula
$\partial \det \mbf{A}(x)/\partial{x} = \det\mbf{A} \Tr(\mbf{A}^{-1}\partial \mbf{A}/\partial x)$.

\end{exercise}

\subsection{Velocity Verlet integrator}

Having established that integrating Hamilton's equations starting from canonically-distributed configurations ensures sampling of the canonical distribution, one should discuss how the integration can be realized in practice. Despite being relatively simple first-order differential equations, Eqs.~(\ref{eq:hamilton}) cannot be integrated analytically except for the simplest problems, so one has to resort to an approximate scheme to evolve the system along a MD trajectory. 
Many more or less complicated \emph{integrators} (algorithms to perform evolve Eqs.~(\ref{eq:hamilton}) over a finite time step $dt$) have been used and proposed, but the simplest and in many ways effective integrator is probably the symmetric-split velocity Verlet algorithm~\cite{verl67pr,tuck+92jcp}. 
In this algorithm, the momentum $\bp$ and the position $\bq$ are propagated according to a linearization of Hamilton's equation:
\begin{equation}
\begin{split}\bp\leftarrow & \bp-\frac{\partial V}{\partial\bq}\frac{dt}{2}\\
\bq\leftarrow & \bq+\frac{\bp}{m}dt\\
\bp\leftarrow & \bp-\frac{\partial V}{\partial\bq}\frac{dt}{2}.
\end{split}
\label{eq:velocity-verlet}
\end{equation}
Note that even though it appears that the force has to be computed twice in Eqs.~\ref{eq:velocity-verlet}, in practice the force computed at the end of one step can be re-used at the beginning of the following step. 
The reason why it is useful to split in two the propagation of the momentum is that in this form the finite-time velocity Verlet propagator is explicitly time-reversible (i.e. integrating a trajectory back in time traces back exactly the trajectory) and symplectic (i.e. preserves exactly phase space volume), see Exercise~\ref{ex:verlet}. 

\subsection{Energy conservation}

The velocity Verlet integrator fulfills exactly two of the properties Hamiltonian dynamics, yet it is not exact. The use of a finite time step entails necessarily an integration error, which -- in a sufficiently large, chaotic system -- will lead to the discrete trajectories to diverge exponentially from the trajectory that would be obtained by exact integration of the dynamics. 
In practice, provided that the time step is sufficiently small, the molecular dynamics trajectory is still sampling an ensemble which is extremely close to the target one, and also exhibits very similar dynamical properties. 
This is at times explained in terms of the existence of a ``shadow Hamiltonian'', which would generate precisely the trajectory that is obtained by finite time step integration and that is very close to the actual Hamiltonian. The hypothetical existence of this shadow Hamiltonian explains why it is important to use an integration algorithm that fulfills the symmetries of Hamiltonian dynamics.

\begin{figure}[tbph]
\begin{centering}
\includegraphics[width=0.7\textwidth]{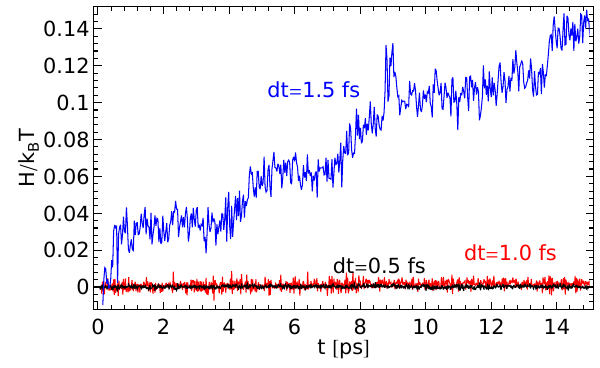}
\par\end{centering}
\caption{\label{fig:traj-nrg}Energy conservation for a simulation of liquid
water at room temperature, using a velocity Verlet integrator with
three different values of the time step.}
\end{figure}

The most straightforward manifestation of integration errors is the fact that the total energy is not conserved along the MD trajectory. 
It is tedious but straightforward to write $V\left(q'\right)+p'/2m$ in terms of a Taylor expansion around the initial value of $q$ and $p$, finding that the leading error term in the expansion is $\mathcal{O}\left(dt^{3}\right)$.
One can exploit the fact that energy conservation is violated because of the finite time step integration to monitor the accuracy of the trajectory -- under the assumption that a simulation with poor energy conservation will contain sizable errors in average and dynamical observables. 
As shown in Figure~\ref{fig:traj-nrg}, for the small values of the time step $dt$, the total energy fluctuates around a constant value, with fluctuations getting smaller as $dt$ is decreased.
For too large values, instead, $H$ exhibits a systematic \emph{drift} -- a sign of a very substantial violation of energy conservation, that should be avoided. 
On a longer time scale, it is common to observe a drift even with reasonable choices of $dt$. In the case of constant-temperature simulations, discussed further below, a small drift is generally acceptable.

\begin{exercise}[label=ex:verlet,title={Velocity Verlet}]
Show that the velocity Verlet integrator~\eqref{eq:velocity-verlet} is time-reversible and symplectic.
Consider a one-dimensional system, compute explicitly the value of the phase space point evolved in time by a finite time step $dt$, and use the definition of the Jacobian determinant from~\eqref{eq:jacobian}.
\end{exercise}

\section{Efficiency of sampling}

In order to compute ensemble averages by generating a sequence of configurations, it is not sufficient that these configurations are distributed according to the probability distribution associated with the ensemble. 
The trajectory $\left\{ A(i)\right\} $ must also satisfy an ergodic hypothesis, i.e. it must be true that 
\begin{equation}
\lim_{M\rightarrow\infty}\frac{1}{M}\sum_{i}A\left(\bq({i})\right)=\int\D{\bq}\,\Prob\left(\bq\right)A\left(\bq\right).
\label{eq:ergodic}
\end{equation}
To see how a set of configurations that satisfy the requirements given in the previous sections could break the assumption~(\ref{eq:ergodic}), imagine a situation in which the configuration space is divided into two disconnected regions so that transitions between any pair of points satisfy detailed balance, but there is zero probability of having a transition between the two regions. 
A single trajectory starting on one of the two areas would never visit the other half of configuration space, and hence the trajectory average would differ from the ensemble average. 

This is far from being a purely academic concern: in a practical case, one does not only require that Eq.~\eqref{eq:ergodic} holds in the $M\rightarrow\infty$ limit, but also would like convergence to be \emph{fast}, to evaluate averages accurately within the limited time available for the simulation.
To see how to get a quantitative measure of the efficiency of sampling, consider obtaining a large number of independent trajectories with $M$ samples each, and compute the average of these independent means: 
\[
\avg{ \frac{1}{M}\sum_{i}A({i}) } = 
\frac{1}{M}\sum_{i}\avg{A({i})} =\avg{ A } .
\]
An analogous expression can be derived for a continuous trajectory, with the summation being replaced by an integral over time. 
Unsurprisingly, the average of the means is the target average value. In technical terms, the mean is an \emph{unbiased estimator} of the average. 
Here we intend $\avg{\cdot} $ to represent an ensemble average, so there are no ergodicity concerns, and the fact that $\avg{A({i})} =\avg{A} $ follows from the fact that at each instant in time the samples are by hypothesis distributed according to the target distribution. 

To assess quantitatively the convergence of the mean to the average, it is useful to introduce the autocorrelation function, 
\begin{equation}
c_{AA}\left(t\right)=\avg{ \left(A({0})-\avg{A}\right)\left(A({t})-\avg{A} \right) }/\sigma^{2}\left(A\right),\qquad\sigma^{2}\left(A\right)=\avg{A^{2}}- \avg{A}^{2},\label{eq:autocorrelation}
\end{equation}
where $t$ indicates a lag between two points at which the function is evaluated. In practice, Eq.~\eqref{eq:autocorrelation} can be estimated as an average over the trajectory, 
\begin{equation}
c_{AA}(t) = \frac{1}{M_t}\sum_i A(i) A(i+t),
\end{equation}
where we assumed $\avg{A}=0$ and $\sigma^2=1$ to simplify the expression.
$c_{AA}(t)$ describes how quickly the trajectory loses memory of fluctuations away from the mean. It starts off at 1 for $t=0$, and (except for cases with pathological behavior) decays to zero for $t\rightarrow\infty$. 
The time (or better, the number of steps, in this discrete formulation) that is necessary to lose the memory of the initial state of $A$ (the autocorrelation time) can be estimated as the integral of the autocorrelation function, i.e 
\[
\tau_{A}=\frac{1}{2}\sum_{t=-\infty}^{\infty}c_{AA}\left(t\right)
\]
It can be seen (Exercise~\ref{ex:error-mean}) that there is a very direct relation between how quickly the trajectory forgets about past fluctuations of an observable and how rapidly the error in the mean 
\begin{equation}
\epsilon_{A}^{2}\left(M\right)=  \avg{\left(\frac{1}{M}\sum_{i}A(i)-\avg{ A} \right)^{2}} \label{eq:error-mean}
\end{equation}
decreases. 
If one considers a trajectory containing $M$ samples, the error in the mean amounts to $\epsilon_{A}^{2}\left(M\right)\approx\sigma^{2}\left(A\right)2\tau_{A}/M$. In other terms, only samples that are spaced by at least $\tau_A$ simulation time contribute substantial information to the mean. 
The autocorrelation time can therefore be taken as a rigorous measure of the ergodicity of a trajectory, and in general, one should manipulate the sampling strategy to minimize $\tau_{A}$ for the observables of interest.

\begin{exercise}[label={ex:error-mean}, title={Autocorrelation and error}]
Show that the error in the mean \eqref{eq:error-mean} of an observable $A$ for a trajectory containing $M$ samples is related to the autocorrelation time by  $\epsilon_{A}^{2}\left(M\right)\approx\sigma^{2}\left(A\right)2\tau_{A}/M$.
You need to exploit the fact that the trajectory is stationary, i.e. that the statistical properties of samples do not depend on their absolute position in the sequence.
\end{exercise}

\section{Langevin dynamics}
\label{sec:langevin-dynamics}
The molecular dynamics approach described in Section~\ref{sec:molecular-dynamics} samples (apart from finite time step errors) the constant \emph{energy} (microcanonical) ensemble. 
In other terms, even though it does conserve the canonical ensemble, and so a collection of independent trajectories starting from uncorrelated points consistent with the finite-temperature Boltzmann distribution would yield correct averages, a \emph{single} trajectory is highly non-ergodic (it does not allow fluctuations of $H$!), and there is no guarantee that it would yield averages consistent with the target constant-temperature conditions. 
Fortunately, it is relatively simple to modify Hamilton equations~(\ref{eq:hamilton}) to allow for energy fluctuations, so as to obtain an ergodic sampling of the canonical ensemble. 
These changes to Hamiltonian dynamics are generally referred to as \emph{thermostats}. 

A simple and elegant approach to achieve ergodic Boltzmann sampling MD is based on the Langevin equation~\cite{lang08cras}. 
Langevin dynamics was initially obtained as a model for Brownian motion and consists in the introduction of viscous friction and noisy force terms on top of Hamilton's equations. In one dimension, Langevin equations read
\begin{equation}
\dot{q}=\frac{p}{m},\quad\dot{p}=-\frac{\partial V}{\partial q}-\gamma p+\sqrt{2m\gamma/\beta}\xi,\quad\left\langle \xi\left(t\right)\xi\left(0\right)\right\rangle =\delta\left(t\right),\label{eq:langevin}
\end{equation}
where $\beta=1/k_{B}T$, $\gamma$ is a friction term, and $\xi$ is a noisy term (technically, the time derivative of a Wiener process) which is completely uncorrelated in time. 
It is not obvious how to define the meaning of Eq.~(\ref{eq:langevin}), particularly for what concerns the noisy force $\xi$, which varies discontinuously from time to time. 
In order to give a more precise meaning to the Langevin equation, it is useful to introduce a few concepts from the theory of random processes and stochastic differential equations~\cite{gard03book}.

\begin{figure}[tbph] \caption{\label{fig:proc-slice}A collection of
sample paths for a random process. Also shown is how the one-time
probability density $\Prob\left(\bx,t\right)$ can be constructed as
the distribution of the points of all the sample paths at a given time.
} \centering{}\includegraphics[width=0.7\textwidth]{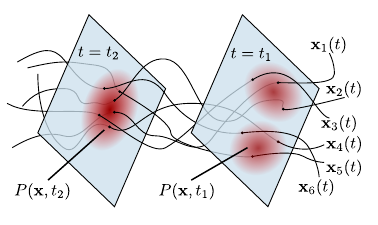}
\end{figure}

\subsection{Random processes and the Fokker-Planck equation}

Consider a system whose state is described by the value of a vector $\bx$ (e.g. the momentum and position $\left(\bp,\bq\right)$), which can evolve in time according to an unknown law, possibly characterized by a degree of random behavior. 
We now assume that we have collected several realizations of this process. 
We will refer to each trajectory as a \emph{sample path} $\bx\left(t\right)$. We let $\Omega$ be the set of all such paths, and label each path according to some index $\omega$. 
One can then describe the random process in terms of the distribution of the points in phase space at a given time, and hence construct a probability density\footnote{The integral here is just used to mean some ``averaging'' procedure
to be performed over all the realizations of the random process.} (figure~\ref{fig:proc-slice})
\[
\Prob\left(\bx,t\right)\propto\int\delta\left(\bx_{\omega}\left(t\right)-\bx\right)\D{\omega}.
\]

This probability, however, does not characterize the random process completely, since one only has knowledge on the ``snapshots'' of the collection of sample paths at different times. 
No information regarding the identity of the paths in the different snapshots has been preserved. 
One could compute the \emph{joint} probability for a sample path to be at $\bx_{1}$ at time $t_{1}$, and at $\bx_{2}$ at time $t_{2}$,\footnote{We  assume times to be ordered according to $t_{1}\ge t_{2}\ge\ldots$.}
\begin{equation}
\Prob\left(\bx_{1},t_{1};\bx_{2},t_{2}\right)\propto\int\delta\left(\bx_{\omega}\left(t_{1}\right)-\bx_{1}\right)\delta\left(\bx_{\omega}\left(t_{2}\right)-\bx_{2}\right)\D{\omega}.\label{eq:th-joint-prob}
\end{equation}
One could generalize this definition and define a hierarchy of $n$-point probability densities.
Fortunately, it is often justified to make a number of assumptions on the form of the joint probability~(\ref{eq:th-joint-prob}) so as to bring it in a more treatable form. 
A first simplification requires the process to be \emph{stationary} in time, i.e. that the two-times joint probabilities only depend on the time difference,
\begin{equation}
\Prob\left(\bx_{1},t_{1};\bx_{2},t_{2}\right)=\Prob\left(\bx_{1},t_{1}-t_{2};\bx_{2},0\right).\label{eq:th-uniform}
\end{equation}
Let us now introduce the \emph{conditional probability} $\Prob\left(\bx_{1},t_{1}|\bx_{2},t_{2}\right)$, which is defined as the probability of the system being at $\bx_{1}$ at a given time $t_{1}$, given that it was as $\bx_{2}$ at
time $t_{2}$. 
Its relation to the joint probability is 
\begin{equation}
\Prob\left(\bx_{1},t_{1}|\bx_{2},t_{2}\right)=\Prob\left(\bx_{1},t_{1};\bx_{2},t_{2}\right)/\Prob\left(\bx_{2},t_{2}\right).\label{eq:th-conditional}
\end{equation}
A random process is said do be Markovian if the joint conditional probability densities only depend on the most recent time frame, e.g.:
\begin{equation}
\Prob\left(\bx_{1},t_{1};\ldots;\bx_{k},t_{k}|\bx_{k+1},t_{k+1};\ldots;\bx_{n},t_{n}\right)=\Prob\left(\bx_{1},t_{1};\ldots;\bx_{k},t_{k}|\bx_{k+1},t_{k+1}\right).\label{eq:th-markov}
\end{equation}

This ansatz means that at each time the model has no memory of the
past history and that further evolution is (probabilistically) determined
uniquely by knowledge of the status at a given instant\footnote{This might seem to be a very crude assumption, but it holds true at least approximately for a large number of physically-relevant problems.}. 
The description of the stochastic process is thus enormously simplified.
Using the definition of conditional probability~(\ref{eq:th-conditional}), one can write
\[
\begin{split}\Prob\left(\bx_{1},t_{1};\bx_{2},t_{2};\bx_{3},t_{3}\right)= & \Prob\left(\bx_{1},t_{1}|\bx_{2},t_{2};\bx_{3},t_{3}\right)\Prob\left(\bx_{2},t_{2};\bx_{3},t_{3}\right)=\\
= & \Prob\left(\bx_{1},t_{1}|\bx_{2},t_{2}\right)\Prob\left(\bx_{2},t_{2}|\bx_{3},t_{3}\right)\Prob\left(\bx_{3},t_{3}\right),
\end{split}
\]
i.e. any joint probability can be broken down to a product of the initial, single-time distribution and a series of conditional probabilities.
If the process is also stationary, the conditional probability will depend only on the time difference, and hence its evolution is completely determined by the initial probability distribution and by the unique two-point conditional probability $\Prob\left(\bx,t|\bx_{0},0\right)$. 

One can see that under mild conditions on the form of $\Prob\left(\bx,t|\bx_{0},0\right)$ (e.g. that it arises from a Markovian, stationary process, with continuous sample paths), the most general way to describe the time evolution of $\Prob\left(\bx,t|\bx_{0},0\right)$ is given by the Fokker-Planck equation~\cite{risk72zpa,gard03book}:
\begin{equation}
{\displaystyle \begin{array}{rll}
\frac{\partial}{\partial t}\Prob\left(\bx,t|\bx_{0},0\right)= & -\sum_{i}\frac{\partial}{\partial x_{i}}\left[a_{i}\left(\bx,t\right)\Prob\left(\bx,t|\bx_{0},0\right)\right]+ & \leftarrow\text{drift}\\
 & +\frac{1}{2}\sum_{ij}\frac{\partial^{2}}{\partial x_{i}\partial x_{j}}\left[D_{ij}\left(\bx,t\right)\Prob\left(\bx,t|\bx_{0},0\right)\right] & \leftarrow\text{diffusion}
\end{array}}\label{eq:fokker-planck}
\end{equation}
Showing the link between a Fokker-Planck equation and Langevin dynamics would require one to first define the meaning of a stochastic differential equation, which can be done for instance by It\={o} calculus (see e.g. Ref.~\cite{gard03book}). 
Here we consider the stochastic differential equation 
\begin{equation}
\dot{\bx}=\mathbf{a}\left(\bx,t\right)+\mathbf{B}\left(\bx,t\right)\boldsymbol{\xi},\label{eq:fp-sde}
\end{equation}
with $\mathbf{B}\left(\bx,t\right)\mathbf{B}\left(\bx,t\right)^{T}=\mathbf{D}\left(\bx,t\right)$, to be essentially a short-hand for the associated Fokker-Planck equation~(\ref{eq:fokker-planck}).

\begin{exercise}[label={ex:liouville},title={Liouville equation}]

Consider the case of Eq.~\eqref{eq:fokker-planck} with $\mathbf{D}=0$ and $\mathbf{B}=0$ (effectively corresponding to the Liouville equation). Take an arbitrary test function $f\left(\bx\right)$, and define $\left\langle \square\right\rangle =\int\D\bx\,\square\,\Prob\left(\bx,t|\bx_{0},0\right)$.

\Question Show the link between the deterministic partial differential equation $\dot{\bx}=\mathbf{a}(\bx,t)$ and the Liouville equation. Compare $\avg{\partial f(\bx)/\partial t}$ and $\partial\avg{ f(\bx)}/\partial t$.

\Question Derive the Liouville equation for the special case of Hamilton's dynamics, where $\bx=(\bp,\bq)$ and $\mathbf{a}$ has no explicit time dependence. This is the Liouville formulation of classical mechanics. 

\end{exercise}

\subsection{Free-particle limit of the Langevin equation}

The free-particle limit of the Langevin equation can be easily integrated using its Fokker-Planck form. In one dimension, it reads just $\dot{p}=-\gamma p+\sqrt{2m\gamma/\beta}\xi$, that corresponds to the Fokker-Planck equation

\begin{equation}
\dot{\Prob}\left(p,t\middle|p_{0},0\right)=\gamma\frac{\partial}{\partial p}\left(p\Prob\right)+\frac{m\gamma}{\beta}\frac{\partial^{2}\Prob}{\partial p^{2}}.\label{eq:fp-langevin}
\end{equation}

First, it is easy to find the stationary probability by taking $\dot{\Prob}=0$.
One integral can be done straight away, and the integration constant has to be zero for $\Prob$ to be positive definite. 
One is left to solve
\[
\frac{\partial \Prob}{\partial p}=-\frac{\beta}{m}p\Prob\qquad\Rightarrow\qquad \Prob\left(p\right)\propto e^{-\beta\frac{p^{2}}{2m}};
\]
at equilibrium, the momenta of a system evolving under a free-particle Langevin equation are canonically distributed. 
Note that the friction $\gamma$ does not enter the stationary solution, as it only governs the relaxation dynamics and not the equilibrium properties. 
The finite-time solution with a boundary condition $\Prob\left(p,0\middle|p_{0},0\right)=\delta\left(p-p_{0}\right)$ is
\[
\Prob\left(p,t\middle|p_{0},0\right)\propto\frac{1}{\sqrt{1-e^{-2\gamma t}}}\exp-\frac{\beta}{2m}\frac{\left(p-p_{0}e^{-\gamma t}\right)^{2}}{1-e^{-2\gamma t}},
\]
as it can be checked by direct substitution into~(\ref{eq:fp-langevin}).
This expression provides an explicit finite-time propagator to obtain a sequence of momenta consistent with the (free-particle) Langevin equation: starting from $p_{0}$, the sample path at time $t$ is a Gaussian centered in $p_{0}e^{-\gamma t}$, with variance $\frac{m}{\beta}\left(1-e^{-2\gamma t}\right)$: given the initial momentum $p\left(0\right)$ one can obtain 
\[
p\left(t\right)=e^{-\gamma t}p\left(0\right)+\sqrt{m/\beta}\sqrt{1-e^{-2\gamma t}}\xi,\quad\left\langle \xi\right\rangle =0,\left\langle \xi^{2}\right\rangle =1,
\]
where $\xi$ is a Gaussian variate with zero mean and unit variance.
This expression provides a practical strategy to introduce Langevin dynamics in a molecular dynamics integrator. The most obvious way to do this involves bracketing the velocity-Verlet propagator between two free-particle Langevin propagators~\cite{buss-parr07pre}

\begin{equation}
\begin{split}
\bp \leftarrow & e^{-\gamma dt/2}\bp+\sqrt{m/\beta}\sqrt{1-e^{-\gamma dt}}\boldsymbol{\xi}_1\\
\bp\leftarrow & \bp-\frac{\partial V}{\partial\bq}\frac{dt}{2}\\
\bq\leftarrow & \bq+\frac{\bp}{m}dt\\
\bp\leftarrow & \bp-\frac{\partial V}{\partial\bq}\frac{dt}{2}\\
\bp \leftarrow & e^{-\gamma dt/2}\bp+\sqrt{m/\beta}\sqrt{1-e^{-\gamma dt}}\boldsymbol{\xi}_2.
\end{split}
\label{eq:obabo}
\end{equation}
However, more accurate and stable combinations have been proposed such as the BAOAB integrator discussed in Ref.~\cite{leim+13jcp}.

\begin{exercise}[label={ex:ho-langevin},title={Langevin dynamics of the harmonic oscillator}]

Solve the Fokker-Planck equation for a harmonic oscillator subject to Langevin dynamics. 
Use mass-scaled coordinates, $p\leftarrow p/\sqrt{m}$ and $q\leftarrow q\sqrt{m}$, so that the equations of motion read
\[
\begin{split}
\dot{q} =& p \\
\dot{p} =& -\omega^2 q -\gamma p + \sqrt{2\gamma/\beta} \xi.
\end{split}
\]

\Question Write the equations of motion in matrix form, in terms of the phase-space vector $(p,q)$.
\Question Write the Fokker-Planck equation associated with this stochastic differential equation.
\Question Find the stationary distribution by setting $\partial \Prob/\partial t=0$, and show that it amounts to canonical sampling for $p$ and $q$. 

\end{exercise}

\begin{figure}[tbph]
\caption{\label{fig:ho-tau}
Autocorrelation time for different observables
for a harmonic oscillator of frequency $\omega$, as a function of
the Langevin friction $\gamma$. Both the friction and the autocorrelation
times are expressed in terms of the intrinsic time scale of the oscillator.}
\centering{}\includegraphics[width=0.7\textwidth]{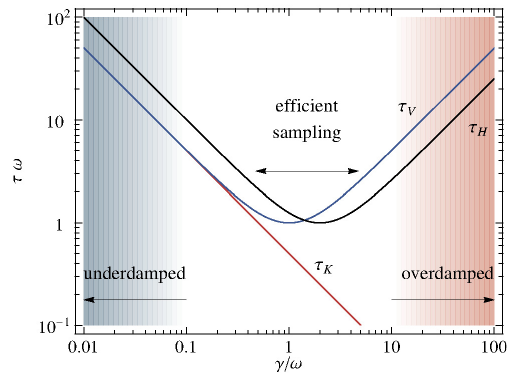}
\end{figure}

\subsection{Sampling efficiency for a harmonic oscillator}

As discussed in Exercise~\ref{ex:ho-langevin}, the Langevin dynamics of a harmonic oscillator corresponds to a matrix generalization of the case of a free particle (a Ornstein-Uhlenbeck process), that can be solved analytically. 
As a consequence, it is possible to compute any static or dynamic quantity describing the stochastic dynamics of such oscillator. 
In particular, it is possible to evaluate autocorrelation times of different observables as a function of the frequency of the oscillator $\omega$ and the friction $\gamma$ -- so that the impact of the Langevin term on the ergodicity of sampling can be assessed quantitatively. 
The autocorrelation time for the potential $V$, the kinetic energy $K$ and the total energy $H$ read respectively: 
\begin{equation}
\tau_{V}=\frac{1}{\gamma}+\frac{\gamma}{\omega^{2}},\quad\tau_{K}=\frac{1}{\gamma},\quad\tau_{H}=\frac{2}{\gamma}+\frac{\gamma}{2\omega^{2}}.\label{eq:ho-correlationtimes}
\end{equation}

Apart from the kinetic energy, for any value of $\gamma$ these quantities grow as $1/\omega^{2}$ -- which is reasonable since $1/\omega$ corresponds to a characteristic time scale for the dynamics of the oscillator. 
It is therefore more convenient to assess the efficiency using the a-dimensional quantity $\kappa=2/\omega\tau$:
\[
\kappa_{V}=2\left(\frac{\omega}{\gamma}+\frac{\gamma}{\omega}\right)^{-1},\quad\kappa_{K}=2\frac{\gamma}{\omega},\quad\kappa_{H}=2\left(\frac{2\omega}{\gamma}+\frac{\gamma}{2\omega}\right)^{-1}.
\]
As shown in Figure~\ref{fig:ho-tau}, there is an optimal range of frictions close to critical damping $\gamma=\omega$ for which the correlation time is minimum, and the dynamics is most ergodic. Lower values of the friction would yield under-damped dynamics, with the oscillator going back and forth with very slow changes in amplitude.
Higher values lead to an over-damped regime, in which the dynamics becomes sluggish, and configuration space exploration is greatly slowed down. 
Figure~\ref{fig:ho-tau} can also be read keeping the friction constant and varying the frequency: oscillators with frequency much different from $\gamma$ would be samples sub-optimally. This poses a problem when one wants to apply a Langevin thermostat to a real system, in which many different time scales will be present at the same time.
To obtain the most ergodic sampling one would need to use a different friction for each normal mode in the system.

\newpage
\section*{Answers to exercises}
\begin{Answer}[ref={ex:canonical-ho}]
The (classical) canonical partition function for a harmonic oscillator reads
\[
Z=\int \D{q}\D{p} e^{-\beta H(p,q)} = \frac{2\pi}{\omega\beta}.
\]

Similarly, one can compute the expectation value of $q^2$, that reads
$1/m\beta\omega^2$, and the kinetic energy that reads simply 
$\avg{p^2/2m} = 1/2\beta $.

\end{Answer}

\begin{Answer}[ref={ex:jacobian}]
Clearly, $J_{ij}\left(0\right)=\frac{\partial x_{i}\left(0\right)}{\partial x_{j}\left(0\right)}=
\delta_{ij}$, so $\CJ\left(0\right)=1$.
One needs then just to show that $\CJ'\left(t\right)=0$ to prove that the volume of the phase space element is conserved by Hamiltonian dynamics.
One can use Jacobi's formula to get $\CJ'\left(t\right)=\CJ\left(t\right)\Tr\left(\mbf{J}^{-1}\mbf{J}'\right).$
Then, one can see that $\left(\mbf{J}^{-1}\right)_{ij}=\partial x_{i}\left(0\right)/\partial x_{j}\left(t\right)$
since 
\[
\sum_{k}\frac{\partial x_{i}\left(0\right)}{\partial x_{k}\left(t\right)}\frac{\partial x_{k}\left(t\right)}{\partial x_{j}\left(0\right)}=\frac{\partial x_{i}\left(0\right)}{\partial x_{j}\left(0\right)}=\delta_{ij},
\]
as the left-hand side is just a chain-rule sum. Considering also that
$J'_{ij}=\partial\dot{x}_{i}\left(t\right)/\partial x_{j}\left(0\right)$,
one gets
\[
\Tr\left(\mbf{J}^{-1}\mbf{J}'\right)=\sum_{ij}\frac{\partial x_{i}\left(0\right)}{\partial x_{j}\left(t\right)}\frac{\partial\dot{x}_{j}\left(t\right)}{\partial x_{i}\left(0\right)}=\sum_{ijk}\frac{\partial x_{i}\left(0\right)}{\partial x_{j}\left(t\right)}\frac{\partial x_{k}\left(t\right)}{\partial x_{i}\left(0\right)}\frac{\partial\dot{x}_{j}\left(t\right)}{\partial x_{k}\left(t\right)}.
\]
The sum over $i$ corresponds to $\mbf{J}\mbf{J}^{-1}=\mbf{1}$,
so 
\[
\Tr\left(\mbf{J}^{-1}\mbf{J}'\right)=\sum_{k}\frac{\partial\dot{x}_{k}\left(t\right)}{\partial x_{k}\left(t\right)}=\frac{\partial}{\partial\bp}\cdot\dot{\bp}+\frac{\partial}{\partial\bq}\cdot\dot{\bq}=\frac{\partial}{\partial\bp}\cdot\left(-\frac{\partial H}{\partial\bq}\right)+\frac{\partial}{\partial\bq}\cdot\frac{\partial H}{\partial\bp}=0
\]
\end{Answer}

\begin{Answer}[ref=ex:verlet]
Write explicitly the propagated position and momentum, $q'$ and $p'$:
\[
q'=q+\frac{p}{m}dt-V'\left(q\right)\frac{dt^{2}}{2m}\qquad p'=p-V'\left(q\right)\frac{dt}{2}-V'\left(q'\right)\frac{dt}{2}.\label{eq:vv-propagator}
\]
It is simple to write the time-reversed step, starting from $\left(q',-p'\right)$ and evolving for $dt$. For instance, from Eqs.~(\ref{eq:vv-propagator}) one can obtain $V'\left(q'\right)\frac{dt}{2}=p-p'-V'\left(q\right)\frac{dt}{2}$ and $V'\left(q\right)\frac{dt}{2}=p+\frac{m}{dt}\left(q-q'\right)$.
Using these one can see that tracing back in time the Velocity Verlet step brings the trajectory back to exactly $(q,-p)$. 
In order to prove that the velocity Verlet step is also exactly symplectic it is sufficient to compute the elements of the Jacobian matrix from Eq.~(\ref{eq:vv-propagator}): $J_{qq}=1-V''\left(q\right)\frac{dt^{2}}{2m}$, $J_{qp}=\frac{dt}{m}$, $J_{pp}=1-V''\left(q'\right)\frac{dt^{2}}{2m}$, $J_{pq}=V''\left(q\right)\frac{dt}{2}-V''\left(q'\right)\frac{dt}{2}\left(1-V''\left(q\right)\frac{dt^{2}}{2m}\right)$
and see that the determinant is $\det\mbf{J}=J_{qq}J_{pp}-J_{qp}J_{pq}=1$.
\end{Answer}

\begin{Answer}[ref={ex:error-mean}]
Start by writing out explicitly the square in Eq.~\eqref{eq:error-mean}
\[
\epsilon_{A}^{2}\left(M\right)=
\frac{1}{M^{2}}\avg{
\sum_{i,j=0}^{M-1}\left(A(i)-\avg{A} \right)\left(A(j)-\avg{A} \right)}
,
\]
and re-arrange the summation so that it runs over the variables $\Delta=j-i$ and $k$. 
This leads to 
\begin{align*}
\epsilon_{A}^{2}\left(M\right) & =\frac{1}{M^{2}}\left<
 \sum_{\Delta=-\left(M-1\right)}^{-1}\sum_{k=\left|\Delta\right|}^{M-1}
 \left(A({k})-\avg{A} \right)\left(A({k+\Delta})-\avg{A}\right)+\right..\\
 & \left.+\sum_{\Delta=0}^{M-1}\sum_{k=0}^{M-1-\left|\Delta\right|}\left(A({k})-\avg{ A} \right)\left(A({k+\Delta})-\avg{A}\right)\right> 
\end{align*}
Then, the crux is assuming that the process that generates the trajectory
is \emph{stationary}, i.e. that the relation between two points in
the sequence only depends on the difference between their position
in the sequence and not on their ``absolute'' location within the
sequence. So the dependence on $k$ becomes immaterial when the ensemble
average is brought inside the summation, and one gets 
\[
\begin{split}\epsilon_{A}^{2}\left(M\right) & =\frac{1}{M^{2}}\sum_{\Delta=-\left(M-1\right)}^{M-1}\left\langle \left(A({0})-\avg{ A} \right)\left(A({\Delta})-\avg{A} \right)\right\rangle \left(M-\left|\Delta\right|\right)=\\
 & =\frac{\sigma^{2}\left(A\right)}{M}\sum_{\Delta=-\left(M-1\right)}^{M-1}c_{AA}\left(\Delta\right)\left(1-\frac{\left|\Delta\right|}{M}\right),
\end{split}
\]
where we have introduced the autocorrelation function.
Taking the limit for $M\rightarrow\infty$ of this sum one obtains the
final result, $\epsilon_{A}^{2}\left(M\right)\approx\sigma^{2}\left(A\right)2\tau_{A}/M$. 

\end{Answer}

\begin{Answer}[ref={ex:liouville}]
While in the general case showing the connection between Eqs.~\ref{eq:fp-sde} and~\ref{eq:fokker-planck} requires the use of stochastic calculus, the deterministic case for which $\mathbf{B}=0$ can be discussed more easily. 
From any given sample path $\mathbf{x}(t)$ and function $f$, one can obtain that 
\[
\frac{\partial}{\partial t}f\left(\bx\right)=
\sum_{k} \frac{\partial f}{\partial x_k} \dot{x}_k =
\nabla f\cdot\dot{\bx}=\nabla f\left(\bx\right)\cdot\mathbf{a}\left(\bx,t\right),
\]
by application of the chain rule and then inserting the differential form of the evolution equation. Then, computing the average over the probability distribution of multiple trajectories, 
\begin{equation}
\begin{split}\begin{split}\left\langle \frac{\partial}{\partial t}f\left(\bx\right)\right\rangle \end{split}
 & =\sum_{i}\int\D\bx\,\frac{\partial f}{\partial x_{i}}\Prob\left(\bx,t|\bx_{0},0\right)a_{i}\left(\bx,t\right)=\\
 & =-\int\D{\bx}f\left(\bx\right)\sum_{i}\frac{\partial}{\partial x_{i}}\left[a_{i}\left(\bx,t\right)\Prob\left(\bx,t|\bx_{0},0\right)\right],
\end{split}
\label{eq:fp-drift-a}
\end{equation}
integrating by parts and knowing that the boundary term must be zero if $\Prob$ is to be normalizable. At the same time, by exchanging the integral and the time derivative one gets 
\begin{equation}
\frac{\partial}{\partial t}\left\langle f\left(\bx\right)\right\rangle =\int\D{\bx}f\left(\bx\right)\frac{\partial}{\partial t}\Prob\left(\bx,t|\bx_{0},0\right).\label{eq:fp-drift-b}
\end{equation}
Since the right hand sides of~(\ref{eq:fp-drift-a}) and~(\ref{eq:fp-drift-b}) are equal for any test function $f$, the drift part of the Fokker-Planck equation~(\ref{eq:fokker-planck}) follows. 
Taking the case of Hamiltonian dynamics, that has no explicit dependence on time, and that in this context can be formulated as 
\[
\frac{\partial}{\partial t}\left(\bp,\bq\right)=\mathbf{a}\left(\bp,\bq\right)=\left(-\nabla V\left(\bq\right),\bp/m\right),
\]
one gets the Liouville formulation of classical mechanics in terms of a probability density of trajectories,
\[
\frac{\partial}{\partial t}\Prob\left(\left(\bp,\bq\right),t|\left(\bp_{0},\bq_{0}\right),0\right)=\nabla V\cdot\nabla_{\bp}\Prob-\frac{\bp}{m}\cdot\nabla_{\bq}\Prob.
\]

\end{Answer}

\begin{Answer}[ref={ex:ho-langevin}]

The Langevin equation for the harmonic oscillator can be written in a matrix form as 
\[
\frac{\partial}{\partial t}\left(\begin{array}{c}
q\\
p
\end{array}\right)=-\left(\begin{array}{cc}
0 & -1\\
\omega^{2} & \gamma
\end{array}\right)\left(\begin{array}{c}
q\\
p
\end{array}\right)+\left(\begin{array}{cc}
0 & 0\\
0 & \sqrt{2\gamma/\beta}
\end{array}\right)\left(\begin{array}{c}
0\\
\xi
\end{array}\right)
\]
that corresponds to the Fokker-Planck equation
\[
\frac{\partial}{\partial t}\Prob\left(\left(p,q\right),t\middle|\left(p_{0},q_{0}\right),0\right)=\left[-p\frac{\partial}{\partial q}+\omega^{2}q\frac{\partial}{\partial p}+\gamma p\frac{\partial}{\partial p}+\gamma\right]\Prob+\frac{\gamma}{\beta}\frac{\partial^{2}\Prob}{\partial p^{2}}.
\]
It is easy to check by direct substitution that the Boltzmann distribution
for the oscillator $\Prob\left(p,q\right)\propto\exp-\beta\left(\frac{1}{2}\omega^{2}q^{2}+\frac{1}{2}p^{2}\right)$
is stationary. 

\end{Answer}

\setchapterpreamble[u]{\margintoc}
\chapter{The basics of path integrals}
\labch{basics}

The purpose of this chapter is to introduce the reader to the very basics of Feynman path integrals. At first, we will draw their connection with the usual Schr\"odinger picture of quantum mechanics and show the connection between time evolution amplitudes and the classical action. This will help the reader to get familiarized with many of the tools and mathematical manipulations typically used in this formalism. We will then show how this toolbox can be used in order to address problems in statistical mechanics.

\section{The time propagator and path integrals}

We start from the usual expression of the time-dependent Schr\"odinger equation, given by
\begin{equation}
i \hbar \frac{\partial}{\partial t} \ket{\Psi(t)} = \hat{H} \ket{\Psi(t)},
\end{equation}
where $\hat{H}$ is the Hamiltonian operator of the system under consideration and $\Psi(t)$ its corresponding time-dependent wave function. This equation allows for the formal solution
\begin{equation}
\ket{\Psi(t)} = e^{-i \hat{H} (t-t_0)/\hbar} \ket{\Psi(t_0)} = \hat{U}(t) \ket{\Psi(t_0)},
\end{equation}
where we have identified the unitary time-evolution operator $\hat{U}(t)=e^{-i \hat{H} (t-t_0)/\hbar}$.

We can now write these equations in the position space $q$ by calculating
\begin{eqnarray}
\braket{q}{\Psi(t)} &=& \Psi(q, t) = \bra{q}\hat{U}(t)\ket{\Psi(t_0)} \nonumber \\
&=& \int dq_0 \bra{q}\hat{U}(t)\ket{q_0}\braket{q_0}{\Psi(t_0)} \nonumber \\
&=& \int dq_0 \bra{q}\hat{U}(t)\ket{q_0} \Psi(q_0, t_0) \nonumber \\
&=& \int dq_0 K(q, q_0, t) \Psi(q_0, t_0) \label{eq:kernel},
\end{eqnarray}
where we have used the resolution of the identity in the position representation $\int dq_0 \ket{q_0}\bra{q_0}=1$. The kernel $K(q, q_0, t)=\bra{q}\hat{U}(t)\ket{q_0}$, often called the ``time evolution amplitude" \cite{klei-shab95pra, Feynman:book}, is the object at the center of the path-integral formulation of quantum mechanics. 

To have an idea how these objects look like and start getting used to some of the mathematical manipulations that we will use in the following sections, we can evaluate this kernel for the free particle in one dimension, described by the Hamiltonian
\begin{equation}
\hat{H} = \frac{\hat{p}^2}{2 m},
\end{equation}
where $\hat{p}$ are the momentum operators (in position representation expressed as $-i \hbar \partial/\partial q$). We set $t_0=0$ for convenience. In this case

\begin{eqnarray}
K^{\text{free}}(q, q_0, t) &=& \bra{q} e^{-\frac{it}{\hbar}\frac{\hat{p}^2}{2m}} \ket{q_0} \nonumber \\
		  &=& \int_{-\infty}^{\infty} dp  \bra{q} e^{-\frac{it}{\hbar}\frac{\hat{p}^2}{2m}} \ket{p} \braket{p}{q_0} \nonumber \\
		  &=& \int_{-\infty}^{\infty} dp  \braket{q}{p} e^{-\frac{it}{\hbar}\frac{p^2}{2m}} \braket{p}{q_0} \nonumber \\
		  &=& \frac{1}{2 \pi \hbar} \int_{-\infty}^{\infty} dp \, e^{\frac{i p q}{\hbar}}e^{-\frac{it}{\hbar}\frac{p^2}{2m}}  e^{-\frac{i p q_0}{\hbar}}   \nonumber \\ &=&                  
		  \frac{1}{2 \pi \hbar} \int_{-\infty}^{\infty} dp \, e^{-\frac{i}{\hbar}p(q_0-q)}e^{-\frac{i}{\hbar}\frac{p^2t}{2m}}
\label{eq:free-particle-1}.
\end{eqnarray}

Up to now, we have used the resolution of the identity in the momentum representation $\int dp \ket{p}\bra{p}=1$ and the well known inner product between position and momenta, given by $\braket{q}{p} = e^{ipq/\hbar}/\sqrt{2 \pi \hbar}$.

\begin{exercise}[label={ex:inner-prod},title={Inner product of position and momenta}]

Show a derivation of the result of $\braket{q}{p}$ and also derive $\braket{p}{q}$.

\end{exercise}

Now, in order to move forward, we need to solve the integral in Eq. \ref{eq:free-particle-1}. This can be done by ``completing the square". In order to do that, we add and subtract from the exponent the term $i(q-q_0)^2m/2t\hbar$, thus recognizing that
\begin{equation}
p(q_0-q)+\frac{p^2 t}{2m} = \frac{1}{2m}\left[p-m\frac{(q-q_0)}{t}\right]^2 t - \frac{m(q-q_0)^2}{2t},  \label{eq:free-particle-1a}
\end{equation}
and then shift the momenta $p'=p-m(q-q_0)/t$ to obtain 
\begin{eqnarray}
K^{\text{free}}(q, q_0, t) & = &  \frac{e^{\frac{i}{\hbar} \frac{m(q_0-q)^2}{2t}}}{2 \pi \hbar}  \int_{-\infty}^{\infty} dp' \, e^{-\frac{i}{\hbar}\frac{p'^2t}{2m}} \nonumber \\
		  		    & = & \frac{e^{\frac{i}{\hbar} \frac{m(q_0-q)^2}{2t}}}{2 \pi \hbar}  \int_{-\infty}^{\infty} dp' \nonumber \\
		  		    & \times &  \left\{\cos\left[\left(p'\sqrt{\frac{t}{2 m \hbar}}\right)^2\right] - i \sin\left[\left(p'\sqrt{\frac{t}{2 m \hbar}}\right)^2\right] \right\}.
\end{eqnarray}
Finally, these equations can be solved with the Fresnel integral equations to give
\begin{equation}
K^{\text{free}}(q, q_0, t)  = \frac{e^{\frac{i}{\hbar} \frac{m(q_0-q)^2}{2t}}}{\sqrt{2 \pi \hbar i t/m}} \label{eq:free-kernel}.
\end{equation}
The result above is central to the following derivation.

We would now like to calculate the time evolution amplitudes for a particle subject to a time-independent , $V(\hat{q})$. The kernel we want to calculate is
\begin{equation}
K(q, q_0, t) = \bra{q} e^{-\frac{it}{\hbar} \left[ \frac{\hat{p}^2}{2m} + V(\hat{q}) \right]} \ket{q_0}.
\end{equation}

Because $[\hat{p},\hat{q}] \neq 0$, we will need to use the Trotter product formula in order to express the kernel in a closed form. This identity (please see proof in e.g. \cite{Trotter:1959ey, cohen-kato1982}) states that\footnote{For numerical reasons, it is usual to use (and implement) the symmetric form of the Trotter splitting, also known as Strang splitting \cite{Strang:1968}. It reads $e^{\hat{A}+\hat{B}} = \lim_{P\to \infty} \left(e^{\hat{B}/2P}  e^{\hat{A}/P} e^{\hat{B}/2P} \right)^P$. The symmetric splitting converges faster with $P$ (error of $\mathcal{O}(P
^{-2})$) than the usual one (error of $\mathcal{O}(P
^{-1})$). For the derivation presented here, it yields the same final expressions and both are exact at $P \to \infty$.}
\begin{equation}
e^{\hat{A}+\hat{B}} = \lim_{P\to \infty} \left(e^{\hat{A}/P} e^{\hat{B}/P} \right)^P \label{eq:trotter},
\end{equation}
for any two operators $\hat{A}$ and $\hat{B}$, giving 
\begin{equation}
K(q, q_0, t) = \lim_{P \to \infty}\bra{q} \left(e^{-\frac{it}{P\hbar}\frac{\hat{p}^2}{2m}} e^{-\frac{it}{P\hbar}V(\hat{q})} \right)^P\ket{q_0}.
\end{equation}

Having the equation in this factorized form, the next step is to introduce $P-1$ times the resolution of the identity in position representation $\int d q_j \ket{q_j}\bra{q_j}=1$, where $j=1,\dots, P-1$,  
\begin{equation}
K(q, q_0, t) = \lim_{P \to \infty} \int dq_1, \dots, dq_{P-1} \prod_{j=0}^{P-1} \bra{q_{j+1}} e^{-\frac{it}{P\hbar}\frac{\hat{p}^2}{2m}} e^{-\frac{it}{P\hbar}V(\hat{q})} \ket{q_{j}},
\end{equation}
where $q_P=q$. The final work is now to evaluate the elements 
\begin{equation}
\bra{q_{j+1}} e^{-\frac{it}{P\hbar}\frac{\hat{p}^2}{2m}} e^{-\frac{it}{P\hbar}V(\hat{q})} \ket{q_{j}} = \bra{q_{j+1}} e^{-\frac{it}{P\hbar}\frac{\hat{p}^2}{2m}} \ket{q_{j}} e^{-\frac{it}{P\hbar}V(q_j)}.
\end{equation}
Luckily, all we need to finalize our results is use Eq. \ref{eq:free-kernel} to obtain
\begin{equation}
\begin{split}
\bra{q_{j+1}} e^{-\frac{it}{P\hbar}\frac{\hat{p}^2}{2m}} e^{-\frac{it}{P\hbar}V(\hat{q})} \ket{q_{j}} =   \frac{e^{\frac{i}{\hbar} \frac{Pm(q_{j}-q_{j+1})^2}{2t}}}{\sqrt{2 \pi \hbar i t/Pm}} e^{-\frac{it}{P\hbar}V(q_j)} \\ 
= \frac{1}{\sqrt{2 \pi \hbar i t/Pm}} e^{\frac{i}{\hbar} \left[ \frac{Pm(q_{j}-q_{j+1})^2}{2t} -\frac{t}{P}V(q_j) \right]}.
\end{split}
 \end{equation}
 
 By identifying $t/P=\delta t$, we finally write
 \begin{equation}
 K(q, q_0, t) = \lim_{\substack{\delta t \to 0 \\ P \to \infty}} \left(\frac{m}{2 \pi \hbar i \delta t}\right)^{P/2} \int dq_1, \dots, dq_{P-1} e^{\frac{i}{\hbar} \left[ \sum_{j=0}^{P-1} \frac{m \delta t}{2}  \left[\frac{(q_{j}-q_{j+1})}{\delta t}\right]^2 - V(q_j) \delta t \right]} \label{eq:kernel-general}
 \end{equation}

 \begin{exercise}[label={ex:harmonic-propagator},title={Time evolution amplitude of the quantum harmonic oscillator}]

Derive $K(q, q_0, t)$ for a harmonic oscillator potential given by $V(\hat{q})=\frac{m \omega^2}{2} \hat{q}^2$.

\end{exercise}
 
\begin{figure}[tbph]
\begin{centering}
\includegraphics[width=0.6\textwidth]{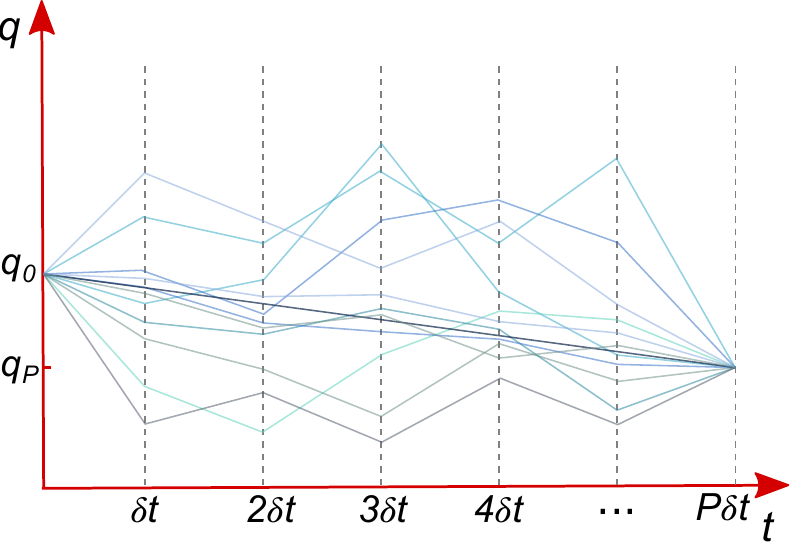}
\par\end{centering}
\caption{\label{fig:time-paths}Discretized possible paths. A particle at point $q_0$ at time $t=0$ tries all possible ways to arrive at $q$ at time $t=P\delta t$. Each path is weighted by the respective action as given by the path-integral time evolution amplitudes $K(q, q_0, t)$.}
\end{figure}

 In the equation above, we identify
 \begin{equation}
 \begin{split}
  S_{P}[q] =  \sum_{j=0}^{P-1} \frac{m \delta t}{2}  \left[\frac{(q_{j}-q_{j+1})}{\delta t}\right]^2 - V(q_j) \delta t   \\ 
  = \sum_{j=0}^{P-1} \frac{m \delta t}{2}  \left[\frac{q(t_j)-q(t_{j+1})}{\delta t}\right]^2 - V(q(t_j)) \delta t  .\label{eq:disc-action}
  \end{split}
 \end{equation}
Such discretized paths are pictorially represented in Fig. \ref{fig:time-paths} The expression above is a Riemann sum and therefore it is easy to see that in the limit of $\delta t \to 0$  (or $P \to \infty$)
 \begin{equation}
 S[q]=\lim_{\substack{\delta t \to 0 \\ P \to \infty}} S_P[q] = \int_0^t dt' \left[ \frac{m }{2}  \dot{q}(t')^2 - V(q(t'))\right]\bigg\rvert_{q(0)=q_0}^{q(t)=q}. \label{eq:action}
 \end{equation}
where $\dot{q}(t')$ denotes the time derivative of the position at $t=t'$.
The integrand in Eq. \ref{eq:action} is easily identifiable as the Lagrangian of the system, and thus $S[q]$ is the classical action (defined as a functional of $q$) along the path from that goes from $q(0)=q_0$ to $q(t)=q_P=q$.

Now, let us go back to examining Eq. \ref{eq:kernel-general}. In the limit $P \to \infty$, one would have an infinite set of points between $q_0$ and $q_P$, representing all possible values of a continuous function $q(t')$ in the given time interval and with the given start and end points.  The term $\int dq_1, \dots, dq_{P-1}$ is equivalent to varying all points of the path, $q(t')$, with fixed end points in the given time interval, and thus to varying the function itself, as discussed in \cite{tuck08book, klei-shab95pra}. One typically writes
\begin{equation}
   \lim_{\substack{\delta t \to 0 \\ P \to \infty}} \left(\frac{m}{2 \pi \hbar i \delta t}\right)^{P/2}  dq_1, \dots, dq_{P-1} =  \mathcal{D}q(t)
\end{equation}
to denote an integration over all possible paths, i.e. a path integral. By finally writing
 \begin{equation}
\langle q | \Psi(t) \rangle = \int dq_0  K(q, q_0, t) \Psi(q_0, t_0) = \int dq_0 \int \mathcal{D}q(t') \,\, e^{\frac{i}{\hbar} S[q(t')]} \Psi(q_0) \label{eq:kernel-general2}
 \end{equation}
it becomes clear that each path is weighted by the complex exponential of the action related to the given path divided by the Planck constant, $\hbar$.

The final expression above lends itself to many interpretations. The time evolution amplitudes calculated in this manner include a sum (integral) over all possible paths that take the system from $q_0$ to $q$, weighted by the complex exponential of the action related to that path, scaled by $1/\hbar$. This term is larger when the variation of the action from one path to another is small, because small oscillations result in more constructive interference of the imaginary term. It is a well known result that the classical path corresponds to a stationary action. Therefore, this term will be dominated by the classical path and by small fluctuations around it. \future{This consideration is the basis for many semiclassical approaches, as will be discussed in \refch{instanton}.}

The minimum action principle of classical mechanics also appears naturally from these expressions. Therefore, the PI formulation of quantum mechanics is a natural way to see how classical and quantum mechanics are related.

\section{Statistical Mechanics: The partition function}

In this section, we will use the tools developed in the previous section in order to address problems in statistical mechanics. In this case, the most important quantity of interest is the partition function in a particular ensemble. We will derive here the formulation for the canonical ensemble, where the quantum partition function is represented by
\begin{equation}
    Z=\Tr[e^{-\beta\hat{H}}],
\end{equation}
where $\beta=1/k_BT$ and $T$ is the temperature. In the basis of the eigenstates of the Hamiltonian, evaluating this trace leads to the usual expression
\begin{equation}
    Z=\sum_i \bra{\psi_i} e^{-\beta\hat{H}} \ket{\psi_i} = \sum_i e^{-\beta E_i}.
\end{equation}

However we will here evaluate this trace in the position representation. This means that we want to calculate 
\begin{equation}
    Z=\int dq \bra{q} e^{-\beta\hat{H}} \ket{q}. \label{eq:part-func}
\end{equation}

\begin{figure}[tbph]
\begin{centering}
\includegraphics[width=0.6\textwidth]{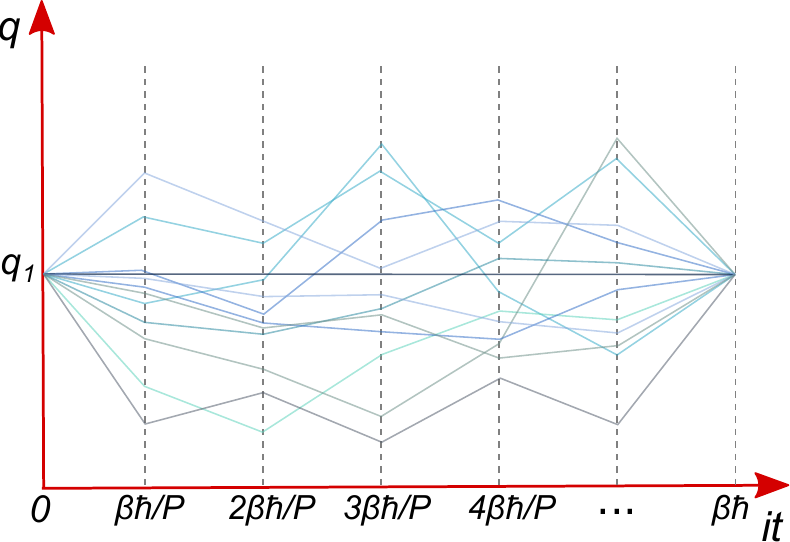}
\par\end{centering}
\caption{\label{fig:im-paths}Discretized possible paths in imaginary time that contribute to the partition function (same end points).}
\end{figure}

At this point we recognize that if $t=-i \tau = -i \beta \hbar$, then $\hat{U}(-i \tau) = e^{-\beta\hat{H}}$. We will refer to $\tau$ as the imaginary time. This establishes an intimate relationship between the path integral formulation of quantum mechanics and quantum statistical mechanics. Statistical mechanics arises as a propagation in imaginary time between $0< \tau <\beta \hbar$. Because of the trace, the initial and final position at these endpoints need to be the same, and in the following we rename $q$ in Eq. \eqref{eq:part-func} as $q_1$. This means that all the tools developed in the last section can also be used here to evaluate the necessary matrix elements in Eq. \eqref{eq:part-func}. 

Using again the Trotter factorization for a Hamiltonian with a generic time-independent potential, $V(\hat{q})$, we can write
\begin{equation}
    Z= \lim_{P \to \infty} \int dq_1, \dots , dq_{P} \prod_{j=1}^{P} \bra{q_{j+1}} e^{-\beta_P\frac{\hat{p}^2}{2m}} e^{-\beta_P V(\hat{q})} \ket{q_j}  \label{eq:part-func-trot}
\end{equation}
where $q_1=q_{P+1}=q$ and $\beta_P=\beta/P$. The equation above can be understood as a product over several high temperature (effective temperature $T^*=PT$) density matrices. As previously, we proceed to evaluate the necessary matrix elements
\begin{equation}
   \bra{q_{j+1}} e^{-\beta_P\frac{\hat{p}^2}{2m}} e^{-\beta_P V(\hat{q})} \ket{q_j}  =  \bra{q_{j+1}} e^{-\beta_P\frac{\hat{p}^2}{2m}}  \ket{q_j} e^{-\beta_P V(q_j)}.
\end{equation}

Using the same techniques as in Eq. \ref{eq:free-particle-1} and \ref{eq:free-particle-1a} (i.e. inserting resolution of identity in momentum space, using the result for the internal product of position and momenta, and ``completing the squares") we get to
\begin{equation}
    \bra{q_{j+1}} e^{-\beta_P\frac{\hat{p}^2}{2m}} e^{-\beta_P V(\hat{q})} \ket{q_j} = \frac{1}{2 \pi \hbar} \sqrt{\frac{2 \pi m}{\beta_P}} e^{-\beta_P \left[V(q_j) + \frac{m}{2 \beta_P^2  \hbar^2} (q_{j+1}-q_{j})^2\right]} \label{eq:matrix-element}
\end{equation}

In the expression above, one can identify the ring-polymer frequency as  $\omega_P=1/(\beta_P \hbar)$. One notes that remarkably, there are no more operators involved in the expressions for these matrix elements, which is the telltale sign of a ``classical world". Plugging this result back into the partition function of Eq. \ref{eq:part-func} under the condition that $q_0=q_P$ we obtain
\begin{equation}
        Z= \lim_{P \to \infty} \left(\frac{1}{2 \pi \hbar}\right)^P \left(\frac{2 \pi m}{\beta_P}\right)^{P/2} \int dq_1 \dots  dq_{P}  e^{-\beta_P \sum_{j=1}^{P} \left\{ \frac{m \omega_P^2}{2} (q_{j+1}-q_{j})^2 + V(q_j)\right\} }  \label{eq:part-func-discrete}
\end{equation}

\begin{exercise}[label={ex:discrete-pf},title={Ring polymer partition function and the harmonic oscillator.}]

Write down the steps leading to Eq. \ref{eq:matrix-element} and derive Eq. \ref{eq:part-func-discrete} for a harmonic potential given by $V(q)=m \omega_0^2 q^2/2$

\end{exercise}

\begin{figure}[tbph]
\begin{centering}
\includegraphics[width=0.7\textwidth]{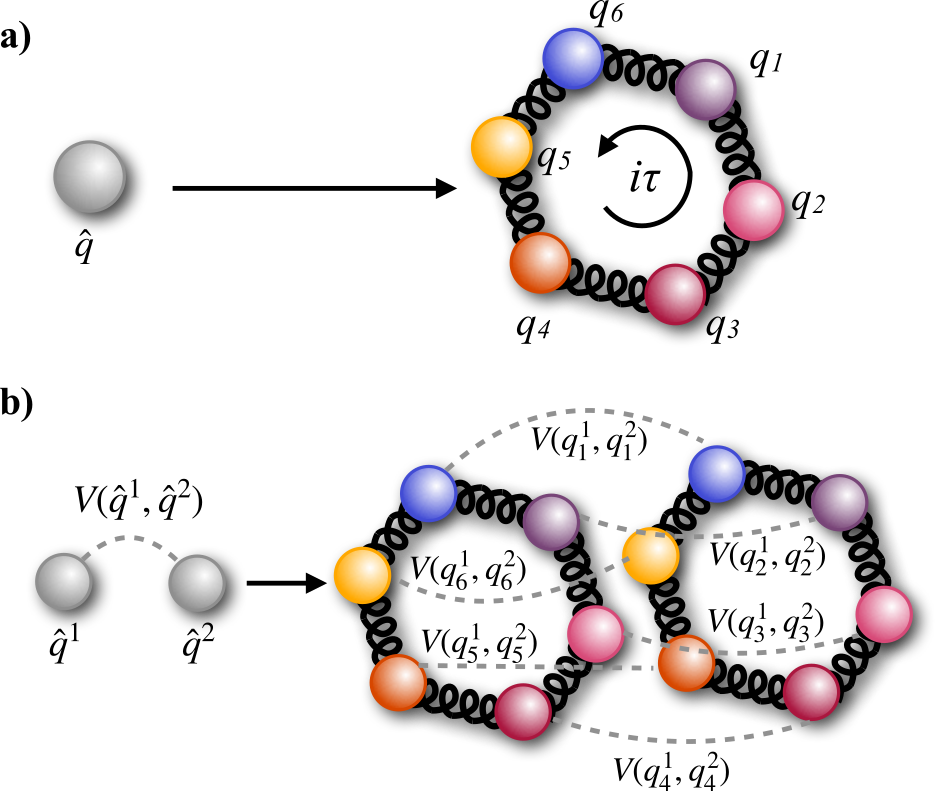}
\par\end{centering}
\caption{\label{fig:ring-polymers} A sketch of classical and quantum particles within the path integral molecular dynamics formalism. In (a), a single classical particle corresponds to one ring polymer, where each bead is a slice in imaginary time (depicted $P = 6$). In (b), two particles interacting through a potential  $V(q^{(1)}, q^{(2)})$ correspond to two ring polymers, where beads with the same index (identified by the same color in the sketch) in each of them interact through the potential $V$.}
\end{figure}

What one can take from Eq. \ref{eq:part-func-discrete} is that the quantum partition function can be recast into a classical object, something that is usually coined as the classical isomorphism. This classical object is a harmonic ring polymer with spring constant $k_P=m \omega_P^2 = m (P k_B T/\hbar)^2$ and number of beads $P$, as pictured in Fig. \ref{fig:ring-polymers}a. Remember that the origin of this term is the quantum kinetic energy, as discussed in the previous section, and the appearance of $\hbar$ on the spring constant is related to the comutation relation between $\hat{p}$ and $\hat{q}$. In analogy to what was discussed in the previous section, it is easy to see that each bead corresponds to a different imaginary time slice. In addition, at high temperatures, the spring constant becomes steadily stiffer (and infinite at the limit of infinitely high temperature), making the position of all ring polymers to collapse at a single point. The length of imaginary time over which one samples ($\beta \hbar$) also gets progressively smaller in this limit, where one can sample with $P=1$. Indeed, it is easy to see that with $P=1$ one automatically recovers the classical limit. The temperature at which a system can be considered classical depends on the characteristics of the potential $V(q)$ and on the particular observable one is interested in. We will discuss further how to estimate such parameters in the next sections.

The extension of Eq. \ref{eq:part-func-discrete} to many particles in multiple dimensions (3D) is straightforward for distinguishable particles. It reads as
\begin{eqnarray}
    Z= \lim_{P \to \infty} \prod_{i=1}^N  \left(\frac{m_i}{2 \pi \hbar^2 \beta_P}\right)^{3P/2} \int \prod_{i=1}^Nd\bm{q}_1^{(i)} \dots d\bm{q}_{P}^{(i)} \times \nonumber \\ \mathrm{e}^{\displaystyle{-\beta_P \left[\sum_{i=1}^N \sum_{j=1}^{P} \frac{m_i \omega_P^2}{2} (\bm{q}^{(i)}_{j+1}-\bm{q}^{(i)}_{j})^2 + \sum_{j=1}^{P} V(\bm{q}^{(1)}_j, \dots, \bm{q}^{(N)}_j)\right]}} \label{eq:mp-part-func}
\end{eqnarray}
under the condition that $\bm{q}^{(i)}_{P+1}=\bm{q}^{(i)}_{1}$. Note that, as pictured in Fig. \ref{fig:ring-polymers}b, the spring term only connects beads that belong to the same atom, while the external potential, $V$ is acting on corresponding beads of the different atoms.

For a system of multiple indistinguishable particles (bosons or fermions), exchange effects cause either a permanent (bosons) or a determinant (fermions) involving the spring terms connecting beads of \textit{different} atoms appear in the evaluation of the canonical partition function \cite{tuck08book, cepe-poll86prl}. This makes the solutions considerably more cumbersome to obtain and, in particular for fermions, also create a sign problem. The assumption of particle distinguishability is very reasonable for most nuclei building up molecules and solids, if one is not dealing with identical particles at extremely low temperatures. We refer the reader to Refs. \cite{Kwon:2012ie, Hirshberg:2019gm, BrieucMarx2020} for more information on this subject. 

\section{Sampling with path integral molecular dynamics}
\label{sec:pimd_sample}

So far, we have discussed how one can obtain an expression for the quantum partition function, $Z$, based on a special classical system, the harmonic ring polymer. What one needs to do now is devise a way to sample the ring polymer potential in a way that allows one to obtain observables of the form
\begin{equation}
    \langle \hat{O} \rangle = \frac{1}{Z} \Tr[e^{-\beta \hat{H}}\hat{O}]. \label{eq:qm-average}
\end{equation}

As discussed in Section \ref{sec:molecular-dynamics}, this sampling can be efficiently achieved in this case by molecular dynamics, if there is a suitable Hamiltonian to generate the equations of motion. The partition function discussed in the previous section is not yet in a format where this can be done. In order to achieve that we need to do some manipulation with the expression, that was first employed in Ref. \cite{ParrinelloRahman1984}. We reintroduce in the integral of Eq. \ref{eq:part-func-discrete} a term associated with an effective kinetic energy in terms of effective momenta of the polymer beads, $p_j$
\begin{eqnarray}
        Z= \lim_{P \to \infty} \left(\frac{1}{2 \pi \hbar }\sqrt{\frac{m}{m'}}\right)^P \int dq_1 \dots  dq_{P} dp_1 \dots dp_{P} \nonumber \\
        \times \exp\left\{-\beta_P  \sum_{j=1}^{P} \left[ \frac{p_j^2}{2m'}+\frac{m \omega_P^2}{2} (q_{j+1}-q_{j})^2 + V(q_j) \right] \right\} , \label{eq:part-func-discrete-withmom}
\end{eqnarray}
with which we can identify 
\begin{equation}
H_P=\sum_{j=1}^{P} \left\{ \frac{p_j^2}{2m'}+\frac{m \omega_P^2}{2} (q_{j+1}-q_{j})^2 + V(q_j) \right\} \label{eq:rp-ham}
\end{equation}
which is typically referred to as the ring polymer Hamiltonian. This Hamiltonian generates classical dynamics in an extended space spanned by all the bead coordinates. In addition, the partition function of Eq. \ref{eq:part-func-discrete-withmom} defines a canonical ensemble at temperature $PT$. It is important to stress that the masses $m'$ are devoid of physical meaning (even if one chooses them equal to the physical mass $m$) and the evaluation of static thermodynamical observables obtained from dynamical equations of motion generated by $H_P$ is completely independent of the value that $m'$ adopts. The molecular dynamics generated in this case is only a sampling tool and the concept of ``time'' when these equations of motion are evolved is, in principle, not physically meaningful. 

The equations of motion generated by $H_P$ (see Eq. \ref{eq:hamilton}) read as
\begin{eqnarray}
    \dot{p_j}= - m \omega_P^2(2q_j - q_{j+1}-q_{j-1}) - \frac{\partial V(q_j)}{\partial q_j} \\
    \dot{q_j}= \frac{p_j}{m'} \label{eq:eom-pimd-naive}
\end{eqnarray}
where the dots indicate taking the time derivative of the physical quantities.

\begin{exercise}[label={ex:rp-freqs},title={Ring polymer internal frequencies}]

If one considers the free ring polymer Hamiltonian, what is the highest frequency of vibration of the ring polymer as a function of $P$ and $T$? 

\end{exercise}

Integrating these equations of motion require some care. A direct integration of Eqs. \ref{eq:eom-pimd-naive} would require small time steps, given that the highest internal frequency of vibration of the ring polymer will be typically much higher than the physical frequencies of motion of the system \cite{Hall:1998km, tuck08book}. One way to circumvent this problem is to work in the normal mode representation of the ring polymer. The free ring polymer Hamiltonian can be written in the following form (assuming $m'=m$)
\begin{equation}
    H_P^0 = \sum_{k=0}^{P-1} \frac{\tilde{p}^2_k}{2m} + \frac{1}{2}m \omega_k^2 \tilde{q}_k^2,
\end{equation}
where $\omega_k=2\omega_P \sin(k\pi/P)$ and 
\begin{equation}
    \tilde{p}_k =  \sum_{j=1}^P p_j C_{jk} \,\,\,\,\, \tilde{q}_k = \sum_{j=1}^P q_j C_{jk}.
\end{equation}

The orthogonal normal mode transformation matrix $\bm{C}$ contains the following elements if $P$ is even 
\cite{craig2006a, ceri+10jcp}
\begin{equation}
    C_{jk}=
    \begin{cases}
    \sqrt{1/P} & k=0 \\
    \sqrt{2/P}\cos (2 \pi jk/P) & 1 \leq k \leq P/2-1 \\
    \sqrt{1/P}(-1)^j & k=P/2 \\
    \sqrt{2/P}\sin (2 \pi jk/P) & P/2+1 \leq k \leq P-1.
    \end{cases}
\end{equation}

For numerical efficiency purposes, the transformation to the normal mode representation is most often implemented via fast Fourier transforms. With this transformation at hand, the integration of the equations of motion for a time step $\Delta t$ can be obtained through the following symplectic algorithm 
\begin{equation}
\begin{gathered}
\begin{array}{cc}
p_j\leftarrow & p_j-\frac{\partial V}{\partial q_j}\frac{\Delta t}{2}\end{array}\\
\tilde{p}_{k}=\sum_{j=1}^{P} p_{j} C_{j k} \quad \tilde{q}_{k}=\sum_{j=1}^{P} q_{j} C_{j k} 
\\ \left(\begin{array}{c}\tilde{p}_{k} 
\\ \tilde{q}_{k}\end{array}\right) \leftarrow\left(\begin{array}{cc}\cos \left(\omega_{k} \Delta t\right) & -m \omega_{k} \sin \left(\omega_{k} \Delta t\right) \\ \frac{1}{m \omega_{k}} \sin \left(\omega_{k} \Delta t\right) & \cos \left(\omega_{k} \Delta t\right)\end{array}\right)\left(\begin{array}{c}\tilde{p}_{k} \\ \tilde{q}_{k}\end{array}\right) \\ 
p_{j}=\sum_{k=0}^{P-1} \tilde{p}_{k} C_{k j} \quad q_{j}=\sum_{k=0}^{P-1} \tilde{q}_{k} C_{k j} \\ 
p_{j} \leftarrow p_{j}-\frac{\partial V}{\partial q_{j}} \frac{\Delta t}{2}
\end{gathered}
\label{eq:pimd-verlet}
\end{equation}

The equations above reduce to the usual Verlet integration scheme if $P=1$. For a free ring polymer, the middle step would be exact for any value of $\Delta t$. For a harmonic potential $V = m \omega^2 q^2/2$, this means that the limitation on the size of the time step will be determined by the highest physical frequency of the systems described by $V$. For anharmonic potentials, nevertheless, the internal modes are not exactly uncoupled and cause a further small limitation on the time step necessary to obtain an accurate time evolution \cite{craig2006a}. Nevertheless, in general this procedure ensures one can run PIMD with similar time steps used for classical-nuclei MD. Another possibility is to transform the representation to the so-called ``staging variables'' \cite{tuck08book} instead of normal modes, which have a similar overall effect. 

Even though the dynamics of the ring polymer is quite non-ergodic~\cite{Hall:1998km}, ensuring ergodicity of the PIMD simulation is necessary in order to sample correctly the canonical ensemble and to be able to evaluate
\begin{equation}
    \langle \hat{O} \rangle = \frac{1}{T} \int_0^T dt O_P(t)  \label{eq:pimd-expectation}
\end{equation}
where $T$ is the total simulation time and $O_P$ is a PIMD estimator for the observable $\hat{O}$. We will briefly discuss how to obtain these estimators in Section \ref{sec:est-simple}. Ergodicity in a PIMD simulation can be ensured by coupling the PIMD momenta to aggressive thermostats~\cite{tuck08book, ceri+10jcp}. \future{Some of these will be discussed in Chapter \ref{ch:gle}.}


\section{Ab initio PIMD}


In this section, a brief explanation about how \textit{ab initio} path integral molecular dynamics can be performed is presented. Intuitively,  assuming that the Born-Oppenheimer approximation is valid, it is natural to conclude that \textit{ab initio} PIMD follows through taking $V(q_j)=E_{BO}(q_j)$ in Eq. \ref{eq:rp-ham} and the corresponding \textit{ab initio} forces of each bead to evolve the equations of motion \ref{eq:eom-pimd-naive}. This conclusion is absolutely correct, but a more detailed derivation is instructive to understand how the Born-Oppenheimer approximation enters the path-integral picture of quantum statistical mechanics. In the following, we summarise the derivation that is also presented in Ref. \cite{MarxHutter:book} and discuss some practical considerations.

Considering $\bm{q}$ and $\bm{r}$ as the position vectors of the nuclei and electrons in the system, respectively, we start from a product basis \textit{ansatz} of the following form
\begin{equation}
\ket{\mu, \bm{q}} = \ket{\mu ; \bm{q}} \otimes \ket{\bm{q}} \label{eq:mixed-basis}
\end{equation}
where a mixed basis is used, meaning that the position representation $\bm{q}$ is employed for the nuclear degree of freedom and the energy representation is used for the electronic degrees of freedom, with $\mu$ representing the index of an (adiabatic) state. The semicolon denotes a parametric dependence. It is to be understood that $\braket{\bm{r}}{\mu; \bm{q}} = \psi_\mu(\bm{r}; \bm{q})$. In addition, we take the states $\ket{\mu; \bm{q}}$ to be the orthogonal adiabatic basis obtained by solving
\begin{equation}
\hat{H}_e  \ket{\mu; \bm{q}} = E_\mu(\bm{q}) \ket{\mu; \bm{q}}
\end{equation} 
where $\hat{H}_e$ is the commonly coined electronic Hamiltonian. In atomic units and for $n$ electrons and $N$ nuclei it is given by 
\begin{equation}
\hat{H}_e = -\sum_{s=1}^n \frac{\nabla_{r_s}^2}{2} + \sum_{s<t}^n \frac{1}{|\bm{r}_s-\bm{r}_{t}|} - \sum_{i, s}^{n,N} \frac{Z_i}{|\bm{r}_s-\bm{q}_i|} +   \sum_{i<i'}^N \frac{Z_i Z_{i'}}{|\bm{q}_{i'}-\bm{q}_i|}. \label{eq:schr-el}
\end{equation}
where $Z$ is the nuclear charge. The full Hamiltonian in position representation is
\begin{equation}
\hat{H} = -\sum_{i=1}^N \frac{\nabla_{q_i}^2}{2 m_i} + \hat{H}_e.
\end{equation}
Note that we write the momentum operator $\hat{p}$ in position representation here for reasons that will become clear soon.

With the basis from Eq.~\ref{eq:mixed-basis}, the partition function is   
\begin{equation}
Z=\int \sum_{\mu} \bra{\bm{q}} \bra{\mu; \bm{q}}e^{-\beta \hat{H}}\ket{ \mu; \bm{q}} \ket{\bm{q}} d\bm{q}. \label{eq:part-func-bo}
\end{equation} 
From here on, we will continue the derivation for a single nucleus in the presence of many electrons for the sake of simplicity. In this case, we can proceed with the Trotter factorization of Eq. \ref{eq:part-func-bo} just like we did in Eq. \ref{eq:part-func-trot}. The Trotter factorization recasts the problem of evaluating Eq.  \ref{eq:part-func-bo} into the evaluation of the following matrix elements
\begin{equation}
\begin{split}
\bra{q_{j+1}}\bra{\mu_{j+1}; q_{j+1}} e^{-\beta_P(-\nabla^2_q/2 m + \hat{H}_e)} \ket{\mu_{j}; q_{j}} \ket{q_j} = \\
\bra{q_{j+1}} \bra{\mu_{j+1}; q_{j+1}} e^{-\beta_P(-\nabla^2_q/2 m)} \ket{\mu_{j}; q_{j}}  \ket{q_j}  e^{-\beta_P E_{\mu_j}(q_j)} \label{eq:matr-el-bo}
\end{split}
\end{equation}
where the Trotter indexes $j$ are such that $\ket{q_{P+1}}=\ket{q_{1}}$ and $\ket{\mu_{P+1}; q_{P+1}}=\ket{\mu_{1}; q_{1}}$, and the notation $\mu_{j}$ relates to the eigenstates of Eq. \ref{eq:schr-el} obtained as solutions of the nuclei fixed at positions $q_j$.

So far, all operations are exact, but dealing with the \textit{nuclear} kinetic energy operator will require some approximations. The adiabatic approximation implies that all terms involving the action of the nuclear gradient operator on the electronic states will be zero. This automatically simplifies the expression above to 
\begin{equation}
\begin{split}
& \bra{q_{j+1}} \bra{\mu_{j+1}; q_{j+1}} e^{-\beta_P(-\nabla^2_q/2 m)} \ket{\mu_{j}; q_{j}}  \ket{q_j}  e^{-\beta_P E_{\mu_j}(q_j)} \approx \\
&  \bra{q_{j+1}} e^{\frac{\beta_P \nabla^2_q}{2 m}} \ket{q_j} \braket{\mu_{j+1}; q_{j+1}}{\mu_{j}; q_{j}} e^{-\beta_P E_{\mu_j}(q_j)}.
\end{split}
\end{equation}

A term involving the projection of the electronic states at different beads (imaginary time slices) still remains. However, without loss of generality, one can assume that the variation in nuclear positions between one imaginary time slice and the next is small and Taylor-expand this variation around $q_{j+1}$, namely
\begin{equation}
\begin{split}
\braket{\mu_{j+1}; q_{j+1}}{\mu_{j}; q_{j}} = \bra{\mu_{j+1}; q_{j+1}} \left[\ket{\mu_{j}; q_{j+1}} + \nabla_q \ket{\mu_{j}; q_{j+1}}(q_j-q_{j+1}) \right] \\ + \mathcal{O}[(q_j-q_{j+1})^2].
\end{split}
\end{equation}
Then, assuming again the adiabatic approximation
\begin{equation}
\begin{split}
\braket{\mu_{j+1}; q_{j+1}}{\mu_{j}; q_{j}} \approx \braket{\mu_{j+1}; q_{j+1}}{\mu_{j}; q_{j+1}} = \delta_{\mu_{j+1}\mu_j}.
\end{split}
\end{equation}

Under the adiabatic approximation, it is then finally possible to treat the matrix element in Eq. \ref{eq:matr-el-bo} just like we did in the previous section, namely
\begin{equation}
\begin{split}
& \bra{q_{j+1}}\bra{\mu_{j+1}; q_{j+1}} e^{-\beta_P(-\nabla^2_q/2 m + \hat{H}_e)} \ket{\mu_{j}; q_{j}} \ket{q_j} \approx \\
& \delta_{\mu_{j+1}\mu_j} e^{-\beta_P E_{\mu_j}(q_j)} \bra{q_{j+1}} e^{\frac{\beta_P \nabla^2_q}{2 m}} \ket{q_j}  = \\
& \delta_{\mu_{j+1}\mu_j}  \left( \frac{m}{2 \pi \hbar^2 \beta_P} \right)^{1/2} e^{-\beta_P [m \omega_P^2 (q_{j+1} -q_j)^2/2 + E_{\mu_j}(q_j) ]}.
\end{split}
\end{equation}

Plugging this back into the Trotter-factorized version of Eq. \ref{eq:part-func-bo}, in which $Z = \lim_{P \to \infty} Z_P$ one obtains
\begin{equation}
\begin{split}
Z_P & = \int dq_1, \dots, dq_P \prod_{j=1}^P \sum_{\mu_j} \delta_{\mu_{j+1}\mu_j}  \left( \frac{m}{2 \pi \hbar^2 \beta_P} \right)^{1/2} e^{-\beta_P [m \omega_P^2 (q_{j+1} -q_j)^2/2 + E_{\mu_j}(q_j) ]} \\
& = \sum_{\mu}  \left( \frac{m}{2 \pi \hbar^2 \beta_P} \right)^{P/2} \int dq_1, \dots, dq_P   e^{-\beta_P \sum_{j=1}^P [m \omega_P^2 (q_{j+1} -q_j)^2/2 + E_{\mu}(q_j) ]}.
\end{split}
\end{equation}

If the system in question is such that the ground electronic state is well separated from all other excited states, then the ground state will be the only state contributing to the expression above, and the Born-Oppenheimer approximation to the path integral formulation of statistical mechanics can be written as 
\begin{equation}
Z  =  \lim_{P \to \infty} \left( \frac{m}{2 \pi \hbar^2 \beta_P} \right)^{P/2} \int dq_1, \dots, dq_P   e^{-\beta_P \sum_{j=1}^P [m \omega_P^2 (q_{j+1} -q_j)^2/2 + E_{0}(q_j) ]}.
\end{equation}
where $E_0(q)$ is the ground state Born-Oppenheimer potential energy surface. The generalization of this expression to many distinguishable nuclei leads to a similar expression as Eq. \ref{eq:mp-part-func}.

Quantum mechanical canonical expectation values for different observables, in this case, can be obtained by performing \textit{ab initio} molecular dynamics, where the force $F$ on atom $i$ are most often obtained on the fly, by calculating
\begin{equation}
F_i = - \nabla_{\bm{q}_i} \bra{0; \bm{q}} \hat{H}_e \ket{0; \bm{q}},
\end{equation}
in which $\mu=0$ is to be understood as the ground electronic state.
In practice, in most applications, one solves an approximation to the electronic time independent Schr\"odinger equation at each bead position (imaginary time slice). 
These approximations can be obtained by any given electronic structure method, be it wave-function based (Hartree-Fock, coupled cluster, configuration interaction, etc.) or electronic-density based (density-functional theory, random-phase approximation, etc.).

\section{Simple path integral estimators for observables} \label{sec:est-simple}

We will end this chapter by discussing how to finally evaluate quantum mechanical thermal expectation values like the ones represented by Eq. \ref{eq:qm-average} in the simplest case when the observable is a pure function of the position. More complex situations will be covered in the next chapter. 

Let us assume that an observable $\hat{O}$ is only a function of the position operator $\hat{O}=O(\hat{q})$. Then, it is true that calculating Eq. \ref{eq:qm-average} in the position representation will yield
\begin{equation}
    \langle \hat{O} \rangle = \frac{1}{Z}\int dq \bra{q} O(\hat{q}) e^{-\beta \hat{H}} \ket{q} = \frac{1}{Z}\int dq \, O(q) \bra{q}  e^{-\beta \hat{H}} \ket{q}.
\end{equation}

In the previous sections we have learned how to obtain $Z$ and the matrix elements $\bra{q}  e^{-\beta \hat{H}} \ket{q}$ in the path integral representation. Therefore, by applying the Trotter splitting techniques, and realizing that evaluating $O(q)$ for any imaginary time slice $q_j$ yields an equivalent expression (cyclic permutation of the ring polymer), one can quickly arrive to the following expression
\begin{equation}
    \langle \hat{O} \rangle = \lim_{P\to\infty}\frac{1}{Z_P} \int dq_1, \dots, dq_P, dp_1, \dots, dp_P\, \frac{1}{P} \sum_{j=1}^P O(q_j)\, e^{-\beta_P H_P} .
\end{equation}
where $H_P$ is given by Eq. \ref{eq:rp-ham} and the expression above lends itself to straightforward sampling with molecular dynamics. In this case, the expression presented in Eq. \ref{eq:pimd-expectation} is evaluated and we can identify the estimator $O_P$ for the observable $O$
\begin{equation}
    \label{eq:simple-estimator}
    O_P(q(t)) = \frac{1}{P}\sum_{j=1}^P O(q_j(t))
\end{equation}

Because of the choice of the position representation for the path integral molecular dynamics formalism, it is more straightforward to obtain expressions for observables that depend only on positions. Thermodynamic relationships can, nevertheless, be used in order to obtain estimators for something like the kinetic energy, expressed only in terms of position-dependent quantities. If it is not possible to do so, estimating a momentum-dependent quantity will involve an open path \cite{morr+07jcp, lin+10prl}. Last but not least, an elegant way to obtain different sorts of observables also involving electronic-structure dependent quantities (dipoles, polarizabilities, etc.) is through adding a perturbation involving the desired observable to the Hamiltonian and evaluating (see Ref.\cite{Shiga:2001gj})
\begin{equation}
    O(q_j) = -\frac{1}{\beta} \sum_{j=1}^P \frac{\partial}{\partial \lambda} \ln \bra{q_j} e^{-\beta_P (\hat{H}+\lambda\hat{O})}\ket{q_{j+1}}\vert_{\lambda=0}.
\end{equation}

As an exercise, we can look at the expectation value for the potential of a harmonic oscillator, and make some considerations about how it converges with $P$ at a given temperature. It is possible to show analytically that the thermal expectation value of the potential of a quantum mechanical harmonic oscillator of frequency $\omega_0$ is
\begin{equation}
    \langle V^{\text{h.o.}} \rangle = \frac{\hbar \omega_0}{4} \coth \left(\frac{\hbar \omega_0}{2 k_B T} \right). \label{eq:texp-analytical}
\end{equation}

Following the expressions above, the PI estimator will be
\begin{equation}
    V_P^{\text{h.o.}}(q_j) = \frac{1}{P} \sum_{j=1}^P \frac{1}{2} m \omega_0^2 q_j^2. \label{eq:estimator-pot}
\end{equation}

The quantum expectation value of $\hat{V}^{\text{h.o.}}$ in the PI representation will then be
\begin{equation}
    \langle \hat{V}^{\text{h.o.}} \rangle = \frac{1}{2} \sum_{j=1}^P m \omega_0^2 \langle q_j^2 \rangle = \frac{k_B T}{2} \sum_{k=0}^{P-1}  \frac{\omega_0^2}{(\omega_k^2+\omega_0^2)} \label{eq:texp-pi}
\end{equation}

\begin{exercise}[label={ex:pi-qho-potential},title={The thermal expectation value of the potential of a quantum harmonic oscillator}]

First, show the result of Eq. \ref{eq:texp-analytical}. You can resort to any picture of quantum mechanics to arrive at that result. 

Then, using the estimator of Eq. \ref{eq:estimator-pot} and the ring polymer normal mode transformation, show the result of Eq. \ref{eq:texp-pi}. 

Finally, plot Eq. \ref{eq:texp-pi} as a function of $P$ for a temperature of 300 K and a harmonic oscillator frequency $\omega_0=3500$ cm$^{-1}$ (this is roughly the frequency of OH stretch modes).
\end{exercise}

For quantities like the potential, one can make the following consideration
in order to gauge the amount of replicas $P$ needed to converge
a simulation. The PI formalism samples the potential in an extended
ring polymer phase space at a temperature $P \times T$. Therefore, $P$ needs
to be high enough to provide enough energy to account for the ZPE
present in the potential, leading to the usual expression $\frac{\hbar \omega_0}{k_BT}< P$. For a
system with many degrees of freedom, one can make such an estimate
by considering the highest vibrational frequency in the system and
the temperature of interest. It is worth noting that other quantities (for
example, heat capacity) can show a different convergence behavior
with $P$.

More advanced estimators and how to reduce the amount of beads required for the simulations will be discussed in the next Chapter.
\setchapterpreamble[u]{\margintoc}

\chapter{Advanced path integral methods}
\labch{pimdadv}

The previous chapter has introduced the fundamental equations and background to the path integral approach to calculate real and imaginary time quantities. In this chapter we will consider methods that can be used to efficiently perform and evaluate properties from path integral simulations.

\section{Momentum dependent observables}
Sec.~\ref{sec:est-simple} introduced the case where the observable of interest is purely a function of position $\hat{O}=O(\hat{q})$ which yields the estimator for the observable given in Eq.~\ref{eq:simple-estimator} as an average over the value of that observable evaluated on each bead of the ring polymer. We will now consider the case where the observable of interest depends on the momentum such as the kinetic energy or the momentum distribution of the system.

\subsection{The primitive kinetic energy estimator}
The previous chapter gives us a prescription for calculating the potential energy, which is depends only on the positions, using the path integral formalism. Hence to derive an estimator for the kinetic energy we can derive an estimator for the total energy of the system, which is the sum of the kinetic and potential energies, and then use this to identify an estimator for the kinetic energy. The total energy can be obtained from the partition function using the standard relation~\cite{tuck08book},
\begin{equation}
    \langle E \rangle = -\frac{1}{Z}\left(\frac{\partial Z}{\partial \beta}\right).
    \label{eq:ener_from_pf}
\end{equation}
Inserting the path integral partition function for a one-dimensional system given in Eq.~\ref{eq:part-func-discrete} yields,
\begin{equation}
    \langle E \rangle = \Bigg\langle\frac{Pk_{B}T}{2} - \sum_{j=1}^{P}\frac{1}{2}m\omega_{P}^{2}(q_{j+1}-q_{j})^2 + \frac{1}{P}\sum_{j=1}^{P}V(q_{j})\Bigg\rangle.
    \label{eq:total_ener_estimator_1d}
\end{equation}
The last term can be recognized as the estimator for the potential energy, $V(q)$, since that is a position dependent operator and hence can be evaluated using Eq.~\ref{eq:simple-estimator} in the previous chapter. The first two terms thus provide an estimator for the kinetic energy $\kappa_{prim}$
\begin{equation}
    \kappa_{prim} = \frac{Pk_{B}T}{2} - \sum_{j=1}^{P}\frac{1}{2}m\omega_{P}^{2}(q_{j+1}-q_{j})^2,
\end{equation}
where ``prim'' denotes that this is the primitive estimator for the kinetic energy in contrast to the more commonly used ``virial'' estimator that will be introduced below. The primitive estimator for the kinetic energy for $N$ particles in three dimensions can be obtained by inserting Eq.~\ref{eq:mp-part-func} into the general expression for the energy in terms of the partition function Eq.~\ref{eq:ener_from_pf} to give,
\begin{equation}
    \kappa_{prim} = \frac{3NPk_{B}T}{2} - \sum_{i=1}^{N}\sum_{j=1}^{P}\frac{1}{2}m\omega_{P}^{2}(\mathbf{q}^{(i)}_{j+1}-\mathbf{q}^{(i)}_{j})^2.
\end{equation}
However, the primitive estimator is a difference of two terms which both grow larger as the number of replicas $P$ is increased. As $P$ is increased this results in larger fluctuations in the instantaneous values of $\kappa_{prim}$ along a path integral trajectory that increase the statistical error bars. Hence as the number of replicas used in the path integral simulations is increased towards the $P \to \infty$ limit in which the exact result is obtained increasingly longer trajectories are needed to average out these fluctuations and obtain an accurate value for the thermodynamic (ensemble averaged) value of the kinetic energy $\left< \kappa_{prim} \right>$\cite{herm-bern82jcp}. To avoid this one can instead derive an alternative kinetic energy estimator, known as the virial estimator\cite{herm-bern82jcp} $\kappa_{virial}$, that gives the same ensemble average $\left< \kappa_{prim} \right> = \left< \kappa_{virial} \right>$ but whose fluctuations do not grow with $P$.

\begin{exercise}[label={ex:primitive_ke},title={Primtive kinetic energy estimator}]

Derive the expression in Eq.~\ref{eq:total_ener_estimator_1d} by inserting Eq.~\ref{eq:part-func-discrete} into Eq.~\ref{eq:ener_from_pf} and then performing the derivative and simplifying.

\end{exercise}

\subsection{The virial kinetic energy estimator}

\begin{figure}[tbph]
\begin{centering}
\includegraphics[width=0.5\textwidth]{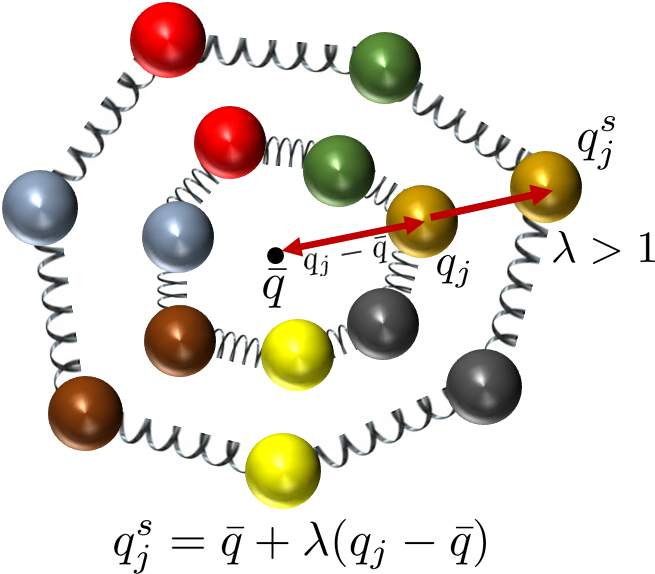}
\par\end{centering}
\caption{Depiction of the coordinate scaled ring polymer used to obtain the virial estimator}
\label{fig:rp_coord_scaling}
\end{figure}

To derive an estimator that possesses more desirable statistical properties than the primitive estimator one can employ a coordinate scaling approach~\cite{Janke1997,yama05jcp}. The coordinate scaling approach provides a quick and physically transparent route to obtain the virial kinetic energy estimator and also provides a more generally powerful tool for constructing other path integral estimators such as those for computing isotopic free energy changes~\cite{ceri-mark13jcp}. To do this we first consider the effect on the partition function of making a coordinate transformation from the ring polymer bead positions $\{ q_{j} \}$ to a set of ring polymer positions that are scaled relative to their centroid position $\{ q^{s}_{j} \}$ via,
\begin{equation}
    q^{s}_{j} = \bar{q} + \lambda(q_j-\bar{q}).
    \label{eq:rp_coord_scaling}
\end{equation}
The centroid of the ring polymer representing each particle is simply the average position (center) of the ring polymer replicas that comprise it,
\begin{equation}
    \bar{q} = \frac{1}{P} \sum_{j=1}^{P} q_{j}.
\end{equation}
Hence for $\lambda<1$ the transformation in Eq.~\ref{eq:rp_coord_scaling} contracts the ring polymer replicas about their centroid while for $\lambda>1$ it dilates them as shown in Fig.~\ref{fig:rp_coord_scaling}.
Applying this change of coordinates to the path integral partition function  (Eq.~\ref{eq:part-func-discrete}) one obtains,
\begin{equation}
     Z = \lambda^{1-P}\left(\frac{m}{2\pi\hbar^{2}\beta_{P}}\right)^{P/2}\int dq^{s}_1 \cdots \int dq^{s}_P \exp{{-\frac{m}{2\hbar^{2}\beta_{P}\lambda^{2}}\sum_{j=1}^{P}(q^{s}_{j+1}-q^{s}_{j})^{2}-\beta_{P}\sum_{j=1}^{P}V(\bar{q}+\lambda^{-1}(q^{s}_{j}-\bar{q}))}} \label{eq:coord-transformed-Z},
\end{equation}
where we have used the Jacobian associated with the transformation,
\begin{equation}
    dq_1^{s}\cdots dq_P^{s} = \lambda^{P-1} dq_1\cdots dq_P.
\end{equation}
From Eq.~\ref{eq:coord-transformed-Z} one can see that if we set $\lambda^{2}=m/\beta_{P}=mP/\beta$ we can remove the inverse temperature ($\beta$) dependence of the spring term leaving the prefactor of that term as $1/2\hbar^{2}$. By using this change of coordinates and choice of $\lambda$ we can thus take the derivative of the scaled coordinate partition function with respect to $\beta$ required to evaluate the total energy without bringing down a expression involving the spring term which produced the statistical convergence issues in the primitive estimator. With this choice of $\lambda$ and making the coordinate transformation yields 
\begin{equation}
    Z = \left[\lambda\left(\frac{1}{2\pi\hbar^{2}}\right)^{P/2}
    \int dq^{s}_1 \cdots \int dq^{s}_P \exp{{-\frac{1}{2\hbar^{2}}\sum_{j=1}^{P}(q^{s}_{j+1}-q^{s}_{j})^{2}-\frac{\beta}{P}\sum_{j=1}^{P}V(\bar{q}+\lambda^{-1}(q^{s}_{j}-\bar{q}))}}\right].
\end{equation}
Inserting this into Eq.~\ref{eq:ener_from_pf} allows us to evaluate the total energy in the temperature scaled coordinates, 
\begin{equation}
    \langle E \rangle = -\Bigg\langle-\frac{1}{2\beta}  - \frac{\beta}{P}\sum_{j=1}^{P}\frac{\partial V(\bar{q}+\lambda^{-1}(q^{s}_{j}-\bar{q}))}{\partial \beta} - \frac{1}{P}\sum_{j=1}^{P}V(\bar{q}+\lambda^{-1}(q^{s}_{j}-\bar{q}))\Bigg\rangle 
    \label{eq:ener_scaled}
\end{equation}
At this point one should be careful in evaluating the derivative with respect to $\beta$ in the middle term since coordinates on which the potential energy, $V$, depend are now a function of $\lambda$ and hence $\beta$:
\begin{align}
    \frac{\beta}{P}\sum_{j=1}^{P}\frac{\partial V(\bar{q}+\lambda^{-1}(q^{s}_{j}-\bar{q}))}{\partial \beta} &=  \frac{\beta}{P}\sum_{j=1}^{P}\frac{\partial V(\bar{q}+\lambda^{-1}(q^{s}_{j}-\bar{q}))}{\partial (\bar{q}+\lambda^{-1}(q^{s}_{j}-\bar{q}))}\frac{\partial (\bar{q}+\lambda^{-1}(q^{s}_{j}-\bar{q}))}{\partial \lambda}\frac{\partial \lambda}{\partial \beta} \\
    &= \frac{\beta}{P}\sum_{j=1}^{P}\frac{\partial V(\bar{q}+\lambda^{-1}(q^{s}_{j}-\bar{q}))}{\partial (\bar{q}+\lambda^{-1}(q^{s}_{j}-\bar{q}))}\left(-\frac{1}{\lambda^2}(q^{s}_{j}-\bar{q})\right)\left(-\frac{\lambda}{2\beta}\right) \\
    &= \frac{1}{2P\lambda}\sum_{j=1}^{P}\frac{\partial V(\bar{q}+\lambda^{-1}(q^{s}_{j}-\bar{q}))}{\partial (\bar{q}+\lambda^{-1}(q^{s}_{j}-\bar{q}))}(q^{s}_{j}-\bar{q}) \label{eq:mid_deriv}
\end{align}
Inserting Eq.~\ref{eq:mid_deriv} into Eq.~\ref{eq:ener_scaled} and then transforming back to the unscaled ring polymer replica coordinates (and noting to account for the Jacobian on the way back) gives an alternative very useful energy estimator,  
\begin{equation}
    \langle E \rangle = \Bigg\langle\frac{1}{2\beta} + \frac{1}{2P}\sum_{j=1}^{P}\frac{\partial V(q_{j})}{\partial q_{j}}(q_{j}-\bar{q}) + \frac{1}{P}\sum_{j=1}^{P}V(q_{j}) \Bigg\rangle.
\end{equation}
From this we can identify the virial estimator for the kinetic energy as
\begin{equation}
    \kappa_{virial} = \frac{k_{B}T}{2} + \frac{1}{2P}\sum_{j=1}^{P}(q_j-\bar{q})\frac{\partial V}{\partial q_j}. \label{eq:ke-virial-1d}
\end{equation}
For $N$ particles and in three-dimensions this gives,
\begin{equation}
    \kappa_{virial} = \frac{3Nk_{B}T}{2} + \frac{1}{2P}\sum_{i=1}^{N}\sum_{j=1}^{P}(\mathbf{q}^{(i)}_{j}-\bar{\mathbf{q}}^{(i)})\frac{\partial V}{\partial \mathbf{q}^{(i)}_{j}}
\end{equation}
where the centroid is defined as,
\begin{equation}
    \bar{\mathbf{q}}^{(i)} = \frac{1}{P} \sum_{j=1}^{P} \mathbf{q}_{j}^{(i)}.
\end{equation}
The virial estimator thus provides an efficient way to evaluate the kinetic energy from a path integral simulation using just the forces and the distance of each replica from the centroid its ring polymer. The name ``virial" arises from the fact that it can also be derived by applying the classical virial theorem~\cite{tuck08book} to the primitive estimator result which was the way it was first obtained \cite{herm-bern82jcp}. Finally, it is important to note that for a given configuration obtained from a path integral simulation $\kappa_{virial}\ne\kappa_{prim}$ but if the path integral simulation is correctly performed and sufficiently averaged then the ensemble averages should be equal i.e. $\left< \kappa_{virial} \right>=\left< \kappa_{prim} \right>$. Indeed, a useful internal consistency check on path integral simulations is to check that the two estimators closely match when averaged over a sufficiently long trajectory. If they do not it often reflects a fundamental error in the simulation being performed such as inaccurate integration of the equations of motion. However, due to its more desirable convergence properties and hence smaller statistical error bars it is preferable to use virial estimator for evaluation of the kinetic energy. 

\subsection{General momentum dependent operators}
\label{sec:genp_op}

\begin{figure}[tbph]
\begin{centering}
\includegraphics[width=0.5\textwidth]{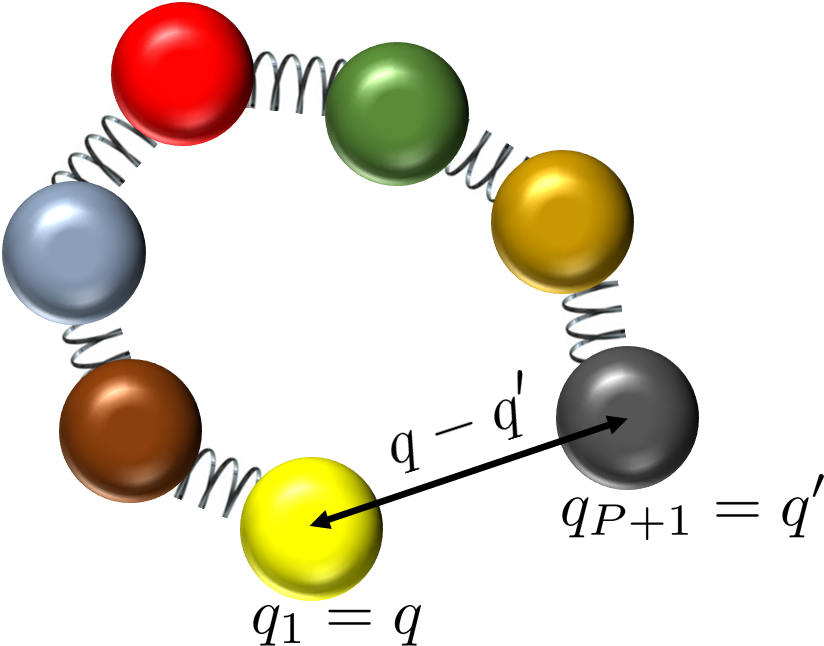}
\par\end{centering}
\caption{Depiction of the open chain ring polymer and its end to end distance. When Fourier transformed the end to end distance can be related to the momentum distribution.}
\label{fig:rp_open_polymer}
\end{figure}

In addition to the kinetic energy one might also be interested in evaluating more general momentum-dependent operators $\hat{O} = O(\hat{p})$. The evaluation of a momentum dependent operator can be done so as to obtain a similar expression to that for position dependent variables,
\begin{align}
    \langle\hat{O}\rangle &= \frac{1}{Z}\text{Tr}\left[e^{-\beta \hat{H}} O(\hat{p})\right]\\
    &= \frac{1}{Z}\int dq \bra{q}e^{-\beta \hat{H}}O(\hat{p})\ket{q}.
\end{align}
Since the observables in this case depends on momentum to evaluate this we first insert a complete set of momentum states followed by a set of position states,
\begin{align}
    \langle\hat{O}\rangle &= \frac{1}{Z}\int dq dp \bra{q}e^{-\beta \hat{H}}O(\hat{p})\ket{p}\bra{p}\ket{q} \\
    &= \frac{1}{Z}\int dq dp~O(p) \bra{q}e^{-\beta \hat{H}}\ket{p}\bra{p}\ket{q} \\
    &= \frac{1}{Z}\int dq dq' dp~O(p) \bra{q}e^{-\beta \hat{H}}\ket{q'}\bra{q'}\ket{p}\bra{p}\ket{q}.
\end{align}
One can then use position-momentum overlap matrix element results given in Eq. \ref{eq:free-particle-1} to obtain,
\begin{align}
    \langle\hat{O}\rangle = \frac{1}{2\pi\hbar Z}\int dqdq'dp~O(p)e^{ip(q-q')/\hbar}\bra{q}e^{-\beta \hat{H}}\ket{q'} \label{eq:momentum_obs}.
\end{align}
From this one can see that evaluating a momentum dependent observable involves evaluating an off-diagonal element of the density matrix i.e. $\bra{q}e^{-\beta \hat{H}}\ket{q'}$ with $q \ne q'$. This is in contrast to the previous section where position dependent operators one only required diagonal elements ($q = q'$). Evaluating the off diagonal element can be done using the same techniques used in deriving Eq.~\ref{eq:kernel-general}. However, it is important to employ the symmetric Trotter splitting in the evaluation of the matrix element in this case since the asymmetric variant is only gives a final result with an error that drops $\mathcal{O}(P^{-1})$ whereas the symmetric gives a final result where the error drops as $\mathcal{O}(P^{-2})$. For diagonal elements due to the symmetry ($q=q'$) the result obtained is identical using either splitting. However, for off-diagonal elements using the symmetric Trotter splitting gives an expression that will converge faster with the number of replicas $P$. Applying the symmetric Trotter splitting gives,
\begin{align}
    \bra{q}e^{-\beta \hat{H}}\ket{q'} &= \lim_{P\to \infty} \bra{q}\left[e^{-\beta \hat{V}/2P}e^{-\beta \hat{K}/P}e^{-\beta \hat{V}/2P}\right]^{P}\ket{q'} \\
    &= \left(\frac{m}{2\pi\hbar^{2}\beta_{P}}\right)^{P/2}\int dq_{2}\cdots dq_{P} \nonumber \\
    &\times\exp{- \sum_{j=1}^{P}\left(\frac{m}{2\hbar^{2}\beta_{P}}\omega_{P}^{2}(q_{j+1}-q_{j})^{2} + \frac{\beta_{P}}{2}(V(q_{j+1})+V(q_{j}))\right)}.
\end{align}
where $q_{1}=q$ and $q_{P+1}=q'$. This path integral expression for an off-diagonal element of the density matrix has many similarities to the diagonal term such as $P$ replicas with adjacent copies coupled by harmonic spring terms. However, it is important to realize that the cyclic condition in this case does not apply i.e. $q_{P+1}\ne q_{1}$. The physical picture is thus one of a open-chain ring polymer as depicted in Fig.~\ref{fig:rp_open_polymer}.

A momentum dependent operator that is of particular interest is the momentum distribution of the particles since this can be extracted from deep inelastic neutron scattering experiments. In classical mechanics the distribution of momentum $n^{cl}(p')$ is independent of the interactions (potential energy) and given by,
\begin{equation}
    n^{cl}(p') = \left(\frac{\beta}{2\pi m}\right)^{3/2}4\pi p'^2 e^{-\beta p'^2/2m}.
\end{equation}
In contrast, in a quantum mechanical system the momentum distribution of particle depends on the potential energy and the resulting confinement of its position which arises as a consequence of the uncertainty principle between positions and momenta i.e. localization in position leads to increased uncertainty in momentum space and vice versa. The quantum mechanical momentum distribution $n(p')$ can be obtained exactly from Eq.~\ref{eq:momentum_obs} by using $O(\hat{p})=\delta(\hat{p}-p'\hat{I}) \rightarrow O(p') = \delta(p-p')$ to give,
\begin{equation}
    n(p') = \frac{1}{2\pi\hbar Z}\int dqdq'~e^{ip'(q-q')/\hbar}\bra{q}e^{-\beta \hat{H}}\ket{q'}.
\end{equation}
This can be recognized as the Fourier transform of the end-to-end distance ($q-q'$) of the open chain ring polymer\cite{cepe-poll86prl,morr+07jcp} as depicted in Fig.~\ref{fig:rp_open_polymer}. The momentum distribution is of particular interest since it can be obtained from deep inelastic neutron scattering experiments~\cite{reit+02prb,reit+04bjp,pant+08prl,flam+12jcp,lin+11prb}. 

\section{Efficient path integral calculations using ring polymer contraction}
The PIMD approach is accompanied by significant extra computational cost owing due the need to make $P$ replicas of the system. A number of efficient methods have been introduced to combat this\cite{mark-ceri18nrc} including the ring polymer contraction approach discussed in this section, higher order path integral approaches (Sec.~\ref{sec:higher_order_PI}) and those that exploit colored noise\future{ (Sec.~\ref{sec:GLE_and_PIMD})}. While the latter two techniques focus on reducing the total number of replicas ring polymer contraction (RPC)\cite{mark-mano08jcp,mark-mano08cpl,Fanourgakis2009,mars-mark16jcp} employs a different approach: instead of reducing the total number of replicas, $P$ it reduces the cost of evaluating the forces on each of them. To do this RPC retains the $P$ replica ring polymer and generates from it a contracted $P'$ replica ring polymer on which the most expensive parts of the potential energy and forces are evaluated. To understand how all these acceleration methods work it is worth first considering how many replicas are required to converge a path integral calculation.

\subsection{Convergence of path integral calculations with numbers of replicas}
\begin{figure}[tbph]
\begin{centering}
\includegraphics[width=0.8\textwidth]{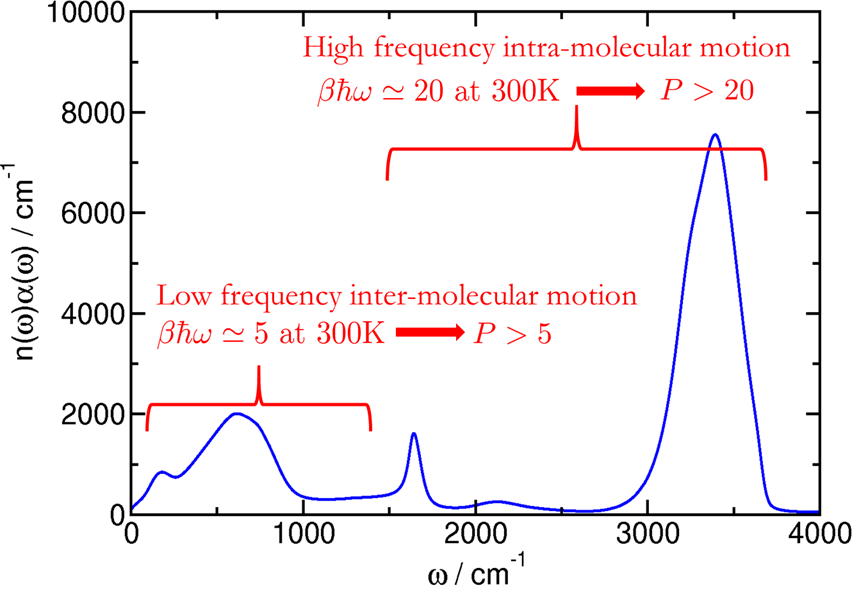}
\par\end{centering}
\caption{Infra red absorption spectrum of liquid water at 300 K showing the different frequencies present in liquid water. To quantize the low frequency intermolecular motion fewer replicas ($P>5$) are needed than to quantize the high frequency intramolecular motion ($P>20$).}
\label{fig:rpc_h20_ir}
\end{figure}

\begin{figure}[tbph]
\begin{centering}
\includegraphics[width=0.9\textwidth]{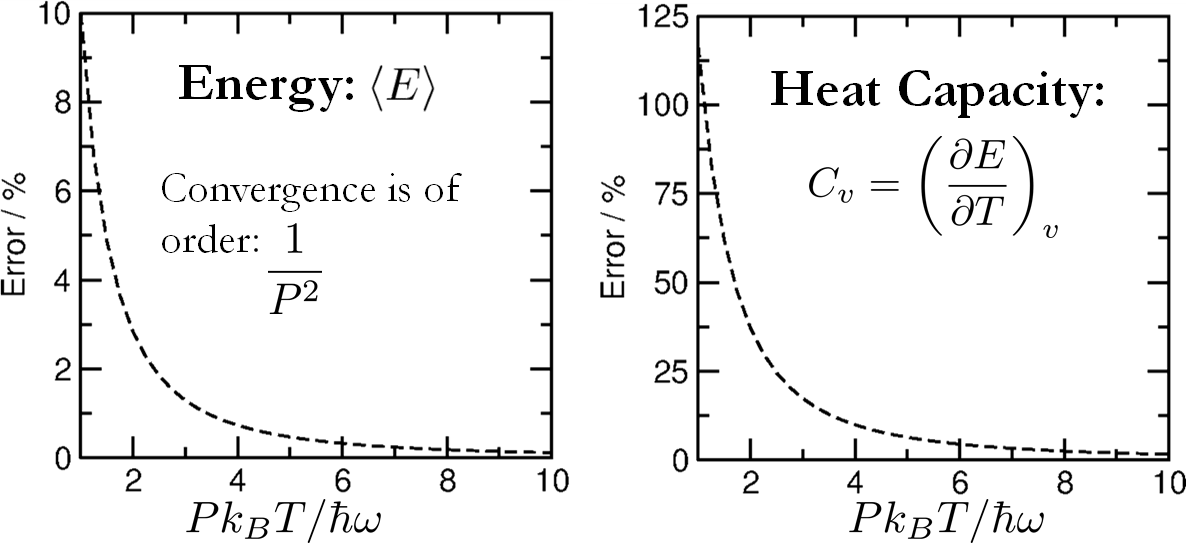}
\par\end{centering}
\caption{Percentage error in the energy (left) and heat capacity (right) for a single harmonic oscillator as a function of $Pk_{B}T/\hbar\omega$.  }
\label{fig:bead_convergence_harmonic}
\end{figure}

As discussed at the end of Chapter 2 the number of replicas required to converge a path integral calculation can be estimated by
\begin{equation}
P >\hbar\omega_{\textrm max}/k_{\textrm B}T
\label{eq:beads_needed}
\end{equation}
where $\omega_{\textrm max}$ is the maximum frequency present in the system of interest. For example the experimental infra-red vibrational absorption spectrum of liquid water is shown in Fig.~\ref{fig:rpc_h20_ir}. This spectrum shows a librational region extending to $\sim$1000cm$^{-1}$ and a region containing the intramolecular motion which extends to wavenumbers approaching 4000 cm$^{-1}$. Using Eq.~\ref{eq:beads_needed} suggests that since $k_{\textrm B}T/hc\simeq 200$ cm$^{-1}$ at room temperature $P$ should be greater than 5 for a rigid model of water (i.e. one where the O-H and H-H intramolecular distances are kept fixed and hence there are no stretching or bending motions), in which only the librational region is modelled. In practice previous studies have found that $P$=6 is typically enough to give a converged result~\cite{Kuharski1984,HernndezdelaPea2004,mill-mano05jcp2}. For fully flexible descriptions of water's potential energy surface where the high frequency intramolecular motions are also present, Eq.\ref{eq:beads_needed} then gives $P>20$ and in practice $P$=32 is found to give a converged result~\cite{Wallqvist1985,paes+06jcp,habe+08jcp}. Finally, it should also be noted for properties that involve computing fluctuations, such as the heat capacity whose estimator is related to the fluctuations of the energy in the simulation, the number of replicas required for convergence is typically considerably higher (by a factor of 2 or more)~\cite{Shiga2005,Shinoda2005,glae-frie02jcp}. The convergence of the energy and heat capacity is shown in Fig.~\ref{fig:bead_convergence_harmonic} for a harmonic oscillator of frequency $\omega$. From this one can see that $Pk_{B}T/\hbar\omega=1$ (i.e. $P = \hbar\omega/k_{B}T$) the error in the energy is $10~\%$ whereas in the heat capacity is $100~\%$. This falls to $2~\%$ and $38~\%$ respectively when $Pk_{B}T/\hbar\omega=2$ (i.e. using the number of replicas determined using $P = 2\hbar\omega/k_{B}T$).

\subsection{Ring polymer contraction}
\label{sub:contraction}
Ring polymer contraction exploits the fact that smoothly varying (low frequency) interactions require less beads for convergence than high frequency rapidly varying ones. This is since the beads are kept close in space due to the strong harmonic spring terms between them and hence any smoothly varying interaction can be approximated with negligible error on a much coarser representation of the imaginary time path, i.e. one with fewer beads. Hence if one can split the forces in a system into components that vary smoothly in space and those which vary rapidly, one can exploit this observation by evaluating the rapidly varying components on all replicas and the smoothly varying ones on a contracted ring polymer comprised of fewer replicas, $P'$. If this splitting is constructed such that the computational cost of the rapidly varying forces is negligible compared to that of the smoothly varying forces, one can decrease the cost of the force evaluations by a factor of $P/P'$.

\begin{figure}[tbph]
\begin{centering}
\includegraphics[width=1.0\textwidth]{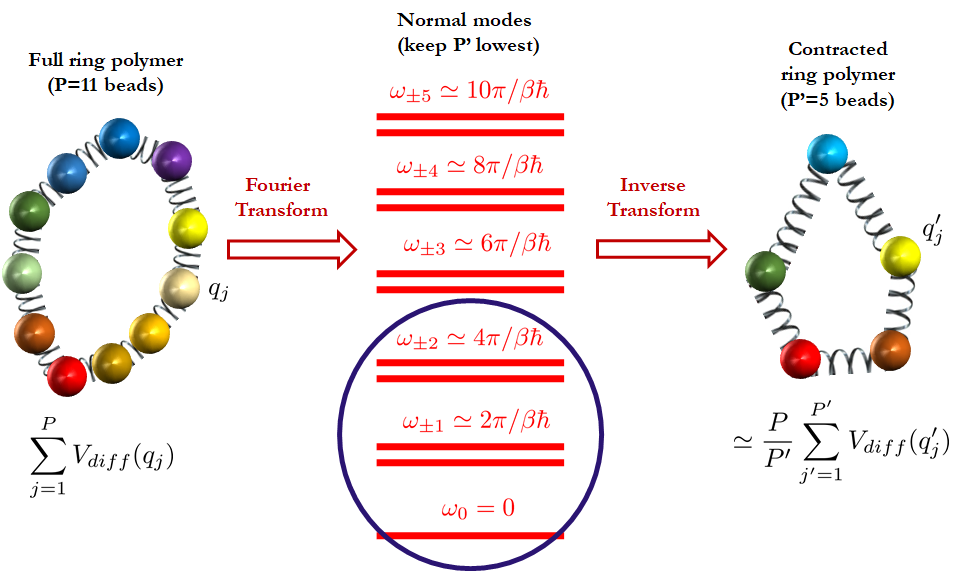}
\par\end{centering}
\caption{Outline of the RPC procedure. In this figure the normal modes shown are labelled using their eigenvalues values in large $P$ limit but in practice the exact values are used.}
\label{fig:rpc_scheme}
\end{figure}

For the moment we will assume that we have been able to find a way to split the full forces $\mathbf{f}_{\textrm{full}}$ into a reference force $\mathbf{f}_{\textrm{ref}}$ which, when taken from the full force $\mathbf{f}_{\textrm{full}}$, leaves a remaining force $\mathbf{f}_{\textrm{diff}}$ that is slowly varying and thus can be well approximated on the contracted polymer
\begin{equation}
\mathbf{f}_{\textrm{diff}} = \mathbf{f}_{\textrm{full}} - \mathbf{f}_{\textrm{ref}}.
\end{equation}
We now wish to construct a set of contracted ring polymer positions with $P'$ replicas ($P'\le P$) on which the difference force can be evaluated. In practice there are many ways one could achieve this following the general form,
\begin{equation}
\label{eq:RPC-r}
\mathbf{q}_{j'}^{(i)} = \sum_{j=1}^P T_{j'j}\,\mathbf{q}_{j}^{(i)}
\end{equation}
where $\mathbf{q}_{j'}^{(i)}$ are the contracted coordinates for particle $i$ and $\mathbf{q}_{j}^{(i)}$ are the full ring polymer coordinates and $T_{j'j}$ is the transformation matrix which maps the full ring polymer coordinates onto the contracted ones. Performing this transformation for each particle transforms the original set of coordinates of the $N$ particle system with each particle consisting of $P$ replicas to the contracted $N$ particle system where each particle only has $P'$ replicas, $\{\mathbf{q}_j^{(i)}\}^{i=1 \ldots N}_{j=1 \ldots P} \to \{\mathbf{q}_{j'}^{(i)}\}^{i=1 \ldots N}_{j'=1 \ldots P'}$. At this stage it is important to identify the specific form of the transformation matrix $T_{j'j}$ which defines the specific RPC scheme to be used. The original, and by far the most commonly employed, contraction scheme~\cite{mark-mano08jcp} involves transforming to the normal mode representation of the free ring polymer, discarding the $P-P'$ highest normal modes and then transforming back to the Cartesian representation as shown schematically in Fig.~\ref{fig:rpc_scheme}. The transformation matrix is given in Ref.~\cite{mark-mano08jcp}. The net effect of this transformation is a Fourier interpolation of the ring polymer (imaginary time path). In the limit where $P'=1$ this contraction scheme reduces each ring polymer to its centroid,
\begin{equation}
\mathbf{q}^{(i)}_{j'= P'=1} = \overline{\mathbf{q}}^{(i)} = \frac{1}{P} \sum_{j=1}^P \mathbf{q}_j^{(i)}
\end{equation}
and when $P=P'$ the ring polymer positions are left unchanged. For intermediate values, the transformation creates a contracted set of $P'$ positions which approximately represent the full ring polymer (effectively taking a lower-order Fourier representation of the imaginary time path).
The slowly varying difference force can then be evaluated on the contracted positions, $\mathbf{f}_{j',\textrm{diff}}^{(i)}$, and then projected back onto the full ring polymer representing each particle:
\begin{equation}
\mathbf{f}_{j,\textrm{diff}}^{(i)} = \frac{P}{P'} \sum_{j'=1}^{P'} T_{j'j} \mathbf{f}_{j',\textrm{diff}}^{(i)}
\end{equation}
With this contracted approximation to the difference force the full ring polymer force can then be obtained. It is vital to note that despite the evaluation of some of the forces on the contracted ring polymer, these forces are exactly projected back onto the full ring polymer. Thus at any instant the positions and forces on the full imaginary time path (ring polymer) representing each particle are known. In addition, although the contracted forces are an approximation to
the full forces on each replica, the dynamics generated from them formally conserve a well defined Hamiltonian~\cite{mark-mano08jcp,mark-mano08cpl}. This allows allows energy conservation to be checked during dynamics to assess integration accuracy and also enables the combination of RPC with reweighting schemes and methods that require a well defined ensemble that is not possible using some other path integral acceleration schemes.

The computational cost savings obtained from RPC depend crucially on the creation of an efficient reference system that accurately captures the rapidly varying parts of the interactions in the full system and thus leaves a slowly varying difference force. The reference force for an RPC approach should thus satisfy the following requirements:
\begin{enumerate}
\item It is computationally quick to evaluate compared to performing the full electronic structure calculation. 
\item It gives a difference force that is smoothly varying in space.
\end{enumerate}
In addition, sometimes other factors may be important - for example if chemical reactivity is of interest such as when performing ab initio path integral simulations a reference should be chosen that does not make assumptions on chemical bonding, i.e. it allows bond breaking and formation as the simulation progresses. Early applications of RPC employing force fields to describe the interactions employed the intramolecular part of the potential energy as the reference force such that the difference force is the intermolecular potential~\cite{mark-mano08jcp}. Further studies then included effects such as short range Coulomb and Lennard-Jones interactions in the reference leaving the difference term to just incorporate long range electrostatics~\cite{mark-mano08cpl} and it was then extended to polarizable force-fields \cite{Fanourgakis2009}. When one wishes to use a ab initio methods the potential energy is not a simple sum of terms like in the case of force fields. However, in these cases one can define a lower level of electronic structure theory as the reference such as density functional tight binding which has been used as a reference with density functional theory as the full potential~\cite{mars-mark16jcp} and density functional theory used a reference for MP2 as the full potential~\cite{kapi+16jcp}.

Finally, it is worth noting that RPC and the reference potentials it required shares many similarities with multiple time-stepping (MTS) schemes~\cite{stre+78mp,tuck+92jcp,lueh+14jcp}. Whereas MTS schemes exploit the slowly varying nature of forces in real time to take larger time steps, RPC takes advantage of the spatially smooth variation of the forces in the imaginary time path integral. Thus when a good reference force for use with RPC can be found it can also usually be used to integrate the equations of motion with a larger time step using MTS schemes giving further acceleration of the path integral calculation.

\section{Higher order path integral methods}
\label{sec:higher_order_PI}

As discussed in Sec.~\ref{sec:genp_op}, the path integral is often constructed using the symmetric Trotter splitting. However, this splitting leads to a global error which decreases as $\mathcal{O}(P^{-2})$. However, by using more complicated ``higher order'' operator splittings the number of beads needed to reach convergence can be further reduced.\cite{DeRaedt1983,taka-imad84jpsj,suzu95pla,chin97pla,Jang2001,pere-tuck11jcp,kapi+16jcp2} These higher order splittings in general lead to faster in convergence in $P$ at the cost of requiring higher derivatives of the potential energy surface and also often require the derivation of more complicated expressions to extract observables. Although a number of higher order splittings have been obtained two of the most widely used in path simulations have been the Takahashi-Imada (TI)~\cite{taka-imad84jpsj,Li1987} and Suzuki-Chin (SC)~\cite{suzu95pla,chin97pla} splittings.

\begin{exercise}[label={ex:commutators},title={Splittings and their leading order errors}]

By expanding both sides of the Boltzmann operator, $e^{-\beta\hat{H}}$, show that the asymmetric Trotter splitting ($e^{-\beta\hat{H}}=e^{-\beta\hat{V}}e^{-\beta\hat{T}}$) has a leading order error of $\beta^2$ and that the symmetric Trotter splitting ($e^{-\beta\hat{H}}=e^{-\beta\hat{v}/2}e^{-\beta\hat{T}}e^{-\beta\hat{v}/2}$) has leading order error of $\beta^3$.

\end{exercise}

The TI splitting is given by,
\begin{equation}
    e^{-\beta \hat{H}} \approx [e^{-\beta_P \hat{T}} e^{-\beta_P \hat{V}_{c}}]^{P}
    \label{eq:ti_split}
\end{equation}
where 
\begin{equation}
    \hat{V}_{c} = \hat{V} + \frac{\beta_{P}^{2}}{24} [[\hat{V},[\hat{T},\hat{V}]].
\end{equation}
Using this splitting the resulting path integral expression obtained by following the steps used in Sec.~\ref{sec:pimd_sample} leads to a path expression like that in Eq.~\ref{eq:part-func-discrete-withmom} but with an additional term added to the ring polymer Hamiltonian given in Eq.~\ref{eq:rp-ham},
\begin{equation}
   V^\text{TI}_P(\mathbf{q})= \frac{1}{24 P^{2} \omega_{P}^{2}}
   \sum_{j=1}^{P} \sum_{i=1}^{N} \frac{\left(\mathbf{f}_i^{(j)}\right)^{2}}{m_{i}} \label{eq:ti_v}
\end{equation}
where $\mathbf{f}_i^{(j)}=-\partial V\left(\mathbf{q}^{(j)}\right)/\partial\mathbf{q}_i^{(j)} $ is the force acting on the $i$-th atom in the $j$-th replica. The TI approach is a 4th order splitting since the resulting expression has a global error that converges as $\mathcal{O}(P^{-4})$ rather than the $\mathcal{O}(P^{-2})$ obtained using the symmetric Trotter splitting. However, this splitting presents two challenges that are not present in the symmetric Trotter case:
\begin{enumerate}
    \item Evaluating the extra TI potential energy term $V^\text{TI}_P(\mathbf{q})$ requires knowledge of the force arising from the external potential on the ring polymer beads.
    \item Evaluating observables, even simple position dependent ones such as the distance between two particles, requires deriving TI specific estimators.
\end{enumerate}
The first of these two problems presents a particular challenge for PIMD since to evolve the system one must know the derivatives of the potential energy which now includes the $V^\text{TI}_P(\mathbf{q})$ term. Since the $V^\text{TI}_P(\mathbf{q})$ contains the forces arising from the external potential evaluating the forces on the ring polymer beads requires evaluating the next derivative of the external potential energy i.e. the Hessian matrix. Although for empirical (analytic) potentials the Hessian can in principle be computed it is not implemented in most software packages and for ab initio surfaces the only way to obtain it is typically via numerical differentiation (i.e. finite difference approximations) using the forces which is computationally extremely expensive. As long as one has access to the forces the above expression is still however tractable for path integral Monte Carlo simulations where moves are accepted or rejected using just the change in the energy which can be evaluated for the TI splitting with just knowledge of the bead positions, potential energy and forces. Since Monte Carlo moves can be assessed (i.e. accepted or rejected) using the TI Hamiltonian and this only differs from the 2nd order Hamiltonian obtained from using Trotter splitting (Eq.~\ref{eq:rp-ham}) by $V^\text{TI}_P(\mathbf{q})$ this has led to the introduction of hybrid Monte Carlo schemes where the 2nd order Hamiltonian is used to perform PIMD simulations to generate new configurations which can then be accepted or rejected using the 4th order TI Hamiltonian \cite{Jang2001,Suzuki2010}. An alternative approach has been to re-weight the configurations obtained from the 2nd order path integral trajectories using the 4th order term \cite{Jang2001,pere-tuck11jcp,mars+14jctc} although for large systems or for systems where the 2nd order evolution leads to a markedly different exploration in phase space both the hybrid Monte Carlo and re-weighting approaches will lead to slow convergence with respect to the length of the PIMD trajectory used which might partially or fully outweigh the savings in using a smaller value of $P$ since one will have to use much longer trajectories~\cite{ceri+12prsa}. 

The SC splitting, although suffering from the same issues regarding the requirement to evaluate the Hessian matrix if it is to be used with PIMD, provides a much more straightforward route to calculating observables and also possesses 4th order accuracy for all elements (whereas TI only gives this accuracy for the trace of the density). The SC splitting is given by,  
\begin{equation}
    e^{-\beta \hat{H}} \approx  [e^{-\beta_P \frac{\hat{V_e}}{3}} e^{-\beta_P \hat{T}} e^{-\beta_P \frac{4\hat{V_o}}{3}} e^{-\beta_P \hat{T}} e^{-\beta_P \frac{\hat{V_e}}{3}}]^{\frac{P}{2}}
\label{eq:sc_split}
\end{equation}
where
\begin{align}
    & \hat{V}_e = \hat{V} + \frac{\alpha}{6} \beta_P^2 [\hat{V},[\hat{T},\hat{V}]], \\
    & \hat{V}_o = \hat{V} + \frac{(1-\alpha)}{12} \beta_P^2 [\hat{V},[\hat{T},\hat{V}]],
\end{align}
and $\alpha$ is a number between zero and one. Using this splitting, inserting the complete sets of positions states and evaluating the resulting matrix elements yields the Hamiltonian in Eq.~\ref{eq:rp-ham} with the external potential energy replaced by the SC potential energy,
\begin{equation}
   V^\text{SC}_P(\mathbf{q})=
   \sum_{j=1}^{P}
     \left(w_j V\left(\mathbf{q}^{(j)}\right) + \sum_{i=1}^{N}\frac{w_j d_j}{m_i \omega_P^2}\left|\mathbf{f}_i^{(j)}\right|^2\right),
     \label{eq:sc_v}
\end{equation}
where the scaling factors,$w_{j}$, for odd and even beads are given by
\begin{equation}
\begin{alignedat}{2}
& w_j=2/3,\quad d_j={\alpha}/{6} \quad && \text{$j$ is even},  \\
& w_j=4/3, \quad d_j={\left(1-\alpha\right)}/{12} \quad && \text{$j$ is odd}. 
\end{alignedat}
\label{eq:sc-wd}
\end{equation}
Here we note that other works have used the bead indexing starting at $j=0$ which results in the opposite labelling of the weighting factors~\cite{kapi+16jcp2} (i.e. $\text{$j$ is even}$ $\to \text{$j$ is odd}$ and vice versa). Again the forces required to evolve PIMD trajectories using the SC Hamiltonian requires the calculation of the Hessian of the external potential,
\begin{equation}
\begin{split}
    \mathbf{f}_{i}^{\text{sc},(j)}  
    &= w_j (\mathbf{f}_{i}^{(j)} + \frac{2 d_j}{\omega_P^2} \tilde{\mathbf{f}}_{i}^{(j)}) =\\
    &=
    w_j \mathbf{f}_{i}^{(j)} + \frac{2 w_j d_j}{\omega_P^2} \sum_{k=1}^{N} \frac{\partial^2 V(\mathbf{q}^{(j)})}{\partial{\mathbf{q}_i^{(j)}} \partial{\mathbf{q}_k^{(j)}} }\frac{\mathbf{f}_{k}^{(j)}}{m_k}.
\end{split}
\end{equation}
However, the main advantage of the SC splitting is that observables can be computed much more straightforwardly than with the TI approach. In particular a general position dependent observable in the SC splitting is given by
\begin{equation}
A(q) = \frac{2}{P}\sum_{j\in \text{odd}} A(\mathbf{q}^{(j)})
\end{equation}
in which the summation is taken over only the odd numbered beads. This reduces the number of beads in the evaluation of observables by half.

This chapter has outlined techniques used to accelerate the convergence of path integral calculations for thermal properties. The next chapter looks at real-time correlation functions and how to use the knowledge that we have gained in these first few chapters to approximate these.

\begin{exercise}[label={ex:coulomb},title={Hydrogen atom using 4th order splittings}]
Evaluate the 4th order correction to the Takahashi–Imada splitting for the Coulomb potential $1/r$. Plot the resulting potential energy including the 4th order term and show that this term removes the singularity.
\end{exercise}

\newpage
\section*{Answers to exercises}

\begin{Answer}[ref={ex:primitive_ke}]

Let's begin with the definition of the partition function, i.e. Eq.~\ref{eq:part-func-discrete}
\begin{equation}
        Z= \lim_{P \to \infty} \left(\frac{1}{2 \pi \hbar}\right)^P \left(\frac{2 \pi m}{\beta_P}\right)^{P/2} \int dq_1 \dots  dq_{P}  e^{f(\beta)}, \label{eq:TMsol_1}
\end{equation}
where we have defined
\begin{equation}
    f(\beta)=-\beta_P \sum_{j=1}^{P}\left[ \frac{m \omega_P^2}{2} (q_{j+1}-q_{j})^2 + V(q_j) \right],
\end{equation}
for future convenience. To determine the average energy we must differentiate Eq.~\ref{eq:TMsol_1} with respect to $\beta$. Note that there is a $\beta$ dependence in both the prefactor and inside the exponential within the integrand. Performing the derivative yields
\begin{eqnarray}
\frac{\partial Z}{\partial \beta} &=&  \lim_{P \to \infty} \bigg[ \left(\frac{1}{2 \pi \hbar}\right)^P \frac{\partial}{\partial \beta}\left(\frac{2 \pi m}{\beta_P}\right)^{P/2} \int dq_1 \dots  dq_{P}  e^{f(\beta)} \nonumber \\ && + \left(\frac{1}{2 \pi \hbar}\right)^P \left(\frac{2 \pi m}{\beta_P}\right)^{P/2} \int dq_1 \dots  dq_{P}  e^{f(\beta)} \frac{\partial f(\beta)}{\partial \beta}\bigg].
\label{eq:TMsol_2}
\end{eqnarray}
We now need to evaluate the expressions $\frac{\partial}{\partial \beta}\left(\frac{2 \pi m}{\beta_P}\right)^{P/2}$ and $\frac{\partial f(\beta)}{\partial \beta}$. The results are
\begin{equation}
\frac{\partial}{\partial \beta}\left(\frac{2 \pi m}{\beta_P}\right)^{P/2} = -\frac{P}{2\beta}\left(\frac{2 \pi m}{\beta_P}\right)^{P/2}.
\end{equation}
and
\begin{equation}
\frac{\partial f(\beta)}{\partial \beta} = \sum^P_{j=1}\left[ \frac{1}{2}m\omega^2_P(q_{j+1}-q_{j})^2 - \frac{1}{P}V(q_j)\right].
\end{equation}
Inserting these expressions into Eq.~\ref{eq:TMsol_2} and factoring gives
\begin{eqnarray}
\frac{\partial Z}{\partial \beta} &=&  \lim_{P \to \infty} \left(\frac{1}{2 \pi \hbar}\right)^P \left(\frac{2 \pi m}{\beta_P}\right)^{P/2} \int dq_1 \dots  dq_{P} \\ &&  \times e^{-\beta_P \sum_{j=1}^{P} \left[ \frac{m \omega_P^2}{2} (q_{j+1}-q_{j})^2 + V(q_j) \right]} \left( -\frac{P}{2\beta} +\sum^P_{j=1} \left[ \frac{1}{2}m\omega^2_P(q_{j+1}-q_{j})^2 - \frac{1}{P}V(q_j)\right]  \right). \nonumber
\end{eqnarray}
The average energy is then given by
\begin{equation}
-\frac{1}{Z}\frac{\partial Z}{\partial \beta} = \left\langle \frac{P}{2\beta} - \sum^P_{j=1}\left[\frac{1}{2}m\omega^2_P(q_{j+1}-q_{j})^2 + \frac{1}{P}V(q_j)\right] \right\rangle,
\end{equation}
where
\begin{equation}
   \langle \cdots \rangle =\lim_{P \to \infty} \frac{1}{Z} \left(\frac{1}{2 \pi \hbar}\right)^P \left(\frac{2 \pi m}{\beta_P}\right)^{P/2} \int dq_1 \dots  dq_{P} e^{-\beta_P \sum_{j=1}^{P} \left[\frac{m \omega_P^2}{2} (q_{j+1}-q_{j})^2 + V(q_j)\right]}(\cdots). 
\end{equation}

\end{Answer}

\begin{Answer}[ref={ex:commutators}]

We begin by expanding the Boltzmann operator $e^{-\beta\hat{H}}$ as
\begin{equation}
e^{-\beta\hat{H}} = 1- \beta\hat{H} +\frac{1}{2}\beta^2\hat{H}^2 - \frac{1}{3!}\beta^3\hat{H}^3 + \mathcal{O}(\beta^4).
\label{eq:TMsol_3}
\end{equation}
We can then expand the asymmetric Trotter Boltzmann operator in a similar fashion
\begin{eqnarray}
e^{-\beta\hat{V}}e^{-\beta\hat{T}} &=& \left[ 1- \beta\hat{V} +\frac{1}{2}\beta^2\hat{V}^2 + \mathcal{O}(\beta^3) \right]\left[ 1- \beta\hat{T} +\frac{1}{2}\beta^2\hat{T}^2 + \mathcal{O}(\beta^3) \right] \nonumber \\
&=& 1- \beta(\hat{V} + \hat{T}) + \frac{1}{2}\beta^2\hat{T}^2 + \frac{1}{2}\beta^2\hat{V}^2 + \beta^2\hat{V}\hat{T} + \mathcal{O}(\beta^3) \nonumber \\
& = & 1- \beta\hat{H} + \frac{1}{2}\beta^2(\hat{T}^2 + \hat{V}^2 + 2\hat{V}\hat{T}) + \mathcal{O}(\beta^3).
\label{eq:TMsol_4}
\end{eqnarray}
Comparing the $\beta^2$ term in Eq.~\ref{eq:TMsol_4} with that in Eq.~\ref{eq:TMsol_3} we see that
\begin{eqnarray}
\hat{H}^2 &=& \hat{T}^2 + \hat{V}^2 + \hat{V}\hat{T} + \hat{T}\hat{V} \label{eq:TMsol_7} \\  &\neq& \hat{T}^2 + \hat{V}^2 + 2\hat{V}\hat{T}. \label{eq:TMsol_8}
\end{eqnarray}
This inequality is due to the fact that $\hat{T}$ and $\hat{V}$ are operators and do not obey the commutative property as ordinary numbers do. This non-commutation of operators is one of the main distinguishing features between quantum and classical mechanics. To quantify the error of assuming commuting operators in the Trotter expansion we can subtract Eq.~\ref{eq:TMsol_8} from Eq.~\ref{eq:TMsol_7} to see that that the error is given by 
\begin{equation}
\text{Error} = \hat{T}\hat{V}-\hat{V}\hat{T} = [\hat{T},\hat{V}]
\end{equation}
meaning that this splitting is correct to only order $\beta$ and has leading order error of $\beta^2[\hat{T},\hat{V}]$. 

Using the same steps for the symmetric Trotter form the expansion gives
\begin{eqnarray}
e^{-\frac{\beta}{2}\hat{V}}e^{-\beta\hat{T}}e^{-\frac{\beta}{2}\hat{V}} &=& \left[ 1- \frac{\beta}{2}\hat{V} +\frac{1}{4\cdot2}\beta^2\hat{V}^2 + \mathcal{O}(\beta^3) \right]\left[ 1- \beta\hat{T} +\frac{1}{2}\beta^2\hat{T}^2 + \mathcal{O}(\beta^3) \right]\nonumber \\ && \times\left[ 1- \frac{\beta}{2}\hat{V} +\frac{1}{4\cdot2}\beta^2\hat{V}^2 + \mathcal{O}(\beta^3) \right] \nonumber \\
&=& 1- \beta\left(\frac{\hat{V}}{2} + \hat{T} +\frac{\hat{V}}{2} \right)\nonumber \\ && + \frac{1}{2}\beta^2\left(\frac{\hat{V}^2}{2} +  \hat{T}^2 + \frac{\hat{V}^2}{4} + \hat{V}\hat{T} + \frac{\hat{V}^2}{4} +\hat{T}\hat{V}  \right) + \mathcal{O}(\beta^3) \nonumber \\
& = & 1- \beta\hat{H} + \frac{1}{2}\beta^2\hat{H}^2 + \mathcal{O}(\beta^3).
\label{eq:TMsol_5}
\end{eqnarray}
We see that the symmetric Trotter splitting (Eq~\ref{eq:TMsol_5}) fixes the commutation problem at the second order that the asymmetric splitting had. However, it is not correct for any terms beyond this and has a leading order error of $\beta^3 \left( [\hat{T},[\hat{T},\hat{V}]] - [\hat{V},[\hat{V},\hat{T}]]\right) $. To correct for these nested commutators one would need further splittings of the Boltzmann operator.

\end{Answer}

\begin{Answer}[ref={ex:coulomb}]

For simplicity we will consider a single particle of mass $m$ in a Coulomb potential which has the form $V(q)=\frac{1}{q}$. From Eq.~\ref{eq:ti_v} we see that in the Takahashi–Imada splitting there is an extra term in the potential that depends on the force squared. The force for the coulomb potential is $-\frac{1}{q^2}$, thus the corrected potential is given as
\begin{equation}
V^\text{TI}_P(q) = -\frac{1}{q} + \frac{\beta^2\hbar^2}{24mP^4}\frac{1}{q^4}
\end{equation}
Since the second term is always positive it counteracts the singularity caused by the bare Coulomb potential at small $q$ values creating a stable minimum, see Fig.~\ref{fig:TI_Coulomb}. This behavior is much more favorable during simulations as the full Takahashi–Imada potential looks attractive at large distances and repulsive at small ones.

 \begin{figure}[H]
 \includegraphics[width=1.0\linewidth]{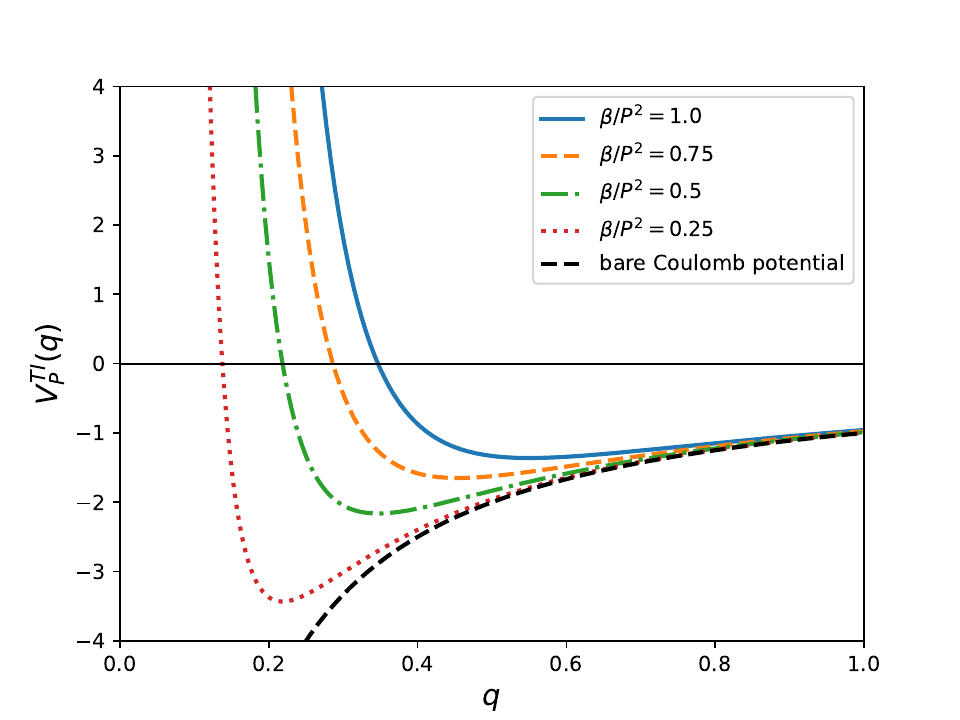}
 \centering
 \caption{Plots of the Takahashi–Imada corrected Coulomb potential for several values of $\beta/P^2$. The bare Coulomb potential is recovered in the classical limit as $\beta\to0$. $m=\hbar=1$ for this figure.}
 \label{fig:TI_Coulomb}
 \end{figure}

\end{Answer}
\setchapterpreamble[u]{\margintoc}

\def\D{\text{d}}

\chapter{Ring polymer molecular dynamics}
\labch{rpmd}

This Chapter provides a brief introduction to the ring polymer molecular dynamics (RPMD) method, which provides a practical way to include nuclear quantum effects in simulations of the dynamical properties condensed phase systems. While path integral molecular dynamics (PIMD) is entirely practical for the calculation of static equilibrium properties, it is simply not feasible to extend it in a rigorous way to the calculation of dynamical properties, because of the infamous sign problem associated with the evaluation of real-time path integrals. RPMD avoids this problem by using the classical trajectories of imaginary time paths to provide a suitable {\em approximation} to real time quantum correlation functions, and hence to the corresponding dynamical observables. For some observables, such as the diffusion coefficients of liquids and the rate coefficients of chemical reactions, the RPMD approximation has proved to be remarkably effective. For others, such as the high frequency infrared and Raman spectra of liquids, it has proved to be rather less so, for reasons we shall endeavour to explain.

\section{Quantum mechanical correlation functions}

Consider a system in the canonical ($NVT$) ensemble with a quantum mechanical Hamiltonian of the form
\begin{equation}
\hat{H}=\sum_{i=1}^N \frac{|\hat{{\bp}}_i|^2}{2m_i}+V(\hat{\bq}),
\label{eq:rpmd1}
\end{equation}
in which $\hat{\bq}\equiv\left\{\hat{\bq}_i\right\}_{i=1}^N$. Many dynamical properties of this system can be related to real time correlation functions of the form
\begin{equation}
c_{AB}(t) = \frac{1}{Z}{\Tr}\left[ e^{-\beta\hat{H}}\hat{A}(0)\hat{B}(t)\right],
\label{eq:rpmd2}
\end{equation}
in which $Z$ is the canonical partition function
\begin{equation}
Z = {\Tr}\left[e^{-\beta\hat{H}}\right]
\label{eq:rpmd3}
\end{equation}
with $\beta=1/\kB T$, and $\hat{O}(t)$ is a Heisenberg-evolved operator at time $t$:
\begin{equation}
\hat{O}(t) = e^{+i\hat{H}t/\hbar}\hat{O}e^{-i\hat{H}t/\hbar}.
\label{eq:rpmd4}
\end{equation}
For example, the diffusion coefficient of a \lq\lq tagged" particle in a liquid can be calculated from its velocity autocorrelation function as 
\begin{equation}
D_i = {1\over 6}\int_{-\infty}^{\infty} c_{{\bv}_i\cdot{\bv}_i}(t)\,\D{t},
\label{eq:rpmd5}
\end{equation}
in which the correlated operators are the velocity operators $\hat{\bv}_i=\hat{\bp}_i/m_i$ of the particle at times 0 and $t$, and the dipole absorption spectrum of a liquid can be calculated from its dipole autocorrelation function as 
\begin{equation}
n(\omega)\alpha(\omega) = {\omega\over 6\hbar c\epsilon_0 V}(1-e^{-\beta\hbar\omega}\,)C_{\boldsymbol{\mu}\cdot\boldsymbol{\mu}}(\omega),
\label{eq:rpmd6}
\end{equation}
where
\begin{equation}
C_{\boldsymbol{\mu}\cdot\boldsymbol{\mu}}(\omega) = \int_{-\infty}^{\infty} e^{-i\omega t}\,c_{\boldsymbol{\mu}\cdot\boldsymbol{\mu}}(t)\,\D{t}.
\label{eq:rpmd7}
\end{equation}
Here $n(\omega)$ is the frequency-dependent refractive index of the liquid, $\alpha(\omega)$ is its Beer-Lambert absorption coefficient, and the correlated operators in $c_{\boldsymbol{\mu}\cdot\boldsymbol{\mu}}(t)$ are the dipole moment operators $\hat{\boldsymbol{\mu}}$ of the liquid at times 0 and $t$.

In these two examples, the experimental observables have been related to standard quantum mechanical correlation functions of the form in Eq.~\eqref{eq:rpmd2}. However, observables can equally well be expressed in terms of other quantum mechanical correlation functions, such as the Kubo-transformed correlation function \cite{KuboBook,zwan+01book}
\begin{equation}
\tilde{c}_{AB}(t) = {1\over \beta Z}\int_0^{\beta} {\Tr}\left[e^{-(\beta-\lambda)\hat{H}}\hat{A}(0)e^{-\lambda\hat{H}}\hat{B}(t)\right]\,\D{\lambda},
\label{eq:rpmd8}
\end{equation}
in which the Boltzmann operator $e^{-\beta\hat{H}}$ has been \lq\lq smeared" around $\hat{A}(0)$. This is not the same as the standard correlation function $c_{AB}(t)$ because $\hat{H}$ does not in general commute with $\hat{A}$. However, it is straightforward to show that the Fourier transforms of $c_{AB}(t)$ and $\tilde{c}_{AB}(t)$,
\begin{equation}
C_{AB}(\omega)=\int_{-\infty}^{\infty} e^{-i\omega t}\,c_{AB}(t)\,\D{t},
\label{eq:rpmd9}
\end{equation}
and
\begin{equation}
\tilde{C}_{AB}(\omega)=\int_{-\infty}^{\infty} e^{-i\omega t}\,\tilde{c}_{AB}(t)\,\D{t},
\label{eq:rpmd10}
\end{equation} 
are related by 
\begin{equation}
C_{AB}(\omega) = D(\omega)\tilde{C}_{AB}(\omega),
\label{eq:rpmd11}
\end{equation}
where 
\begin{equation}
D(\omega) = {\beta\hbar\omega\over (1-e^{-\beta\hbar\omega})}.
\label{eq:rpmd12}
\end{equation}
So standard correlation functions can easily be reconstructed from Kubo-trans\-formed correlation functions, and vice versa. Moreover observables can equally well be written in terms of the Kubo-transformed correlation functions and their frequency spectra. For example, Eq.~\eqref{eq:rpmd5} can be rewritten as
\begin{equation}
D_i = {1\over 6}\int_{-\infty}^{\infty} \tilde{c}_{{\bv}_i\cdot{\bv}_i}(t)\,\D{t},
\label{eq:rpmd13}
\end{equation}
and Eq.~\eqref{eq:rpmd6} becomes
\begin{equation}
n(\omega)\alpha(\omega) = {\beta\omega^2\over 6c\epsilon_0V}\tilde{C}_{\boldsymbol{\mu}\cdot\boldsymbol{\mu}}(\omega),
\label{eq:rpmd14}
\end{equation}
where we have used the fact that $D(0)=1$ (i.e., the fact that the zero frequency components of the standard and Kubo-transformed spectra are the same) to obtain Eq.~\eqref{eq:rpmd13}.

Note in passing that, while the relationship between $n(\omega)\alpha(\omega)$ and $C_{\boldsymbol{\mu}\cdot\boldsymbol{\mu}}(\omega)$ in Eq.~\eqref{eq:rpmd6} involves $\hbar$, that between $n(\omega)\alpha(\omega)$ and $\tilde{C}_{\boldsymbol{\mu}\cdot\boldsymbol{\mu}}(\omega)$ in Eq.~\eqref{eq:rpmd14} does not. This is just one indication that the Kubo-transformed quantum correlation function $\tilde{c}_{\boldsymbol{\mu}\cdot\boldsymbol{\mu}}(t)$ is more of a \lq\lq classical" object than the standard quantum correlation function $c_{\boldsymbol{\mu}\cdot\boldsymbol{\mu}}(t)$. There are many other such indications, including the fact that the Kubo-transformed correlation function has the same symmetry properties as the corresponding classical correlation function \cite{crai-mano04jcp}, and the fact that the Kubo-transformed correlation function plays the same role in quantum mechanical linear response theory as the classical correlation function plays in classical linear response theory \cite{KuboBook,zwan+01book}. Since RPMD is just classical molecular dynamics in an extended phase space \cite{habe+13arpc}, it should therefore come as no surprise that the quantum mechanical correlation function it approximates is $\tilde{c}_{AB}(t)$ rather than $c_{AB}(t)$.

\vfill
\newpage
\begin{exercise}[label={ex:rpmd-1},title={Kubo transform and detailed balance}]
The simplest way to establish the relationship between $C_{AB}(\omega)$ and $\tilde{C}_{AB}(\omega)$ in Eq.~\eqref{eq:rpmd11} is to note that, for a finite volume $V$ (however large), the Hamiltonian in Eq.~(1) has a discrete spectrum. So one can write down expressions for $c_{AB}(t)$, $\tilde{c}_{AB}(t)$, $C_{AB}(\omega)$, and $\tilde{C}_{AB}(\omega)$ in terms of the discrete orthonormal eigenstates $\ket{k}$ and real eigenvalues $E_k$ of $\hat{H}$. For example
\begin{align*}
c_{AB}(t) &= {1\over Z}{\Tr}\left[e^{-\beta\hat{H}}\hat{A}(0)\hat{B}(t)\right]
\nonumber\\
&= {1\over Z}{\Tr}\left[e^{-\beta\hat{H}}\hat{A}\,e^{+i\hat{H}t/\hbar}\hat{B}e^{-i\hat{H}t/\hbar}\right]
\nonumber\\
&= {1\over Z}\sum_{jk} e^{-\beta E_j}\bra{j}\hat{A}\ket{k}e^{iE_kt/\hbar}\bra{k}\hat{B}\ket{j}e^{-iE_jt/\hbar}
\nonumber\\
&={1\over Z}\sum_{jk} e^{-\beta E_j} A_{jk}B_{kj}e^{+i(E_k-E_j)t/\hbar},
\end{align*}
and therefore
\begin{align*}
C_{AB}(\omega) &= \int_{-\infty}^{\infty} e^{-i\omega t} c_{AB}(t)\,\D{t}
\nonumber\\
&= {2\pi\over Z}\sum_{jk} e^{-\beta E_j}A_{jk}B_{kj}\,\delta(\omega-(E_k-E_j)/\hbar).
\end{align*}
\begin{itemize}
\item[(a)] Work out the corresponding expressions for $\tilde{c}_{AB}(t)$ and $\tilde{C}_{AB}(\omega)$, and use the latter of these to verify Eq.~\eqref{eq:rpmd11}.
\item[(b)] Derive the detailed balance conditions
\begin{equation*}
C_{AB}(-\omega) = e^{-\beta\hbar\omega}C_{BA}(\omega)\quad\hbox{and}\quad \tilde{C}_{AB}(-\omega)=\tilde{C}_{BA}(\omega).
\end{equation*}
\end{itemize}
\end{exercise}

\begin{exercise}[label={ex:rpmd-2},title={Symmetry properties of $\tilde{c}_{AB}(t)$}]
Inserting Eq.~\eqref{eq:rpmd4} into Eq.~\eqref{eq:rpmd8} gives
\begin{equation*}
\tilde{c}_{AB}(t) = {1\over \beta Z}\int_0^{\beta} {\Tr}\left[e^{-(\beta-\lambda)\hat{H}}\hat{A}\,e^{-\lambda\hat{H}}e^{+i\hat{H}t/\hbar}\hat{B}\,e^{-i\hat{H}t/\hbar}\right]\,\D{\lambda}.
\end{equation*}
Note that the Boltzmann operators in the trace commute with the evolution operators, and that the trace is invariant to a cyclic permutation of the operators within it.
\begin{itemize}
\item[(a)] Show by changing the order of the operators in the trace, and then changing the integration variable to $\lambda'=\beta-\lambda$, that $\tilde{c}_{AB}(t) = \tilde{c}_{BA}(-t)$.
\item[(b)] Assuming that $\hat{A}$ and $\hat{B}$ are both Hermitian operators, show that $\tilde{c}_{AB}(t)=\tilde{c}_{AB}(t)^*$.
\item[(c)] Assuming that the matrix elements of $\hat{A}$ and $\hat{B}$ in the basis of eigenstates of $\hat{H}$ are real, show that $\tilde{c}_{AB}(t) = \tilde{c}_{AB}(-t)^*$.
\end{itemize}
Combining all three of these results shows that, if assumptions in (b) and (c) are valid, $c_{AB}(t)=c_{BA}(t)$ is a real and even function of $t$.
\end{exercise}

\section{Ring polymer correlation functions}

Without any further ado, let us now introduce the ring polymer correlation functions that approximate the Kubo-transformed quantum mechanical correlation functions in Eq.~\eqref{eq:rpmd8}. This is easiest to do in the special case where the correlated operators $\hat{A}$ and $\hat{B}$ are purely configurational operators, $\hat{A} = A(\hat{\bq})$ and $\hat{B} = B(\hat{\bq})$. In this case, the $P$-bead RPMD approximation to $\tilde{c}_{AB}(t)$ is simply a classical correlation function in an extended phase space \cite{crai-mano04jcp,habe+13arpc}
\begin{equation}
\tilde{c}_{AB}(t) \simeq {1\over (2\pi\hbar)^fZ}\int \D^f{\bp}_0\int \D^f {\bq}_0\, e^{-\beta_PH_P({\bp}_0,{\bq}_0)}A_P({\bq}_0)B_P({\bq}_t),
\label{eq:rpmd15}
\end{equation}
where $\beta_P=\beta/P$, $f=3NP$ is the total number of degrees of freedom, and $H_{P}({\bp},{\bq})$ is the ring polymer Hamiltonian
\begin{equation}
H_{P}({\bp},{\bq}) = \sum_{i=1}^N \sum_{j=1}^P {|{\bp}_{ij}|^2\over 2m_i}+{1\over 2}m_i\omega_P^2|{\bq}_{ij}-{\bq}_{ij+1}|^2+\sum_{j=1}^P V({\bq}_j),
\label{eq:rpmd16}
\end{equation}
with $\omega_P=1/\beta_P\hbar$, ${\bq}_{iP+1}\equiv {\bq}_{i1}$, and ${\bq}_j\equiv\{{\bq}_{ij}\}_{i=1}^N$. The dynamics that takes the initial ring polymer phase space point $({\bp}_0,{\bq}_0)$ to $({\bp}_t,{\bq}_t)$ is the classical dynamics generated by this Hamiltonian,
\begin{equation}
{\D{\bp}_t\over \D{t}} = -{\partial H({\bp}_t,{\bq}_t)\over\partial {\bq}_t}\quad\hbox{and}\quad
{\D{\bq}_t\over \D{t}} = +{\partial H({\bp}_t,{\bq}_t)\over\partial {\bp}_t}, 
\label{eq:rpmd17}
\end{equation}
and the correlated observables $A_P({\bq}_0)$ and $B_P({\bq}_t)$ in Eq.~\eqref{eq:rpmd15} are averages over the beads of the ring polymer necklace at times 0 and $t$:
\begin{equation}
A_P({\bq}) = {1\over P}\sum_{j=1}^P A({\bq}_j)\quad\hbox{and}\quad 
B_P({\bq}) = {1\over P}\sum_{j=1}^P B({\bq}_j).
\label{eq:rpmd18}
\end{equation}

This already explains (for example) how to calculate the RPMD approximation to the dipole autocorrelation function $c_{\boldsymbol{\mu}\cdot\boldsymbol{\mu}}(t)$ in Eq.~\eqref{eq:rpmd7},
\begin{equation}
\tilde{c}_{\boldsymbol{\mu}\cdot\boldsymbol{\mu}}(t) \simeq {1\over (2\pi\hbar)^fZ}\int \D^f{\bp}_0\int \D^f{\bq}_0\, e^{-\beta_PH_P({\bp}_0,{\bq}_0)}\boldsymbol{\mu}_P({\bq}_0)\cdot\boldsymbol{\mu}_P({\bq}_t),
\label{eq:rpmd19}
\end{equation}
in which $\boldsymbol{\mu}_P({\bq})$ is simply the ring polymer average of the classical dipole moment vector at the ring polymer configuration ${\bq}$,
\begin{equation}
\boldsymbol{\mu}_P({\bq}) = {1\over P}\sum_{j=1}^P \boldsymbol{\mu}({\bq}_j).
\label{eq:rpmd20}
\end{equation}
However, Eq.~\eqref{eq:rpmd15} does not explain how to calculate the RPMD approximation to the correlation functions of more general operators, such as the velocity autocorrelation function $c_{{\bv}_i\cdot {\bv}_i}(t)$ in Eq.~\eqref{eq:rpmd5}.

For this, we note that the velocity operator of the tagged particle $i$ is the Heisenberg time derivative of its position operator $\hat{\bq}_i$,
\begin{equation}
\hat{\bv}_i={i\over\hbar}\left[\hat{H},\hat{\bq}_i\right] = {\hat{\bp}_i\over m_i},
\label{eq:rpmd21}
\end{equation} 
and hence that the velocity autocorrelation function $\tilde{c}_{{\bv}_i\cdot{\bv}_i}(t)$ is minus the second time derivative of the position autocorrelation function $\tilde{c}_{{\bq}_i\cdot{\bq}_i}(t)$:
\begin{align}
-{\D^2{}\over \D{t^2}}\tilde{c}_{{\bq}_i\cdot{\bq}_i}(t)
&=-{\D^2{}\over \D{t^2}}{1\over\beta Z}\int_0^{\beta} {\Tr}\left[e^{-(\beta-\lambda)\hat{H}}\hat{\bq}_i\,e^{-\lambda\hat{H}}\cdot e^{+i\hat{H}t/\hbar}\hat{\bq}_i\,e^{-i\hat{H}t/\hbar}\right]\,\D{\lambda}
\nonumber\\
&=-{\D{}\over\D{t}}{1\over\beta Z}\int_0^{\beta} {\Tr}\left[e^{-(\beta-\lambda)\hat{H}}\hat{\bq}_i\,e^{-\lambda\hat{H}}\cdot e^{+i\hat{H}t/\hbar}\hat{\bv}_i\,e^{-i\hat{H}t/\hbar}\right]\,\D{\lambda}
\nonumber\\
&=-{\D{}\over\D{t}}{1\over\beta Z}\int_0^{\beta} {\Tr}\left[e^{-\lambda\hat{H}}\hat{\bv}_i\,e^{-(\beta-\lambda)\hat{H}}\cdot e^{-i\hat{H}t/\hbar}\hat{\bq}_i\,e^{+i\hat{H}t/\hbar}\right]\,\D{\lambda}
\nonumber\\
&={1\over\beta Z}\int_0^{\beta} {\Tr}\left[e^{-\lambda\hat{H}}\hat{\bv}_i\,e^{-(\beta-\lambda)\hat{H}}\cdot e^{-i\hat{H}t/\hbar}\hat{\bv}_i\,e^{+i\hat{H}t/\hbar}\right]\,\D{\lambda}
\nonumber\\
&={1\over\beta Z}\int_0^{\infty} {\Tr}\left[e^{-(\beta-\lambda)\hat{H}}\hat{\bv}_i\,e^{-\lambda\hat{H}}\cdot e^{+i\hat{H}t/\hbar}\hat{\bv}_i\,e^{-i\hat{H}t/\hbar}\right]\,\D{\lambda}
\nonumber\\
&\equiv \tilde{c}_{{\bv}_i\cdot{\bv}_i}(t).
\label{eq:rpmd22}
\end{align}
It follows that we can obtain an RPMD approximation to the velocity autocorrelation function simply by differentiating the RPMD approximation to the position autocorrelation function in the same way:
\begin{align}
-{\D^2{}\over\D{t^2}}\tilde{c}_{{\bq}_i\cdot{\bq}_i}(t)
&=-{\D^2\over\D{t^2}}{1\over (2\pi\hbar)^fZ}\int \D^f{\bp}_0\int \D^f{\bq}_0\,
e^{-\beta_PH_P({\bp}_0,{\bq}_0)} \bar{\bq}_{i,0}\cdot\bar{\bq}_{i,t}
\nonumber\\ 
&=-{\D{}\over\D{t}}{1\over (2\pi\hbar)^fZ}\int \D^f{\bp}_0\int \D^f {\bq}_0\,
e^{-\beta_PH_P({\bp}_0,{\bq}_0)} \bar{\bq}_{i,0}\cdot\bar{\bv}_{i,t}
\nonumber\\
&=-{\D{}\over\D{t}}{1\over (2\pi\hbar)^fZ}\int \D^f{\bp}_t\int \D^f{\bq}_t\,
e^{-\beta_PH_P({\bp}_t,{\bq}_t)} \bar{\bq}_{i,0}\cdot\bar{\bv}_{i,t}
\nonumber\\
&=-{\D{}\over\D{t}}{1\over (2\pi\hbar)^fZ}\int \D^f{\bp}_0\int \D^f{\bq}_0\,
e^{-\beta_PH_P({\bp}_0,{\bq}_0)} \bar{\bq}_{i,-t}\cdot\bar{\bv}_{i,0}
\nonumber\\
&={1\over (2\pi\hbar)^fZ}\int \D^f{\bp}_0\int \D^f{\bq}_0\,
e^{-\beta_PH_P({\bp}_0,{\bq}_0)} \bar{\bv}_{i,-t}\cdot\bar{\bv}_{i,0}
\nonumber\\
&={1\over (2\pi\hbar)^fZ}\int \D^f{\bp}_t\int \D^f{\bq}_t\,
e^{-\beta_PH_P({\bp}_t,{\bq}_t)} \bar{\bv}_{i,0}\cdot\bar{\bv}_{i,t}
\nonumber\\
&={1\over (2\pi\hbar)^fZ}\int \D^f{\bp}_0\int \D^f{\bq}_0\,
e^{-\beta_PH_P({\bp}_0,{\bq}_0)} \bar{\bv}_{i,0}\cdot\bar{\bv}_{i,t}
\nonumber\\
&\equiv \tilde{c}_{{\bv}_i\cdot{\bv}_i}(t),
\label{eq:rpmd23}
\end{align}
where 
\begin{equation}
\bar{\bq}_{i,t}={1\over P}\sum_{j=1}^P {\bq}_{ij,t}\quad\hbox{and}\quad \bar{\bv}_{i,t} ={1\over m_iP}\sum_{j=1}^P {\bp}_{ij,t}
\label{eq:rpmd24}
\end{equation}
are the position and velocity centroids of the tagged particle at time $t$.

In the third and seventh lines of Eq.~\eqref{eq:rpmd23}, we have used Liouville's theorem
\begin{equation}
\D{\bp_0}\D{\bq_0} = \D{\bp_t}\D{\bq_t}
\label{eq:rpmd25}
\end{equation}
and the conservation of the ring-polymer Hamiltonian
\begin{equation}
e^{-\beta_PH_P({\bp}_0,{\bq}_0)} = e^{-\beta_PH_P({\bp}_t,{\bq}_t)}
\label{eq:rpmd26}
\end{equation}
to shift the Boltzmann sampling of the ring polymer phase space variables between times 0 and $t$, 
and in the fourth and sixth lines we have simply shifted the origin of time back and forth between 0 and $-t$ (which is permissible because the correlation function only depends on the difference between the times at which the two correlated observables are evaluated -- it is the same in Lausanne as it is in Oxford, despite the hour's time difference between the two cities). These manipulations play the same role in the present context as the commutativity of the Boltzmann and evolution operators and the invariance of the trace to a cyclic permutation play in the quantum mechanical argument in Eq.~\eqref{eq:rpmd22}. 

The final result in Eq.~\eqref{eq:rpmd23} is clearly very natural: the RPMD approximation to the velocity autocorrelation function of the tagged particle simply involves correlating its centroid (ring-polymer averaged) velocity at time 0 with the same velocity at time $t$. Entirely analogous results are also obtained for other RPMD correlation functions involving the time derivatives of local (configurational) operators. For example, the reactive flux operator that we shall discuss in the next Chapter is the Heisenberg time derivative of a local side operator that projects onto the product side of a dividing surface in configuration space, and the RPMD approximation to the reactive flux autocorrelation function is simply the first time derivative of the corresponding flux-side correlation function. Not all quantum mechanical correlation functions of interest can be treated in this way. For example, it is hard to see how to write the energy flux operators that are correlated in the Green-Kubo formula for thermal conductivity in terms of Heisenberg time derivatives of local operators. However, a number of interesting quantum mechanical correlation functions do involve either purely local operators or their Heisenberg time derivatives, and all of these can in principle be approximated using RPMD.

Exercise~\ref{ex:rpmd-3} shows that the RPMD approximation to a Kubo-transformed correlation function involving local operators becomes exact in the short-time limit. A more detailed analysis that compares the Taylor series expansions of the exact and RPMD correlation functions around $t=0$ reveals that the leading error in the RPMD approximation is $O(t^7)$ when $A(\hat{\bq})$ and $B(\hat{\bq})$ are both linear functions of (some subset of) the coordinate operators, and $O(t^3)$ when they are both non-linear operators (see Ref.~\cite{braa-mano06jcp} and the correction in Ref.~\cite{ross+14jcp}). This accuracy in the short-time limit is central to the accuracy of the RPMD rate theory discussed in the next Chapter, because it ensures that the corresponding transition state theory (which is obtained by taking the limit as $t\to 0_+$ of the RPMD flux-side correlation function) has a direct connection to the semiclassical instanton approximation in the deep quantum tunnelling regime \cite{rich-alth09jcp}. But in other contexts, the precise order of the short-time error in the RPMD approximation is frankly largely irrelevant. For the calculation of more general condensed phase correlation functions, such as $\tilde{c}_{{\bv}_i\cdot{\bv}_i}(t)$ in Eq.~\eqref{eq:rpmd13} and $\tilde{c}_{\boldsymbol{\mu}\cdot\boldsymbol{\mu}}(t)$ in Eq.~\eqref{eq:rpmd14}, what matters far more is how accurate the approximation is at times on the order of the thermal time ($\beta\hbar$) and beyond. This is much harder to establish analytically, but there is one important observation we can make about it. Because RPMD correlation functions are simply classical correlation functions in the extended phase space of the ring polymer, they share all of the desirable features of classical correlation functions. The most important of these is that they are consistent with the underlying equilibrium distribution. Moreover the regression of the spontaneous fluctuations in RPMD is to the quantum mechanical equilibrium distribution rather than the classical equilibrium distribution, by virtue of its connection with PIMD.

Exercise~\ref{ex:rpmd-4} shows that the RPMD correlation function of two local operators has exactly the same symmetry properties as the corresponding quantum mechanical Kubo-transformed correlation function. 
\begin{exercise}[label={ex:rpmd-3},title={Short time limit}]
In the limit as $t\to 0$, the exact Kubo-transformed correlation function in Eq.~\eqref{eq:rpmd8} becomes
\begin{equation*}
\tilde{c}_{AB}(0) = {1\over\beta Z}\int_0^{\beta} {\Tr}\left[ e^{-(\beta-\lambda)\hat{H}}\hat{A}\,e^{-\lambda\hat{H}}\hat{B}\right]\,\D{\lambda}.
\end{equation*}
Develop a $P$-bead imaginary time path integral approximation to this in the case where $\hat{A}=A(\hat{\bq})$ and $\hat{B}=B(\hat{\bq})$ are local operators and show that this coincides with the $t\to 0$ limit of the RPMD correlation function
\begin{equation*}
\tilde{c}_{AB}(0) \simeq {1\over (2\pi\hbar)^fZ}\int \D^f{\bp}\int \D^f{\bq}\, e^{-\beta_PH_P({\bp},{\bq})}A_P({\bq})B_P({\bq}).
\end{equation*}
\end{exercise}

\begin{exercise}[label={ex:rpmd-4},title={Symmetry properties of RPMD}]
When $A(\hat{\bq})$ and $B(\hat{\bq})$ are Hermitian operators (real functions of the Hermitian coordinate operators), it is clear that $A_P({\bq}_0)$ and $B_P({\bq}_t)$ in Eq.~\eqref{eq:rpmd15} will both be real, and hence that the RPMD approximation to $\tilde{c}_{AB}(t)$ will also be real: $\tilde{c}_{AB}(t) = \tilde{c}_{AB}(t)^*$. This is the second of the three symmetry properties of the quantum mechanical Kubo-transformed correlation function that was established in Exercise~1.2.

\begin{itemize}
\item[(a)] Use the manipulations in Eqs.~\eqref{eq:rpmd25} and~\eqref{eq:rpmd26} to show that the RPMD approximation to $\tilde{c}_{AB}(t)$ in Eq.~\eqref{eq:rpmd15} satisfies $\tilde{c}_{AB}(t)=\tilde{c}_{BA}(-t)$.
\item[(b)] See if you can find a separate argument (based on the effect of reversing the initial momentum of a classical trajectory) to show that the correlation function in Eq.~\eqref{eq:rpmd15} also satisfies $\tilde{c}_{AB}(t) = \tilde{c}_{AB}(-t)$.
\end{itemize}
\end{exercise}

\section{What RPMD gets right}

The RPMD approximation to a Kubo-transformed quantum mechanical correlation function is clearly not exact. Among other things, it does not contain any quantum mechanical phase information. This means that it will not be able to capture any subtle quantum mechanical interference effects in the real-time dynamics. However, these effects are often rapidly quenched in condensed phase systems, and rarely play any significant role. The dominant quantum mechanical effects in these systems are zero point energy and tunnelling effects, which the RPMD approximation captures extremely well (as we shall demonstrate in detail in the next Chapter on ring polymer rate theory).

Other things that the RPMD approximation gets right include the high temperature (classical) limit, in which a single ring polymer bead suffices and the RPMD correlation function reduces correctly to a classical correlation function, the short-time limit (as we have already discussed), and the limit of a harmonic interaction potential when the correlated operators $\hat{A}$ and $\hat{B}$ are linear functions of coordinates or momenta \cite{crai-mano04jcp}. It is also faithful to all quantum mechanical symmetries, including time-reversal and detailed balance, as we have shown in Exercises~\ref{ex:rpmd-2} and~\ref{ex:rpmd-4}. And above all, it is consistent with the quantum mechanical equilibrium distribution, by virtue of its connection to PIMD.  Let us now illustrate some of these features with two early applications of RPMD to the dynamics of liquids.

\subsection{Quantum diffusion in liquid para-hydrogen}

Liquid para-hydrogen has become a standard test case for condensed phase quantum dynamics methods for a number of reasons. The de Broglie thermal wave length of a hydrogen molecule at the triple point temperature of 13.8 K is $\lambda=h/\sqrt{2\pi m \kB T}\simeq 3.3$ \AA, which is only just larger than the hard-sphere diameter for the interaction between to hydrogen molecules ($\sigma\simeq 3.0$ \AA). This implies that the exchange of para-hydrogen molecules will not have a significant effect on the properties of the liquid. Furthermore, since the critical point temperature $T_c\simeq 33.1$ K is substantially lower than the rotational temperature of a hydrogen molecule ($\theta_{\mrm{rot}}\simeq 87.6$ K), the vast majority of para-hydrogen molecules will be in their ground rotational state throughout the liquid phase. Since $J=0$ rotational wave functions are spherically symmetric, this implies that the interaction between the molecules can be modelled to a good approximation with an isotropic pair potential, which once again simplifies the calculations.

The first application of RPMD to a condensed phase problem was therefore to liquid para-hydrogen \cite{mill-mano05jcp}. The self-diffusion coefficient of the liquid was calculated by averaging Eq.~\eqref{eq:rpmd13} over all $N$ molecules in the simulation,
\begin{equation}
D = {1\over 6N}\sum_{i=1}^N\int_{-\infty}^{\infty} \tilde{c}_{{\bv}_i\cdot{\bv}_i}(t)\,\D t,
\label{eq:rpmd27}
\end{equation}
with $\tilde{c}_{{\bv}_i\cdot{\bv}_i}(t)$ calculated using Eq.~\eqref{eq:rpmd23}. This was done using liquid densities obtained from earlier PIMD calculations under zero pressure at temperatures of 14 and 25 K, with $P=48$ beads at the lower temperature and $P=24$ at the higher. At each temperature, the calculations were repeated for a range of system sizes, and the established finite size relation \cite{dunw-krem93jcp,yeh-humm04jpcb}
\begin{equation}
D(L) = D(\infty)-\xi{\kB T\over 6\pi\eta L}
\label{eq:rpmd28}
\end{equation}
was used to extrapolate the results to the limit of infinite system size. Here $\eta$ is the shear viscosity of the liquid, $L$ is the side length of the simulation cell, and $\xi=2.837$ is the appropriate geometric coefficient for a cubic simulation box. More details of the simulations are given in the original paper \cite{mill-mano05jcp}.

\begin{figure}[t]
\includegraphics[width=0.7\linewidth]{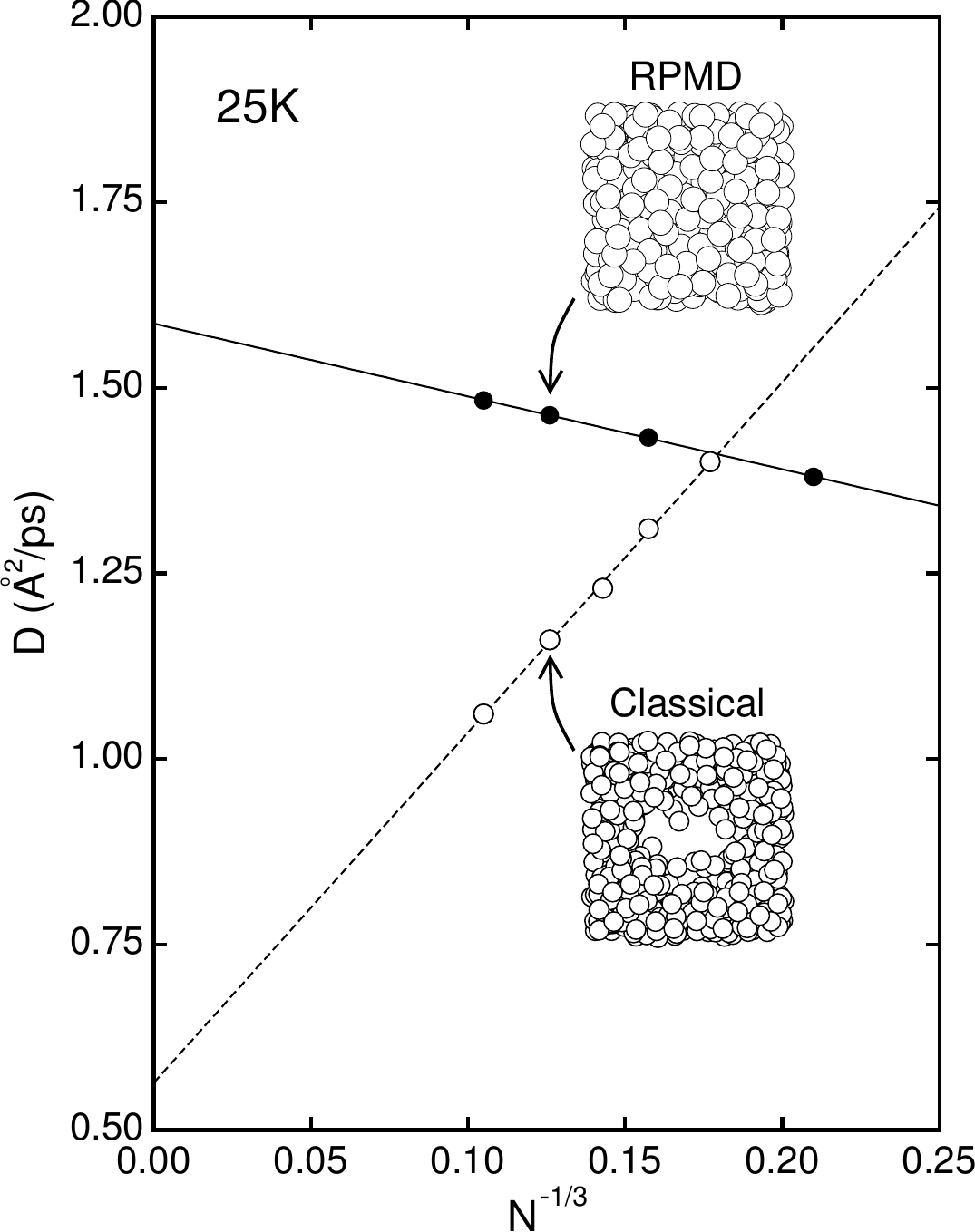}
 \centering
 \caption{System-size scaling of RPMD ($P=24$) and classical ($P=1$) self-diffusion coefficients for liquid para-hydrogen at the $T=25$ K, $V=31.7$ cm$^3$ mol$^{-1}$ state point. The illustrations show typical configurations from the simulations of a system containing $N=500$ molecules~\cite{mill-mano05jcp}. The experimental diffusion coefficient is 1.6 \AA$^2$ ps$^{-1}$, in excellent agreement with the extrapolation of the RPMD simulation to the limit of infinite system size.}
 \label{ph2}
 \end{figure}

The key result of this study is shown in Fig.~\ref{ph2}, which compares the RPMD ($P=24$) and purely classical ($P=1$) extrapolations of the diffusion coefficient at 25 K to the limit of infinite system size ($N^{-1/3}\to 0$). One sees that both the RPMD and the classical results are linear functions of $N^{-1/3}\propto L^{-1}$, although with entirely different slopes and for entirely different reasons. The RPMD diffusion coefficient increases with increasing system size as expected from Eq.~\eqref{eq:rpmd28}, a fit to which gives a shear viscosity of $\eta=1.4\times 10^{-5}$ N s m$^{-2}$ (in reasonable agreement with the experimental shear viscosity of $0.94\times 10^{-5}$ N s m$^{-2}$ at 25 K and saturated vapour pressure). However, the classical diffusion coefficient {\em decreases} with increasing system size and is therefore manifestly inconsistent with Eq.~\eqref{eq:rpmd28}.

The snapshots of the simulations in Fig.~\ref{ph2} explain why the RPMD and classical diffusion coefficients are so different. In these snapshots, the radii of the classical particles have been set equal to the hard-sphere radius for the interaction between two para-hydrogen molecules ($\sigma/2\simeq 1.5$ \AA). The radii of the quantum particles have been \lq\lq swollen" from this by an amount equal to the average radius of gyration of the ring polymer, which was calculated from the estimator
\begin{equation}
\rG^2({\bq}) = {1\over NP}\sum_{i=1}^N\sum_{j=1}^P |{\bq}_{ij}-\bar{\bq}_i|^2,
\label{eq:rpmd29}
\end{equation}
and found to be $\left<\rG({\bq})^2\right>^{1/2}\simeq 0.5$ \AA. As a result of the swelling, the quantum molecules are larger than their classical counterparts, they fill the simulation box, and the snapshot of the RPMD calculation in Fig.~\ref{ph2} looks like a configuration from a liquid simulation. This is to be expected, because the density of the liquid was obtained from a PIMD simulation rather than a classical simulation. Since the classical molecules are smaller, the classical liquid is denser, and in the same size simulation box a bubble opens up in the classical simulation. (In other words, the $NVT$ state point used in the calculation is in the liquid region of the quantum mechanical phase diagram but in the liquid-gas co-existence region of the classical phase diagram.) Since the molecules at the surface of the bubble are less crowded than those in the bulk liquid, they diffuse faster, and since the fraction of these molecules decreases as $N^{-1/3}$ the overall diffusion coefficient obtained from the classical simulation  decreases with increasing system size.

The classical diffusion coefficient extrapolated to the limit of infinite system size (thereby eliminating the effect of the bubble) is 0.60 \AA$^2$ ps$^{-1}$, whereas the RPMD diffusion coefficient is 1.59 \AA$^2$ ps$^{-1}$, in significantly better agreement with the experimental value of 1.6 \AA$^2$ ps$^{-1}$. This is clearly a situation in which the RPMD approximation is physically reasonable: it is consistent with the thermodynamic state point given by a PIMD simulation, and the \lq\lq swelling" of molecules as a result of thermal quantum fluctuations is a well-known physical phenomenon \cite{chan-woly81jcp} that has now been used to explain nuclear quantum effects in a wide range of condensed phase contexts (see, e.g., Refs.~\cite{mark+08jcp} and \cite{mark+11np}). But how do we know that the agreement with experiment in this case is not simply fortuitous (arising, for example, from a cancellation between the errors in the RPMD approximation and the errors in the interaction potential that was used in the simulation)?  

One way to check this is to use the average kinetic energy of the molecules obtained from a PIMD simulation at the same thermodynamic state point to perform an internal consistency check of the RPMD approximation to the velocity autocorrelation function $\tilde{c}_{{\bv}_i\cdot {\bv}_i}(t)$. The average kinetic energy of a tagged molecule $i$ in the liquid can be calculated from a PIMD simulation using the centroid virial estimator \cite{herm-bern82jcp}
\begin{equation}
\TCV({\bq}) = {3\over 2\beta}+{1\over 2P}\sum_{j=1}^{P} ({\bq}_{ij}-\bar{\bq}_i)\cdot
{\partial V({\bq}_{1j},\ldots,{\bq}_{Nj})\over\partial {\bq}_{ij}},
\label{eq:rpmd30}
\end{equation}
which one simply averages during the simulation to obtain $\left<T\right>=\left<\TCV({\bq})\right>$. And as shown in Exercise~1.5, $\left<T\right>$ can also be calculated from the Kubo-transformed velocity autocorrelation function as
\begin{equation}
\left<T\right> = {m_i\over 2}\left[\tilde{c}_{{\bv}_i\cdot{\bv}_i}(0)+\int_0^{\infty}
{2\over (1-e^{+2\pi t/\beta\hbar})}{\D{}\tilde{c}_{{\bv}_i\cdot{\bv}_i}(t)\over\D{t}}\,\D{t}\right].
\label{eq:rpmd31}
\end{equation}
So by inserting the RPMD approximation to $\tilde{c}_{{\bv}_i\cdot{\bv}_i}(t)$ into this equation and comparing the result with $\left<\TCV({\bq})\right>$, we have a way to check the accuracy of the RPMD approximation that is independent of any errors in the interaction potential.

The upshot of this consistency check for liquid para-hydrogen at 25 K is that $\left<\TCV({\bq})\right> = 62.0$ K, whereas the value of $\left<T\right>$ obtained using the RPMD approximation to $\tilde{c}_{{\bv}_i\cdot{\bv}_i}(t)$ is 64.5 K \cite{mill-mano05jcp}. For comparison, the purely classical kinetic energy is $3\kB T/2$, corresponding to a temperature of 37.5 K. So the RPMD approximation is not bad for the velocity autocorrelation function of liquid para-hydrogen: it overestimates the quantum contribution to the kinetic energy at 25 K by less than 10\%. And this is quite a stringent test, because while Eq.~\eqref{eq:rpmd31} only probes the behaviour of $\tilde{c}_{{\bv}_i\cdot{\bv}_i}(t)$ for times on the order of the thermal time, this is $\beta\hbar\simeq 0.3$ ps at 25 K, which is comparable to the decay time of the velocity autocorrelation function \cite{braa+06cpl}. So it provides some reason for confidence in the RPMD approximation to diffusion coefficients in other (less quantum mechanical) contexts, such as the one we shall consider next.

\begin{exercise}[label={ex:rpmd-5},title={Average kinetic energy from $\tilde{c}_{\bv_i\cdot\bv_i}(t)$}]
The average kinetic energy of a tagged molecule $i$ in a liquid is given quantum mechanically by
\begin{equation*}
\left<T\right> = {1\over Z}{\Tr}\left[e^{-\beta\hat{H}}\hat{T}_i\right] = {m_i\over 2Z}{\Tr}\left[e^{-\beta\hat{H}}\hat{\bv}_i\cdot\hat{\bv}_i\right] \equiv {m_i\over 2}c_{{\bv}_i\cdot{\bv}_i}(0),
\end{equation*}
where $c_{{\bv}_i\cdot{\bv}_i}(t)$ is the standard quantum mechanical velocity autocorrelation function.

\begin{itemize}
\item[(a)] Use Eq.~\eqref{eq:rpmd11} to show that this can be re-written in terms of the Kubo-transformed velocity autocorrelation function $\tilde{c}_{{\bv}_i\cdot{\bv}_i}(t)$ as
\begin{equation*}
\left<T\right> = {m_i\over 4\pi} \int_{-\infty}^{\infty}\D{\omega} \int_{-\infty}^{\infty}\D{t}\,
{\beta\hbar\omega\over (1-e^{-\beta\hbar\omega})}e^{-i\omega t}\,\tilde{c}_{{\bv}_i\cdot{\bv}_i}(t).
\end{equation*}
\item[(b)] {\bfseries Harder:} Derive Eq.~\eqref{eq:rpmd31} by evaluating the integral over $\omega$ in this expression.
\end{itemize}
\end{exercise}

\subsection{Competing quantum effects in liquid water}

Soon after the study of liquid para-hydrogen discussed above, we published the first RPMD study of the dynamics of liquid water \cite{mill-mano05jcp2}. In common with previous studies that had used the centroid molecular dynamics (CMD) method, this study found rather large quantum mechanical effects on both the self-diffusion coefficient of the molecules in the liquid and their orientational relaxation times.  A number of subsequent studies seemed to confirm this. However, all of these studies used either rigid-body water models or models with harmonic OH stretching potentials. Moreover these potentials were all obtained by fitting classical molecular dynamics simulations to experimental measurements, which effectively includes nuclear quantum effects in an approximate way within the potential. The use of such potentials in path integral simulations leads to a \lq\lq double counting" of nuclear quantum effects, which is clearly undesirable.

In order to remedy these deficiencies, we have since developed a new interaction potential for water, the q-TIP4P/F model \cite{habe+09jcp}. This is a flexible water model with an anharmonic OH stretching potential that is specifically designed for use in path integral (PIMD, RPMD, and CMD) simulations. Its parameters were obtained by using these simulations to fit a wide variety of structural, thermodynamic, and dynamic properties of the model to experimental measurements (including the OO, OH, and HH radial distribution functions, the density isotherm at atmospheric pressure, the melting point of ice, the dielectric constant of water, and the self diffusion coefficients, orientational relaxation times, and infrared absorption spectra of both light and heavy water).  Somewhat annoyingly, having done all of this, we found that the nuclear quantum effects in the room temperature liquid are not very large after all!

 \begin{figure}[t]
 \includegraphics[width=0.7\linewidth]{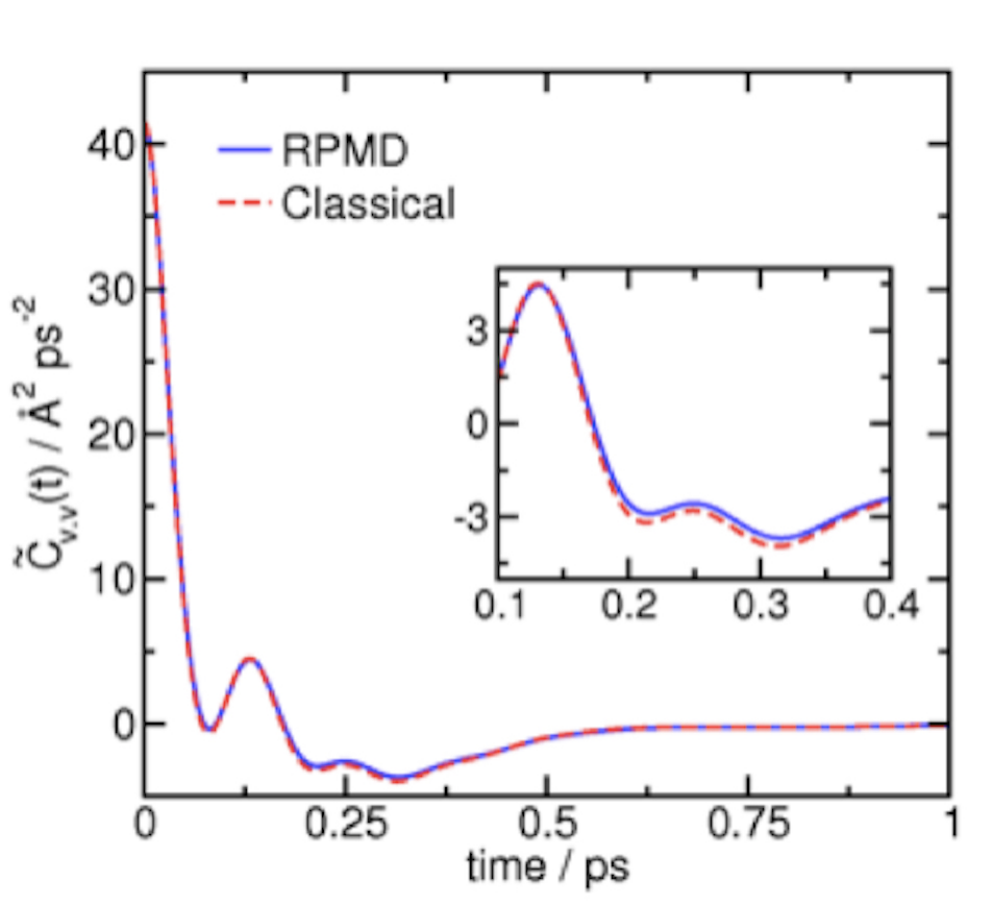}
 \centering
 \caption{RPMD and classical velocity autocorrelation functions of the q-TIP4P/F water model under ambient conditions. The two curves are clearly very similar. Their integrals give $D_\mathrm{qm}/D_\mathrm{ cl}\simeq 1.1$ -- a mere 10\% quantum enhancement in the diffusion coefficient.}
 \label{qTIP4PF}
 \end{figure}

To give just one example, Fig.~\ref{qTIP4PF} compares the RPMD $(P=32)$ and purely classical $(P=1)$ molecular centre-of-mass velocity autocorrelation functions of the q-TIP4P/F water model at room temperature. The two curves are almost identical, and the quantum enhancement in the diffusion coefficient is a mere 10\%. This is significantly smaller than the quantum enhancements that had been found for rigid-body water models in previous RPMD and CMD simulations, which had ranged from 40\% to over 50\%. It is also smaller than the 25\% quantum enhancement that was found in RPMD simulations of a flexible water model with a purely harmonic OH stretch \cite{habe+09jcp}. But {\em why} was the quantum enhancement of the diffusion coefficient for q-TIP4P/F water found to be so small?

 \begin{figure}[htb]
 \includegraphics[width=0.7\linewidth]{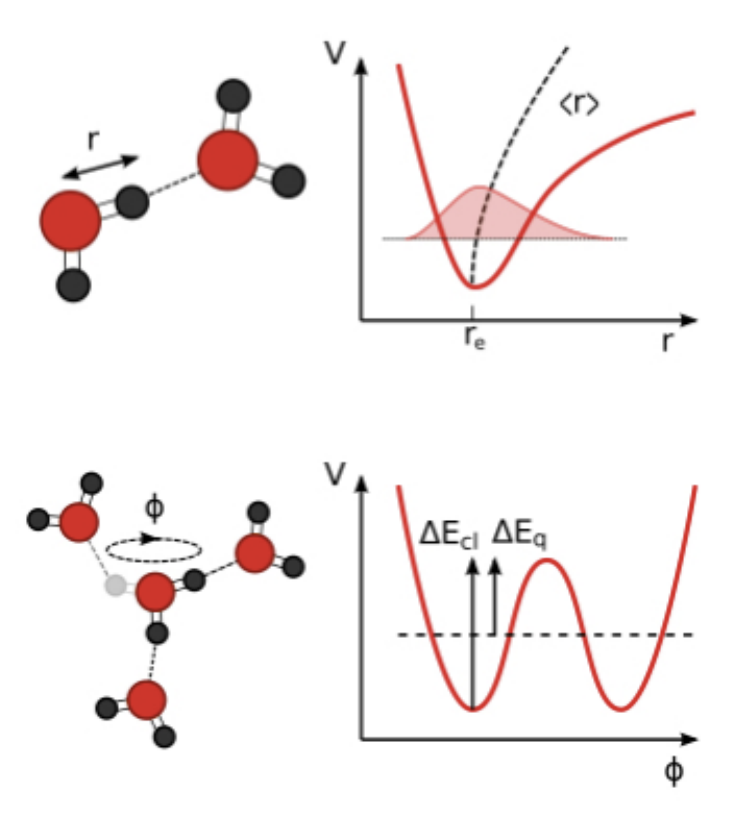}
 \centering
 \caption{Competing quantum effects in liquid water. The upper panel illustrates the effect of zero point energy on the average bond length of an anharmonic OH bond. This stretches the bond, gives the water molecule a larger dipole moment, strengthens the intermolecular interactions between water molecules, and slows down their diffusion in the liquid. The lower panel illustrates the effect of zero point energy on the librational (hydrogen bond breaking and making) modes of the liquid. This decreases the activation energy for hydrogen bond rearrangements and speeds up the diffusion. The two effects are in opposite directions, resulting in a very small {\em net} nuclear quantum effect on the liquid dynamics.}
 \label{water}
 \end{figure}

The analysis of the calculations performed in Ref.~\cite{habe+09jcp} answered this question in terms of {\em competing quantum effects}. First it was noted that the ensemble-averaged OH bond length was larger in the RPMD simulation than in the classical simulation. As is illustrated in the top panel of Fig.~\ref{water},  this is to be expected from the effect of zero point motion in an anharmonic bond. The increase OH bond length leads to a larger molecular dipole moment, which strengthens the intermolecular interactions in the liquid and slows down the diffusion.\footnote{This is certainly true for a water model like q-TIP4P/F, in which the electrostatic interactions between the atoms are modelled using simple point charges. And it also expected to be true in {\em ab initio} water, because the ground electronic state of a water molecule has a significant ionic contribution in the vicinity of its equilibrium geometry.} On the other hand, the zero point energy in the librational (hydrogen bond making and breaking) modes of the liquid leads to a lower activation energy for hydrogen bond rearrangements, as illustrated in the bottom panel of Fig.~\ref{water}. This intermolecular effect speeds up the diffusion and counteracts the intramolecular effect, resulting in a very small {\em net} quantum effect on the self-diffusion coefficient.

In order to verify this picture of competing quantum effects, we turned off the intramolecular quantum effect by removing the molecular flexibility of the q-TIP4P/F model and confining each water molecule to its equilibrium geometry \cite{habe+09jcp}. This was found to give a quantum/classical self-diffusion coefficient ratio of $D_\mathrm{qm}/D_\mathrm{cl}=1.43$, in agreement with those obtained in earlier RPMD and CMD studies of rigid body water models. When there is no competition from the intramolecular effect, the intermolecular effect (the disruption of the hydrogen bonding network due to zero point energy in the librational modes of the liquid) leads to quite a significant quantum enhancement of the diffusion coefficient. But when the competition is present the net enhancement is very small. Note also that the anharmonicity of the OH bond is an essential ingredient in the intramolecular effect, as illustrated in Fig.~\ref{water}. The average length of a harmonic OH bond does {\em not} increase when zero point motion is included -- it is the same in the vibrational ground state as it is at the equilibrium OH bond length. This explains why the quantum enhancement in the diffusion coefficient is also significant for flexible water models with harmonic OH stretching potentials, which like rigid-body models are blind to competing quantum effects.

The notion of competing quantum effects is now very well established. There have since been many more examples of how they influence the structural, thermodynamic, and dynamic properties of molecular liquids and solids. Examples from our group include studies of a surface-specific isotope effect in mixtures of light and heavy water \cite{liu+13jpcc}, nuclear quantum effects in water exchange around aqueous lithium and fluoride ions \cite{wilk+15jcp}, nuclear quantum effects in H$^+$ and OH$^-$ diffusion along confined water wires \cite{ross+16jpcl}, and nuclear quantum effects in the extended jump model of water reorientation and hydrogen-bond dynamics \cite{wilk+17jpcl}. There have since also been many other studies by other groups that have confirmed the notion of competing quantum effects. Since this notion was first highlighted in the above RPMD study of a flexible water model \cite{habe+09jcp}, it is definitely something that RPMD \lq\lq gets right".

One final remark is that it does require a method that is consistent with the quantum mechanical equilibrium distribution (i.e., with PIMD) to correctly identify quantum mechanical effects in the dynamics of liquids. A hilarious counter-example is that when the classical Wigner model (or linearised semiclassical initial value representation, LSC-IVR) is used to calculate the self-diffusion coefficient of the q-TIP4P/F water model it finds a quantum enhancement factor of $D_\mathrm{qm}/D_\mathrm{cl}=3$ at room temperature \cite{habe-mano09jcp}. This is completely bogus, and arises because the classical dynamics in the LSC-IVR does not conserve the Wigner transform of the Boltzmann operator. The initially-quantised phase space distribution in this method contains around $10\,\kB T$ of zero point energy in each intramolecular OH stretching mode, which is not conserved during the subsequent dynamics. As a result, this zero point energy \lq\lq leaks out" into the intermolecular modes of the liquid over the $\sim 1$ ps timescale of the velocity autocorrelation, massively (and entirely spuriously) speeding up the diffusion \cite{habe-mano09jcp}. The zero point energy leakage does not happen in RPMD because its dynamics is consistent with the (PIMD) equilibrium distribution, which is undoubtedly the single most important thing that the method gets right.

\section{What RPMD gets wrong}

The examples given above show that RPMD provides an entirely reasonable way to include nuclear quantum effects in the diffusion coefficients of liquids. It is equally reasonable for the orientational correlation times of molecular liquids \cite{habe+09jcp}, and as we shall see in the next Chapter it provides an absolutely superb way to estimate the quantum mechanical rate coefficients of chemical reactions in complex systems. The fundamental reason for this is that all of these properties can be related to the {\em zero frequency} components of various (Kubo-transformed) time correlation spectra,
$\tilde{C}(0) = \int_{-\infty}^{\infty} \tilde{c}(t)\,\D{t}$. These zero frequency components are largely unaffected by the fictitious high frequency dynamics of the internal modes of the ring polymer, which are simply there to ensure consistency with the quantum mechanical Boltzmann distribution and have nothing whatsoever to do with real-time quantum dynamics. However, higher frequency observables, such as the components of dipole absorption spectra $\tilde{C}_{\boldsymbol{\mu}\cdot\boldsymbol{\mu}}(\omega)$ with $\beta\hbar\omega\gg 1$, may well be affected by the fictitious dynamics of the ring polymer internal modes, and so these provide an obvious place to look for a breakdown of RPMD.

Another cause for concern is that, even for correlation functions involving local operators, RPMD is only exact in the limit of a harmonic potential when one or other of the correlated operators is a linear function of $\hat{\bq}$ \cite{crai-mano04jcp}. This suggests that the approximation will generally be worse for correlation functions involving non-linear operators than for those involving linear operators. A systematic way to investigate this is to consider the RPMD approximations to coherent and incoherent dynamic structure factors, the latter of which (for example) is the time-to-frequency Fourier transform of a correlation function of the form
\begin{equation}
F_{\mathrm{s}}(k,t) = {1\over ZN}\sum_{i=1}^N {\Tr}\left[e^{-\beta\hat{H}}e^{-i{\bk}\cdot\hat{\bq}_i(0)}
e^{+i{\bk}\cdot\hat{\bq}_i(t)}\right].
\label{eq:rpmd32}
\end{equation}  
For small $k\equiv |{\bk}|$, the correlated operators in Eq.~(1.40) are approximately linear functions of $\hat{\bq}$, and one might expect the RPMD approximation to work well. But for larger $k$, they become increasingly non-linear, and one might expect the approximation to break down. These expectations were confirmed early on in the RPMD literature, as we shall describe below. But first, let us demonstrate that the RPMD approximation does indeed break down in the high frequency region for the dipole absorption spectrum of liquid water, and discuss what can be done to mitigate this.

\subsection{The vibrational spectrum of liquid water}

\begin{figure}[t]
 \includegraphics[width=0.7\linewidth]{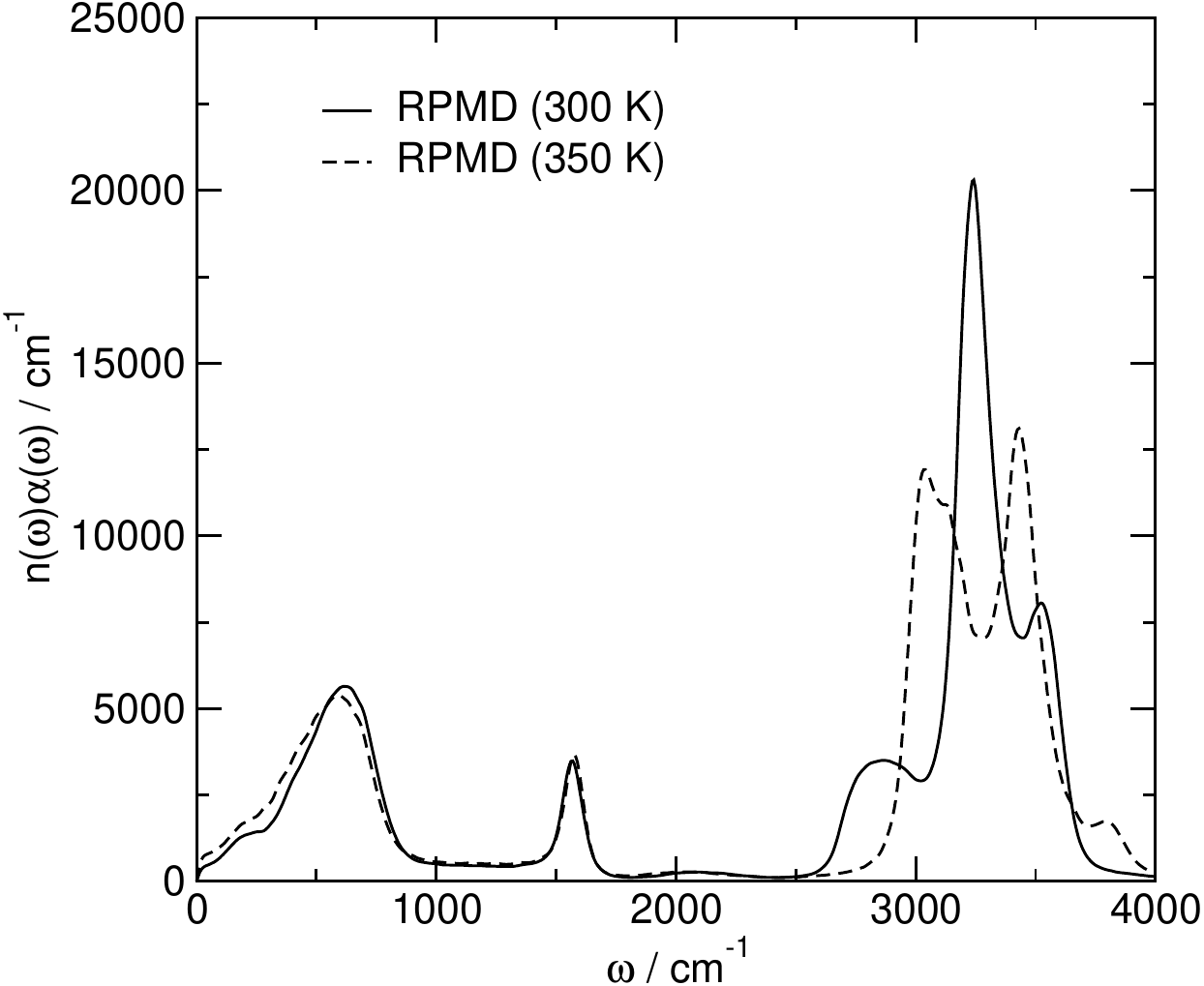}
 \centering
 \caption{RPMD dipole absorption spectra of the TTM3-F model of liquid water at 300 and 350 K \cite{habe+08jcp}. The low frequency librational and intramolecular bending bands are physically reasonable, but the high frequency \ce{OH} stretching band is contaminated by \lq\lq spurious resonances" associated with the internal modes of the ring polymer.}
 \label{TTM3F}
 \end{figure}

The RPMD approximation to the dipole absorption spectrum of liquid water, $n(\omega)\alpha(\omega)$ in Eq.~(1.14), is shown at two different temperatures in Fig.~\ref{TTM3F}. These calculations used a flexible and polarizable Thole-type potential energy model (TTM3-F) that was specifically developed for spectroscopic simulations. The spectra exhibit three main bands: a broad librational band below 1000 cm$^{-1}$, a narrow intramolecular HOH bending band at around 1600 cm$^{-1}$, and a broader OH stretching band at around 3300 cm$^{-1}$. The experimental spectra are almost identical to these simulated spectra in the librational and bending regions, but the experimental stretching band consists of a single broad (inhomogeneously broadened) peak in which the symmetric and antisymmetric OH stretching modes are not resolved.  There is therefore clearly something wrong with the RPMD simulation in the OH stretching region, and this is highlighted by its temperature dependence. Increasing the temperature from 300 to 350 K causes a slight red shift to the librational band as a result of anharmonicity, and has no discernible effect on the intramolecular bending band. However, it wildly changes the structure of the resonances that are seen in the simulated OH stretching region, which is clearly physically unreasonable. 

The only possible explanation for this breakdown of the RPMD approximation is that the spurious resonances in the OH stretching region are due to interactions with the internal modes of the ring polymer, which borrow intensity from the physical OH stretching modes and contaminate the simulated spectrum. The lowest internal modes of a {\em free} ring polymer have frequencies of 
\begin{equation}
\omega_k = {2P\over\beta\hbar}\sin\left({k\pi\over P}\right) \simeq {2\pi k\over \beta\hbar}
\label{eq:rpmd33}
\end{equation}
for $k=1,2,\ldots$, which occur at integer multiples of $\sim 1300$ cm$^{-1}$ at 300 K and $\sim 1517$ cm$^{-1}$ at 350 K. In combination with a physical mode with a frequency of $\omega$, the internal mode frequencies are shifted to $\sqrt{\omega^2+\omega_k^2}$, and it is relatively easy to find such combinations with frequencies in the OH stretching region. For example, the interaction between the second internal mode of the ring polymer at 300 K and the water bending mode gives a combined oscillation with a wavenumber of $\sqrt{2600^2+1600^2}=3052$ cm$^{-1}$, and the same calculation at 350 K gives an oscillation with a wavenumber of $\sqrt{3033^2+1600^2}=3429$ cm$^{-1}$.

This explanation for the spurious resonances was confirmed in Ref.~\cite{habe+08jcp} by repeating the calculation with the partially-adiabatic (PA) CMD method, in which the masses associated with the internal modes of the ring polymer are reduced so as to shift their frequencies beyond the spectral range of interest. This was found to remove the resonances and leave a single broad peak in the OH stretching region, in much better agreement with experiment. Moreover this broad peak was found to blue shift slightly on increasing the temperature from 300 to 350 K. This is again consistent with experiment, and is to be expected from the greater disruption of the hydrogen bonding network and the concomitant strengthening of the intramolecular OH bonds at higher temperatures \cite{habe+08jcp}. 

Unfortunately, while it does remove the spurious resonances, replacing RPMD with (PA)-CMD does not completely solve the problem of computing condensed phase vibrational spectra, because CMD suffers from a \lq\lq curvature problem" when it is applied to systems with both bond stretching and librational (or rotational) modes \cite{witt+09jcp,ivan+10jcp}. As the temperature is lowered, the ring polymer spreads out around the librational modes, and the centroid potential of mean force that is used in CMD becomes softer in the stretching modes. This results in a spurious red-shift in the computed stretching frequencies. The red shift is not especially pronounced in the OH stretching band of liquid water at room temperature, but it is a serious issue in ice at lower temperatures. 

\subsection{An aside on thermostatted RPMD}

An alternative which both mitigates the resonance problem of RPMD and avoids the curvature problem of CMD is simply to apply a Langevin thermostat to the internal modes of the ring polymer during the dynamics, {\em without} adjusting the masses associated with these modes so as to shift their frequencies. The resulting thermostatted (T)-RPMD method is identical to RPMD except for the equation of motion for $\D{\bp}_t/\D{t}$ in Eq.~\eqref{eq:rpmd17}, which becomes \cite{ross+14jcp}
\begin{equation}
{\D{\bp}_t\over\D{t}} =-{\partial H({\bp}_t,{\bq}_t)\over\partial {\bq}_t}-\boldsymbol{\gamma}{\bp}_t+\sqrt{2\boldsymbol{m}\boldsymbol{\gamma}\over \beta_P}\boldsymbol{\xi}(t)
\label{eq:rpmd34}
\end{equation}
where $\boldsymbol{m}$ is a diagonal mass matrix, $\boldsymbol{\gamma}$ is a real, symmetric, positive semi-definite friction matrix that is diagonal in the atomic index $i$, and $\boldsymbol{\xi}(t)$ is a vector of uncorrelated normal deviates with unit variance and zero mean ($\left<\xi_{ij}(t)\right>=0$ and $\left<\xi_{ij}(0)\xi_{i'j'}(t)\right>=\delta_{ii'}\delta_{jj'}\delta(t)$ for $i=1,\ldots,N$ and $j=1,\ldots,P$). Provided $\boldsymbol{\gamma}$ only acts on the internal modes of the ring polymer and has no effect on the centroid, one can show that {\em all} of the established properties of RPMD (its symmetry properties, its behaviour in the short time limit, and all of the situations in which it becomes exact) are unaltered by replacing Eq.~\eqref{eq:rpmd17} with Eq.~\eqref{eq:rpmd34} \cite{ross+14jcp}. 

\begin{figure}[t]
\includegraphics[width=0.9\linewidth]{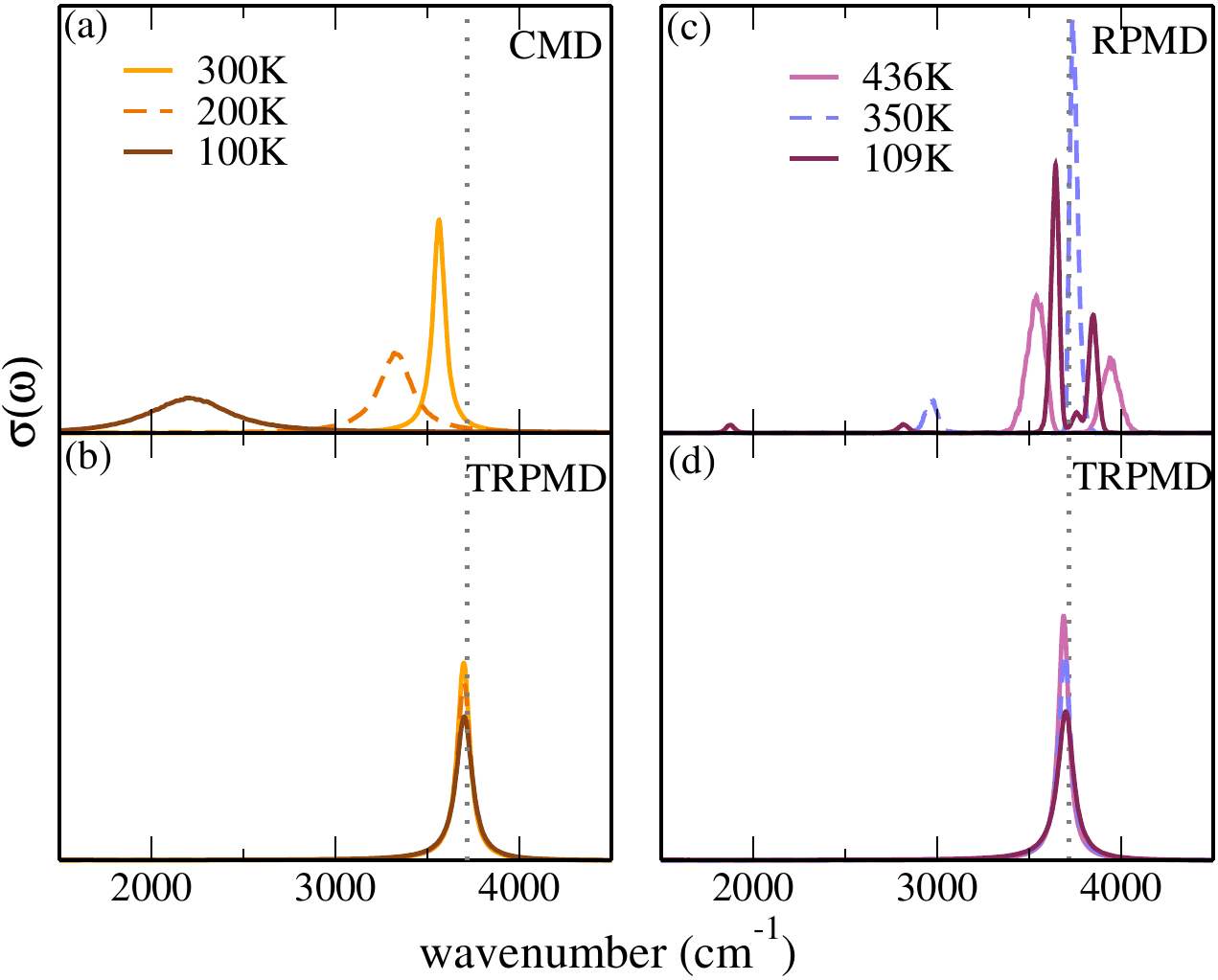}
 \centering
 \caption{Dipole absorption cross sections for a harmonic OH model. Panels (a) and (b) compare the CMD and TRPMD methods at 100, 200 and 300 K, while panels (c) and (d) compare the RPMD and TRPMD methods at 109, 350, and 436 K. The dotted grey line indicates the correct harmonic vibrational frequency of the model, $\omega_{\ce{ OH}}=3716$ cm$^{-1}$.}
 \label{TRPMD}
\end{figure}

In practice, it is convenient to choose $\boldsymbol{\gamma}$ to be diagonal in the ring polymer normal mode representation, and to choose its diagonal elements in this representation so as to give optimum canonical sampling of the free ring polymer configuration space \cite{ross+14jcp}. This gives $\gamma_{ik,ik}=\omega_k$ where $\omega_k$ is the internal mode frequency in Eq.~\eqref{eq:rpmd33}. Note that, since $\omega_0=0$, this prescription automatically detaches the thermostat from the centroid. A simple example of how well this works is provided in Fig.~\ref{TRPMD}, which compares the CMD, RPMD, and TRPMD dipole absorption spectra of a harmonic OH molecule that is free to rotate in 3D space so as to bring out the curvature problem of CMD \cite{ross+14jcp}. As a result of this curvature problem, the CMD spectrum exhibits an uncontrolled red shift in the OH stretching region that becomes particularly pronounced at lower temperatures. The RPMD spectrum exhibits spurious resonances associated with the internal modes of the ring polymer, which are especially clear at the three temperatures we have chosen to show in the figure. The TRPMD spectrum is immune to both of these problems, and consists of just a single peak close to the correct vibrational frequency of the model (3716 cm$^{-1}$). This is clearly a big improvement over both CMD and RPMD, and it strongly suggests that TRPMD is the best of the three methods for calculating vibrational spectra. However, it is still not perfect. The thermostat in TRPMD damps the oscillations of the internal modes of the ring polymer but does not remove them completely, and this results in a broadening of the calculated OH stretching peak that becomes more pronounced at lower temperatures. We have therefore decided to include this discussion of TRPMD in this section on what RPMD \lq\lq gets wrong". Given the flexibility in the choice of $\boldsymbol{\gamma}$ in Eq.~\eqref{eq:rpmd34}, it is possible that one could do better than the simple prescription $\gamma_{ik,ik}=\omega_k$, and there has been some more recent work on this (including work on the use of a generalized Langevin equation \cite{ross+18jcp}). But it is probably fair to say that the problem of computing accurate vibrational spectra with TRPMD has still not entirely been solved.

\subsection{Incoherent dynamic structure factors} 
\label{NLOP}

Finally, let us return to the \lq\lq non-linear operator problem" and use the self part of the intermediate scattering function in Eq.~\eqref{eq:rpmd32} to illustrate how the RPMD approximation breaks down as the correlated operators $\hat{A}$ and $\hat{B}$ become increasingly non-linear functions of the coordinate operator $\hat{\bq}$.

The incoherent dynamic structure factor $S_{\mathrm{inc}}(k,\omega)$ is the time-to-frequency Fourier transform of the intermediate scattering function,
\begin{equation}
S_{\mathrm{inc}}(k,\omega) = {1\over 2\pi}\int_{-\infty}^{\infty}\,e^{-i\omega t}F_{\mathrm{s}}(k,t)\,\D{t}.
\label{eq:rpmd35}
\end{equation}
In view of Eqs.~\eqref{eq:rpmd11} and~\eqref{eq:rpmd12}, this can be calculated equivalently as
\begin{equation}
S_{\mathrm{inc}}(k,\omega) = {\beta\hbar\omega\over (1-e^{-\beta\hbar\omega})}\tilde{S}_{\mathrm{inc}}(k,\omega),
\label{eq:rpmd36}
\end{equation}
where  $\tilde{S}_{\mathrm{inc}}(k,\omega)$ is the incoherent relaxation spectrum
\begin{equation}
\tilde{S}_{\mathrm{inc}}(k,\omega) = {1\over 2\pi}\int_{-\infty}^{\infty}\,e^{-i\omega t}\tilde{F}_{\mathrm{s}}(k,t)\,\D{t},
\label{eq:rpmd37}
\end{equation}
and the self relaxation function $\tilde{F}_{\mathrm{s}}(k,t)$ is the Kubo-transformed version of $F_{\mathrm{s}}(k,t)$,
\begin{equation}
\tilde{F}_{\mathrm{s}}(k,t) = {1\over \beta ZN}\sum_{i=1}^N\int_0^{\beta}\, {\Tr}\left[e^{-(\beta-\lambda)\hat{H}}e^{-i{\bk}\cdot\hat{\bq}_i(0)}e^{-\lambda\hat{H}}e^{+i{\bk}\cdot\hat{\bq}_i(t)}\right]\,\D{\lambda}.
\label{eq:rpmd38}
\end{equation}  
Since this is just a Kubo-transformed correlation function involving local operators, it can be approximated in RPMD using Eq.~\eqref{eq:rpmd15}. We shall call this direct RPMD approximation to $\tilde{F}_{\mathrm{s}}(k,t)$ \lq\lq RPMD-F" in what follows. (Note that, in an isotropic liquid, the result is independent of the direction of the wave vector ${\bk}$; $\tilde{F}_{\mathrm{s}}(k,t)$ just depends on $k=|{\bk}|$.)

 \begin{figure}[t]
 \includegraphics[width=0.8\linewidth]{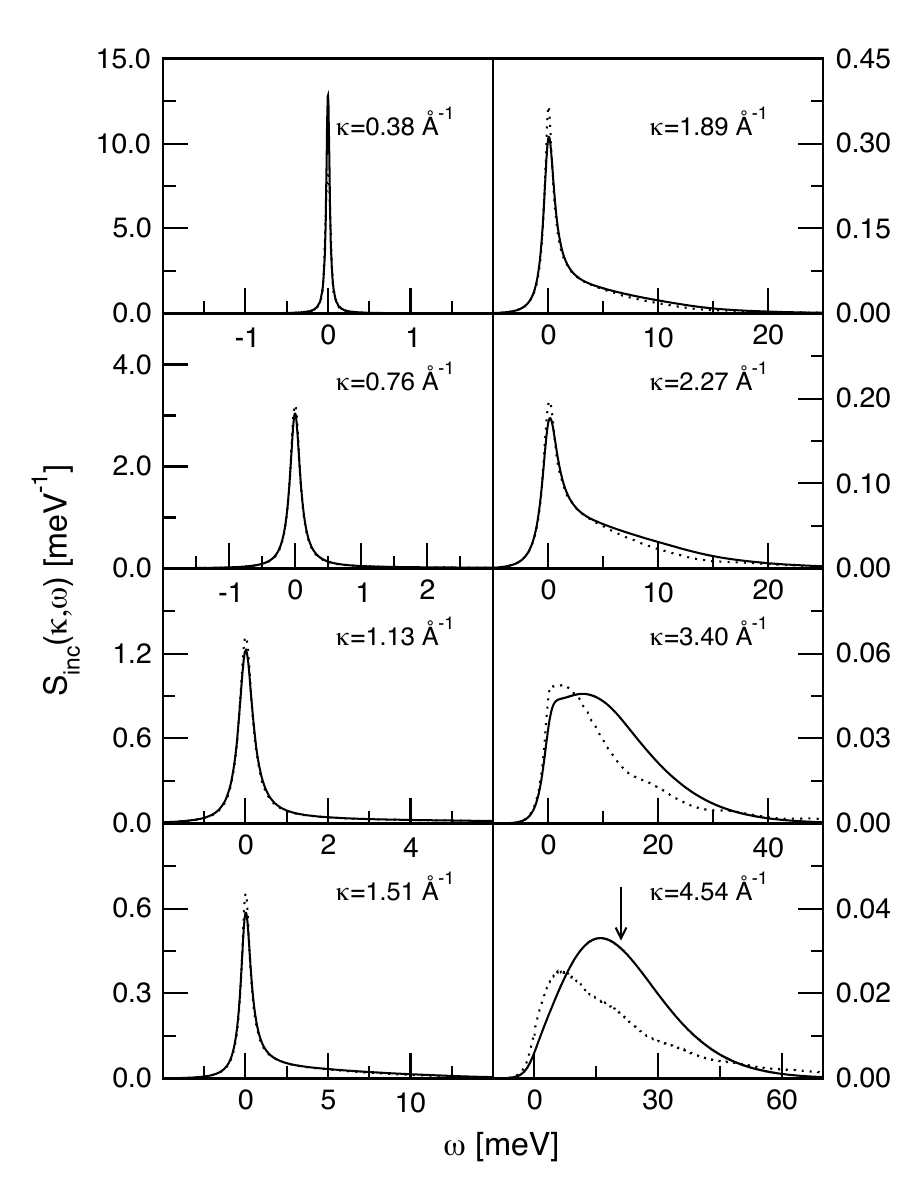}
 \centering
 \caption{Comparison of the RPMD-F (dotted) and RPMD-G (solid) approximations to the incoherent dynamic structure factor of liquid para-hydrogen at 14 K, for various values of the wave number $k$ \cite{crai-mano06cp}.}
 \label{Sinc}
 \end{figure}

An alternative way to calculate the incoherent dynamic structure factor is to use a Gaussian (second order cumulant) approximation to the intermediate scattering function \cite{rahm+62pr},
\begin{equation}
F_{\mathrm{s}}(k,t) \simeq e^{-\gamma(t)k^2},
\label{eq:rpmd39}
\end{equation}
where
\begin{equation}
\gamma(t) = -i{\hbar t\over 2m}+{1\over 3}\int_0^t\,(t-t')c_{{\bv}\cdot{\bv}}(t')\,\D{t}'
\label{eq:rpmd40}
\end{equation}
and $c_{{\bv}\cdot{\bv}}(t)$ is the standard (non-Kubo-transformed) velocity autocorrelation function
\begin{equation}
c_{{\bv}\cdot{\bv}}(t) = {1\over N}\sum_{i=1}^N c_{{\bv}_i\cdot{\bv}_i}(t).
\label{eq:rpmd41}
\end{equation}
This can be calculated from the RPMD approximation to $\tilde{c}_{{\bv}_i\cdot{\bv}_i}(t)$ by inverting the Kubo-transform in the frequency domain as in Eq.~\eqref{eq:rpmd36}, and then Fourier transforming back to the time domain. We shall call this indirect (Gaussian) RPMD approximation to $F_{\mathrm{s}}(k,t)$ \lq\lq RPMD-G" in what follows. It has a clear advantage over the RPMD-F approximation in that the velocity operators that are correlated in $\tilde{c}_{{\bv}_i\cdot{\bv}_i}(t)$ are Heisenberg time derivatives of {\em linear} position operators, and so this approach entirely avoids the non-linear operator problem. It does rest on the accuracy of the Gaussian approximation in Eq.~\eqref{eq:rpmd39}, but this is known to be exact in a number of important limits (including both the short- and long-time limits) and for a number of important model systems (including a perfect gas, a harmonic solid, and a Brownian liquid), so it is often likely to be very reliable.
 
Fig.~\ref{Sinc} compares the RPMD-F and RPMD-G approximations to the incoherent dynamic structure factor of liquid para-hydrogen at 14 K, for various values of the wave number $k$ \cite{crai-mano06cp}. The RPMD-G approximation is expected to be reliable for this problem for all values of $k$ considered in the figure, and in fact it has been shown to give excellent agreement with inelastic neutron scattering experiments along two very different kinematic lines (probing both diffusive and impulsive dynamics) in the $(k,\omega)$ plane \cite{crai-mano06cp}. The figure can therefore be taken to illustrate the breakdown of the RPMD-F approximation as $k$ increases and the correlated operators $e^{\pm i{\bk}\cdot \hat{\bq}_i}$ in Eq.~\eqref{eq:rpmd32} become increasingly non-linear functions of $\hat{\bq}$. One sees that for small $k$ (below about 2 \AA$^{-1}$) both approximations are in good agreement. Here the non-linearity of the correlated operators is not strong enough to cause any significant issues in the RPMD-F approximation to $\tilde{F}_{\mathrm{s}}(k,t)$. However, the accuracy of the approximation clearly begins to deteriorate beyond $k=2$ \AA$^{-1}$, and by the time $k>3$ \AA$^{-1}$ it is really rather poor. Fig.~\ref{Sinc} therefore shows that, while RPMD provides a reasonable approximation to condensed phase correlation functions involving mildly non-linear operators, it is significantly less accurate when the correlated operators are strongly non-linear functions of $\hat{\bq}$.

This is the interpretation of Fig.~\ref{Sinc} that was given in the early paper in which these calculations were first performed \cite{crai-mano06cp}. However, it is interesting to note that there is another interpretation, based on the observation that the peak in the incoherent dynamic structure factor moves to higher $\omega$ with increasing $k$. According to this second (and arguably more general, since it is consistent with the breakdown of RPMD for high frequency vibrational spectra) interpretation, RPMD works best when {\em both} $\omega$ and $k$ are small -- i.e., when the dynamics being captured is in the hydrodynamic regime.

\vfill
\newpage
\section{Summary}

Hopefully this Chapter has explained what RPMD is (a simple way to approximate quantum mechanical correlation functions in condensed phase systems from classical trajectories in an extended phase space), what it is good for (the calculation of diffusion coefficients and orientational relaxation times in molecular liquids), and what it is not (the calculation of high-frequency vibrational spectra and correlation functions involving strongly non-linear operators). \future{We have not discussed what RPMD is absolutely superb for (the calculation of chemical reaction rate coefficients in complex systems), because that is the subject of another Chapter.}

\newpage
\section*{Answers to exercises}

\begin{Answer}[ref=ex:rpmd-1]

\begin{itemize}

\item[(a)]
\begin{align*}
\tilde{c}_{AB}(t)
&= {1\over \beta Z}\int_0^{\beta} {\Tr}\left[e^{-(\beta-\lambda)\hat{H}}\hat{A}\,e^{-\lambda\hat{H}}e^{+i\hat{H}t/\hbar}\hat{B}e^{-i\hat{H}t/\hbar}\right] \D{\lambda} \\
&= {1\over \beta Z}\int_0^{\beta} \sum_{jk} e^{-\beta E_j}A_{jk}B_{kj}
e^{-\lambda(E_k-E_j)}e^{+i(E_k-E_j)t/\hbar}. \\
\\
\therefore\quad
\tilde{C}_{AB}(\omega) &= \int_{-\infty}^{\infty} e^{-i\omega t}\tilde{c}_{AB}(t)\,\D{t}\\
&= {2\pi\over \beta Z}\int_0^{\beta} \sum_{jk} e^{-\beta E_j}A_{jk}B_{kj}
e^{-\lambda(E_k-E_j)}\delta(\omega-(E_k-E_j)/\hbar) \\
&= {1\over\beta}\int_{0}^{\beta} e^{-\lambda\hbar\omega}\,\D{\lambda} \times
{2\pi\over Z}\sum_{jk} e^{-\beta E_j}A_{jk}B_{kj}\delta(\omega-(E_k-E_j)/\hbar)\\
&\equiv {(1-e^{-\beta\hbar\omega})\over \beta\hbar\omega}\times C_{AB}(\omega).
\end{align*}

\item[(b)] 
\begin{align*}
C_{AB}(-\omega) &= {2\pi\over Z}\sum_{jk} e^{-\beta E_j}A_{jk}B_{kj}\delta(\omega+(E_k-E_j)/\hbar)\\
&= {2\pi\over Z}\sum_{kj} e^{-\beta E_k-\beta\hbar\omega}B_{kj}A_{jk}\delta(\omega-(E_j-E_k)/\hbar)\\
&\equiv e^{-\beta\hbar\omega}\times C_{BA}(\omega).\\
\\
\therefore\quad
\tilde{C}_{AB}(-\omega) &= {(e^{\beta\hbar\omega}-1)\over\beta\hbar\omega}C_{AB}(-\omega)
={(1-e^{-\beta\hbar\omega})\over\beta\hbar\omega}C_{BA}(\omega) = \tilde{C}_{BA}(\omega).
\end{align*}

\end{itemize}
\end{Answer}

\begin{Answer}[ref=ex:rpmd-2]
\begin{itemize}

\item[(a)]
\begin{align*}
\tilde{c}_{BA}(-t) &= {1\over\beta Z}\int_0^{\beta} {\Tr}\left[e^{-(\beta-\lambda)\hat{H}}\hat{B}\,e^{-\lambda\hat{H}}e^{-i\hat{H}t/\hbar}\hat{A}\,e^{+i\hat{H}t/\hbar}\right]\,\D{\lambda}\\
&= {1\over\beta Z}\int_0^{\beta} {\Tr}\left[e^{-\lambda\hat{H}}\hat{A}\,e^{-(\beta-\lambda)\hat{H}}e^{+i\hat{H}t/\hbar}\hat{B}\,e^{-i\hat{H}t/\hbar}\right]\,\D{\lambda}\\
&= {1\over\beta Z}\int_0^{\beta} {\Tr}\left[e^{-(\beta-\lambda')\hat{H}}\hat{A}\,e^{-\lambda'\hat{H}}e^{+i\hat{H}t/\hbar}\hat{B}\,e^{-i\hat{H}t/\hbar}\right]\,\D{\lambda}' = \tilde{c}_{AB}(t).
\end{align*}

\item[(b)] $\hat{H}^{\dagger}=\hat{H}$. If $\hat{A}$ and $\hat{B}$ are also Hermitian, then recycling the last step of part (a) gives
\begin{align*}
\tilde{c}_{AB}(t)^* &= {1\over\beta Z}\int_0^{\beta} {\Tr}\left[\left(e^{-(\beta-\lambda)\hat{H}}\hat{A}\,e^{-\lambda\hat{H}}e^{+i\hat{H}t/\hbar}\hat{B}\,e^{-i\hat{H}t/\hbar}\right)^{\dagger}\right]\,\D{\lambda}\\
&= {1\over\beta Z}\int_0^{\beta} {\Tr}\left[e^{+i\hat{H}t/\hbar}\hat{B}\,e^{-i\hat{H}t/\hbar}e^{-\lambda\hat{H}}\hat{A}\,e^{-(\beta-\lambda)\hat{H}}\right]\,\D{\lambda}\\
&= {1\over\beta Z}\int_0^{\beta} {\Tr}\left[e^{-\lambda\hat{H}}\hat{A}\,e^{-(\beta-\lambda)\hat{H}}e^{+i\hat{H}t/\hbar}\hat{B}\,e^{-i\hat{H}t/\hbar}\right]\,\D{\lambda} = \tilde{c}_{AB}(t).
\end{align*}

\item[(c)] Assuming that the matrix elements $A_{jk}$ and $B_{kj}$ are real, the result that $\tilde{c}_{AB}(t) = \tilde{c}_{AB}(-t)^*$ follows directly from the first part of Exercise 1.1(a) and the fact that $(-it)^*=+it$:
\begin{align*}
\tilde{c}_{AB}(t)
&= {1\over \beta Z}\int_0^{\beta} {\Tr}\left[e^{-(\beta-\lambda)\hat{H}}\hat{A}\,e^{-\lambda\hat{H}}e^{+i\hat{H}t/\hbar}\hat{B}e^{-i\hat{H}t/\hbar}\right] \D{\lambda} \\
&= {1\over \beta Z}\int_0^{\beta} \sum_{jk} e^{-\beta E_j}A_{jk}B_{kj}
e^{-\lambda(E_k-E_j)}e^{+i(E_k-E_j)t/\hbar}. 
\end{align*}

\end{itemize}
\end{Answer}

\begin{Answer}[ref=ex:rpmd-3]

A $(P+1)$-point trapezium rule discretisation of the integral over $\lambda$ in Eq.~(1.31), with the weights $w_0=w_P=1/2$ and $w_{j}=1$ for $j=1,\ldots,P-1$, gives
\begin{align*}
\tilde{c}_{AB}(0) &= {1\over\beta Z}\int_0^{\beta} {\Tr}\left[ \hat{A}\,e^{-\lambda\hat{H}}\hat{B}\,e^{-(\beta-\lambda)\hat{H}}\right]\,\D{\lambda}
\nonumber\\
&\simeq {1\over PZ} \sum_{j=0}^P w_j {\Tr}\left[ \hat{A}\,e^{-j\beta_P\hat{H}}\hat{B}\,e^{-(P-j)\beta_P\hat{H}}\right]
\nonumber\\
&={1\over Z}\int \D[f]{\bq}\, \bra{{\bq}_1}e^{-\beta_P\hat{H}}\ket{{\bq}_2}\cdots \bra{{\bq}_P}e^{-\beta_P\hat{H}}\ket{{\bq}_1}\,A({\bq}_1){1\over P}\sum_{j=0}^P w_jB({\bq}_{j+1})
\nonumber\\
&={1\over (2\pi\hbar)^fZ}\int \D[f]{\bp}\int \D[f]{\bq}\,e^{-\beta_PH_P({\bp},{\bq})} A({\bq}_1){1\over P} \sum_{j=1}^P B({\bq}_j)
\nonumber\\
&={1\over (2\pi\hbar)^fZ}\int \D[f]{\bp}\int \D[f]{\bq}\,e^{-\beta_PH_P({\bp},{\bq})} {1\over P}\sum_{i=1}^P A({\bq}_i){1\over P} \sum_{j=1}^P B({\bq}_j)
\nonumber\\
&\equiv{1\over (2\pi\hbar)^fZ}\int \D[f]{\bp}\int \D[f]{\bq}\,e^{-\beta_PH_P({\bp},{\bq})} A_P({\bq})B_P({\bq}),
\end{align*}
where we have used the fact that ${\bq}_{P+1}\equiv{\bq}_1$ in the fourth line and the fact that cyclic permutations of the ring polymer beads do not change either $H_P({\bp},{\bq})$ or $B_P({\bq})$ (so averaging over all $P$ of them is the same as using any one of them) to obtain the fifth.
\end{Answer}

\begin{Answer}[ref=ex:rpmd-4]

\begin{itemize}
\item[(a)] The manipulations described in Eqs.~(1.29) and (1.30) and the following text can be used to show that the RPMD approximation to $\tilde{c}_{AB}(t)$ satisfies $\tilde{c}_{AB}(t)=\tilde{c}_{BA}(-t)$ as follows:
\begin{align*}
\tilde{c}_{AB}(t) &= {1\over (2\pi\hbar)^fZ}\int \D[f]{\bp}_0\int \D[f]{\bq}_0\, e^{-\beta_PH_P({\bp}_0,{\bq}_0)}A_P({\bq}_0)B_P({\bq}_t)
\nonumber\\
&= {1\over (2\pi\hbar)^fZ}\int \D[f]{\bp}_t\int \D[f]{\bq}_t\, e^{-\beta_PH_P({\bp}_t,{\bq}_t)}A_P({\bq}_0)B_P({\bq}_t)
\nonumber\\
&= {1\over (2\pi\hbar)^fZ}\int \D[f]{\bp}_0\int \D[f]{\bq}_0\, e^{-\beta_PH_P({\bp}_0,{\bq}_0)}A_P({\bq}_{-t})B_P({\bq}_0)
\nonumber\\
&= \tilde{c}_{BA}(-t).
\end{align*}
\item[(b)] The fact that the RPMD approximation to $\tilde{c}_{AB}(t)$ also satisfies $\tilde{c}_{AB}(t)=\tilde{c}_{AB}(-t)$ when $\hat{A}$ and $\hat{B}$ are local operators follows from the fact that the classical trajectories of RPMD satisfy ${\bq}_t[-{\bp}_0,{\bq}_0]={\bq}_{-t}[{\bp}_0,{\bq}_0]$, and the fact that the ring polymer Hamiltonian $H_P({\bp}_0,{\bq}_0)$ is an even function of ${\bp}_0$ (so ${\bq}_t[{\bp}_0,{\bq}_0]$ and ${\bq}_{t}[-{\bp}_0,{\bq}_0]={\bq}_{-t}[{\bp}_0,{\bq}_0]$ have equal Boltzmann weights in the correlation function).
\end{itemize}
\end{Answer}

\begin{Answer}[ref=ex:rpmd-5]
\begin{itemize}
\item[(a)] 
\begin{align*}
\left<T\right> &= {m_i\over 2}c_{{\bv}_i\cdot{\bv}_i}(0)
\nonumber\\
&={m_i\over 4\pi}\int_{-\infty}^{\infty}\D{\omega}\,C_{{\bv}_i\cdot{\bv}_i}(\omega)
\nonumber\\
&={m_i\over 4\pi}\int_{-\infty}^{\infty}\D{\omega}\,{\beta\hbar\omega\over (1-e^{-\beta\hbar\omega})}\tilde{C}_{{\bv}_i\cdot{\bv}_i}(\omega)
\nonumber\\
&={m_i\over 4\pi}\int_{-\infty}^{\infty}\D{\omega}\int_{-\infty}^{\infty}\D{t}\,{\beta\hbar\omega\over (1-e^{-\beta\hbar\omega})}e^{-i\omega t}\,\tilde{c}_{{\bv}_i\cdot{\bv}_i}(t).\\
\end{align*}
\item[(b)] This is indeed harder! Try to do the integral over $\omega$ yourself. When you get stuck and have given up all hope of ever being able to do it, see:\\
\\
{\em Sum rule constraints on Kubo-transformed correlation functions.}\\
B. J. Braams, T. F. Miller III and D. E. Manolopoulos,\\
Chem. Phys. Lett. 418, 179-184 (2006).
\end{itemize}
\end{Answer}

\setchapterpreamble[u]{\margintoc}
\chapter{Colored-noise methods} \labch{gle}

The methods we have discussed this far rely on performing classical canonical sampling of the path integral Hamiltonian, or on performing classical molecular dynamics based on it. 
Methods to accelerate the convergence of PIMD with the number of replicas, presented in \refch{pimdadv}, relied on reducing the computational cost of evaluating the forces, or on constructing a faster-converging version of the discretized path integral, in both cases relying on classical sampling of an extended system to compute quantum mechanical observables. 
An alternative approach involves using non-equilibrium sampling, using a generalized Langevin equation (GLE) to enforce frequency-dependent fluctuations, that reflect the quantum mechanical distribution of a harmonic oscillator. 
In this Chapter we give a pedagogic introduction to the definition and properties of GLEs, and to their use in molecular simulations -- which extends beyond the realm of modeling nuclear quantum effects. 
We then show how they can be used to enforce quantum fluctuations, and be combined with path integrals to obtain a method with systematic, but faster, convergence to the exact quantum distribution. 

\section{Generalized Langevin Equations}

The Langevin equation, as introduced in Section~\ref{sec:langevin-dynamics}, can be seen as a model of the coupling between a physical system and its environment. A more careful treatment of the effects of the system-bath coupling leads to a non-Markovian, history-dependent generalized Langevin equation (GLE) arises~\cite{kubo66rpp,bern-fors71arpc,hene72jpa,fox87jsp,zwan+01book,lucz05chaos}
\begin{equation}
\begin{split}
 \dot{q}&=p/m\\
 \dot{p}&=-V'(q)-\int_{-\infty}^t K(t-s) p(s)\mathrm{d} s +\zeta(t).
\end{split}
\label{eq:nonmark-sde}
\end{equation}
Comparing this equation with the conventional, Markovian case, Eq.~\ref{eq:langevin}, one sees that the friction memory kernel $K(t)$ describes dissipation, and corresponds to the parameter $\gamma$ in Eq.~\eqref{eq:langevin}, while $H(t)=\left<\zeta(t)\zeta(0)\right>$ describes the intensity of the noisy force -- more precisely, their time correlation function. In order to achieve canonical sampling a fluctuation-dissipation theorem must hold, requiring that the noise and friction memory kernels are related by $H(t)=k_B T K(t)$~\cite{zwan+01book}. 
The Markovian Langevin equation is recovered for $H(t)\propto K(t)\propto \delta(t)$. 
Eq.~\eqref{eq:nonmark-sde} is obtained when the dynamical variables associated with the bath are integrated out, leaving an effective,  history-dependent dynamics for the system only. 
GLEs of this form have been used in the past to model an open system coupled to a physical bath~\cite{kant08prb}, and play an important role in the theory of coarse graining.

From the point of view of using Eqs.~\eqref{eq:nonmark-sde} as a sampling device, to control the equilibration properties of molecular dynamics~\cite{otto+12jfa}, and the convergence of statistical averages, this form has many advantages. 
One of the crucial aspects is best understood by considering the (conventional) Langevin dynamics of a multi-dimensional system. Writing it using mass-scaled variables
($q\leftarrow \sqrt{m}q$, $p\leftarrow p/\sqrt{m}$), the dynamics can be expressed in the compact form
\begin{equation}
\begin{split}
\dot{\mbf{q}} = &\mbf{p}\\
\dot{\mbf{p}} = & -\partial V/\partial \mbf{q} -\gamma \mbf{p} + \sqrt{2\gamma/\beta}\boldsymbol{\xi}.
\end{split}
\end{equation}
Here $\boldsymbol{\xi}$ is a vector of uncorrelated Gaussian random variates, 
$\left<\xi_i(t)\xi_j(0)\right>=\delta_{ij}\delta(t)$. Now consider an orthogonal 
transformation of the coordinates, $\tilde{\mbf{q}}=\mbf{O}\mbf{q}$ and
$\tilde{\mbf{p}}=\mbf{O}\mbf{p}$. Because of the orthogonality of $\mbf{O}$, it is true that $\partial V/\partial\tilde{\mbf{q}}=\mbf{O}\partial V/\partial{\mbf{q}}$,  and that $\left<\mbf{O}\boldsymbol{\xi}(t) \boldsymbol{\xi}(0)^T\mbf{O}^T\right>=\delta(t) \mbf{1}$.
Thanks to the Gaussian statistics of $\boldsymbol{\xi}$, also the transformed noise $\tilde{\boldsymbol{\xi}}=\mbf{O}\boldsymbol{\xi}$ has identical Gaussian statistics.
Hence, the equations of motion in the transformed coordinates are completely  equivalent to those in the original coordinates. The invariance of Langevin dynamics under an orthogonal transformation means for instance that  if a Langevin dynamics with friction $\gamma$ is applied to a multi-dimensional harmonic system, the very same statistical and dynamical properties would be observed if  the equations of motion were written in the Cartesian basis or in normal-modes coordinates. 
For example, the predictions for the autocorrelation time of the potential or kinetic energy (as in Eqs.~\eqref{eq:ho-correlationtimes}) apply for each normal mode separately, even though the dynamics is integrated in Cartesian coordinates without explicit knowledge of the systems's vibrational frequencies or Hessian eigenvectors.

A similar argument applies if one considers a multi-dimensional system, with identical non-Markovian dynamics having independent Gaussian $\zeta$ applied to the different degrees of freedom. 
Thus, an analysis of the behavior of Eq.~\eqref{eq:nonmark-sde} for a 1D harmonic oscillator is sufficient to predict the behavior of a large assembly of coupled harmonic oscillators, for example a harmonic crystal -- and, at least approximately, the behavior of an anharmonic system with a dynamics that spans both slow and fast time scales. 
However, the non-Markovian nature of the equations of motion means that it is considerably more complex to derive analytical estimates of the sampling properties, and that the practical implementation of the equations of motion would be riddled with difficulties.

To circumvent this inconvenience, we introduce $n$ fictitious degrees of freedom $\mbf{s}$, and write a Markovian Langevin dynamics in an extended phase space~\cite{marc-grig83jcp}

\begin{equation}
 \begin{split}
  \dot{q}=&p\\
 \!\left(\! \begin{array}{c}\dot{p}\\ \dot{\mbf{s}} \end{array}\!\right)\!=&
 \left(\!\begin{array}{c}-V'(q)\\ \mbf{0}\end{array}\!\!\right)
 \!-\!\left(\!
 \begin{array}{cc}
 a_{pp} & \mbf{a}_p^T \\ 
 \bar{\mbf{a}}_p & \mbf{A}
 \end{array}\!\right)\!
 \left(\!\begin{array}{c}p\\ \mbf{s}\end{array}\!\right)\!+\!
 \left(\!
 \begin{array}{cc}
 b_{pp} & \mbf{b}_p^T \\ 
\bar{\mbf{b}}_p & \mbf{B} \end{array}\!\right)\!
  \left(\!\begin{array}{c}\multirow{2}{*}{$\boldsymbol{\xi}$ }\\ \\\end{array}\!\right),
\end{split}
\label{eq:mark-sde}
\end{equation}
Here, $\boldsymbol{\xi}$ is a vector of $n+1$ uncorrelated Gaussian random numbers, with
$\left<\xi_i\left(t\right)\xi_j\left(0\right)\right>=\delta_{ij}\delta\left(t\right)$.
The conventional Langevin equation Eq.~(\ref{eq:langevin}) is recovered when $n=0$.
It is easy to see, by integrating out the additional degrees of freedom in a Mori-Zwanzig fashion, that Equations~\eqref{eq:mark-sde} are statistically equivalent to Eqs.~\eqref{eq:nonmark-sde} with $K(t)=2a_{pp} \delta(t)-\mbf{a}_p^T e^{-\left|t\right|\mbf{A}}\bar{\mbf{a}}_p$ and an analogous (albeit more cumbersome) expression for the noise correlation function $H(t)$~\cite{ceri+10jctc,ceri10phd}. 

The GLE has hence been reformulated as a linear, Markovian stochastic differential equation, that can be thought as a matrix generalization of white-noise Langevin dynamics (an Ornstein-Uhlenbeck process~\cite{gard03book}). 
To distinguish expressions where the matrices are restricted to act on $(p,\mbf{s})$ and expressions where matrices act on the full state vector $\mbf{x}=\left(q,p,\mbf{s}\right)^T$, we use the same labelling introduced in Refs.~\cite{ceri+09prl2,ceri10phd,ceri+10jctc}:
\newcommand\arS{\rule{0pt}{12pt}}
\begin{equation}
\begin{array}{ccccc}
      &   q   &    p   &   \mbf{s}  & \arS \\ \cline{2-4}
\multicolumn{1}{c|}{q} & m_{qq} & m_{qp} & \multicolumn{1}{c|}{\mbf{m}_q^T} & \arS \\\cline{3-4}
\multicolumn{1}{c|}{p} & \multicolumn{1}{c|}{\bar{m}_{qp}} &  m_{pp} &  \multicolumn{1}{c|}{\mbf{m}_p^T} & \arS \\\cline{4-4}
\multicolumn{1}{c|}{\mbf{s}} &  \multicolumn{1}{c|}{\bar{\mbf{m}}_q}  &  \multicolumn{1}{c|}{\bar{\mbf{m}}_p} &  \multicolumn{1}{c|}{\mbf{M}} & \arS \\\cline{2-4}
\end{array}
\hspace{-8pt}\begin{array}{cc}
\arS \\ \arS \\
\left.\rule{0pt}{12pt}\right\}\!\mbf{M}_p \\
\end{array}
\hspace{-8pt}\begin{array}{cc}
\arS \\
\left.\rule{0pt}{20pt}\right\}\!\mbf{M}_{qp} 
\end{array}
\label{eq:notation}
\end{equation}

The form of Eqs.~\eqref{eq:mark-sde} is very general, and comprises as special cases many related GLE implementations~\cite{stel+14prb,bacz-bond13jcp}, that  have on their side a more transparent relation to a physical model of the bath. 
Here we are only interested in obtaining the maximum flexibility with the most compact formulation possible, and will therefore derive all of our results in the general case of arbitrary $\mbf{A}_p$ and $\mbf{B}_p$ matrices. 
$\mbf{A}_p$ and $\mbf{B}_p$ determine the static covariance matrix $\mbf{C}_p$, that reflects the fluctuations  of $p$ and $\mbf{s}$ in the free-particle limit. 
The three matrices must satisfy the relation $\mbf{A}_p\mbf{C}_p+\mbf{C}_p\mbf{A}_p^T=\mbf{B}_p\mbf{B}_p^T$.
Furthermore, a sufficient condition to fulfill fluctuation-dissipation theorem and achieve canonical sampling is that $\mbf{C}_p=k_B T$. In practice if one wants to sample configurations consistent with Boltzmann statistics the  diffusion matrix $\mbf{B}_p$ is determined by the value of $\mbf{A}_p$ via $\mbf{B}_p\mbf{B}_p^T=k_B T\left(\mbf{A}_p+\mbf{A}_p^T\right)$.

\begin{exercise}[label=ex:gle-exponential,title={An exponential-memory GLE}]
Consider the case of a friction matrix
\begin{equation*}
\mbf{A}_p = \frac{1}{\tau} \left( 
\begin{array}{cc}
    0 & -\sqrt{\gamma\tau}  \\
\sqrt{\gamma\tau} & 1
\end{array}
\right)
\end{equation*}
\begin{enumerate}
\item What noise matrix $\mbf{B}_p$ should be used to obtain a GLE that samples the canonical ensemble at temperature $T$?
\item Write explicitly the equations of motion associated with $\mbf{A}_p$ and $\mbf{B}_p$, in terms of $q$, $p$, $s$
\item Compute the friction autocorrelation kernel $K(t)$ that is associated with this GLE.
\end{enumerate}
\end{exercise}

Just as for plain Langevin dynamics, the case of a generalization to multiple degrees of freedom, in which equivalent (same-parameters) but independent (uncorrelated) GLEs are applied to each coordinate is invariant to a unitary transformation that is applied simultaneously to the physical degrees of freedom and their associated extended momenta $\mbf{s}$.
Thus, predictions for the statistical and dynamical properties of a one-dimensional harmonic oscillator will be realized on each vibrational mode of a real system, regardless of whether the GLE is integrated in the normal-modes or in the Cartesian basis. 
Contrary to the explicitly non-Markovian case, it is now possible to solve the harmonic case analytically, since the dynamics for $\mbf{x}=\left(q,p,\mbf{s}\right)^T$ reads simply 
\begin{equation}
\begin{split}
 \dot{\mbf{x}}= &-\mbf{A}_{qp} \mbf{x} + \mbf{B}_{qp}\boldsymbol{\xi}, \\\text{where }&\, 
\mbf{A}_{qp}=\left(\begin{array}{ccc}
0         & -1/m &\mbf{0} \\
\omega^2  & \multicolumn{2}{c}{\multirow{2}{*}{ $\mbf{A}_p$ }} \\
\mbf{0} & & \\
\end{array}\right)\quad
\mbf{B}_{qp}=\left(\begin{array}{ccc}
0         & 0 &\mbf{0} \\
0  & \multicolumn{2}{c}{\multirow{2}{*}{ $\mbf{B}_p$ }} \\
\mbf{0} & & \\
\end{array}\right)
\end{split}
\label{eq:harmonic}
\end{equation}
which is itself just an Ornstein-Uhlenbeck process. 

\section{Equilibrium GLE sampling}

Refs.~\cite{ceri10phd,ceri+10jctc} report the expressions for a number of static and dynamic properties of the dynamics of a harmonic oscillator of frequency $\omega$ as a function of the parameters $\mbf{A}_p$ and $\mbf{B}_p$. 
These include expressions for the correlation times~\eqref{eq:ho-correlationtimes}, that are cumbersome but straightforward, basically requiring diagonalisation of small matrices of size $n+2$.
It is therefore possible to evaluate the sampling properties of the GLE exactly, without  statistical error and without having to run test calculations. 
Furthermore, the dynamics of a multi-dimensional harmonic system will comply with these predictions without having to know explicitly the vibrational modes.
This makes it possible to design GLE dynamics that address several practical problems that are often encountered when using MD to sample the constant-temperature canonical ensemble. 

\begin{figure}[btp]
\caption{\label{fig:ho1d-optimal} Sampling efficiency for the potential energy, $\kappa_V$,
for different optimal sampling GLE thermostat, that have been designed to yield constant mormalized efficiency
over different ranges of frequency (two, four, six orders of magnitude from bottom to top). 
Red lines are obtained from matrices with $n=2$, blue lines are the best fit for matrices
with $n=4$, and the black line is the most balanced choice of white noise, shown as reference.
}
\centering\includegraphics[width=1.0\linewidth]{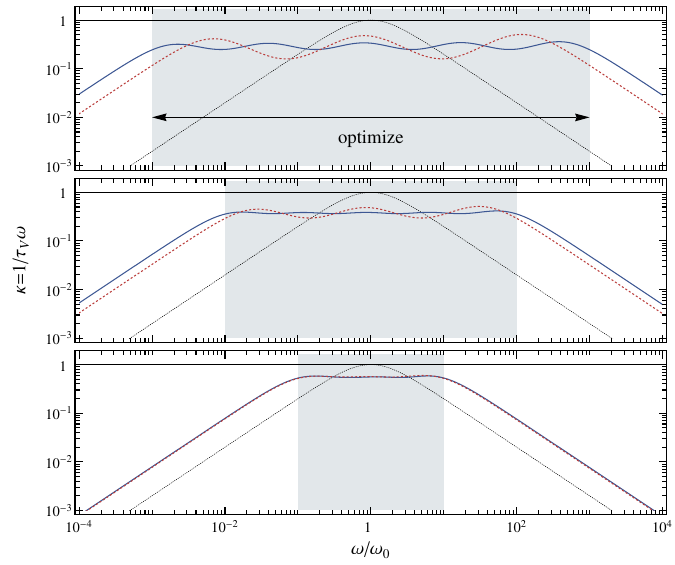}
\end{figure}

\subsection{Enhanced sampling efficiency}

Depending on the number of additional degrees of freedom $n$, the GLE thermostat involves a large number of parameters, and it would be completely impractical to choose them by trial and error, based on test simulations for a specific system. 
For this approach to be useful, the parameters have to be decided \emph{a priori}, based on the analytical estimates that can be computed in the harmonic limit.
Let us focus for instance on the task of optimising the sampling efficiency for the potential energy of the harmonic oscillator, measured in terms of $\kappa_V=1/\tau_V\omega$, a normalized measure which is $1$ in the optimal case and smaller for sub-optimal sampling (see Section~\ref{sec:langevin-dynamics}). 

The strategy for designing a transferable ``optimal sampling'' GLE is quickly explained. The only piece of information that is needed is a rough estimate of the range of frequencies that is needed for the problem at hand (e.g. 1 to 4000 cm$^{-1}$  for a liquid containing O--H bonds). 
Then, one can start from a tentative (e.g. random) $\mbf{A}_p$ matrix, and compute the value of $\kappa_V$ for a number of frequencies $\omega\in\left[1,4000\right]$cm$^{-1}$. 
The elements of $\mbf{A}_p$ can then be modified, so as to iteratively optimize the values of $\kappa_V$ across the chosen frequency range, aiming for a large, constant value of the sampling efficiency. 
We refer the reader interested in the details of the fitting procedure to Refs.~\cite{ceri10phd,ceri+10jctc}, and to the open-source GLE fitting code.\footnote{A development version is available at \url{https://github.com/cosmo-epfl/gle4md}} 
One important aspect is the need of parametrising $\mbf{A}_p$ (and in some cases $\mbf{B}_p$) in a way that enforces automatically some mathematical constraints that must be fulfilled in order to get a well-behaved stochastic dynamics (e.g. the eigenvalues of $\mbf{A}_p$ must have positive-definite real part). 
An objective function is then introduced that evaluates how much the GLE generated by the tentative set of parameters deviates from the desiderata (e.g. large and constant  $\kappa_V$). A minimum is then found by iteratively changing the parameters, for instance using the Nelder-Mead downhill simplex algorithm~\cite{neld-mead65cj}. 

The response of the GLE as a function of frequency is typically very smooth (in fact, it 
takes considerable effort to obtain fits with sharp changes as a function of $\omega$), so
it is not worth to target specifically the vibrational density of states of a given system. 
Asking for equally efficient sampling of all vibrational modes over a broad range of frequencies
yields a very transferable set of parameters, that can be used to obtain efficient sampling
for many different systems without the need of time-consuming preliminary tests.

\begin{figure}[btp]
\caption{\label{fig:ho1d-smart} Sampling efficiency for the potential energy
$\kappa_V$ as a function of the frequency,
for a white-noise thermostat optimised for $\omega=0.4$cm$^{-1}$ (black), for a
optimal sampling GLE optimised between 0.4 and 4000cm$^{-1}$ (red) and for
smart sampling GLE optimised between 0.4 and 4000cm$^{-1}$ (blue). The upper panel
shows the velocity-velocity correlation spectrum for a flexible TIP4P water 
model~\cite{habe+09jcp}, as reference.}
\centering\includegraphics[width=0.75\linewidth]{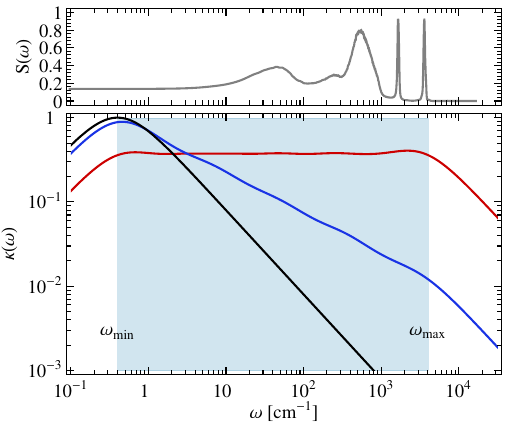}
\end{figure}

\begin{figure}[btp]
\centering\includegraphics[width=0.75\linewidth]{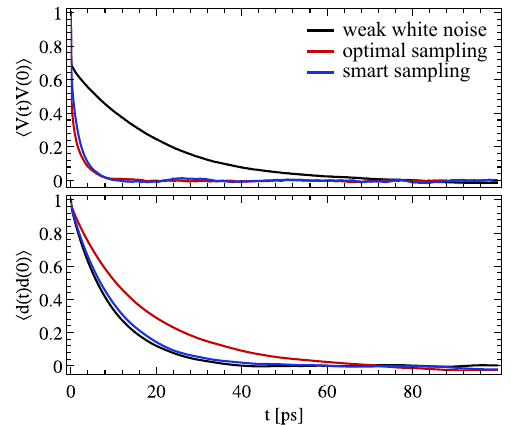}
\caption{\label{fig:smart-water} Autocorrelation functions of the potential energy (top)
and the components of the total dipole moment of the simulation cell (bottom, $\left<\mbf{d}(0)\cdot\mbf{d}(t)\right>$),
for a simulation of a flexible water model~\cite{habe+09jcp}.
The different curves correspond to simulations performed with a white-noise 
Langevin thermostat optimised for $\omega=0.4$cm$^{-1}$ (black), for a
optimal sampling GLE optimised between 0.4 and 4000 cm$^{-1}$ (red) and for
smart sampling GLE optimised between 0.4 and 4000 cm$^{-1}$ (blue).
Note that smart-sampling has comparable performance to a very weak Langevin thermostat
for hard-to-compute properties depending on slow collective rearrangement of atoms,
while being dramatically more efficient for faster-converging properties such as
the potential energy.
}
\end{figure}

\subsection{Smart sampling GLE \label{sub:smart}}

If one observes carefully the $\kappa_V(\omega)$ curves optimised over different frequency ranges
(Figure~\ref{fig:ho1d-optimal}), it becomes apparent that there is a trade-off between the breadth of the 
range and the constant sampling efficiency that can be achieved. At the extremes of the 
fitted range the GLE is orders of magnitude more effective than the best choice of white noise,
however the GLE constant $\kappa_V$ is about 50\%\ of the ideal value of 1, and tends to
be lower for the broader fitting ranges.

While an ``optimal sampling'' GLE guarantees that no vibrational mode is severely 
over or under-damped, one could argue that treating evenly all vibrational modes
is not the best choice possible. Slow, collective modes are the most challenging, 
while fast vibrations will be sampled several times during the simulation, and so 
achieving optimal sampling efficiency is less crucial. A smarter sampling strategy
would be to consider an estimate of the maximum simulation time one can afford 
 $t_\text{max}$, and to ensure that vibrations with frequency $\omega_\text{min}=2\pi/t_\text{max}$
(the slowest one can hope to observe) are sampled with maximum efficiency. 
All the faster modes, up to the maximum frequency present $\omega_\text{max}$, should be sampled as 
efficiently as possible, \emph{without negatively affecting the sampling of slower modes}.
Empirically, it seems to be possible to obtain a decay of $\kappa_V(\omega)\sim 1/\sqrt{\omega}$ 
above $\omega_\text{min}$ -- rather than the $1/\omega$ decay that would be 
expected for white noise (see Figure~\ref{fig:ho1d-smart}) -- but a formal treatment of this problem is not yet available. 

As shown in Figure~\ref{fig:smart-water}, in a practical case (a MD simulation of liquid water at room temperature) this leads to a shorter correlation time for a hard-to-compute property such as the cell dipole moment, and a slightly longer correlation time for the potential energy, that has a short correlation time and is easy to converge anyway. 
This example demonstrates on one hand that a ``smart-sampling'' GLE  helps ensuring that as much statistics as possible is extracted from demanding MD simulations, and on the other hand illustrates the philosophy of GLE thermostatting: a set of parameters is optimized based on very general considerations and  analytical estimates in the harmonic limit, and then it is applied to a real, complex simulation yielding results that are compatible with the initial set of desiderata.

\begin{exercise}[label=ex:gle-canonical,title={Parameters for a smart-sampling GLE}]
You are setting up calculations for a first-principles molecular dynamics simulation of pure molten silica. 
You have a CPU grant worth one million CPU hours, and you can run one fs of dynamics in 30s, using 1024 cores. 
Estimate the minimal and maximum frequency of the range you should use for a ``smart sampling'' GLE simulation to get the best possible statistics from a single MD trajectory. \emph{Hint: look up an IR spectrum of silica, or quartz, to get an idea of the fastest vibrational modes.}
\end{exercise}

\subsection{Stabilizing multiple time-step dynamics}

As a final demonstration of the flexibility of GLE thermostats, let us consider the problem of stabilising multiple time step (MTS) dynamics. 
The general idea behind MTS is similar to that underlying ring-polymer contraction (see Section~\ref{sub:contraction}). Whenever it is possible to decompose the inter-atomic forces in a fast (and inexpensive) component and one slowly-varying (and expensive) one, it is possible and advantageous to introduce a multiple time step procedure~\cite{stre+78mp,tuck+92jcp}, whereby the fast component of the force is evaluated often, and the slow component is evaluated once every several steps, reducing dramatically the cost of the simulation while maintaining an accurate description of the physics. 
These methods are common when using empirical force fields, where the slowly-varying component of the force typically corresponds to long-range electrostatics. 
More recently, attempts have been made to develop similar schemes for ab initio MD~\cite{stee13jcp,lueh+14jcp}, and with machine-learning potentials~\cite{ross+20jctc} although the partitioning of the forces in slow and fast components is less obvious.
Multiple time step concepts also underlie the integration of PIMD using an exact propagator for the free ring polymer, discussed in Section~\refch{basics}.

In either case, it is known that even with an effective splitting there is a limit to the ratio between the slower and the faster time steps, because the fast degrees of freedom in the system enter in resonance with the small errors in the seldom-updated slow component~\cite{schl+98jcp}. 
It is also well-known that such ``resonance barrier'' can be overcome by coupling the system to a strong thermostatting bath~\cite{bart-schl98jcp}. 
However, as we have seen in Section~\ref{sec:langevin-dynamics}, over-damped thermostatting reduces the sampling efficiency of diffusive modes, so that despite the larger outer time step, little or no advantage is obtained in terms of statistical efficiency. 

Ideally, one would like to use a thermostat that is active on the fast degrees of freedom, while leaving the slow components of the dynamics unaffected. 
In fact, methods have been devised that implement this concept, based on approximate knowledge of the vibrational patterns associated with fast molecular motion~\cite{izag+99jcp}.
Coloured-noise dynamics can also be used to this aim, with the considerable advantage that no prior knowledge of the dynamics is needed, and that one can simply specify a cutoff frequency below which the GLE dynamics aims at disturbing minimally the system's dynamics~\cite{morr+11jcp}. 

\section{Non-equilibrium GLE sampling}

Sampling configurations consistent with constant-temperature equilibrium
conditions is perhaps the most straightforward application of our GLE
framework. Eq.~\eqref{eq:mark-sde} is however considerably more general,
and in principle one could very easily realise a stochastic dynamics for which
$k_BT(\mbf{A}_p+\mbf{A}_p^T)\ne\mbf{B}_p\mbf{B}_p^T$, which does
not satisfy the fluctuation-dissipation theorem and therefore does not
guarantee sampling of the canonical ensemble. 

A simulation based on this dynamics could be regarded as a model of 
the coupling of the physical system with several baths at different 
temperature, each coupled preferentially to different frequency ranges. 
In fact, one can exploit the possibility of solving the dynamics 
in the harmonic limit (and once more the invariance of a multi dimensional
GLE to a unitary rotation of the coordinates) to predict the stationary
distribution of a $N$-dimensional harmonic system in terms of a frequency-dependent
effective temperature $T^\star(\omega)$. 
Furthermore, by fitting the parameters in $\mbf{A}_p$
and  $\mbf{B}_p$ one can tune the fluctuations of potential $\left<\omega^2 q^2\right>(\omega)$ 
and kinetic energy $\left<p^2\right>(\omega)$ to obtain the desired frequency dependence, leading to a number of useful effects.

\subsection{$\delta$-thermostat and $f$-thermostat}

\begin{figure}[btp]
\caption{\label{fig:delta-ice} Velocity-velocity correlation spectra for a flexible ice model~\cite{habe+09jcp}, computed out of a series of simulations using $\delta$-thermostats targeted at different frequencies $\omega_0$. }
\centering\includegraphics[width=0.75\linewidth]{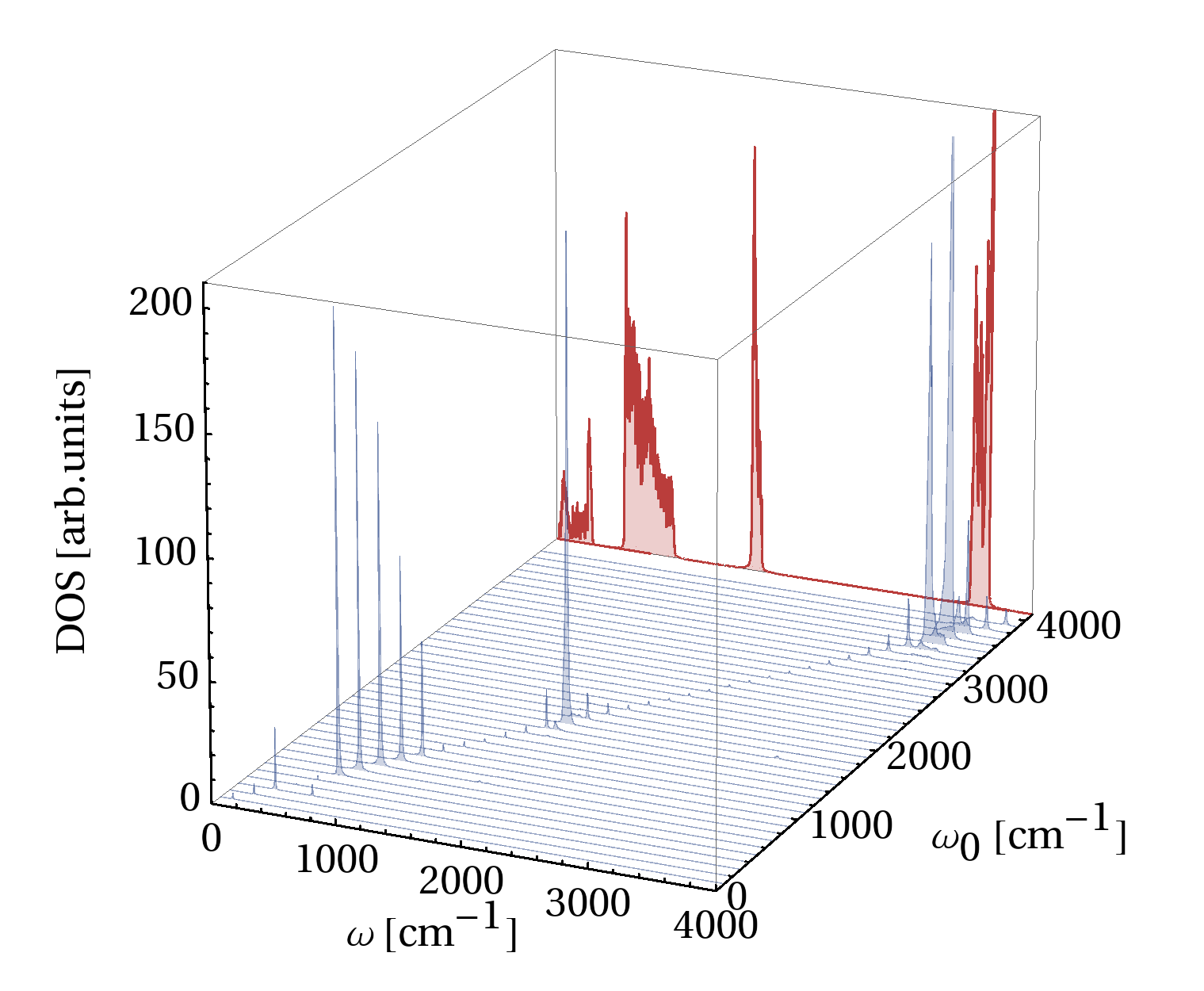}
\end{figure}

Perhaps, the simplest example of a ``non-equilibrium'' GLE is one that enforces a $\delta$-like dependency of the effective temperature as a function of frequency~\cite{ceri-parr10pcs}.
In a harmonic system, this GLE will set a narrow range of frequencies to a 
finite effective temperature, and all other normal modes to a near-zero $T^\star$. 
Figure~\ref{fig:delta-ice} shows the effect of applying $\delta$-thermostats
targeting different frequencies $\omega_0$ to a quasi-harmonic system, namely 
an empirical force field model of ice. Despite the anharmonicity of the system, 
the dynamics does automatically excite the normal modes corresponding to
the desired target. Once more, no information on the intrinsic vibrations of the system is used, and the GLEs can be integrated in the Cartesian basis.
Modified $\delta$ thermostats that excite a few modes while setting the others at a finite, constant temperature have also been  used to model pump-probe experiments, and to estimate energy relaxation in liquids~\cite{dett+17jctc,dett+19jpcl}.

Another application of
this idea is the (approximate) evaluation of the density of states of a positive-definite matrix $\mbf{M}$. 
The idea is to use $\mbf{M}$ to define the quadratic potential of an artificial harmonic dynamics. When $\delta$-thermostats centred on different frequencies $\omega$ are used on top of such potential, the mean kinetic energy obtained from the trajectory can be related to the density of eigenstates in the vicinity of $\epsilon=\omega^2$. 
This concept can be generalized to construct an $f$-thermostat~\cite{nava+14pre}, which can be used to evaluate selected elements of positive functions of positive-definite matrices. In practice one designs a GLE that enforces $\omega$-dependent fluctuations of momentum that are related to the target matrix function $f$ by $\left<p^2\right>(\omega)=f(\omega^2)$. 
Then, thanks to the rotational invariance of GLE dynamics, one sees that $\left<\mbf{p}\mbf{p}^T\right>\approx f(\mbf{M})$, even though the artificial dynamics is \emph{not} performed in the basis of the eigenvalues of $\mbf{M}$, and in fact there is no explicit knowledge of the spectrum of the matrix.
If $\mbf{M}$ is sparse, the dynamics can be propagated with linear scaling effort, and so selected elements of $f(\mbf{M})$ can be computed with linear complexity, avoiding the diagonalisation of $\mbf{M}$.

\begin{exercise}[label=ex:gle-fermi,title={A Fermi thermostat}]
What is the $T^\star(\omega)$ curve that the GLE must enforce in order to obtain $\left<\mbf{p}\mbf{p}^T\right>=f_{\epsilon,\beta}(\mbf{M})$, where $f_{\epsilon,\beta}(x)$ is a Fermi function with chemical potential $\epsilon$ and inverse temperature $\beta$? 
\end{exercise}

\subsection{The quantum thermostat}

The methods discussed this far demonstrate nicely the effects that can be obtained 
by using GLEs that do not fulfil the fluctuation-dissipation relation between 
friction and noise memory kernels. Furthermore, they suggest a possible application to 
the evaluation of physical effects by \emph{ab initio} molecular dynamics, beyond
conventional Boltzmann sampling. 

Consider a harmonic oscillator of frequency $\omega$, sampled canonically at inverse temperature
$\beta=1/k_BT$. Its phase-space distributions for position and momentum are Gaussians
$\rho(p)\propto \exp -p^2/2\sigma^2_p$ and $\rho(q)\propto \exp -q^2/2\sigma^2_q$, regardless
of whether the oscillator is described classically or according to quantum mechanics.
The classical and quantum cases only differ for the mean fluctuations:
for a classical oscillator $\sigma_p^2=m/\beta$ and $\sigma_q^2=1/\beta m \omega^2$,
while for a quantum oscillator $\sigma_p^2=m\frac{\hbar\omega}{2}\coth \frac{\beta\hbar\omega}{2}$
and $\sigma_q^2=\frac{\hbar}{2m\omega}\coth \frac{\beta\hbar\omega}{2}$. 
One sees that quantum fluctuations at temperature $T$ correspond to the 
fluctuations of a classical oscillator at the effective temperature 
\begin{equation}
\label{eq:qt-tstar}
T^\star(\omega)=\frac{\left<{p^2}\right>}{mk_B}
=\frac{m\left<{q^2}\right>}{k_B}
=\frac{\hbar\omega}{2k_B}\coth \frac{\hbar\omega}{2k_B T}.
\end{equation}

If one could perform a simulation of a compound in which different  normal modes 
are thermalised at the effective, frequency-dependent temperature $T^\star(\omega)$, then 
the phase-space distribution and thermodynamic properties of the system
would correspond to the distinguishable-particles quantum description of the
nuclear degrees of freedom -- at least in the harmonic limit. 

To achieve this while using conventional thermostatting, one would need to know the 
normal modes frequencies and phonon displacement patterns, and apply tailored white-noise 
thermostats at different temperatures working in the normal modes representation. 
GLE thermostatting, on the other hand, makes it possible to obtain the desired
distribution without the need of knowing the normal modes of the system being studied.
One only needs to fit a set of parameters that enforces the quantum fluctuations for any frequency within a range that encompasses the vibrational modes relevant for the system at hand, and then apply the same GLE to each Cartesian degree of freedom. 
The quantum $T^\star(\omega)$ is then enforced automatically, giving quantum fluctuations at the cost of conventional molecular dynamics.

\begin{figure}[btp]
\caption{\label{fig:quantum-qdw} Panel b) shows the expectation values of potential, kinetic and total energy
for a proton in a 1D quartic double-well potential with minima separated by 1\AA, as a function of the height 
of the barrier. Dotted lines correspond to reference quantum mechanical results, the continuous blue line
corresponds to the classical mean total energy, and the red dots are the average values obtained
from a quantum-thermostat simulation. As shown in panel a), the different barrier heights span
different regimes, going from a quasi-classical limit for small barriers, to strongly quantised but quasi-harmonic
conditions at very high barriers, with an intermediate regime in which tunnelling is non-negligible.}
\centering\includegraphics[width=1.0\linewidth]{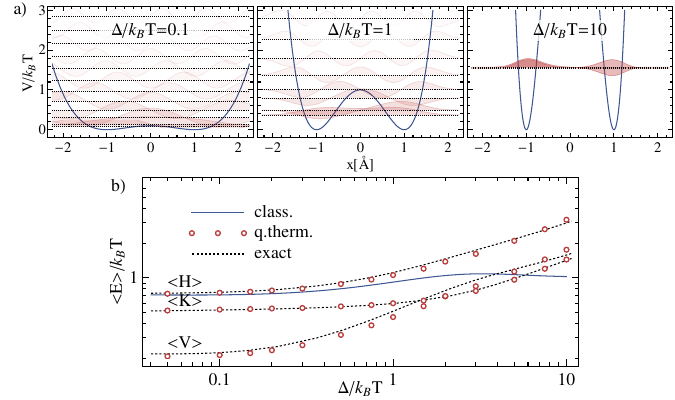}
\end{figure}

This ``quantum thermostat'' (QT) idea~\cite{ceri+09prl2,ceri+10jctc} works surprisingly well also for strongly anharmonic potentials (Figure~\ref{fig:quantum-qdw}). 
In fact, the main limitation when applying it to real systems does not depend much on failure to describe strongly anharmonic behaviour, but rather on the consequences of weak anharmonic coupling in a multi-dimensional system, giving rise to zero-point energy leakage~\cite{ceri+10jctc}.
In practice the quantum thermostat tries to keep normal modes of different frequencies  at different temperatures, and so in the presence of anharmonic coupling energy will tend to flow from high-frequency/high-temperature to low-frequency/classical temperature modes.
This energy flow was not accounted for when designing $T^\star(\omega)$, and so there will be a (significant) deviation between the desired quasi-harmonic quantum fluctuations and the actual fluctuations. 

This is a common problem in semi-classical methods to treat quantum nuclear effects~\cite{habe-mano09jcp}, and has been also recognized in other stochastic approaches to obtain approximate quantum effects, such as the quantum thermal bath (QTB) method~\cite{damm+09prl,bedo+14prb}, which can be construed as a special case of a QT in which the friction is instantaneous, and the memory kernel of the noise is consistent with the quantum frequency-dependent fluctuations~\eqref{eq:qt-tstar}.\footnote{Strictly speaking, a quantum thermal bath only ensures the correct fluctuations for a harmonic oscillator in the weak-coupling limit. However, for realistic values of the friction term, the distribution is very close to the correct one.  }
A possible solution to the problem of zero-point energy leakage by exploiting the tunability of the GLE thermostats, enforcing a strong coupling across the whole frequency range so as to counterbalance the energy transfer associated with anharmonic couplings.
This approach improves significantly the performance of the quantum thermostat when applied to anharmonic problems~\cite{ceri+10jctc,ceri10phd}, and makes it possible to describe qualitatively the role of NQEs in several real applications.
Alternatively, it has been proposed to use the relationship between the force-velocity and the velocity-velocity correlation function to detect ZPE leakage, and to introduce a correction term to the target $T^\star(\omega)$~\cite{mang+19jctc}. At least for homogeneous systems, this ``adaptive quantum thermal bath'' approach can improve substantially the accuracy of these out-of-equilibrium GLE approaches without having to perform simulations in an overdamped regime. 

\section{Combining GLE and PIMD}
\label{sec:GLE_and_PIMD}

The approximations behind the quantum thermostat and related semi-classical methods are essentially uncontrolled, and can be regarded as inexpensive techniques to assess qualitatively the importance of NQEs, more than to reach quantitative conclusions for systems in which a fully-converged reference calculation is impossible.
However, coloured-noise is not only useful for approximate calculations: it can be used together with path integral molecular dynamics to obtain a systematically (and quickly!) converging method to quantitatively evaluate nuclear quantum effects.

\subsection{Accelerating the convergence of configurational properties: PI+GLE}

The quantum thermostat yields exact (apart from the very small errors in the fit  of $T^\star(\omega)$) quantum fluctuations in the harmonic limit, but exhibits uncontrolled errors in real, anharmonic problems. 
Path integral molecular dynamics (PIMD) on the contrary can be converged systematically, but involve a very large computational overhead that is largely due to the high-frequency, strongly quantised  vibrations, that are typically very close to harmonic. 
These considerations suggest that a hybrid technique, combining PIMD and correlated noise, could help achieve faster convergence while still allowing for quantitative accuracy and controlled error. 
The crux is designing a GLE thermostat that enforces exact quantum fluctuations in the harmonic limit \emph{for any number of replicas}, even in cases where PIMD alone would be far from converged. 
Then, such a PI+GLE method would be always exact for harmonic problems, and naturally converge to (Boltzmann-sampled) PIMD when the number of beads is large enough to have a converged result in the absence of a non-canonical GLE. 

\begin{figure}[btp]
\caption{\label{fig:pigle-gp} The frequency-dependent temperature (expressed as $g_P(x)=T^*(2x/\beta\hbar)/PT$, 
with $x=\beta\hbar\omega/2$) for the PI+GLE technique using different numbers of beads.}
\centering\includegraphics[width=0.75\linewidth]{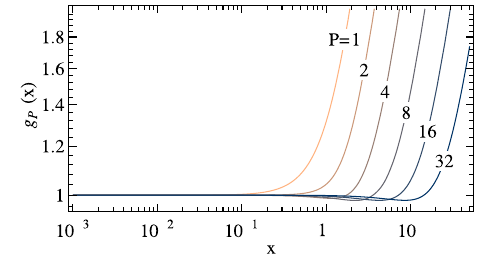}
\end{figure}

In order to work out the properties of the GLE that would achieve this goal, one can proceed in the same way as with the quantum thermostat, only considering that now, in the presence of a harmonic potential of frequency $\omega$, the dynamics will be characterised by frequencies $\omega_k=2\omega_P\sin k\pi/P$ (recall that $\omega_P=Pk_BT/\hbar$, see \ref{sec:est-simple} for a discussion of the calculation of path integral estimators for the harmonic oscillator). 
These are the frequencies that will be picked up by the colored-noise dynamics, so, introducing a frequency-dependent \emph{configurational} temperature $T^*(\omega)=\left<q^2\right>(\omega)m\omega^2/k_B$ (momentum fluctuations are not important per se in a PIMD framework), one gets the requirement for having quantum fluctuations of the beads to be 
\begin{multline}
\frac{m\omega^2}{k_B T}\left<q^2\right>=\frac{m\omega^2}{P{k_B T}}\sum_i\left<q_i^2\right>=
\frac{m\omega^2}{P{k_B T}}\sum_k\left<\tilde{q}_k^2\right>=\\
\frac{1}{P} \sum_k \frac{T^*(\sqrt{\omega^2+\omega_k^2})/T}{1+\omega_k^2/\omega^2} = 
\frac{\hbar\omega}{2k_B T}\coth \frac{\hbar\omega}{2k_B T}.
\label{eq:tw-pigle}
\end{multline}
Since the frequencies of the free ring polymer $\omega_k$ are shifted by the physical frequency of the oscillator $\omega$, Eq.~\eqref{eq:tw-pigle} 
must be seen as a functional equation that defines the $T^*(\omega)$ curve -- if any --
that satisfies it for any oscillator frequency $\omega$. In Ref.\cite{ceri+11jcp} 
it is discussed how to numerically solve this functional equation, and the resulting 
target $T^*(\omega)$ for different number of beads are provided in the supporting
information of the same reference.
Note incidentally that it is essential that one can tell precisely what will be the 
ring-polymer dynamics in a harmonic potential: the discussion should be modified
if one used anything other than the physical masses for the normal modes propagation. 

\begin{exercise}[label=ex:pigle-limit,title={Low and high-frequency limits of PI+GLE}]
Consider the limit of \eqref{eq:tw-pigle} for $\omega\rightarrow 0$ and for $\omega\rightarrow \infty.$ Keeping in mind that in $x\coth x$ tends to $1$ for $x\rightarrow 0$, and to $x$ for $x\rightarrow \infty$, show what are the limits of $T^\star(\omega)$ for $P=2$.
\end{exercise}

\begin{figure}[btp]
\caption{\label{fig:pigle-qdw} Probability density for a hydrogen atom in a 
quartic double-well potential with the minima
separated by 0.6\AA{} and a barrier height of 1000K. 
All simulations were performed with a target temperature of 300 K. 
The exact quantum mechanical result (dashed black line) was obtained by
numerical solution of the Schr{\"o}dinger equation, the contributions 
of the various eigenstates being averaged with the appropriate Boltzmann weight. 
The four panels compare this exact result with
the PIMD (blue line) and PI+GLE (red line) results with increasing bead numbers.}
\centering\includegraphics[width=0.75\linewidth]{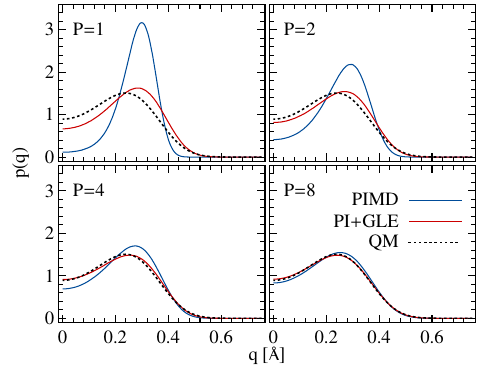}
\end{figure}

Figure~\ref{fig:pigle-gp} shows the frequency-dependent effective temperature 
relative to the temperature $P T$ that would be used in a conventional PIMD simulation,
as a function of the number of replicas. As the number of replicas increases, 
the curve is just equal to one up to larger and larger ``quantumness'' parameter
$x=\beta\hbar\omega/2$. For a given frequency and temperature, 
as the number of beads is increased, PI+GLE will behave more and more as a conventional
PIMD with Boltzmann sampling of the ring-polymer Hamiltonian. 
This implies that PI+GLE is bound to converge to the exact quantum averages, just because
in the large $P$ limit it converges to PIMD. As shown in Fig.~\ref{fig:pigle-qdw},
even for a strongly anharmonic quantum problem, the convergence is considerably accelerated.

\begin{figure}[btp]
\caption{\label{fig:pigle-water} The average value of the potential energy, the virial kinetic energy 
and the constant-volume heat capacity for a simulation of
a flexible water model~\cite{habe+09jcp} at T = 298 K, plotted as a function of the number of beads. The results
obtained with conventional PIMD and PI+GLE are compared, and the value of V obtained with
the original quantum thermostat10 (QT) is also reported.}
\centering\includegraphics[width=0.75   \linewidth]{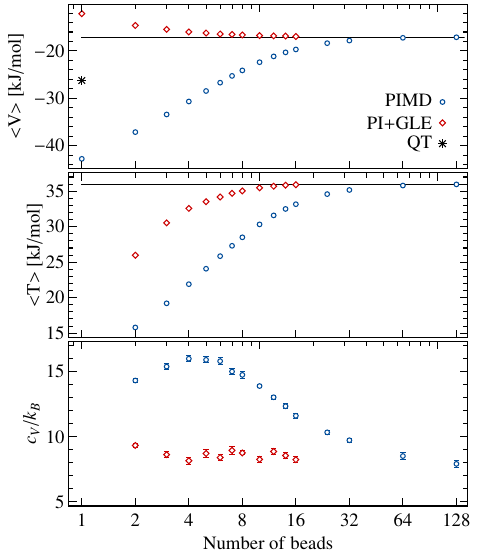}
\end{figure}

Furthermore, PI+GLE makes it possible to systematically converge results also in the case
of real, anharmonic multidimensional problems. As the number of replicas is increased,
the $T^*(\omega)$ curve becomes flatter and flatter, and so there is a less pronounced 
effective temperature gradient between the high and low frequency modes, so zero-point
energy leakage (using here the term in a loose sense) is a lesser concern than in the 
case of the quantum thermostat. Even though it is still important to enforce effective
coupling to the thermostat, one can avoid the strong overdamping that must be used with
the quantum thermostat, with considerable advantages in terms of sampling efficiency for 
for the slow, diffusion-like modes. 
Figure~\ref{fig:pigle-water} shows the convergence with number of beads of potential and kinetic energy 
for a quantum simulation of an empirical water model~\cite{habe+09jcp} at room temperature,
comparing plain PIMD and PI+GLE. Colored noise accelerates dramatically the convergence of 
observables to the quantum expectation values, and the possibility of converging results
systematically makes it possible to assess the error. 
A careful examination of Figure~\ref{fig:pigle-water} shows that the mean kinetic energy 
$\left<T\right>$ converges somewhat more slowly than $\left<V\right>$. This is due to a
specific shortcoming of PI+GLE, that will be addressed in the next section.

\subsection{Including imaginary-time correlations: PIGLET}

As noted above, when using PI+GLE the quantum kinetic energy seems to converge more slowly than 
the average potential, or other structural quantities such as radial distribution functions. 
If one considers carefully the expression for the kinetic energy estimator~\eqref{eq:ke-virial-1d}
in the harmonic limit, it becomes apparent why:
\begin{equation}
\begin{split}
\left<T\right>=&\frac{1}{2\beta} + \frac{1}{2P}\omega^2 \sum_{i=0}^{P-1}\left<q_i^2\right> - \frac{1}{2}\omega^2\left<\bar{q}^2\right>=\\
=&\left<V\right>+\frac{1}{2\beta} - \frac{1}{2}\omega^2\left<\bar{q}^2\right>.
\end{split}
\label{eq:piglet-tv}
\end{equation}
In a quantum mechanical description, the average potential and kinetic energy of a harmonic
oscillator of frequency $\omega$ read
\begin{equation}
\left<V\right>=\left<T\right>=\frac{\hbar\omega}{4}\coth {\beta\hbar\omega\over 2}.
\label{eq:piglet-tv-tgt}
\end{equation}
In order to obtain the correct quantum value for $\left<T\right>$, it is not sufficient
to design the GLE so that the fluctuations of $q$ are consistent with 
$\left<V\right>=\frac{\hbar\omega}{4}\coth {\beta\hbar\omega\over 2}$, but it is also
necessary to make sure that $\frac{1}{2}\omega^2\left<\bar{q}^2\right>=\frac{1}{2\beta}$.

This points at a general limitation of the basic PI+GLE idea: only the ``marginal'' distribution
of the beads is bound to be exact in the harmonic limit, which warrants accelerated convergence of
any observable that depends only on $q$ but does not necessarily help converging more complex
estimators that also depend on the correlations between different beads.
Fortunately, it is relatively easy to extend the PI+GLE idea to include further correlations. 
In fact, PIMD can be very effectively propagated in the free-particle normal modes representation,
i.e. by transforming the coordinates to $(\tilde{q}_k,\tilde{p}_k)$ and writing the equations
of motion in that base (see Eq.~\eqref{eq:pimd-verlet} and Ref.~\cite{ceri+10jcp}). The physical potential acts in the same way on all the beads, so in
the harmonic limit it only amounts to a diagonal perturbation of the free ring polymer, that 
changes the vibrational frequencies as discussed above, but does not change the eigenvectors. 
Hence, in a multi-dimensional context it is possible to transform individual degrees of freedom in
the \emph{free particle} normal modes representation, without the need to diagonalise the physical
potential, and it is possible to apply GLEs with different temperature curves $T_k^*(\omega)$ 
onto different \emph{free particle} coordinates.
In practice, this makes it possible to enforce multiple constraints on the distribution of the 
ring polymer, including bead-bead correlations as well as the marginal distribution that guarantees
fast convergence of structural properties. 

\begin{figure}[btp]
\caption{\label{fig:piglet-water} The quantum contribution to the potential energy, and to the 
kinetic energy of hydrogen and oxygen atoms as computed by the centroid virial estimator for
 a simulation of a flexible water model~\cite{habe+09jcp} at T = 298 K, plotted as a function 
 of the number of beads. Note the much accelerated convergence rate of the kinetic energy
 when using PIGLET compared to PI+GLE.}
\centering\includegraphics[width=0.8\linewidth]{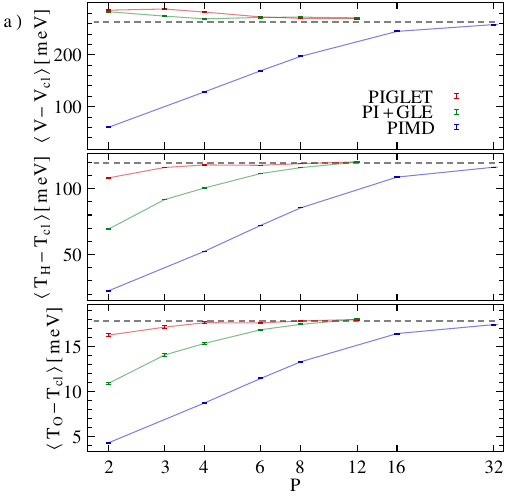}
\end{figure}

Let us elaborate on this idea, focussing on the objective of accelerating the convergence of the 
centroid-virial kinetic energy estimator. Eqs.~\eqref{eq:piglet-tv} and~\eqref{eq:piglet-tv-tgt}
clearly require that the centroid must be distributed classically, so $T_0^*(\omega)=PT$, as in 
conventional PIMD. There is then just one more condition to be enforced, namely the marginal 
distribution of the $q_i$s, which now has to be determined assuming a classical distribution for the centroid mode:
\begin{equation}
\!\!\!\!\!\frac{m\omega^2}{P{k_B T}}\sum_k\left<\tilde{q}_k^2\right>=
1+
\frac{1}{P} \sum_{k>0} \frac{T^*(\sqrt{\omega^2 + \omega_k^2})/T}{1+\omega_k^2/\omega^2} = 
\frac{\hbar\omega}{2k_B T}\coth \frac{\hbar\omega}{2k_B T}.
\label{eq:tw-piglet}
\end{equation}
This functional equation can be solved similarly to Eq.~\eqref{eq:tw-pigle}, now singling out 
the $k=1$ term to devise a fixed-point iteration that converges to the desired, universal
$T^*(\omega)$ curve. 

Figure~\ref{fig:piglet-water} shows clearly that PIGLET is considerably more efficient than 
PI+GLE in converging the quantum kinetic energy of atoms, even for an anharmonic problem such as liquid water.
The convergence of structural properties for the two methods is very similar, highlighting 
the fact that it is possible to manipulate bead-bead correlations without disrupting the 
efficient convergence of the marginal distribution of individual beads. 

The centroid-virial estimator does not exhaust the list of physical observables that depend
on bead-bead correlations: imaginary-time correlation functions provide moment constraints 
that can be used to improve the reliability of real-time approximate quantum dynamics~\cite{egor+97jcp,habe+07jcp},
scaled-coordinate estimators make it possible to obtain directly the heat capacity~\cite{yama05jcp},
displaced path estimators can be used to compute the particle momentum distribution~\cite{lin+10prl},
and free-energy perturbation estimators give access to isotope fractionation ratios~\cite{ceri-mark13jcp}.
In all of these cases one could try to figure out which constraints ought to be enforced
on the ring polymer distribution to obtain exact expectation values in the harmonic case,
and use them to determine a number of $T_k^*(\omega)$ curves to be used to design
GLEs for the different ring-polymer normal modes. 

In some cases it is possible that a given combination of PIMD and GLEs accelerates convergence
of estimators it has not been designed specifically for. For instance, PI+GLE does yield faster
convergence of the centroid-virial kinetic energy (albeit not as fast as PIGLET), and PIGLET
appears to speed up the convergence of the ``thermodynamic'' free-energy perturbation estimator
of isotope fractionation~\cite{ceri-mark13jcp}. The possibility of obtaining systematic convergence
by increasing the number of beads means one can empirically assess the accuracy for a given estimator
and physical problem by performing test simulations with increasing number of beads, much like
one would converge a plane waves cutoff, a $k$-points mesh, or conventional PIMD.
It is also possible to combine GLE thermostatting with high-order path integration~\cite{suzu95pla,chin97pla,pere-tuck11jcp}, but such a SC+GLE scheme yields only marginal improvements over a GLE method based on Trotter PIMD~\cite{kapi+16jcp2}.

\section{Dynamical properties from GLE trajectories}

Thermostats that fulfill the classical fluctuation-dissipation theorem generate a canonical probability distribution of configurations, irrespective of the GLE parameters. 
The dynamics, however, is modified by the coupling between the Hamiltonian dynamics and the stochastic terms. 
As usual, we consider the harmonic limit of the problem, for which one can obtain the time correlation functions as a function of $\mbf{A}_p$ and $\mbf{B}_p$, and use it to understand the effect of the thermostat on the dynamics.

\begin{figure}[btp]
\caption{\label{fig:gle-dynamics} Momentum-momentum (full lines) and position-position (dashed lines) correlation functions for a harmonic oscillator of frequency $\omega_0$ subject to a friction $\gamma$. 
}
\centering\includegraphics[width=0.8\linewidth]{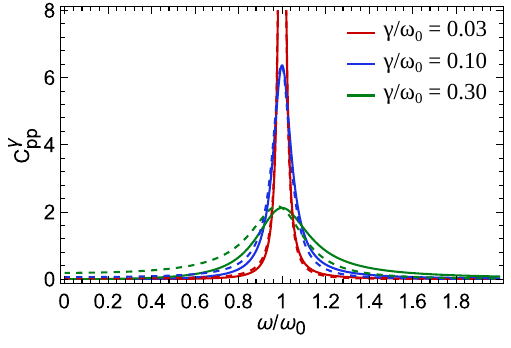}
\end{figure}

For instance, the Fourier transform of the momentum-momentum correlation function for a harmonic oscillator of frequency $\omega_0$, that is directly linked to the vibrational density of states, reads:
\begin{equation}
C_{pp}(\omega_0,\omega) = \frac{1}{[\mathbf{C}_{qp}(\omega_0)]_{pp}} \left[\frac{\mathbf{A}_{qp}(\omega_0)}{\mathbf{A}^2_{qp}(\omega_0) +\omega^2} \mathbf{C}_{qp}(\omega_0)\right]_{pp} \label{eq:vvac-gle},
\end{equation}
where the stationary covariance matrix $\mathbf{C}_{qp}$ can be obtained by solving the equation
$\mathbf{A}_{qp}\mathbf{C}_{qp}+
\mathbf{C}_{qp}\mathbf{A}_{qp}^T=
\mathbf{B}_{qp}\mathbf{B}_{qp}^T$ (see Ref.~\cite{ceri10phd}).
In the white-noise limit, this reduces to 
\begin{equation}
C^{(\gamma)}_{pp}(\omega_0,\omega) = 
\frac{2 \gamma  \omega ^2}{\pi  \left(\gamma ^2
   \omega ^2+\omega ^4-2 \omega _0^2 \omega
   ^2+\omega _0^4\right)}
 \label{eq:vvac-gamma-decomp}.
\end{equation}
As shown in Figure~\ref{fig:gle-dynamics}, the Langevin term leads to a broadening of the peak associated with the correlation function of a harmonic oscillator.

\begin{figure}[bp]
\caption{\label{fig:qt-purify} Velocity-velocity correlation functions for a simulation of liquid water at $T=300$~K. 
The figure compares a simulation from classical MD (NVE) with one using a quantum thermostat (QT), whose strong coupling nature disrupts the low-frequency part of the spectrum. 
Applying the deconvolution procedure (QT$\rightarrow$NVE) recovers the classical behavior of the low-frequency part of the correlation spectrum, and one can see a red shift of the stretch peak which is a typical manifestation of quantum nuclear effects. Adapted from Ref.~\cite{ross+18jcp}.
}
\centering\includegraphics[width=1.0\linewidth]{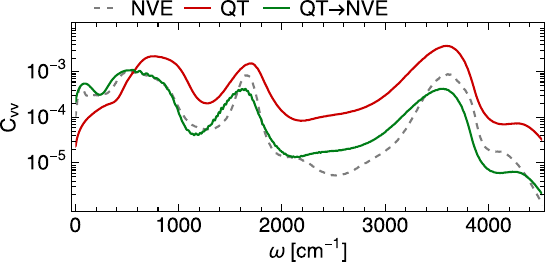}
\end{figure}

When applied to a realistic system, this broadening results in a loss of resolution of the observables, such as vibrational spectra, that reflect its dynamical behavior.
If the system can be modelled as an assembly of harmonic oscillators with density of states $g(\omega)$, its correlation spectrum in the presence of a GLE can be expressed as the convolution of $g$ with the broadened response of a single oscillator
\begin{equation}
C_{vv}^\text{GLE}(\omega)=\int \D\omega' g(\omega') C(\omega',\omega). \label{eq:cpp-convolution}
\end{equation}
As discussed in Ref.~\cite{ross+18jcp}, this ansatz makes it possible to recover the unperturbed $g(\omega)$ by applying an appropriate deconvolution algorithm to the correlation function computed from a GLE trajectory. 
In particular, it was found that the Iterative Image Space Reconstruction Algorithm (ISRA), that enforces positive-definiteness of the solution\cite{daub-mueh86ieee,arch-titt95ss}, is much more effective and stable than a regularized inversion of Eq.~\eqref{eq:cpp-convolution}. 
This approach can be used to extract usable dynamical information from thermostatted trajectories, even for non-equilibrium schemes such as the quantum  thermostat,\cite{ross+18jcp} or the quantum thermal bath.\cite{mauger2021arxiv}

\begin{figure}[btp]
\caption{\label{fig:piglet-purify} IR (top) and Raman (bottom) spectra of liquid water at 300K, computed using a machine-learning potential energy, dipole and polarizability surfaces. Simulations were performed using a PIGLET thermostat and $P=6$ beads, and the centroid trajectories were deconvoluted based on the analytical estimate of the GLE-induced broadening\cite{ross+18jcp}. Adapted from Ref.~\cite{kapi+20jcp}.
}
\centering\includegraphics[width=0.8\linewidth]{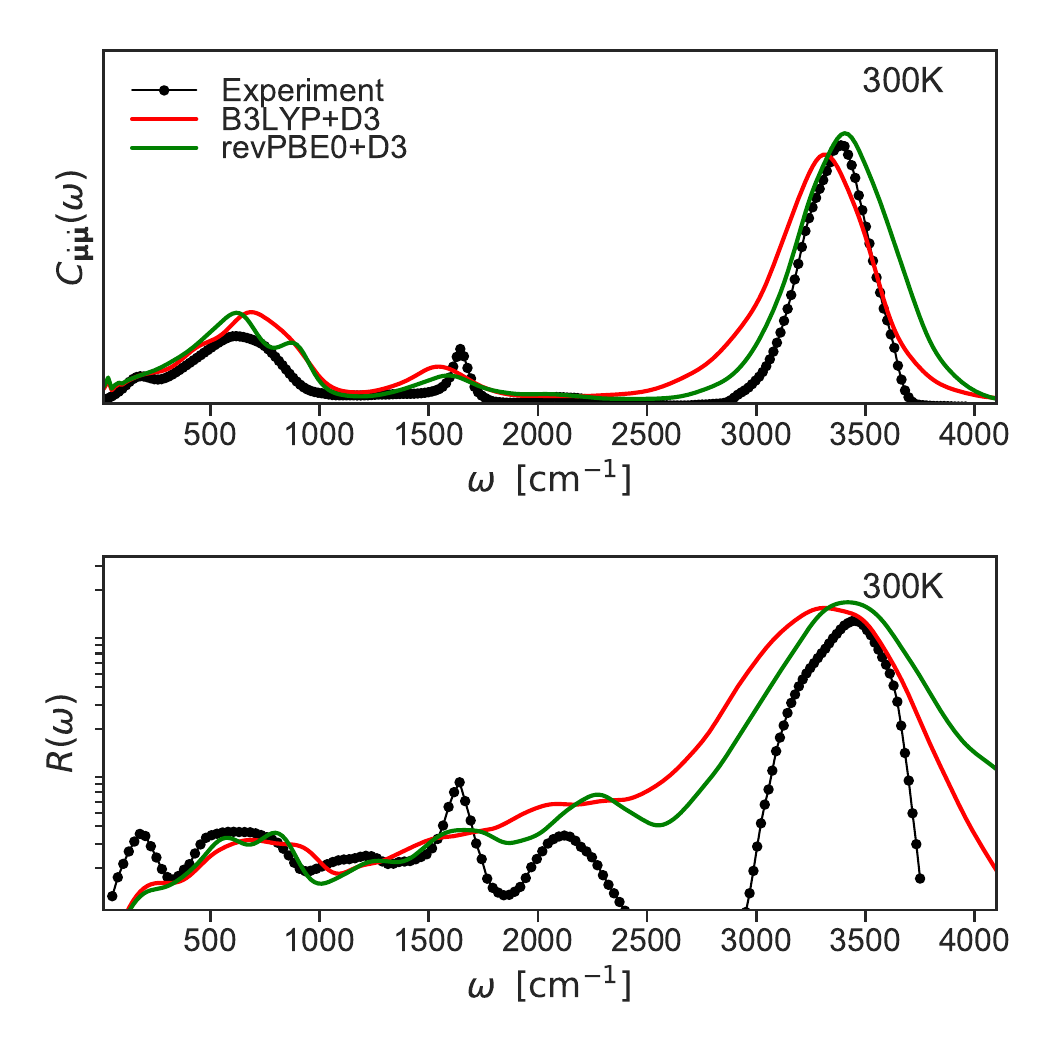}
\end{figure}

A similar idea can also be applied to PIGLET trajectories, to obtain estimates of the quantum correction to vibrational correlations that are similar in accuracy to those that can be obtained from (T)RPMD or CMD. 
Essentially one would run a PIGLET trajectory (which uses a classical thermostat for the centroid) and then apply the deconvolution procedure to CMD-like correlation functions computed for the centroid, according to the GLE parameters used for the $k=0$ normal mode. 
The deconvolution eliminates the broadening that is introduced by the rather aggressive thermostat, and recovers spectra in good qualitative agreement with those obtained by conventional CMD or (T)RPMD. Fig.~\ref{fig:piglet-purify} for an example of the IR and Raman spectra of water, obtained using this scheme in combination with machine-learning models of energy and dielectric response\cite{kapi+20jcp}. 

\textbf
\newpage
\section*{Exercise answers}

\begin{Answer}[ref=ex:gle-exponential]
\begin{enumerate}
\item We have to find $\mbf{B}_p$ such that
$$
\mbf{B}_p\mbf{B}_p^T = k_BT (\mbf{A}_p+\mbf{A}_p^T) = \left( 
\begin{array}{cc}
    0 & 0  \\
0 & 2k_B T/\tau
\end{array}
\right)
$$
which is satisfied by 
$$
\mbf{B}_p = \left( 
\begin{array}{cc}
    0 & 0 \\
0 & \sqrt{2k_B T/\tau}
\end{array}
\right)
$$
\item The combination of the Hamiltonian evolution of $(q,p)$ and the GLE associated to $(p,s)$ yields
\begin{equation*}
\begin{split}
\dot{q} = & p \\
\dot{p} = & -\partial V/\partial q + s \sqrt{\gamma/\tau } \\
\dot{s} = & -p\sqrt{\gamma/\tau } -s/\tau + \xi\sqrt{2k_BT/\tau}   \\
\end{split}
\end{equation*}
where $\xi$ is an uncorrelated Gaussian noise. 
\item By just substituting the entries of $\mbf{A}_p$ into the expression for the memory kernel associated with the non-Markovian friction one obtains
$$
K(t) = \frac{\gamma}{\tau} e^{-t/\tau}
$$
\end{enumerate}
\end{Answer}

\begin{Answer}[ref=ex:gle-canonical]
First, you need to realize that most of the experimental spectra of silica also contain a signal for \ce{Si-O-H} groups, so quartz is a better reference. 
The highest vibrational mode can then be seen to be around 1100 cm$^{-1}$. 
Given the amount of CPU time available, and the cost of simulating one fs worth of trajectory, the overall budget is a little short of what is needed to simulate 60 ps of trajectory. 
This is then the longest time scale which can be explored given the budget, and that for which the best possible sampling efficiency should be obtained. It corresponds to a frequency of 0.5~cm$^{-1}$.
Note how both of these estimates need not be exceedingly precise: changing either by a factor of two would affect minimally the accuracy over the accessible frequency range. 
\end{Answer}

\begin{Answer}[ref={ex:gle-fermi}]
Given that the eigenvalues of the positive-definite matrix $\mbf{M}$ are equal to the square of the dynamical frequencies, and that $\left<p^2\right>=T$ (assuming unit mass and $k_B=1$) one must have  $T^\star(\omega) = f_{\epsilon,\beta}(\omega^2)$.
\end{Answer}

\begin{Answer}[ref={ex:pigle-limit}]
First, consider the case of $\omega\rightarrow 0$. 
The term in the sum is zero for all $k$ except $k=0$ (for which $\omega_0^2/\omega^2=1$). Given that the right-hand side of the equation tends to one, 
\begin{equation}
T^\star(\omega) = P T,
\end{equation}
i.e. for low frequencies the target temperature should be that of a conventional PIMD simulation. 
In the high-$\omega$ limit, and for $P=2$, Eq. \eqref{eq:tw-pigle} reads
\begin{equation}
\frac{1}{2}\left[T^\star(\omega) + \frac{\omega^2 T^\star\left(\sqrt{\omega^2+4\omega_P^2}\right) }{\omega^2 + 4\omega_P^2}\right] \approx \frac{\hbar\omega}{2k_B}.
\end{equation}
Expanding the second term around $\omega$ one finds that the left-hand side is $T^\star(\omega)+\mcal{O}(1/\omega)$, showing that in the high-frequency limit $T^\star(\omega)\approx \hbar\omega/2k_B$.
\end{Answer}
\def\std{\ooalign{$-$\cr\hfil $\circ$\hfil \cr}}
\newcommand{\elecspin}[1]{\hat {\mathbf{S}}_#1}
\newcommand{\prop}[1]{e^{-i #1 t}}
\newcommand{\iprop}[1]{e^{+i #1 t}}
\setchapterpreamble[u]{\margintoc}
\chapter{Adiabatic Ring Polymer Rate Theory}
\labch{rates-adiabatic}

This Chapter and the next provide an introduction to the quantum mechanical theory of chemical reaction rates, and how these rates can be approximated using path integral methods. Although the theory will be presented   using one-dimensional notation, the methods we shall describe are easily generalised to treat arbitrarily complex reactions in their full dimensionality. In particular, they do not make any assumptions about the dominance of a single reaction pathway: all contributing pathways are sampled with the appropriate Boltzmann weights. This sets the present methods apart from the ring polymer instanton methods, which are only applicable to relatively simple reactions that are dominated by a single tunneling pathway.

Here we shall consider electronically adiabatic reactions that proceed on a single Born-Oppenheimer potential energy surface, and devote Chapter~\ref{ch:rates-nonadiabatic} to the case of non-adiabatic reactions. Since both cases have been discussed in a recent review article \cite{lawr-mano20fd}, we shall keep the presentation here to a minimum, with a focus on using exercises to illustrate the material. More details can be found in Ref.~\cite{lawr-mano20fd}, and in the other references we shall give as we go along. (See also these Oxford lecture notes \cite{manonotes}, which show how the flux-side correlation function formulation of reaction rate theory in Eq.~\ref{RateEq2} can be derived from quantum scattering theory.)

\section{Quantum Rate Theory}\label{QRT}

The simplest possible model for a (bimolecular) chemical reaction is one-dimensional barrier transmission problem with a Hamiltonian of the form
\begin{equation}
\hat{H} = {\hat{p}^2\over 2m}+V(\hat{q}),\label{RateEq1}
\end{equation}
in which the potential $V(q)$ tends to zero as $q\to-\infty$ (the reactant asymptote) and to a constant as $q\to\infty$ (the product asymptote).  The exact quantum mechanical thermal rate constant for this reaction can be written as \cite{mill+83jcp}
\begin{equation}
k(T) = {1\over Q_r(T)}\lim_{t\to\infty} \bar{c}_{fs}(t),\label{RateEq2}
\end{equation}
where $Q_r(T)=\sqrt{mk_{\text{B}}T/2\pi\hbar^2}$ is the reactant partition function per unit length (it would become the reactant partition function per unit volume in three-dimensional space) and $\bar{c}_{fs}(t)$ is a flux-side correlation function
\begin{equation}
\bar{c}_{fs}(t) = {\text{tr}}\left[e^{-\beta\hat{H}/2}\hat{F}e^{-\beta\hat{H}/2}e^{+i\hat{H}t/\hbar}\hat{\theta}\,e^{-i\hat{H}t/\hbar}\right],\label{RateEq3}
\end{equation}
with $\beta=1/k_{\text{B}}T$. Here the flux operator
\begin{equation}
\hat{F} = {i\over\hbar}\left[\hat{H},\hat{\theta}\right]\label{RateEq4}
\end{equation} 
is the Heisenberg time derivative of a projection operator onto the product side of a dividing surface between the reactants and products at $q=q^{\ddagger}$,
\begin{equation}
\hat{\theta} = \theta(\hat{q}-q^{\ddagger}),\label{RateEq5}
\end{equation}
and $\theta(x)$ is a Heavyside step function.

\begin{figure}[htb!]
 \resizebox{1.2\columnwidth}{!} {\includegraphics{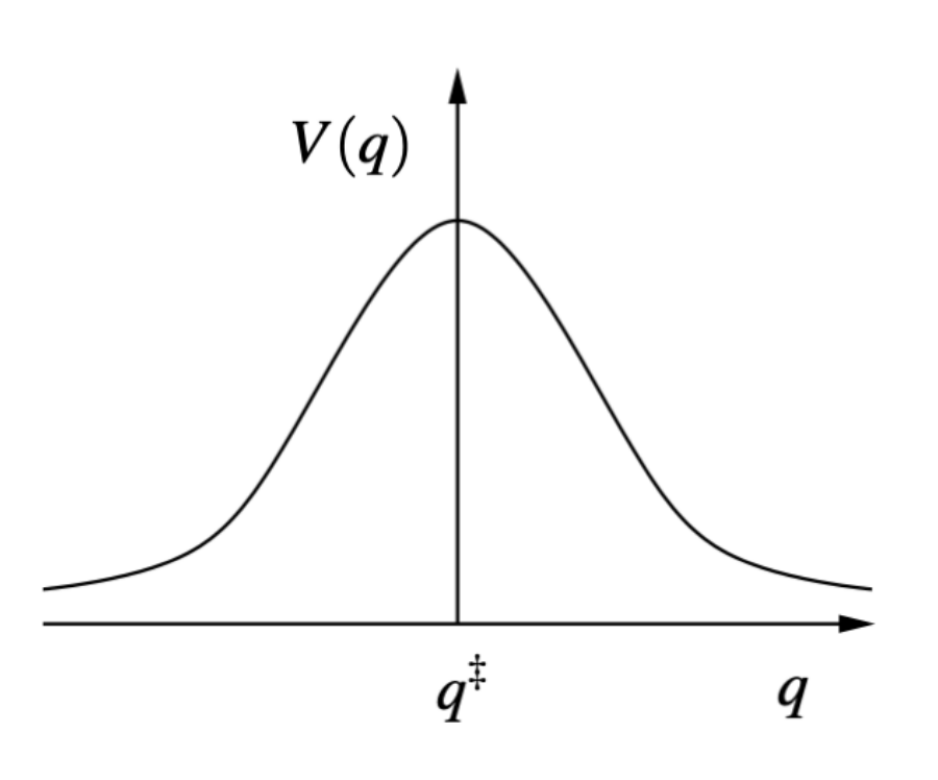}}
 \centering
 \caption{A model one-dimensional barrier transmission problem.}
 \label{RateFig1}
\end{figure}

\begin{exercise}[label={ex:rate-exercise1},title={Quantum rate theory}]

\Question 
Show that $k(T)$ in Eq.~\ref{RateEq2} has the correct dimensions for a bimolecular rate  constant in one-dimensional space, in which concentrations are measured in terms of the number of particles per unit length.
\Question 
Show that the flux operator in Eq.~\ref{RateEq4} can be written equivalently as
\begin{equation}
\hat{F} = {1\over 2m}\left[\hat{p}\,\delta(\hat{q}-q^{\ddagger})+\delta(\hat{q}-q^{\ddagger})\,\hat{p}\right].\label{RateEq6}
\end{equation}
\Question 
Use the definition of $\hat{F}$ in Eq.~\ref{RateEq4} to prove that 
\begin{equation}
\lim_{t\to 0} \bar{c}_{fs}(t)=0.\label{RateEq7}
\end{equation}
\Question 
Show that $k(T)$ in Eq.~\ref{RateEq2} can be written equivalently as
\begin{equation}
k(T) = {1\over Q_r(T)}\int_0^{\infty} \bar{c}_{ff}(t)\,{\mathrm{d}}t,\label{RateEq8}
\end{equation}
where $\bar{c}_{ff}(t)$ is the flux-flux correlation function
\begin{equation}
\bar{c}_{ff}(t) = {\text{tr}}\left[e^{-\beta\hat{H}/2}\hat{F}e^{-\beta\hat{H}/2}e^{+i\hat{H}t/\hbar}\hat{F}\,e^{-i\hat{H}t/\hbar}\right].\label{RateEq9}
\end{equation}
\Question 
Show that $\bar{c}_{ff}(t)$ is an even function of $t$, and hence that the rate constant can also be written as
\begin{equation}
k(T) = {1\over 2Q_r(T)}\int_{-\infty}^{\infty} \bar{c}_{ff}(t)\,{\mathrm{d}}t.\label{RateEq10}
\end{equation}
\Question 
Show that this the same as
\begin{equation}
k(T) = {1\over 2Q_r(T)}\int_{-\infty}^{\infty} \tilde{c}_{ff}(t)\,{\mathrm{d}}t,\label{RateEq11}
\end{equation}
where $\tilde{c}_{ff}(t)$ is the Kubo-transformed correlation function
\begin{equation}
\tilde{c}_{ff}(t) = {1\over\beta}\int_0^{\beta} {\mathrm{d}}\lambda\,{\text{tr}}\left[e^{-(\beta-\lambda)\hat{H}}\hat{F}e^{-\lambda\hat{H}}e^{+i\hat{H}t/\hbar}\hat{F}\,e^{-i\hat{H}t/\hbar}\right].\label{RateEq12}
\end{equation}
\Question  {\bfseries{(Harder!)}} Obtain an explicit expression for $\bar{c}_{fs}(t)$ in the case of free-particle motion [where $V(q)=0$], and plot it as a function of $t/\beta\hbar$ in the range $[-3,+3]$.

\end{exercise}

\section{Classical Rate Theory}\label{CRT}

In the classical limit, the trace in Eq.~\ref{RateEq3} becomes a phase-space average, the operators within the trace become functions of the phase space variables, and the quantum evolution operators are replaced by classical time evolution. The classical limit of the rate constant is thus \cite{manonotes}
\begin{equation}
k^{\text{cl}}(T) = {1\over Q_r(T)}\lim_{t\to\infty} c^{\text{cl}}_{fs}(t),\label{RateEq13}
\end{equation}
where $c^{\text{cl}}_{fs}(t)$ is the classical flux-side correlation function
\begin{equation}
c^{\text{cl}}_{fs}(t) = {1\over 2\pi\hbar}\int {\mathrm{d}}p_0\int {\mathrm{d}}q_0\,e^{-\beta H(p_0,q_0)}\delta(q_0-q^{\ddagger}){p_0\over m}\theta(q_t-q^{\ddagger}).\label{RateEq14}
\end{equation} 
Here $H(p,q)=p^2/2m+V(q)$ is the classical Hamiltonian and $q_t\equiv q_t(p_0,q_0)$ is the position at time $t$ of a classical trajectory generated by this Hamiltonian that starts out at the initial phase space point $(p_0,q_0)$ at time 0. As in the quantum case, the rate constant is obtained from a Boltzmann average of the correlation between the flux through the dividing surface at time 0 [$\delta(q_0-q^{\ddagger})p_0/m$] and the population on the product side of the dividing surface in the long-time limit [$\lim_{t\to\infty}\theta(q_t-q^{\ddagger})$], divided by the reactant partition function per unit length (which is the same in both cases).

\newpage
\begin{exercise}[label={ex:rate-exercise2},title={Classical rate theory}]
\Question 
Show that
\begin{equation}
\lim_{t\to 0_+} c^{\text{cl}}_{fs}(t) = {1\over 2\pi\beta\hbar} e^{-\beta V(q^{\ddagger})},\label{RateEq15}
\end{equation}
and hence that classical rate theory has the well-defined transition state theory limit
\begin{equation}
k_{\text{cl}}^{\text{TST}}(T) = {1\over Q_r(T)}\lim_{t\to 0_+} c_{fs}^{\text{cl}}(t) \equiv {k_{\text{B}}T\over h}{1\over Q_r(T)}e^{-V(q^{\ddagger})/k_{\text{B}}T}.\label{RateEq16}
\end{equation}
What is the analogous expression in quantum rate theory?
\Question 
Show that the classical transition state theory rate constant can be interpreted as
\begin{equation}
k_{\text{cl}}^{\text{TST}}(T) = {1\over 2}\left<|\dot{q}|\right>_{\text{cl}} e^{-\beta V(q^{\ddagger})},\label{RateEq17}
\end{equation}
where
\begin{equation}
{1\over 2}\left<|\dot{q}|\right>_{\text{cl}} = {\displaystyle{{1\over 2}\int_{-\infty}^{\infty} {\mathrm{d}}p\,{|p|\over m}e^{-\beta p^2/2m}}\over
\displaystyle{\int_{-\infty}^{\infty} {\mathrm{d}}p\, e^{-\beta p^2/2m}}}\label{RateEq18}
\end{equation}
is the thermal expectation value of the forward classical flux through the dividing surface and $e^{-\beta V(q^{\ddagger})}$ is the probability that a thermal fluctuation will bring the classical system to the top of the reaction barrier.

\end{exercise}
\section{RPMD Rate Theory}\label{RPMDRT}

As discussed in the lecture on RPMD, ring polymer molecular dynamics is simply classical molecular dynamics in the extended phase space of the ring polymer expression for the quantum mechanical partition function. So ring polymer rate theory is simply classical rate theory in this extended phase space.

For the present one-dimensional barrier transmission problem, the RPMD rate constant can be written in a variety of different ways, depending on the choice of the dividing surface between reactants and products in ring polymer coordinate space \cite{crai-mano05ajcp,crai-mano05bjcp}. The most general expression is 
\begin{equation}
k^{\text{RPMD}}(T) = {1\over Q_r(T)}\lim_{t\to\infty} c^{\text{RPMD}}_{fs}(t),\label{RateEq19}
\end{equation}
where the RPMD flux-side correlation function is defined as
\begin{equation}
c_{fs}^{\text{RPMD}}(t) = {1\over (2\pi\hbar)^P}\int {\mathrm{d}}{\mathbf{p}}_0\int {\mathrm{d}}{\mathbf{q}}_0\,e^{-\beta_PH_P({\mathbf{p}}_0,{\mathbf{q}}_0)}\dot{\theta}[s({\mathbf{q}}_0)]\theta[s({\mathbf{q}}_t)],\label{RateEq20}
\end{equation}
with
\begin{equation}
\dot{\theta}[s({\mathbf{q}})] = \delta[s({\mathbf{q}})]\sum_{j=1}^P {\partial s({\mathbf{q}})\over\partial q_j}\cdot {p_j\over m}.\label{RateEq21}
\end{equation}
Here $s({\mathbf{q}})=0$ is a dividing surface between reactants and products in the ring polymer coordinate space, and ${\mathbf{q}}_t\equiv {\mathbf{q}}_t({\mathbf{p}}_0,{\mathbf{q}}_0)$ is the coordinate at time $t$ of a trajectory generated by the classical ring polymer Hamiltonian
\begin{equation}
H_P({\mathbf{p}},{\mathbf{q}}) = \sum_{j=1}^P \left[{p_j^2\over 2m}+{1\over 2}m\omega_P^2(q_j-q_{j-1})^2+V(q_j)\right]\label{RateEq22}
\end{equation}
from the initial phase space point $({\mathbf{p}}_0,{\mathbf{q}}_0)$ at time $t=0$. [In Eq.~\ref{RateEq22}, $\omega_P=1/(\beta_P\hbar)$ with $\beta_P=\beta/P$, and $q_0\equiv q_P$, as is usual in RPMD.]

\begin{figure}[b!]
 \resizebox{1.6\columnwidth}{!} {\includegraphics{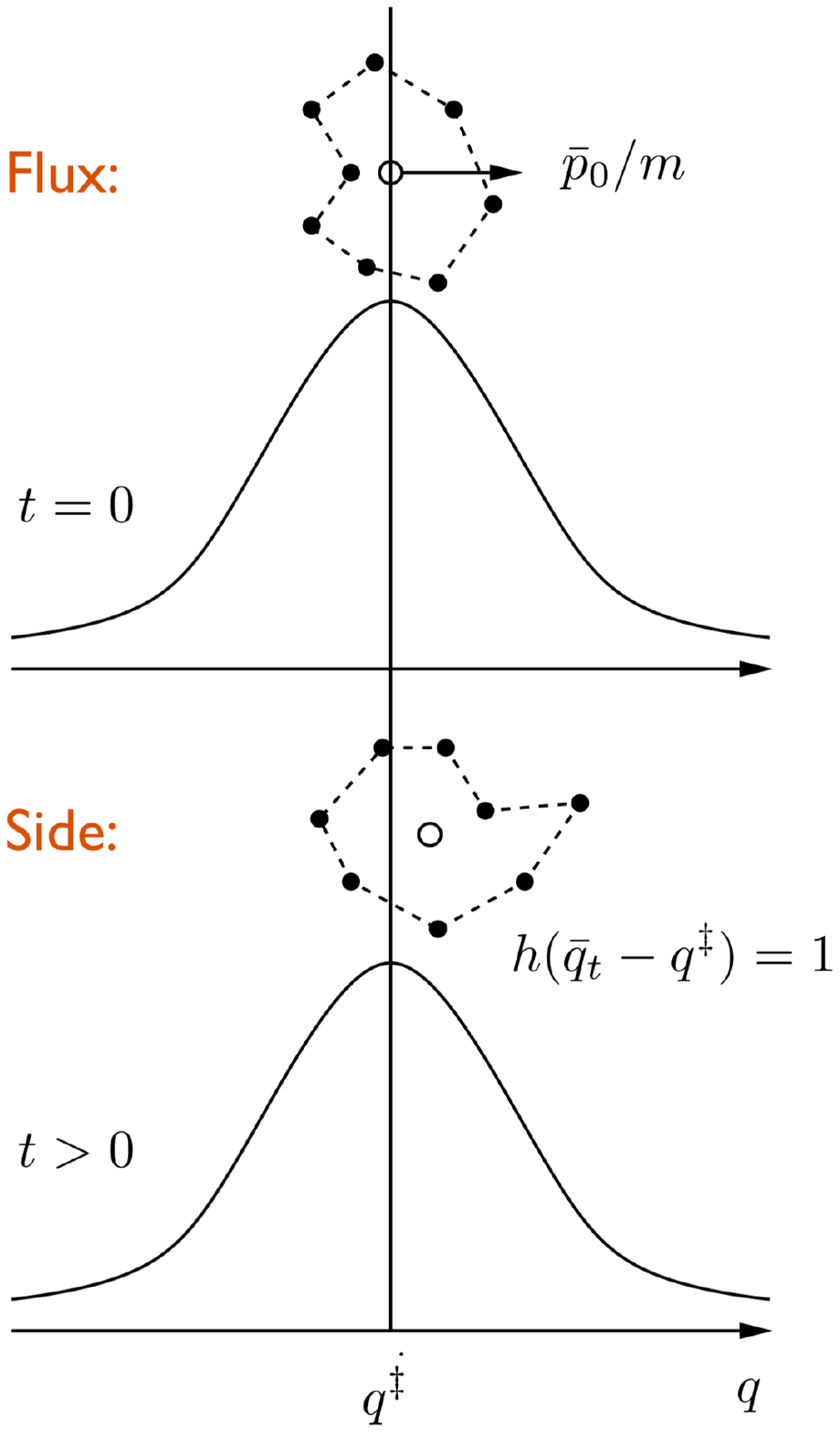}}
 \centering
 \caption{Illustration of RPMD rate theory.}
 \label{RateFig3}
 \end{figure}

Any choice of the dividing surface $s({\mathbf{q}})=0$ will give the same RPMD rate constant, provided it separates reactants from products. We are therefore free to choose the dividing surface for convenience, and the most convenient choice is to base it on the centroid 
\begin{equation}
\bar{q} = {1\over P}\sum_{j=1}^P q_j. \label{RateEq23}
\end{equation}
In particular, if we set $s({\mathbf{q}})=\bar{q}-q^{\ddagger}$, we find that Eq.~\ref{RateEq20} reduces to a form that resembles the classical flux-side correlation function in Eq.~\ref{RateEq14},
\begin{equation}
c_{fs}^{\text{RPMD}}(t) = {1\over (2\pi\hbar)^P}\int {\mathrm{d}}{\mathbf{p}}_0\int {\mathrm{d}}{\mathbf{q}}_0\,e^{-\beta_PH_P({\mathbf{p}}_0,{\mathbf{q}}_0)}\delta(\bar{q}_0-q^{\ddagger}){\bar{p}_0\over m}\theta(\bar{q}_t-q^{\ddagger}),\label{RateEq24}
\end{equation}
where $\bar{p}_0$ is the centroid momentum $\bar{p}=(1/P)\sum_{j=1}^P p_j$ evaluated at time zero. With this choice, the (thermal) flux of ring polymer centroids through the dividing surface at $t=0$ is correlated with the population of ring polymer centroids on the product side of the dividing surface at time $t$, as illustrated schematically in Fig.~\ref{RateFig3}.

The more general expression for the flux-side correlation function in Eq.~\ref{RateEq20} can be used to establish a  connection between (the optimum transition state theory approximation to) the RPMD rate and the semiclassical instanton approximation to the rate \cite{mill75jcp}, and so used to explain why RPMD works in the deep quantum tunnelling regime \cite{rich-alth09jcp}. However, the simplified expression in Eq.~\ref{RateEq24} is the one that been used in most  practical applications of RPMD rate theory, so it is this that we shall focus on in the following exercise.

\begin{exercise}[label={ex:rate-exercise3},title={RPMD rate theory}]

{\textbf {\em Preamble:}} The simplest way to do several parts of this exercise is to note that the free ring polymer Hamiltonian
\begin{equation}
h_P({\mathbf p},{\mathbf q}) = \sum_{j=1}^P \left[{p_j^2\over 2m}+{1\over 2}m\omega_P^2(q_j-q_{j-1})^2\right]\label{RateEq25}
\end{equation}
can be simplified by transforming the ring polymer momenta and coordinates into the normal mode representation
\begin{equation}
\tilde{p}_k = \sum_{j=1}^P p_jC_{jk}\quad\hbox{and}\quad
\tilde{q}_k = \sum_{j=1}^P q_jC_{jk},\label{RateEq26}
\end{equation}
where in the case of even $P$ the elements of the orthogonal transformation matrix are
\begin{equation}
C_{jk} = \begin{cases}
\sqrt{1/P}, & k=0\\
\sqrt{2/P}\cos(2\pi jk/P), & 1\le k\le P/2-1\\
\sqrt{1/P}(-1)^j, & k=P/2\\
\sqrt{2/P}\sin(2\pi jk/P), & P/2+1\le k\le P-1.\\
\end{cases}\label{RateEq27}
\end{equation}
This gives 
\begin{equation}
h_P(\tilde{\mathbf p},\tilde{\mathbf q}) = \sum_{k=0}^{P-1} \left[{\tilde{p}_k^2\over 2m}+{1\over 2}m\omega_k^2\tilde{q}_k^2\right],
\label{RateEq28}
\end{equation}
where $\omega_k=2\omega_P\sin(k\pi/P)$ for $k=0,\ldots,P-1$. Since the transformation is orthogonal, we have that 
\begin{equation}
\int {\mathrm d}{\mathbf p} \equiv \int {\mathrm d}\tilde{\mathbf p} \quad\hbox{and} \int {\mathrm d}{\mathbf q} \equiv \int {\mathrm d}\tilde{\mathbf q},\label{RateEq29}
\end{equation}
and that ${\mathbf p}\cdot{\mathbf p}=\tilde{\mathbf p}\cdot\tilde{\mathbf p}$ and ${\mathbf q}\cdot{\mathbf q}=\tilde{\mathbf q}\cdot\tilde{\mathbf q}$, both of which can help with calculations. For example, the fact that ${\mathbf q}\cdot{\mathbf q}=\tilde{\mathbf q}\cdot\tilde{\mathbf q}$ implies that the ring polymer Hamiltonian of a system with a harmonic potential $V(q) = {1\over 2}m\omega^2q^2$ can be written in the normal mode representation as
\begin{equation}
H_P(\tilde{\mathbf p},\tilde{\mathbf q}) = \sum_{k=0}^{P-1} \left[{\tilde{p}_k^2\over 2m}+{1\over 2}m(\omega_k^2+\omega^2)\tilde{q}_k^2\right].
\label{RateEq30}
\end{equation}

\Question 
Starting from the expression for $c_{fs}^{\mathrm RPMD}(t)$ in Eq.~\ref{RateEq24}, show that
\begin{equation}
\lim_{t\to 0_+} c^{\mathrm RPMD}_{fs}(t) = {1\over 2}\left<|\dot{q}|\right>_{\mathrm cl} Q(q^{\ddagger}),
\label{RateEq31}
\end{equation}
where ${1\over 2}\left<|\dot{q}|\right>_{\mathrm cl}$ is the forward classical flux through the dividing surface defined in Eq.~\ref{RateEq18} and
\begin{equation}
Q(q^{\ddagger}) = {1\over (2\pi\hbar)^P}\int {\mathrm d}{\mathbf p}\int {\mathrm d}{\mathbf q}\, e^{-\beta_PH_P({\mathbf p},{\mathbf q})}\delta(\bar{q}-q^{\ddagger})\label{RateEq32}
\end{equation}
is a centroid-constrained partition function. It follows from this that ring polymer rate theory has a well-defined transition state theory limit,
\begin{equation}
k^{\mathrm QTST}(T) = {1\over Q_r(T)}\lim_{t\to 0_+} c_{fs}^{\mathrm RPMD}(t) = {1\over 2}\left<|\dot{q}|\right>_{\mathrm cl}{Q(q^{\ddagger})\over Q_r(T)},\label{RateEq33}
\end{equation}
which is known in the literature as the \lq\lq centroid density quantum transition state theory" approximation to the rate constant \cite{voth+89jcp}.

\Question Show that if $q_r$ is in the reactant region where the potential $V(q)$ is zero, then $Q(q_r)=Q_r(T)$ (the reactant partition function per unit length).\\ 

\noindent
{\textbf {\em Hint:}} The calculation is straightforward for $P=1$, and for $P>1$ you can use the identity $\prod_{k=1}^{P-1} 2\sin(k\pi/P) = P$.

\Question Show that the factor of $\displaystyle{Q(q^{\ddagger})\over Q_r(T)}$ in Eq.~\ref{RateEq33} can be calculated as
\begin{equation}
{Q(q^{\ddagger})\over Q_r(T)} = e^{-\beta F(q^{\ddagger})} = \exp\left[-\beta{\int_{q_r}^{q^\ddagger}} F'(q)\,{\mathrm d}q\right],\label{RateEq34}
\end{equation}
where 
\begin{equation}
F'(q) = \left<{1\over P}\sum_{j=1}^P {{\mathrm d}V(q_j)\over{\mathrm d}q_j}\right>_{\bar{q}=q}\label{RateEq35}
\end{equation}
with
\begin{equation}
\left<(\cdots)\right>_{\bar{q}=q} = 
{\displaystyle{\int {\mathrm d}{\mathbf p}\int {\mathrm d}{\mathbf q}\, (\cdots)e^{-\beta_PH_P({\mathbf p},{\mathbf q})}\delta(\bar{q}-q)}\over
\displaystyle{\int {\mathrm d}{\mathbf p}\int {\mathrm d}{\mathbf q}\, e^{-\beta_PH_P({\mathbf p},{\mathbf q})}\delta(\bar{q}-q)}}\label{RateEq36}
\end{equation}
is a standard PIMD average with the centroid of the ring polymer constrained at $q$.

\Question
Show that the RPMD rate constant can be written as
\begin{equation}
k^{\mathrm RPMD}(T) = k^{\mathrm QTST}(T)\lim_{t\to\infty}\kappa(t),\label{RateEq37}
\end{equation}
where
\begin{equation}
\kappa(t) = \sqrt{2\pi\beta m}\left<{\bar{p}_0\over m}\theta(\bar{q}_t-q^{\ddagger})\right>_{\bar{q}_0=q^{\ddagger}}\label{RateEq38}
\end{equation}
is a time-dependent transmission coefficient that allows for recrossing of the transition state dividing surface.\\

Once $k^{\mathrm QTST}(T)$ has been calculated, one thus simply runs RPMD trajectories initialised in the constrained ensemble with $\bar{q}_0=q^{\ddagger}$ and counts those that go on to form products with weights of $\sqrt{2\pi\beta m}\,(\bar{p}_0/m)$ to obtain $k^{\mathrm RPMD}(T)$. If $k^{\mathrm QTST}(T)$ is calculated by thermodynamic integration as in Eq.~\ref{RateEq34}, the overall scheme is known as the Bennett-Chandler method \cite{benn77acs,chan78jcp}. This works well even for strongly activated reactions in which reaching the transition state is a rare event [i.e, for which $Q(q^{\ddagger})/Q_r(T)\sim e^{-\beta V(q^{\ddagger})}$ is exponentially small], and it is the basis of all practical RPMD rate theory calculations.

\Question
{{\textbf {(Harder!)}}} Show that, at temperatures above the instanton crossover temperature $T_{\mathrm c}=\hbar\omega_b/2\pi k_{\mathrm B}$, RPMD rate theory gives the exact quantum mechanical result 
\begin{equation}
k(T)Q_r(T) = {1\over (2\pi\beta\hbar)}{(\beta\hbar\omega_b/2)\over \sin(\beta\hbar\omega_b/2)}\label{RateEq39}
\end{equation}
for the thermal rate constant of a parabolic barrier $V(q)=-{1\over 2}m\omega_b^2q^2$,
in the limit as the number of beads $P\to\infty$.\\

This suggests that RPMD rate theory will also correctly capture the \lq\lq shallow tunnelling" through the parabolic tip of the reaction barrier in more general situations. Below $T_{\mathrm c}$, the parabolic approximation to the reaction barrier breaks down, and the tunnelling enters the \lq\lq instanton" regime. Here the analysis becomes more complicated, but  RPMD is also expected to provide a good approximation to the quantum mechanical rate constant by virtue of its connection to semiclassical instanton theory. In particular, Richardson and Althorpe have shown that there is a direct connection between the optimum transition state theory approximation to the RPMD rate constant (obtained by optimising the dividing surface $s({\mathbf q})=0$ in Eq.~\ref{RateEq20}) and the rate constant given by the \lq\lq Im F" version of semiclassical instanton theory \cite{rich-alth09jcp}.
\end{exercise}

\section{Additional Comments and Further Reading}

Everything we have said above is straightforward to generalise to gas phase bimolecular reactions in their full dimensionality, without making any reduced dimensionality approximations or assuming (as is commonly done in more approximate theories) that the rotations and vibrations of the transition state can be treated separately. The resulting theory automatically includes zero-point energy and tunnelling effects, both of which are known to be important in reactions involving hydrogen atoms \cite{lawr-mano20fd}. The implementation of RPMD rate theory for gas phase atom-diatom reactions is described in~\cite{coll+09jcp}, and the generalisation to polyatomic reactions in~\cite{sule+11jcp}. This generalisation has since been implemented in a freely available computer program \cite{sule+13cpc} that has already been used to calculate the rates of chemical reactions containing as many as nine atoms. We expect that there will be many more applications as machine-learned {\em ab initio} potential energy surfaces for polyatomic reactions become more widely available. 

RPMD rate theory can also be applied to condensed phase reactions, and indeed the first such application (to a system-bath model for condensed phase proton transfer) was reported in the first RPMD rate theory paper \cite{crai-mano05ajcp}. For a unimolecular reaction, $Q_r(T)$ becomes the full reactant partition function
\begin{equation}
Q_r(T) = {\text{tr}}\left[e^{-\beta\hat{H}}\theta[-s(\hat{\mathbf{q}})]\right]\label{RateEq40}
\end{equation}
rather than the reactant partition function per unit volume, and the RPMD rate is typically calculated after a \lq\lq plateau time" $t_{\text{p}}$ rather than in the limit as $t\to \infty$: 
\begin{equation}
k^{\text{RPMD}}(T) = {c_{fs}^{\text{RPMD}}(t_{\text{p}})\over Q_r(T)}.\label{RateEq41}
\end{equation}
Here the plateau time is supposed to be longer than the time taken for a typical reactive trajectory initiated at the transition state to leave the barrier region, but short compared with the timescale for the onset of the reverse reaction \cite{chan78jcp}. When this separation of time scales is difficult to achieve, one can use the more general expression \cite{lawr+19jcp}
\begin{equation}
k^{\text{RPMD}}(T) = {Q_r(T)^{-1}c_{fs}^{\text{RPMD}}(t_{\text{p}})\over 1-\left[Q_r(T)^{-1}+Q_p(T)^{-1}\right]\int_0^{t_{\text{p}}} c_{fs}^{\text{RPMD}}(t)\,{\mathrm{d}}t},\label{RateEq42}
\end{equation}
in which $Q_p(T)$ is the product partition function
\begin{equation}
Q_p(T) = {\text{tr}}\left[e^{-\beta\hat{H}}\theta[s(\hat{\mathbf{q}})]\right]\label{RateEq43}
\end{equation}
and the denominator in Eq.~\ref{RateEq42} accounts for the onset of the reverse reaction. Here $t_{\text{p}}$ simply has to be sufficiently large for the right-hand side of Eq.~\ref{RateEq42} to have reached a plateau value.

This condensed phase version of RPMD rate theory has also seen numerous applications, ranging from early studies of proton transfer in a polar solvent \cite{coll+08jcp}, and the hopping of muonium atoms between the cavities in hexagonal ice \cite{mark+08jcp}, to a fully atomistic simulation of enzyme-catalysed hydride transfer in aqueous dihydrofolate reductase \cite{boek+11pnas}. We expect that it too will see many more applications as accurate machine-learned potentials for reactive condensed phase systems become more widely available.

\section*{Exercise answers}
\begin{Answer}[ref=ex:rate-exercise1]

\begin{enumerate}

\item
The second order rate equation for a bimolecular reaction is
\begin{equation*}
-{{\mathrm{d}}c_{\text{A}}\over {\mathrm{d}}t} = kc_{\text{A}}c_{\text{B}}.
\end{equation*}
In one dimension where concentrations are in particles per unit length, this gives $[k]=$ LT$^{-1}$. From Eq.~\ref{RateEq4}, $[\hat{F}]={\text{T}}^{-1}$, and since $\hat{\theta}$ is dimensionless, $[c_{fs}(t)]={\text{T}}^{-1}$. The dimensions of $Q_r(T)$ (the reactant partition function per unit length) are clearly L$^{-1}$, and so Eq.~\ref{RateEq2} gives $[k]= $LT$^{-1}$ as required.

\item
Since $\hat{\theta} = \theta(\hat{q}-q^{\ddagger})$ commutes with $V(\hat{q})$,
\begin{equation*}
\hat{F} = {i\over\hbar}\left[\hat{H},\hat{\theta}\right]
={i\over 2m\hbar}\left[\hat{p}^2,\hat{\theta}\right] = {i\over 2m\hbar}\left(\hat{p}\left[\hat{p},\hat{\theta}\right]+\left[\hat{p},\hat{\theta}\right]\hat{p}\right).
\end{equation*}
In the position representation, $\hat{p}=-i\hbar\displaystyle{\partial\over \partial q}$,  $\hat{\theta} = \theta(q-q^{\ddagger})$, and therefore $\left[\hat{p},\hat{\theta}\right]=-i\hbar\,\delta(q-q^{\ddagger})$, giving
\begin{equation*}
 \hat{F} = {1\over 2m}\left[\hat{p}\,\delta(\hat{q}-q^{\ddagger})+\delta(\hat{q}-q^{\ddagger})\,\hat{p}\right].
 \end{equation*}
 
\item
 From Eqs.~\ref{RateEq3} and \ref{RateEq4},
 \begin{align*}
\bar{c}_{fs}(0) &= {\text{tr}}\left(e^{-\beta\hat{H}/2}\hat{F}e^{-\beta\hat{H}/2}\hat{\theta}\right)\\
 &= {i\over\hbar}{\text{tr}}\left(e^{-\beta\hat{H}/2}\left[\hat{H},\hat{\theta}\right]e^{-\beta\hat{H}/2}\hat{\theta}\right)\\
 &= {i\over\hbar}{\text{tr}}\left(e^{-\beta\hat{H}/2}\hat{H}\hat{\theta}\,e^{-\beta\hat{H}/2}\hat{\theta}-e^{-\beta\hat{H}/2}\hat{\theta}\hat{H}e^{-\beta\hat{H}/2}\hat{\theta}\right)\\
 &= {i\over\hbar}{\text{tr}}\left(\hat{H}e^{-\beta\hat{H}/2}\hat{\theta}\,e^{-\beta\hat{H}/2}\hat{\theta}-\hat{H}e^{-\beta\hat{H}/2}\hat{\theta}\,e^{-\beta\hat{H}/2}\hat{\theta}\right)\\
 &=0,
 \end{align*}
 where we have used the fact that $e^{-\beta\hat{H}/2}$ commutes with $\hat{H}$ in the first term and the fact that the trace is invariant to a cyclic permutation of the operators within it in the second.
 
\item
 The flux-side and flux-flux correlation functions in Eqs.~\ref{RateEq3} and~\ref{RateEq9} are clearly related by
 \begin{align*}
 {{\mathrm{d}}\over {\mathrm{d}}t}\bar{c}_{fs}(t) &= {{\mathrm{d}}\over {\mathrm{d}}t}{\text{tr}}\left[e^{-\beta\hat{H}/2}\hat{F}e^{-\beta\hat{H}/2}e^{+i\hat{H}t/\hbar}\hat{\theta}\,e^{-i\hat{H}t/\hbar}\right]\\
 &={\text{tr}}\left[e^{-\beta\hat{H}/2}\hat{F}e^{-\beta\hat{H}/2}e^{+i\hat{H}t/\hbar}{i\over\hbar}\left[\hat{H},\hat{\theta}\right]\,e^{-i\hat{H}t/\hbar}\right]\\
&= {\text{tr}}\left[e^{-\beta\hat{H}/2}\hat{F}e^{-\beta\hat{H}/2}e^{+i\hat{H}t/\hbar}\hat{F}\,e^{-i\hat{H}t/\hbar}\right]\\
&=\bar{c}_{ff}(t).
\end{align*}
So
\begin{align*}
\int_0^{\infty} \bar{c}_{ff}(t)\,{\mathrm{d}}t &= \int_0^{\infty} {{\mathrm{d}}\bar{c}_{fs}(t)\over {\mathrm{d}}t}\,{\mathrm{d}}t = \lim_{t\to\infty} \bar{c}_{fs}(t)-\bar{c}_{fs}(0)=\lim_{t\to\infty}\bar{c}_{fs}(t),
\end{align*}
where we have used the result from part~(c) in the final step. Hence
\begin{align*}
k(T) &= {1\over Q_r(T)}\lim_{t\to\infty} \bar{c}_{fs}(t) = {1\over Q_r(T)}\int_0^{\infty} \bar{c}_{ff}(t)\,{\mathrm{d}}t.
\end{align*}

\item
The proof that $\bar{c}_{ff}(t)$ is an even function of $t$ uses the commutivity of the evolution and Boltzmann operators and the invariance of the trace to a cyclic permutation. From Eq.~\ref{RateEq9},
\begin{align*}
\bar{c}_{ff}(-t) &= {\text{tr}}\left[e^{-\beta\hat{H}/2}\hat{F}\,e^{-\beta\hat{H}/2}e^{-i\hat{H}t/\hbar}\hat{F}\,e^{+i\hat{H}t/\hbar}\right]\\
&= {\text{tr}}\left[e^{+i\hat{H}t/\hbar}\hat{F}\,e^{-i\hat{H}t/\hbar}e^{-\beta\hat{H}/2}\hat{F}\,e^{-\beta\hat{H}/2}\right]\\
&= {\text{tr}}\left[e^{-\beta\hat{H}/2}\hat{F}\,e^{-\beta\hat{H}/2}e^{+i\hat{H}t/\hbar}\hat{F}\,e^{-i\hat{H}t/\hbar}\right]\\
&= \bar{c}_{ff}(t).
\end{align*}
Given this, it is clear that
\begin{equation*}
k(T) = {1\over Q_r(T)}\int_0^{\infty} \bar{c}_{ff}(t)\,{\mathrm{d}}t = {1\over 2Q_r(T)}\int_{-\infty}^{\infty} \bar{c}_{ff}(t)\,{\mathrm{d}}t.
\end{equation*}

\item
The symmetrically thermalised correlation function $\bar{c}_{ff}(t)$ and the Kubo-transformed correlation function $\tilde{c}_{ff}(t)$ can both be related to the standard quantum mechanical correlation function
\begin{equation*}
c_{ff}(t) = {\text{tr}}\left[e^{-\beta\hat{H}}\hat{F}\,e^{+i\hat{H}t/\hbar}\hat{F}\,e^{-i\hat{H}t/\hbar}\right],
\end{equation*}
which is an analytic function of its argument throughout the strip $0\le {\mathrm{Im}}(t)\le \beta\hbar$ in the complex $t$ plane. [It is undefined outside this strip because the matrix elements of the Boltzmann operator are undefined at negative temperatures unless the spectrum of the Hamiltonian is bounded from above, which is not the case for the Hamiltonian $\hat{H}=\hat{p}^2/2m+V(\hat{q})$.] The relationships are 
\begin{align*}
\bar{c}_{ff}(t) &= {\text{tr}}\left[e^{-\beta\hat{H}/2}\hat{F}e^{-\beta\hat{H}/2}e^{+i\hat{H}t/\hbar}\hat{F}\,e^{-i\hat{H}t/\hbar}\right]\\
&={\text{tr}}\left[e^{-\beta\hat{H}}\hat{F}e^{+i\hat{H}(t+i\beta\hbar/2)}\hat{F}e^{-i\hat{H}(t+i\beta\hbar/2)/\hbar}\right]\\
&=c_{ff}(t+i\beta\hbar/2)
\end{align*}
and
\begin{align*}
\tilde{c}_{ff}(t) &= {1\over\beta}\int_0^{\beta} {\mathrm{d}}\lambda\,{\text{tr}}\left[e^{-(\beta-\lambda)\hat{H}}\hat{F}e^{-\lambda\hat{H}}e^{+i\hat{H}t/\hbar}\hat{F}\,e^{-i\hat{H}t/\hbar}\right]\\
&={1\over\beta}\int_0^{\beta} {\mathrm{d}}\lambda\,{\text{tr}}\left[e^{-\beta\hat{H}}\hat{F}e^{+i\hat{H}(t+i\lambda\hbar)}\hat{F}e^{-i\hat{H}(t+i\lambda\hbar)/\hbar}\right]\\
&={1\over\beta}\int_0^{\beta} {\mathrm{d}}\lambda\,c_{ff}(t+i\lambda\hbar).
\end{align*}

It follows from these relationships and the analyticity of $c_{ff}(t)$ that the time integrals from $-\infty$ to $\infty$ of all three correlation functions are the same. For any $\lambda$ between 0 and $\beta$,
\begin{align*}
\int_{-\infty}^{\infty} c_{ff}(t+i\lambda\hbar)\,{\mathrm{d}}t &= \int_{-\infty+i\lambda\hbar}^{\infty+i\lambda\hbar} c_{ff}(\tau)\,{\mathrm{d}}\tau\\ &= \int_{-\infty}^{\infty} c_{ff}(\tau)\,{\mathrm{d}}\tau\\ &\equiv \int_{-\infty}^{\infty} c_{ff}(t)\,{\mathrm{d}}t,
\end{align*}
where we have changed the integration variable to $\tau=t+i\lambda\hbar$ in the first line, deformed the integration contour down to the real $\tau$ axis in the second, and noted that the boundary terms vanish because $c_{ff}(\tau)\to 0$ as ${\mathrm{Re}}(\tau)\to \pm\infty$. (If this were not the case the rate constants in Eqs.~\ref{RateEq10} and \ref{RateEq11} would be undefined.)

Hence 
\begin{align*}
\int_{-\infty}^{\infty} \bar{c}_{ff}(t)\,{\mathrm{d}}t = \int_{-\infty}^{\infty} \tilde{c}_{ff}(t)\,{\mathrm{d}}t,
\end{align*}
and we can equally well write the rate constant in terms of either correlation function.

\item
In the case of free particle motion, $\hat{H}={\hat{p}^2/2m}$, and using Eq.~\ref{RateEq6} for the flux operator gives
\begin{align*}
\bar{c}_{fs}(t) &= {\text{tr}}\left[e^{-\beta\hat{H}/2}\hat{F}e^{-\beta\hat{H}/2}e^{+i\hat{H}t/\hbar}\hat{\theta}\,e^{-i\hat{H}t/\hbar}\right]\\
&={\text{tr}}\left[\hat{F}e^{+i\hat{H}\tau/\hbar}\hat{\theta}\,e^{-i\hat{H}\tau^*/\hbar}\right]\\
&={1\over 2m}{\text{tr}}\left[\left[\hat{p}\,\delta(\hat{q}-q^{\ddagger})+\delta(\hat{q}-q^{\ddagger})\hat{p}\right]e^{+i\hat{p}^2\tau/2m\hbar}\theta(\hat{q}-q^{\ddagger})e^{-i\hat{p}^2\tau^*/2m\hbar}\right],
\end{align*}
where $\tau=t+i\beta\hbar/2$. Evaluating the trace in the basis of coordinate eigenstates gives
\begin{align*}
\bar{c}_{fs}(t) &= {1\over 2m}\int_{q^{\ddagger}}^{\infty} {\mathrm{d}}q \left[\bigl<q^{\ddagger}\bigr|e^{+i\hat{p}^2\tau/2m\hbar}\bigl|q\bigr>\bigl<q\bigr|e^{-i\hat{p}^2\tau^*/2m\hbar}\hat{p}\bigl|q^{\ddagger}\bigr>+{\text{c}.c.}\right]\\
&={1\over m}{\mathrm{Re}}\int_{q^{\ddagger}}^{\infty}{\mathrm{d}}q\, \bigl<q^{\ddagger}\bigr|e^{+i\hat{p}^2\tau/2m\hbar}\bigl|q\bigr>\bigl<q\bigr|e^{-i\hat{p}^2\tau^*/2m\hbar}\hat{p}\bigl|q^{\ddagger}\bigr>.
\end{align*}
The required coordinate matrix elements are
\begin{align*}
\bigl<q^{\ddagger}\bigr|e^{+i\hat{p}^2\tau/2m\hbar}\bigl|q\bigr> &= \sqrt{m\over -i2\pi\hbar\tau}e^{-im(q^{\ddagger}-q)^2/2\hbar\tau}\\
\bigl<q\bigr|e^{-i\hat{p}^2\tau^*/2m\hbar}\hat{p}\bigl|q^{\ddagger}\bigr> &= \sqrt{m\over +i2\pi\hbar\tau^*}{m(q-q^{\ddagger})\over\tau^*}e^{+im(q-q^{\ddagger})^2/2\hbar\tau^*}
\end{align*}
and therefore
\begin{align*}
\bar{c}_{fs}(t) &= {m\over 2\pi\hbar|\tau|}{\mathrm{Re}}\int_{q^{\ddagger}}^{\infty} {\mathrm{d}}q\,{(q-q^{\ddagger})\over\tau^*}e^{-\beta m(q-q^{\ddagger})^2/2|\tau|^2}\\
&= {m\over 2\pi\hbar|\tau|}{\mathrm{Re}} \left({\tau\over\beta m}\right)\\
&= {1\over 2\pi\beta\hbar}{(t/\beta\hbar)\over\sqrt{(t/\beta\hbar)^2+1/4}}.
\end{align*}

Note that this is a real, odd, and continuous function of $t$, which passes through zero at $t=0$ and reaches its long-time plateau value within a few multiples of the thermal time $\beta\hbar$. These features are not unique to the free motion case: they are generic features of the exact quantum mechanical flux-side correlation $\bar{c}_{fs}(t)$ when the location of the dividing surface ($q^{\ddagger}$) is chosen appropriately.
 \end{enumerate}
\end{Answer}
\begin{figure}[htb!]
 \resizebox{1.6\columnwidth}{!} {\includegraphics{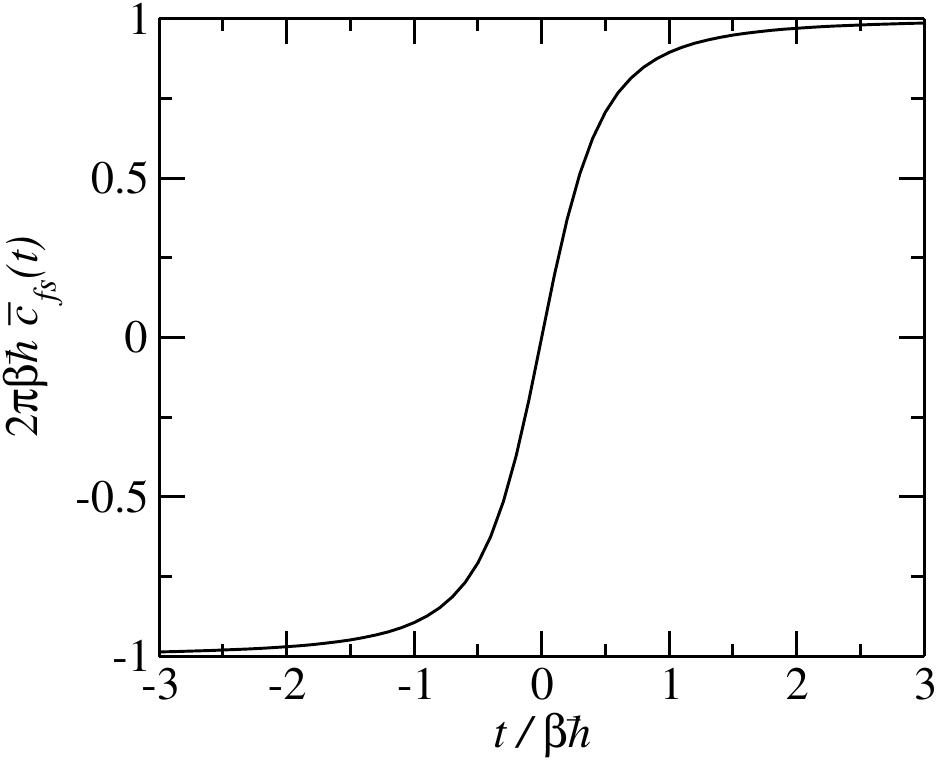}}
 \centering
 \caption{Quantum mechanical $\bar{c}_{fs}(t)$ for free motion.}
 \label{RateFig2}
 \end{figure}

\vspace{0.5cm}

\begin{Answer}[ref=ex:rate-exercise2]

\begin{enumerate}
\item
To obtain the short time limit of $c^{\text{cl}}_{fs}(t)$ in Eq.~\ref{RateEq14}, first note that
\begin{align*}
\lim_{t\to 0_+} \delta(q_0-q^{\ddagger})\theta(q_t-q^{\ddagger})=\delta(q_0-q^{\ddagger})\theta(p_0).
\end{align*} 
So 
\begin{align*}
\lim_{t\to 0_+} c^{\text{cl}}_{fs}(t) &= {1\over 2\pi\hbar}\int {\mathrm{d}}p_0\int {\mathrm{d}}q_0\,e^{-\beta H(p_0,q_0)}\delta(q_0-q^{\ddagger}){p_0\over m}\theta(p_0)\\
&={1\over 2\pi\hbar}\int_0^{\infty} {\mathrm{d}}p_0\,{p_0\over m}e^{-\beta p_0^2/2m}\,\int_{-\infty}^{\infty} {\mathrm{d}}q_0\,e^{-\beta V(q_0)}\delta(q_0-q^{\ddagger})\\
&={1\over 2\pi\hbar}\cdot{1\over\beta}\cdot e^{-\beta V(q^{\ddagger})},
\end{align*}
which gives the standard (textbook) transition state theory result for our model one-dimensional barrier transmission problem:
\begin{equation*}
k_{\text{cl}}^{\text{TST}}(T) = {1\over Q_r(T)}\lim_{t\to 0_+}c_{fs}^{\text{cl}}(t) \equiv {k_{\text{B}}T\over h}{1\over Q_r(T)}e^{-V(q^{\ddagger})/k_{\text{B}}T}.
\end{equation*}

There is no direct analogue of this result in the quantum case because $\lim_{t\to 0_+}\bar{c}_{fs}(t)=0$, as we have shown in Exercises~\ref{ex:rate-exercise1} (3) and (7).

\item
The two elementary integrals in this question give
\begin{equation*}
{1\over 2}\left<|\dot{q}|\right>_{\text{cl}} = {\displaystyle{{1\over 2}\int_{-\infty}^{\infty} {\mathrm{d}}p\,{|p|\over m}e^{-\beta p^2/2m}}\over
\displaystyle{\int_{-\infty}^{\infty} {\mathrm{d}}p\, e^{-\beta p^2/2m}}} = {\displaystyle{1/\beta}\over \displaystyle{\sqrt{2\pi m/\beta}}}={1\over\sqrt{2\pi\beta m}},
\end{equation*}
and combining this with the fact that $Q_r(T)=\sqrt{m/2\pi\beta\hbar^2}$ gives
\begin{equation*}
k_{\text{cl}}^{\text{TST}}(T) = {1\over 2\pi\beta\hbar Q_r(T)}e^{-\beta V(q^{\ddagger})} \equiv {1\over 2}\left<|\dot{q}|\right>_{\text{cl}}e^{-\beta V(q^{\ddagger})},
\end{equation*}
the interpretation of which is given in the question.

\end{enumerate}
\end{Answer}

\begin{Answer}[ref=ex:rate-exercise3]
\begin{enumerate}
\item 
Beginning as we did in the answer to Exercise~\ref{ex:rate-exercise2} 1, we have
\begin{align*}
\lim_{t\to 0_+}c_{fs}^{\text{RPMD}}(t) &= {1\over (2\pi\hbar)^P}\int {\mathrm{d}}{\mathbf{p}}_0\int {\mathrm{d}}{\mathbf{q}}_0\,e^{-\beta_PH_P({\mathbf{p}}_0,{\mathbf{q}}_0)}\delta(\bar{q}_0-q^{\ddagger}){\bar{p}_0\over m}\theta(\bar{p}_0)\\
&\equiv {1\over 2}\left<|\dot{\bar{q}}|\right>_{\text{RPMD}}Q(q^{\ddagger}),
\end{align*}
where we have defined
\begin{equation*}
{1\over 2}\left<|\dot{\bar{q}}|\right>_{\text{RPMD}} = {\displaystyle{{1\over 2}\int {\mathrm{d}}{\mathbf{p}}\,{|\bar{p}|\over m}e^{-\beta_P |{\mathbf{p}}|^2/2m}}\over
\displaystyle{\int {\mathrm{d}}{\mathbf{p}}\,  e^{-\beta_P |{\mathbf{p}}|^2/2m}}}
\end{equation*}
and
\begin{equation*}
Q(q^{\ddagger}) = {1\over (2\pi\hbar)^P}\int {\mathrm{d}}{\mathbf{p}}\int {\mathrm{d}}{\mathbf{q}}\, e^{-\beta_PH_P({\mathbf{p}},{\mathbf{q}})}\delta(\bar{q}-q^{\ddagger}).
\end{equation*}
To show that this is the same as Eq.~\ref{RateEq31}, we therefore have to show that ${1\over 2}\left<|\dot{\bar{q}}|\right>_{\text{RPMD}} = {1\over 2}\left<|\dot{q}|\right>_{\text{cl}}$. This can be done by transforming both the numerator and the denominator of ${1\over 2}\left<|\dot{\bar{q}}|\right>_{\text{RPMD}}$ into the ring polymer normal mode representation, and noting that $\beta_P=\beta/P$ and $\bar{p}=\tilde{p}_0/\sqrt{P}$: 
\begin{align*}
{1\over 2}\left<|\dot{\bar{q}}|\right>_{\text{RPMD}} &= {\displaystyle{{1\over 2}
\int {\mathrm{d}}\tilde{\mathbf{p}}\,{|\bar{p}|\over m}e^{-\beta_P|\tilde{\mathbf{p}}|^2/2m}}\over
\displaystyle{\int {\mathrm{d}}\tilde{\mathbf{p}}\,e^{-\beta_P |\tilde{\mathbf{p}}|^2/2m}}}\\
&={\displaystyle{{1\over 2}\int_{-\infty}^{\infty} {\mathrm{d}}\tilde{p}_0\,{|\bar{p}|\over m}e^{-\beta_P\tilde{p}_0^2/2m}}\over
\displaystyle{\int_{-\infty}^{\infty} {\mathrm{d}}\tilde{p}_0\,e^{-\beta_P\tilde{p}_0^2/2m}}}\\
&={\displaystyle{{1\over 2}\int_{-\infty}^{\infty} {\mathrm{d}}\bar{p}\,{|\bar{p}|\over m}e^{-\beta\bar{p}^2/2m}}\over
\displaystyle{\int_{-\infty}^{\infty} {\mathrm{d}}\bar{p}\,e^{-\beta\bar{p}^2/2m}}}={1\over 2}\left<|\dot{q}|\right>_{\text{cl}}.
\end{align*}
\item

Since $q_r$ is in the reactant region where $V(q)=0$, we can replace the ring-polymer Hamiltonian $H_P({\mathbf{p}},{\mathbf{q}})$ with the free ring polymer Hamiltonian $h_P({\mathbf{p}},{\mathbf{q}})$ in the expression for $Q(q_r)$ [i.e., in Eq.~\ref{RateEq32} with $q^{\ddagger}\to q_r$]. And since $\bar{q}=\tilde{q}_0/\sqrt{P}$ and $\delta(\tilde{q}_0/\sqrt{P}-q_r)=\sqrt{P}\delta(\tilde{q}_0-q_r\sqrt{P})$, this gives
\begin{align*}
Q(q) &= {1\over (2\pi\hbar)^P}\int {\mathrm{d}}{\mathbf{p}}\int {\mathrm{d}}{\mathbf{q}}\, e^{-\beta_Ph_P({\mathbf{p}},{\mathbf{q}})}\delta(\bar{q}-q)\\
&= {1\over (2\pi\hbar)^P}\int {\mathrm{d}}\tilde{\mathbf{p}}\int {\mathrm{d}}\tilde{\mathbf{q}}\, e^{-\beta_Ph_P(\tilde{\mathbf{p}},\tilde{\mathbf{q}})}\sqrt{P}\delta(\tilde{q}_0-q\sqrt{P})\\
&= {1\over (2\pi\hbar)^P}\left({2\pi m\over\beta_P}\right)^{P/2}\sqrt{P}\left({2\pi\over\beta_Pm}\right)^{(P-1)/2}\prod_{k=1}^{P-1}{1\over \omega_k}\\
&=\sqrt{m\over 2\pi\beta\hbar^2}\, P\prod_{k=1}^{P-1} {1\over 2\sin(k\pi/P)}\\
&= Q_r(T),
\end{align*}
where we have used $\beta_P=\beta/P$, $\omega_k=(2/\beta_P\hbar)\sin(k\pi/P)$, and the identity $\prod_{k=1}^{P-1} 2\sin(k\pi/P)=P$ to get to the bottom line.

\item

First note that, since $\bar{q}=\tilde{q}_0/\sqrt{P}$,
\begin{align*}
{{\mathrm{d}}\over {\mathrm{d}}\bar{q}} = \sqrt{P}{\partial\over\partial\tilde{q}_0} =\sqrt{P}\sum_{j=1}^P {\partial q_j\over\partial\tilde{q}_0}{\partial\over\partial q_j} = \sqrt{P}\sum_{j=1}^P C_{j0}{\partial\over\partial q_j}=\sum_{j=1}^P {\partial\over\partial q_j},
\end{align*}
where we have used the orthogonality of the bead to normal mode transformation to write
\begin{align*}
q_{j} = \sum_{k=0}^{P-1} C_{jk}\tilde{q}_k
\end{align*}
and then used the expression for $C_{j0}$ in Eq.~\ref{RateEq27}. Note also that the harmonic spring terms in the ring polymer Hamiltonian
\begin{align*}
\sum_{j=1}^P {1\over 2}m\omega_P^2(q_j-q_{j-1})^2 = \sum_{k=0}^{P-1} {1\over 2}m\omega_k^2\tilde{q}_k^2
\end{align*}
are independent of $\tilde{q}_0$ (and therefore of $\bar{q})$ because $\omega_0=0$.

In view of these observations, the derivative of the free energy $F(q) = -\displaystyle{1\over\beta}\ln {Q(q)\over Q_r(T)}$ can be calculated as 
\begin{align*}
F'(q) &= -{1\over \beta Q(q)}{{\mathrm{d}}Q(q)\over {\mathrm{d}}q}\\
&= -{1\over 2\pi\hbar\beta Q(q)}\int {\mathrm{d}}{\mathbf{p}}_0\int {\mathrm{d}}{\mathbf{q}}_0\,e^{-\beta_PH_P({\mathbf{p}}_0,{\mathbf{q}}_0)} {{\mathrm{d}}\over {\mathrm{d}}q}\delta(q-\bar{q})\\
&= -{1\over 2\pi\hbar\beta Q(q)}\int {\mathrm{d}}{\mathbf{p}}_0\int {\mathrm{d}}{\mathbf{q}}_0\,\delta(q-\bar{q}){{\mathrm{d}}\over{\mathrm{d}}\bar{q}}\,e^{-\beta_PH_P({\mathbf{p}}_0,{\mathbf{q}}_0)}\\
&= {1\over 2\pi\hbar\beta Q(q)}\int {\mathrm{d}}{\mathbf{p}}_0\int {\mathrm{d}}{\mathbf{q}}_0\,\delta(q-\bar{q})\left[\beta_P\sum_{j=1}^P {{\mathrm{d}}V(q_j)\over {\mathrm{d}}q_j}\right]e^{-\beta_PH_P({\mathbf{p}}_0,{\mathbf{q}}_0)}\\
&= {1\over 2\pi\hbar Q(q)}\int {\mathrm{d}}{\mathbf{p}}_0\int {\mathrm{d}}{\mathbf{q}}_0\,\left[{1\over P}\sum_{j=1}^P {{\mathrm{d}}V(q_j)\over {\mathrm{d}}q_j}\right]e^{-\beta_PH_P({\mathbf{p}}_0,{\mathbf{q}}_0)}\delta(q-\bar{q})\\
&\equiv \left<{1\over P}\sum_{j=1}^P {{\mathrm{d}}V(q_j)\over {\mathrm{d}}q_j}\right>_{\bar{q}=q},
\end{align*}
 and in view of the result in part (b), the factor of $Q(q^{\ddagger})/Q_t(T)$ in $K^{\text{QTST}}(T)$ can be calculated as
\begin{align*}
{Q(q^{\ddagger})\over Q_t(T)} = e^{-\beta F(q^{\ddagger})} = \exp\left[-\beta{\int_{q_r}^{q^\ddagger}} F'(q)\,{\mathrm{d}q}\right],
\end{align*}
where $q_r$ is in the reactant region such that $Q(q_r)=Q_r(T)$.

\item

Making use of the results we have already established, we have
\begin{align*}
\kappa(t) &= {c_{fs}^{\text{RPMD}}(t)\over\displaystyle{{1\over 2}\left<|\dot{q}|\right>_{\text{cl}}Q(q^{\ddagger})}}\\
&=\sqrt{2\pi\beta m}\,{c_{fs}^{\text{RPMD}}(t)\over Q(q^{\ddagger})}\\
&=\sqrt{2\pi\beta m}\,
{\displaystyle{ \int {\mathrm{d}}{\mathbf{p}}_0\int {\mathrm{d}}{\mathbf{q}}_0\,e^{-\beta_PH_P({\mathbf{p}}_0,{\mathbf{q}}_0)}{\bar{p}_0\over m}\delta(\bar{q}_0-q^{\ddagger})\theta(\bar{q}_t-q^{\ddagger}) }\over\displaystyle{  \int {\mathrm{d}}{\mathbf{p}}_0\int {\mathrm{d}}{\mathbf{q}}_0\,e^{-\beta_PH_P({\mathbf{p}}_0,{\mathbf{q}}_0)}\delta(\bar{q}_0-q^{\ddagger}) }}\\
&\equiv \sqrt{2\pi\beta m}\left<{\bar{p}_0\over m}\theta(\bar{q}_t-q^{\ddagger})\right>_{\bar{q}_0=q^{\ddagger}}.
\end{align*}

Choosing $q^{\ddagger}=0$, which is at the top of the parabolic barrier potential $V(q)=-{1\over 2}m\omega_b^2q^2$ and therefore the most natural position for the transition state dividing surface, we have
\begin{align*}
k^{\text{RPMD}}(T)Q_r(T) &= \lim_{t\to\infty} c_{fs}^{\text{RPMD}}(t)\\
&= \lim_{t\to\infty}{1\over (2\pi\hbar)^P}\int {\mathrm{d}}{\mathbf{p}}\int {\mathrm{d}}{\mathbf{q}}\, e^{-\beta_PH_P({\mathbf{p}},{\mathbf{q}})}{\bar{p}\over m}\delta(\bar{q})\theta[\bar{q}(t)],
\end{align*} 
where $\bar{q}(t)$ is the centroid of a ring polymer that has evolved from the initial phase space point $({\mathbf{p}},{\mathbf{q}})$ for time $t$. Transforming this to the normal mode representation using $\bar{p}=\tilde{p}_0/\sqrt{P}$ and $\bar{q}=\tilde{q}_0/\sqrt{P}$ gives
\begin{align*}
k^{\text{RPMD}}(T)Q_r(T) = \lim_{t\to\infty}{1\over (2\pi\hbar)^P}\int {\mathrm{d}}\tilde{\mathbf{p}}\int {\mathrm{d}}\tilde{\mathbf{q}}\, e^{-\beta_PH_P(\tilde{\mathbf{p}},\tilde{\mathbf{q}})}{\tilde{p}_0\over m}\delta(\tilde{q}_0)\theta[\tilde{q}_0(t)],
\end{align*}
where
\begin{align*}
H_P(\tilde{\mathbf{p}},\tilde{\mathbf{q}}) = \sum_{k=0}^{P-1} \left[{\tilde{p}_k^2\over 2m}+{1\over 2}m(\omega_k^2-\omega_b^2)\tilde{q}_k^2\right].
\end{align*}
Since the Hamiltonian is diagonal in the normal modes, $k^{\text{RPMD}}(T)Q_r(T)$ separates into a product of $P$ independent phase space integrals
\begin{align*}
k^{\text{RPMD}}(T)Q_r(T) = \prod_{k=0}^{P-1} I_k,
\end{align*}
where
\begin{align*}
I_0 &= \lim_{t\to\infty}{1\over 2\pi\hbar}\int {\mathrm{d}}\tilde{p}_0\int {\mathrm{d}}\tilde{q}_0\, e^{-\beta_P[\tilde{p}_0^2/2m-m\omega_b^2\tilde{q}_0^2/2]}{\tilde{p}_0\over m}\delta(\tilde{q}_0)\theta[\tilde{q}_0(t)]\\
&= \lim_{t\to\infty}{1\over 2\pi\hbar}\int {\mathrm{d}}\tilde{p}_0\, e^{-\beta_P\tilde{p}_0^2/2m}{\tilde{p}_0\over m}\theta\left[{\tilde{p}_0\sinh(\omega_bt)\over m\omega_b}\right]\\
&={1\over 2\pi\hbar}\int {\mathrm{d}}\tilde{p}_0\,e^{-\beta_P\tilde{p}_0^2/2m}{\tilde{p}_0\over m}\theta[\tilde{p}_0]\\
&={1\over 2\pi\beta_P\hbar},
\end{align*}
and
\begin{align*}
I_k &= {1\over 2\pi\hbar} \int {\mathrm{d}}\tilde{p}_k\int {\mathrm{d}}\tilde{q}_k\,e^{-\beta_P[\tilde{p}_k^2/2m+m(\omega_k^2-\omega_b^2)\tilde{q}_k^2/2]}\\
&={1\over 2\pi\hbar}\cdot\left({2\pi m\over\beta_P}\right)^{1/2}\cdot\left({2\pi\over \beta_Pm(\omega_k^2-\omega_b^2)}\right)^{1/2}\\
&={1\over\beta_P\hbar(\omega_k^2-\omega_b^2)^{1/2}},
\end{align*}
for $k\ge 1$, {provided} $\omega_k>\omega_b$. So if we are above the instanton crossover temperature where $\omega_1=(2/\beta_P\hbar)\sin(\pi/P)\simeq 2\pi/\beta\hbar>\omega_b$ (or $T>T_{\text{c}}=\hbar\omega_b/2\pi k_{\text{B}}$), we have
\begin{align*}
k^{\text{RPMD}}(T)Q_r(T) &= {1\over 2\pi\beta_P\hbar}\prod_{k=1}^{P-1} {1\over \beta_P\hbar(\omega_k^2-\omega_b^2)^{1/2}}\\
&={1\over 2\pi\hbar}\,P\prod_{k=1}^{P-1}{1\over [4\sin^2(k\pi/P)-(\beta\hbar\omega_b/P)^2]^{1/2}}.
\end{align*}
In order to show that this converges on the exact quantum mechanical result given in the question when we increase the number of ring polymer beads, we still have to prove that
\begin{align*}
\lim_{P\to\infty} P\prod_{k=1}^{P-1}{1\over [4\sin^2(k\pi/P)-(\beta\hbar\omega_b/P)^2]^{1/2}} = {\sin(\beta\hbar\omega_b/2)\over (\beta\hbar\omega_b/2)}.
\end{align*}
The easiest way to do this is to note that the $P$-bead path integral approximation to the partition function of a simple harmonic oscillator with $V(q)={1\over 2}m\omega^2q^2$ converges on the exact quantum mechanical result in the limit as $P\to\infty$, which implies using a similar argument to the one we have just given that \cite{crai-mano05ajcp}
\begin{align*}
\lim_{P\to\infty} P\prod_{k=0}^{P-1}{1\over [4\sin^2(k\pi/P)^2+(\beta\hbar\omega/P)^2]^{1/2}} = {1\over 2\sinh(\beta\hbar\omega/2)}.
\end{align*}
Eliminating the first term in the product with $k=0$ gives
\begin{align*}
\lim_{P\to\infty} P\prod_{k=1}^{P-1}{1\over [4\sin^2(k\pi/P)^2+(\beta\hbar\omega/P)^2]^{1/2}} = {\beta\hbar\omega/2\over 2\sinh(\beta\hbar\omega/2)},
\end{align*}
and setting $\omega=i\omega_b$ immediately gives the desired result. Hence RPMD does indeed give the exact quantum mechanical rate constant for a parabolic barrier.

\end{enumerate}
\end{Answer}

\setchapterpreamble[u]{\margintoc}
\chapter{Non-adiabatic Ring Polymer Rate Theory}\labch{rates-nonadiabatic}  

Chapter~\ref{ch:rates-adiabatic} has shown how RPMD rate theory can be used to calculate the rates of electronically adiabatic reactions that proceed on a single Born-Oppenheimer potential energy surface.
This Chapter extends the methodology to electronically non-adiabatic reactions in the Fermi Golden Rule limit, and more general non-adiabatic reactions that lie between the Born-Oppenheimer and Fermi Golden Rule extremes. 
The extension for non-adiabatic reactions in the Fermi Golden Rule limit was originally suggested by Wolynes in the 1980s, and so predates RPMD rate theory by almost twenty years. However, the treatment of reactions that lie between the Golden Rule and Born-Oppenheimer limits is a much harder problem that is still a topic of ongoing research. In Section~\ref{Interpolation} we describe a simple interpolation formula that can be used to combine the results of RPMD rate theory and Wolynes theory calculations for reactions in the intermediate regime. This formula is not perfect, but it is simple to implement, and there are good reasons to believe that it is the best one will ever be able to do with ring polymer methods for general electronically non-adiabatic reactions.

\section{Fermi's Golden Rule}\label{FGR}

\def\bra#1{\left<{#1}\right|}
\def\ket#1{\left|{#1}\right>}

Consider an electronically non-adiabatic reaction (such as an electron transfer reaction) that involves a transition between two diabatic electronic states $\ket{0}$ and $\ket{1}$. Sticking with one-dimensional notation for simplicity, the generic Hamiltonian for this reaction is
\begin{equation}
\hat{H} = \hat{H}_0\ket{0}\bra{0}+\hat{H}_1\ket{1}\bra{1}+{\Delta}\left(\ket{0}\bra{1}+\ket{1}\bra{0}\right),
\label{RateEq44}
\end{equation}
where
\begin{equation}
\hat{H}_i = {\hat{p}^2\over 2m}+V_i(\hat{q})\label{RateEq45}
\end{equation}
and we have taken the diabatic electronic coupling matrix element $\Delta$ to be a constant, independent of the nuclear coordinate $\hat{q}$ (the Condon approximation). 

The exact quantum mechanical rate constant for the transition from state $\ket{0}$ to state $\ket{1}$ can be written in the same way as in Eqs.~\ref{RateEq2} and \ref{RateEq3},
\begin{equation}
k(T) = {1\over Q_r(T)}\lim_{t\to\infty} \bar{c}_{fs}(t),\label{RateEq46}
\end{equation}
where 
\begin{equation}
\bar{c}_{fs}(t) = {\text{tr}}\left[e^{-\beta\hat{H}/2}\hat{F}e^{-\beta\hat{H}/2}e^{+i\hat{H}t/\hbar}\hat{\theta}\,e^{-i\hat{H}t/\hbar}\right].\label{RateEq47}
\end{equation}
However the reactant partition function is now 
\begin{equation}
Q_r(T) = {\text{tr}}\left[e^{-\beta\hat{H}}\ket{0}\bra{0}\right] = {\text{tr}}\bra{0}e^{-\beta\hat{H}}\ket{0},\label{RateEq48}
\end{equation}
(where \lq\lq Tr" denotes a trace over both the nuclear and the electronic degrees of freedom whereas \lq\lq tr" denotes a trace over just the nuclear degrees of freedom as in Chapter.~\ref{ch:rates-adiabatic}, the projection operator onto products is 
\begin{equation}
\hat{\theta} = \ket{1}\bra{1},\label{RateEq49}
\end{equation}
and the corresponding reactive flux operator is
\begin{equation}
\hat{F}={i\over\hbar}\left[\hat{H},\hat{\theta}\right] = {i\Delta\over\hbar}\left(\ket{0}\bra{1}-\ket{1}\bra{0}\right).\label{RateEq50}
\end{equation}

The arguments in parts 4,5 and 6 of Exercise~\ref{ex:rate-exercise1} also apply in the present context and show that the rate in Eq.~\ref{RateEq46} can be written equivalently as
\begin{equation}
k(T) = {1\over 2Q_r(T)}\int_{-\infty}^{\infty} c_{ff}(t)\,{\mathrm{d}}t,\label{RateEq51}
\end{equation}
where $c_{ff}(t)$ is the standard quantum mechanical flux-flux correlation function
\begin{equation}
{c}_{ff}(t) = {\text{tr}}\left[e^{-\beta\hat{H}}\hat{F}\,e^{+i\hat{H}t/\hbar}\hat{F}\,e^{-i\hat{H}t/\hbar}\right].\label{RateEq52}
\end{equation}
This equivalent expression provides a more convenient way to calculate the rate in the non-adiabatic limit where $\Delta$ can be treated as a perturbation. As shown in the following exercise, the leading order contribution to the rate constant in the limit as $\Delta\to 0$ is the Golden Rule rate
\begin{equation}
k(T) = {\Delta^2\over\hbar^2 Q_r(T)}\int_{-\infty}^{\infty}c_0(t)\,{\mathrm{d}}t,\label{RateEq53}
\end{equation}
where
\begin{equation}
Q_r(T)={\text{tr}}\left[e^{-\beta\hat{H}_0}\right]\label{RateEq54}
\end{equation}
and
\begin{equation}
c_0(t) = {\text{tr}}\left[e^{-\beta\hat{H}_0}e^{-i\hat{H}_0t/\hbar}e^{+i\hat{H}_1t/\hbar}\right].\label{RateEq55}
\end{equation}

\vspace{0.5cm}

\begin{figure}[htb!]
 \resizebox{1.6\columnwidth}{!} {\includegraphics{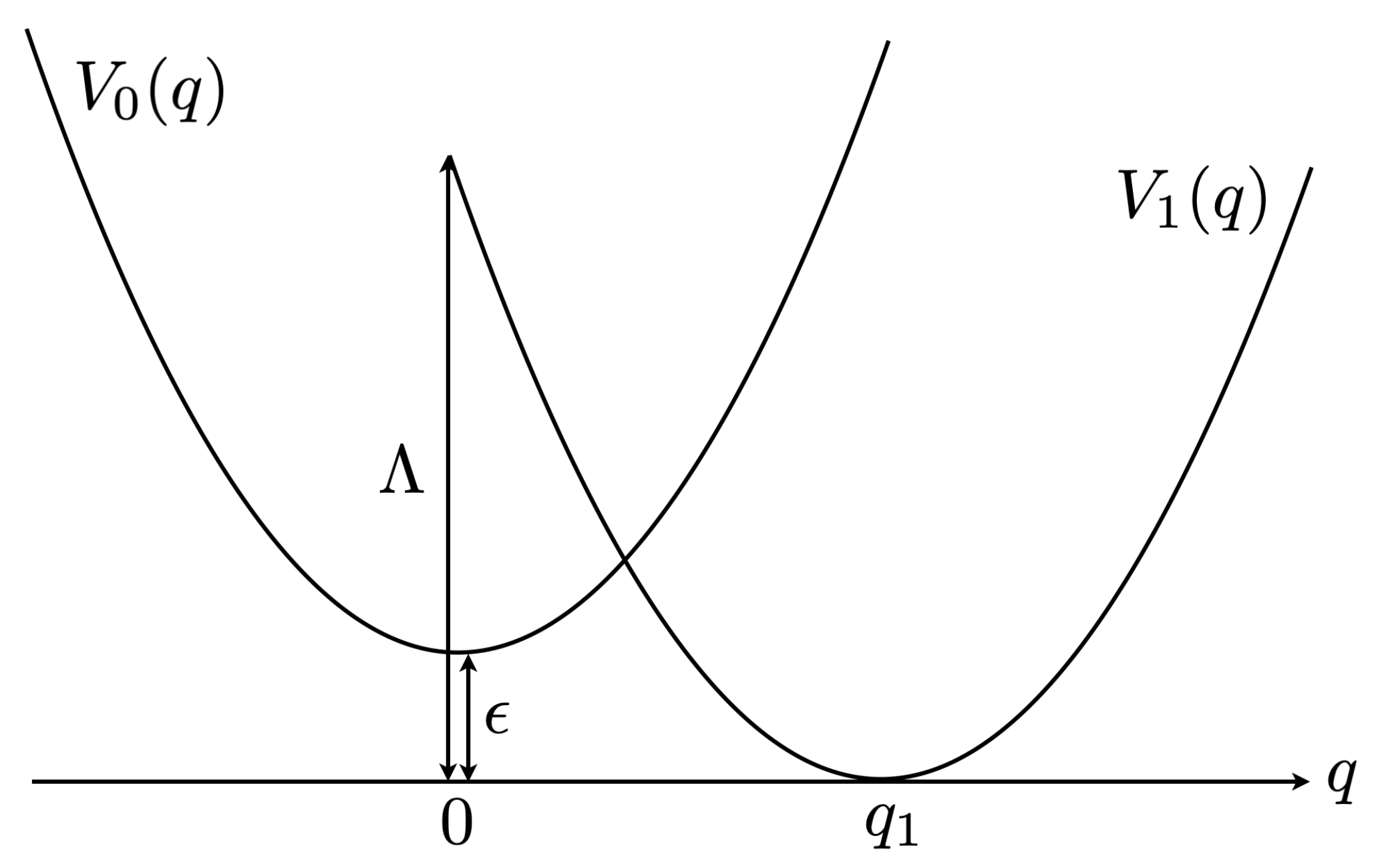}}
 \centering
 \caption{A one-dimensional spin-boson model.}
 \label{RateFig4}
 \end{figure}

\begin{exercise}[label={ex:rate-exercise4},title={Fermi's Golden Rule}]
\Question
Show that the flux-flux correlation function in Eq.~\ref{RateEq52} can be written as
\begin{equation}
c_{ff}(t) = {\Delta^2\over\hbar^2}\left[c_0(t)+c_1(t)+O(\Delta^2)\right],\label{RateEq56}
\end{equation}
where $c_0(t)$ is defined in Eq.~\ref{RateEq55} and
\begin{equation}
c_1(t) = {\text{tr}}\left[e^{-\beta\hat{H}_1}e^{-i\hat{H}_1t/\hbar}e^{+i\hat{H}_0t/\hbar}\right].\label{RateEq57}
\end{equation}
This does not immediately lead us to Eq.~\ref{RateEq53} because $c_0(t)$ and $c_1(t)$ are not the same. However, one can show that their time integrals from $-\infty$ to $+\infty$ are the same for any problem for which the quantum mechanical Golden Rule rate is well defined, thereby completing the derivation of Eq.~\ref{RateEq53}.

\Question
The quantum mechanical Golden Rule rate is only well defined if one or other (or both) of the Hamiltonians $\hat{H}_0$ and $\hat{H}_1$ has a continuous spectrum. This is the case (for example) for a one-dimensional electronic predissociation model with $V_0(q)={1\over 2}kq^2$ and $V_1(q)=Ae^{-q/q_1}-\varepsilon$. Supposing that $\hat{H}_0$ has a discrete spectrum and $\hat{H}_1$ a continuous spectrum as is the case in this predissociation model, show that the time integrals from $-\infty$ to $+\infty$ of $c_0(t)$ and $c_1(t)$ are the same.

\Question
In the classical limit, Eq.~\ref{RateEq54} becomes
\begin{equation}
Q_r(T) = {1\over 2\pi\hbar}\int {\mathrm{d}}p\int {\mathrm{d}}q\,e^{-\beta[p^2/2m+V_0(q)]},\label{RateEq58}
\end{equation}
Eq.~\ref{RateEq55} becomes
\begin{equation}
c_0(t) = {1\over 2\pi\hbar}\int {\mathrm{d}}p\int {\mathrm{d}}q\,e^{-\beta[p^2/2m+V_0(q)]}e^{-i[V_0(q)-V_1(q)]t/\hbar},\label{RateEq59}
\end{equation}
and the Golden Rule rate in Eq.~\ref{RateEq53} becomes well defined even when the classical motion in both potentials  is bounded (because classical densities of states are continuous).\\

Consider such a problem with $V_0(q)={1\over 2}kq^2$ and $V_1(q)={1\over 2}k(q-q_1)^2-\varepsilon$, as illustrated in Fig.~\ref{RateFig4}. Show that for this problem the classical Golden Rule rate is just the Marcus theory rate 
\begin{equation}
k(T) = {\Delta^2\over \hbar}\sqrt{\pi\beta\over\Lambda}e^{-\beta(\Lambda-\varepsilon)^2/4\Lambda},\label{RateEq60}
\end{equation}
where $\Lambda={1\over 2}kq_1^2$.

\end{exercise}

\section{Wolynes Theory}

In a pioneering paper published over thirty years ago, Wolynes made a steepest descent approximation to the time integral in Eq.~\ref{RateEq53} to obtain \cite{woly87jcp}
\begin{equation}
k(T)\simeq {\Delta^2\over \hbar Q_r(T)}\sqrt{2\pi\over -\beta F''(\lambda_{\mathrm{sp}})}e^{-\beta F(\lambda_{\mathrm{sp}})},\label{RateEq61}
\end{equation}
where $F(\lambda)=-\ln[c_0(i\lambda\hbar)]/\beta$ is an effective free energy such that
\begin{equation}
e^{-\beta F(\lambda)} = {\text{tr}}\left[e^{-(\beta-\lambda)\hat{H}_0}e^{-\lambda\hat{H}_1}\right],\label{RateEq62}
\end{equation}
and $\lambda_{\mathrm{sp}}$ can be found from the saddle point condition $F'(\lambda_{\mathrm{sp}})=0$. Noting that $e^{-\beta F(0)}={\text{tr}}\left[e^{-\beta\hat{H}_0}\right]=Q_r(T)$, we see that Eq.~\ref{RateEq61} can be re-written as
\begin{equation}
k(T)\simeq {\Delta^2\over\hbar}\sqrt{2\pi\over -\beta F''(\lambda_{\mathrm{sp}})}\exp\left[-\beta\int_0^{\lambda_{\mathrm{sp}}}F'(\lambda)\,{\mathrm{d}}\lambda\right]. \label{RateEq63}
\end{equation}
Since this only involves free energy derivatives, it is ideally suited to an imaginary time path integral calculation.

A $P$-bead path integral discretisation of Eq.~\ref{RateEq62} can be constructed in the same way as the standard Trotter product discretisation of an ordinary quantum mechanical partition function $Q={\text{tr}}\left[e^{-\beta\hat{H}}\right]$. The only difference is that we must now pay attention to which beads of the ring polymer necklace are on electronic state $\ket{0}$ and which on electronic state $\ket{1}$. This is straightforward to do for $\lambda$ in the range $0\le\lambda\le\beta$, and when one does it one finds that the result can be written in the form
\begin{equation}
e^{-\beta F(\lambda_l)} = {1\over (2\pi\hbar)^P}\int {\mathrm{d}}{\mathbf{p}}\int {\mathrm{d}}{\mathbf{q}}\,e^{-\beta_PH_P^{(l)}({\mathbf{p}},{\mathbf{q}})},\label{RateEq64}
\end{equation}
where $\lambda_l=l\beta_P$ for $l=0,\ldots,P$. Here
\begin{equation}
H_P^{(0)}({\mathbf{p}},{\mathbf{q}}) = h_P({\mathbf{p}},{\mathbf{q}})+\sum_{j=0}^P V_0(q_j),\label{RateEq65}
\end{equation}
\begin{equation}
H_P^{(P)}({\mathbf{p}},{\mathbf{q}}) = h_P({\mathbf{p}},{\mathbf{q}})+\sum_{j=0}^P V_1(q_j),\label{RateEq66}
\end{equation}
and
\begin{equation}
H_P^{(l)}({\mathbf{p}},{\mathbf{q}}) = h_P({\mathbf{p}},{\mathbf{q}})+\bar{V}(q_1)+\sum_{j=2}^l V_1(q_j)+\bar{V}(q_{l+1})+\sum_{j=l+2}^P V_0(q_j)\label{RateEq67}
\end{equation}
for $l=1,\ldots,P-1$, where $\bar{V}(q)={1\over 2}\left[V_0(q)+V_1(q)\right]$ is the average potential and $h_P({\mathbf{p}},{\mathbf{q}})$ is the free ring polymer Hamiltonian in Eq.~\ref{RateEq25}.

\begin{figure}[htb!]
 \resizebox{1.7\columnwidth}{!} {\includegraphics{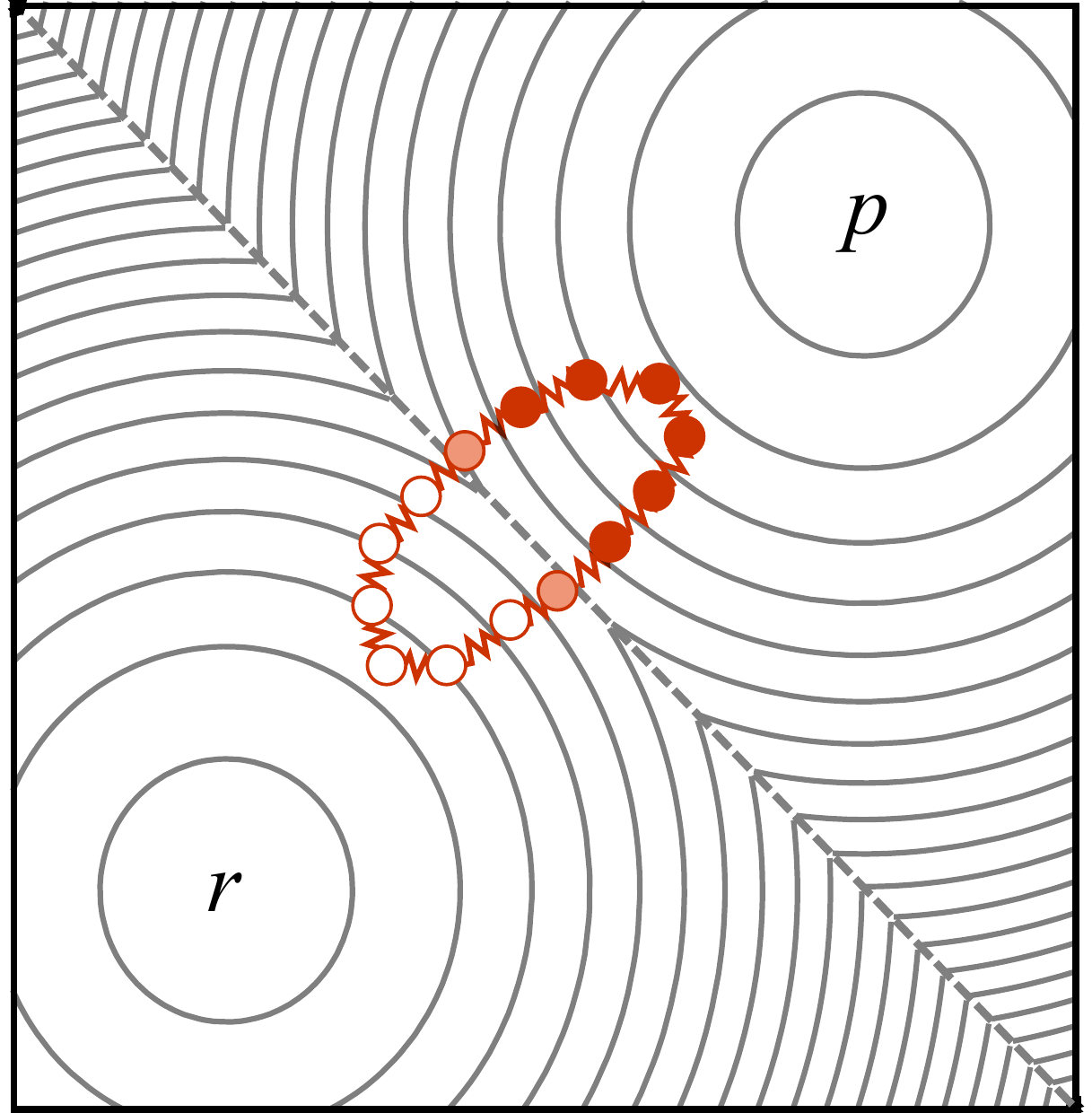}}
 \centering
 \caption{ Illustration of the path integral used in Wolynes theory when $l=P/2$ and $\lambda_l = \beta/2$. The hollow ring polymer beads are on the reactant surface ($r$), the filled beads on the product surface ($p$), and the shaded beads on the average of the two.}
 \label{RateFig5}
 \end{figure}

Using this discretisation, the derivatives of $F(\lambda)$ that are needed to evaluate Eq.~\ref{RateEq63} can be written as \cite{lawr-mano18jcp}
\begin{equation}
-\beta F'(\lambda_l) = \left<\delta V(q_1)\right>_l\label{RateEq68}
\end{equation}
and
\begin{equation}
-\beta F''(\lambda_l) = \left<\delta V(q_1)\delta V(q_{l+1})\right>_l-\left<\delta V(q_1)\right>_l^2, \label{RateEq69}
\end{equation}
where $\delta V(q) = V_0(q)-V_1(q)$ and
\begin{equation}
\left<\cdots\right>_l = {\displaystyle{\int {\mathrm{d}}{\mathbf{p}}\int {\mathrm{d}}{\mathbf{q}}\, e^{-\beta_PH_P^{(l)}({\mathbf{p}},{\mathbf{q}})}\left(\cdots\right)}\over\displaystyle{\int {\mathrm{d}}{\mathbf{p}}\int {\mathrm{d}}{\mathbf{q}}\, e^{-\beta_PH_P^{(l)}({\mathbf{p}},{\mathbf{q}})}}}\label{RateEq70}
\end{equation}
is a canonical average in the system with the ring polymer Hamiltonian $H_P^{(l)}({\mathbf{p}},{\mathbf{q}})$. For example, when $l=P/2$ and $\lambda_l=\beta/2$, this system has the ring polymer Hamiltonian illustrated in Fig.~\ref{RateFig5}.

\begin{exercise}[label={ex:rate-exercise5},title={Wolynes Theory}]
\Question
Show that, in the classical limit, Wolynes theory gives the same rate as Marcus theory for the spin-boson model in Fig.~\ref{RateFig4}  [with $V_0(q)={1\over 2}kq^2$ and $V_1(q)={1\over 2}k(q-q_1)^2-\varepsilon$].

\Question
Show that when $l=1,\ldots,P-1$, a Trotter product discretisation of 
\begin{equation}
e^{-\beta F(\lambda_l)} = {\text{tr}} \left[e^{-\lambda_l\hat{H}_1}e^{-(\beta-\lambda_l)\hat{H}_0}\right]
\end{equation}
can be written in the form of Eq.~\ref{RateEq64} with the Hamiltonian $H_P^{(l)}({\mathbf{p}},{\mathbf{q}})$ in Eq.~\ref{RateEq67}. 

\Question
Verify the expressions for $-\beta F'(\lambda_l)$ and $-\beta F''(\lambda_l)$ in Eqs. \ref{RateEq68} and \ref{RateEq69}.
\end{exercise}

\section{Interpolation Formula}\label{Interpolation}

\begin{figure}[htb!]
 \resizebox{2.5\columnwidth}{!} {\includegraphics{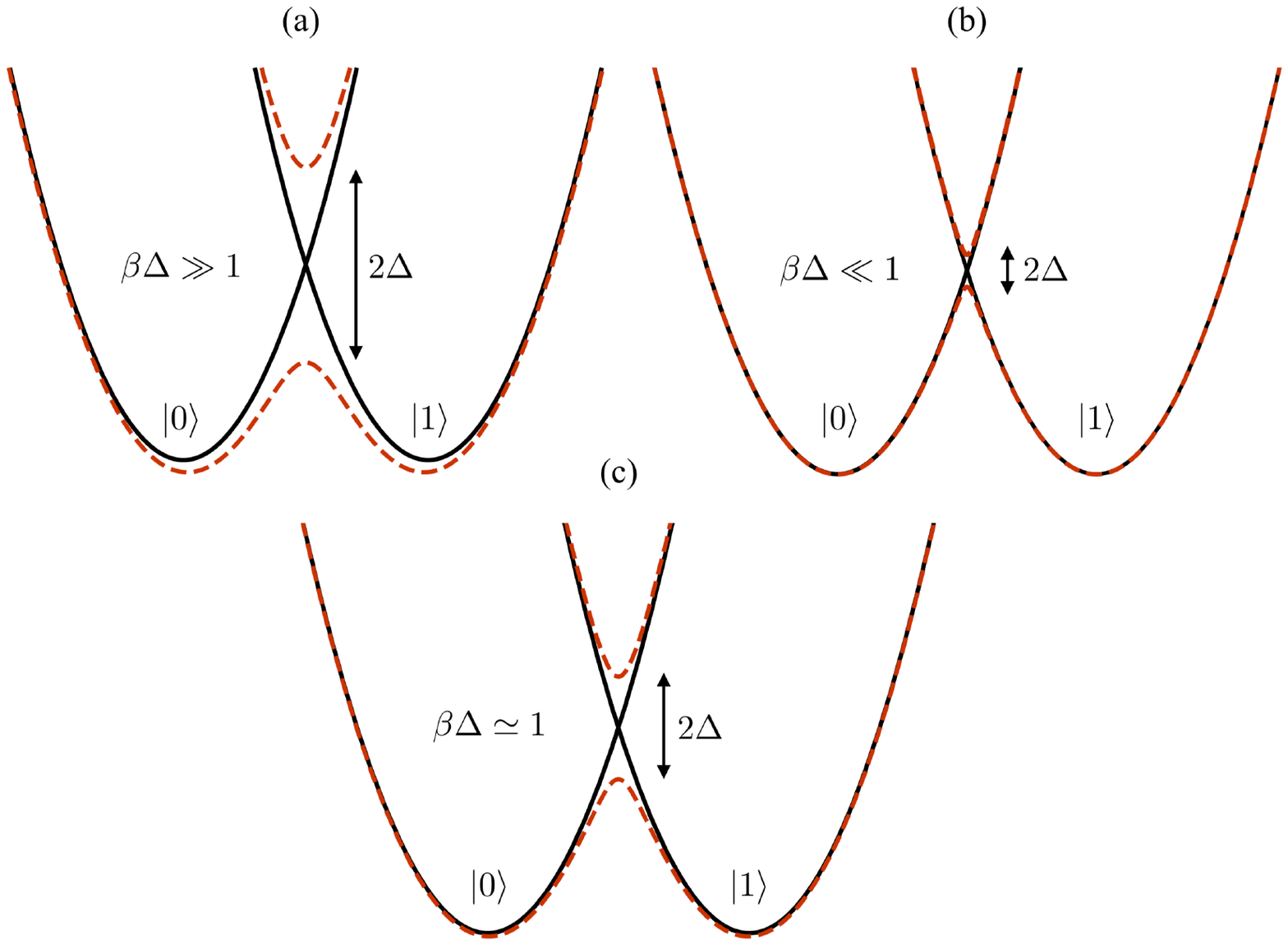}}
 \centering
 \caption{Various regimes for an electronically non-adiabatic reaction. The diabatic electronic potential energies $V_0(q)$ and $V_1(q)$ are shown in solid black, and the adiabatic energies $V_{\pm}(q) = \bar{V}(q)\pm{1\over 2}\sqrt{\delta V(q)^2+4\Delta^2}$ in dashed red, both as functions of the diabatic energy gap $\delta V(q)=V_0(q)-V_1(q)$. (a) The adiabatic limit where $\beta\Delta\gg 1$, the upper electronic adiabat is thermally inaccessible, and the Born-Oppenheimer rate on the lower adiabat is all that is required. (b) The non-adiabatic limit where $\beta\Delta\ll 1$ and Fermi's Golden Rule (second order perturbation theory in $\Delta$) gives the correct rate. (c) The intermediate regime where $\beta\Delta\simeq 1$ and neither approximation is justified (on its own).}
 \label{RateFig6}
 \end{figure}

Finally, consider an electronically non-adiabatic reaction in which the electronic coupling matrix element $\Delta$ is too large to be treated as a perturbation and yet too small for the reaction to be assumed to proceed solely on the ground adiabatic Born-Oppenheimer potential energy surface, as illustrated in Fig.~\ref{RateFig6}(c). 

\begin{figure}[htb!]
 \resizebox{1.5\columnwidth}{!} {\includegraphics{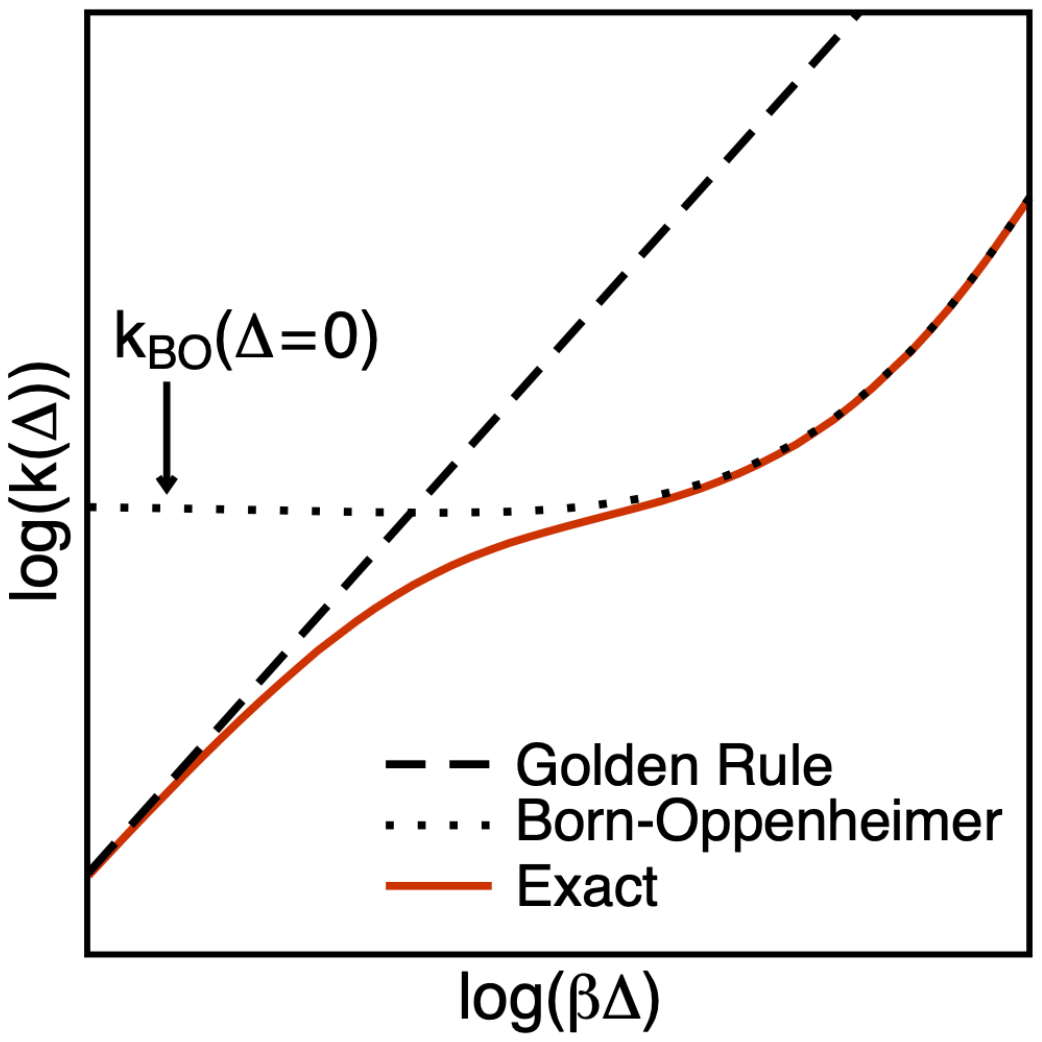}}
 \centering
 \caption{Illustration of the transition from non-adiabatic (Fermi Golden Rule) to adiabatic (Born-Oppenheimer) behaviour of an electron transfer rate constant (red curve), as a function of $\beta\Delta$. Note the logarithmic scales on both axes.}
 \label{RateFig7}
 \end{figure}

Neither RPMD rate theory nor Wolynes theory can be used to solve this problem. However, an effective solution can be constructed by combining the two methods, as illustrated in Fig.~\ref{RateFig7}. This shows the Fermi Golden Rule rate, $k_{\text{GR}}(\Delta)$, the Born-Oppenheimer rate, $k_{\text{BO}}(\Delta)$, and the exact quantum mechanical rate, $k(\Delta)$, of a typical electronically non-adiabatic reaction as a function of the electronic coupling strength. The Golden Rule rate is exact for small $\Delta$ but increases as $\Delta^2$, resulting in an over-estimation of the exact rate at large $\Delta$. The Born-Oppenheimer rate is exact for large $\Delta$ but decreases monotonically to a plateau value at small $\Delta$. The plateau value is the adiabatic rate on the cusped potential $V_{\text{min}}(q)=\min[V_0(q),V_1(q)]$, and the behaviour of $k_{\text{BO}}(\Delta)$ for larger $\Delta$ can be understood in terms of the decrease in the ground adiabatic reaction barrier (and hence the Born-Oppenheimer activation energy) with increasing $\Delta$.

These observations suggest making a Pad\'e-like approximation to $k(\Delta)$,
\begin{equation}
k(\Delta) \simeq {k_{\text{BO}}(0)k_{\text{GR}}(\Delta)\over k_{\text{BO}}(\Delta)+k_{\text{GR}}(\Delta)},\label{RateEq72}
\end{equation}
where $k_{\text{BO}}(0)$ is the adiabatic rate on the cusped potential. As $\Delta\to 0$, the denominator in this expression becomes dominated by $k_{\text{BO}}(\Delta\to0)$, which then cancels with the $k_{\text{BO}}(0)$ in the numerator to leave $k(T) \simeq k_{\text{GR}}(\Delta)$, the correct non-adiabatic rate constant. And at large $\Delta$, the denominator becomes dominated by $k_{\text{GR}}(\Delta)$, which then cancels with the $k_{\text{GR}}(\Delta)$ in the numerator to leave $k(T)\simeq k_{\text{BO}}(\Delta)$, the correct adiabatic rate constant. Moreover both limits are approached from below, consistent with the behaviour of the exact rate constant in Fig.~\ref{RateFig7}: $k(T)$ approaches $k_{\text{GR}}(T)$ from below as $\Delta\to 0$ and $k_{\text{BO}}(T)$ from below at large $\Delta$. One can also show that  Eq.~\ref{RateEq72} satisfies detailed balance (i.e., that $k_f(\Delta)/k_b(\Delta)=K$, where $k_f(\Delta)$ and $k_b(\Delta)$ are the forward and backward rate constants and $K$ is the equilibrium constant for the reaction), and that it reduces to the well-known Zusmann formula in the appropriate (strong friction and classical nuclear motion) regime \cite{lawr+19jcp}.

Since RPMD rate theory can be used to calculate both $k_{\text{BO}}(\Delta)$ and $k_{\text{BO}}(0)$ in Eq.~\ref{RateEq72}, and Wolynes theory to calculate $k_{\text{GR}}(\Delta)$, there is no problem in implementing this interpolation formula with well-established imaginary time path integral techniques. And when one does this, one finds that the error in the interpolation formula is no larger than the error in RPMD rate theory in the Born-Oppenheimer limit or the error in Wolynes theory in the Golden Rule limit, for a wide variety of electron transfer and other electronically non-adiabatic reactions \cite{lawr+19jcp}. The interpolation formula is also physically motivated and appealingly simple. I doubt there will ever be a better or more practical way to use path integral methods to calculate electronically non-adiabatic reaction rates in the intermediate coupling regime.

\section{Additional Comments and Further Reading}\label{Non-adiabatic comments}

Since Wolynes theory is a purely imaginary time path integral method, it is just as easy to implement for complex reactions in the Golden Rule limit as RPMD rate theory is in the Born-Oppenheimer limit. Indeed the theory was used to study a fully atomistic model of aqueous ferrous-ferric electron transfer soon after it was first published \cite{bade+90jcp}, and an atomistic model of biochemical electron transfer just one year later \cite{zhen+91cp}. 

One possible issue with Wolynes theory is that its path integral implementation can only be used directly when the saddle point $\lambda_{\mathrm{sp}}$ lies in the range $0\le \lambda_{\mathrm{sp}}\le \beta$. This seems to exclude calculations in the Marcus inverted regime for reasons explained in the answer to Exercise~\ref{ex:rate-exercise5} 1. However, we have recently shown that the theory can be analytically continued into the inverted regime using path integral information calculated in the normal regime, giving a rate that is just as accurate in the inverted regime as it is in the normal regime \cite{lawr-mano18jcp}. We have also proposed an alternative \lq\lq linear golden rule" (LGR) approximation to the rate constant that uses the same ingredients as Wolynes theory but avoids the need for analytic continuation \cite{lawr-mano20ajcp}, and shown that both this and Wolynes theory provide a realistic description of nuclear quantum effects in condensed phase electron transfer reactions \cite{lawr-mano20bjcp}. This was originally established for the Wolynes theory description of aqueous ferrous-ferric electron transfer by Chandler and coworkers over thirty years ago \cite{bade+90jcp}, and it is just as true today as it was then (despite a recent misguided suggestion to the contrary, which we shall save the authors the embarrassment of citing here).

The interpolation formula discussed in Sec.~\ref{Interpolation} has so far only been validated for some simple scattering and spin-boson models: it not yet been used in an atomistic simulation. This is a pity, because the formula is both very simple and entirely general, and it is expected to work extremely well \cite{lawr+19jcp}. It should be straightforward to apply the formula in conjunction with RPMD rate theory and Wolynes theory to any condensed phase non-adiabatic reaction with a Hamiltonian of the form in Eq.~\ref{RateEq44}, and also to allow for non-Condon effects by letting $\Delta\to \Delta(\hat{q})$. If this were combined with machine learning two-state \lq\lq empirial valence bond" Hamiltonians of the form in Eq.~\ref{RateEq44} from {\em ab intio} data, I imagine it could make quite an interesting research project.

\newpage
\section*{Exercise answers}

\begin{Answer}[ref=ex:rate-exercise4]

\begin{enumerate}
\item
Inserting Eq.~\ref{RateEq50} into Eq.~\ref{RateEq51} gives
\begin{align*}
c_{ff}(t) = {\Delta^2\over\hbar^2}{\text{tr}}
\Bigl[\, &\bra{0}e^{-i\hat{H}t/\hbar}e^{-\beta\hat{H}}\ket{0}\bra{1}e^{+i\hat{H}t/\hbar}\ket{1}\\
+         &\bra{1}e^{-i\hat{H}t/\hbar}e^{-\beta\hat{H}}\ket{1}\bra{0}e^{+i\hat{H}t/\hbar}\ket{0}\\
-          &\bra{1}e^{-i\hat{H}t/\hbar}e^{-\beta\hat{H}}\ket{0}\bra{1}e^{+i\hat{H}t/\hbar}\ket{0}\\
-          &\bra{0}e^{-i\hat{H}t/\hbar}e^{-\beta\hat{H}}\ket{1}\bra{0}e^{+i\hat{H}t/\hbar}\ket{1}\,\Bigr].
\end{align*}
Combining this with the fact that $\bra{i}e^{-\alpha\hat{H}}\ket{i} = e^{-\alpha\hat{H}_i}+O(\Delta^2)$ and
$\bra{i}e^{-\alpha\hat{H}}\ket{j} = -\alpha\Delta+O(\Delta^2)$ for $i\not=j$ (both of which can be verified by using Eq.~\ref{RateEq44} for $\hat{H}$ and expanding the exponential operators as Taylor series in $-\alpha\hat{H}$) immediately gives
\begin{align*}
c_{ff}(t) = {\Delta^2\over\hbar^2}\left[c_0(t)+c_1(t)+O(\Delta^2)\right],
\end{align*}
where $c_0(t)$ and $c_1(t)$ are defined in Eqs.~\ref{RateEq55} and~\ref{RateEq57}.

\item
When $\hat{H}_0$ has discrete eigenstates $\hat{H}_0\ket{\nu}=E_{\nu}\ket{\nu}$ and $\hat{H}_1$ has a continuous spectrum, we can evaluate the time integral of $c_0(t)$ as
\begin{align*}
\int_{-\infty}^{\infty} c_0(t)\,{\mathrm{d}}t &= \int_{-\infty}^{\infty} {\text{tr}}\left[e^{-\beta\hat{H}_0}e^{+i\hat{H}_1t/\hbar}e^{-i\hat{H}_0t/\hbar}\right]\,{\mathrm{d}}t\\
&=\int_{-\infty}^{\infty} \sum_{\nu} \bra{\nu}e^{-\beta E_{\nu}}\int {\mathrm{d}}E\, \delta(E-\hat{H}_1)\,e^{+iEt/\hbar}\,e^{-iE_{\nu}t/\hbar} \ket{\nu}\,{\mathrm{d}}t\\
&= 2\pi\hbar\sum_{\nu} \bra{\nu}e^{-\beta E_{\nu}}\int {\mathrm{d}}E\, \delta(E-\hat{H}_1)\,\delta(E-E_{\nu})\ket{\nu}\\
&= 2\pi\hbar\sum_{\nu} e^{-\beta E_{\nu}}\bra{\nu}\delta(E_{\nu}-\hat{H}_1)\ket{\nu},
\end{align*}
and the time integral of $c_1(t)$ as
\begin{align*}
\int_{-\infty}^{\infty} c_1(t)\,{\mathrm{d}}t &= \int_{-\infty}^{\infty} {\text{tr}}\left[e^{-\beta\hat{H}_1}e^{-i\hat{H}_1t/\hbar}e^{+i\hat{H}_0t/\hbar}\right]\,{\mathrm{d}}t\\
&=\int_{-\infty}^{\infty} \sum_{\nu} \bra{\nu}\int {\mathrm{d}}E\, \delta(E-\hat{H}_1)\,e^{-\beta E}\,e^{-iEt/\hbar}\,e^{+iE_{\nu}t/\hbar} \ket{\nu}\,{\mathrm{d}}t\\
&= 2\pi\hbar\sum_{\nu} \bra{\nu}\int {\mathrm{d}}E\, \delta(E-\hat{H}_1)\,e^{-\beta E}\,\delta(E-E_{\nu})\ket{\nu}\\
&= 2\pi\hbar\sum_{\nu} e^{-\beta E_{\nu}}\bra{\nu}\delta(E_{\nu}-\hat{H}_1)\ket{\nu},
\end{align*}
both of which give the same result.\\

It also follows from this argument that we can use a standard formula from quantum scattering theory,
\begin{align*}
\delta(E-\hat{H}_1) = -{1\over\pi}{\mathrm{Im}}\, \hat{G}_1^+(E),
\end{align*}
where
\begin{align*}
\hat{G}_1^+(E) = \lim_{\eta\to 0_+} \left(E+i\eta-\hat{H}_1\right)^{-1},
\end{align*}
to calculate the Golden Rule rate constant in Eq.~\ref{RateEq53} as \cite{lawr-mano18jcp}
\begin{align*}
k(T) = -{2\Delta^2\over \hbar Q_r(T)}\sum_{\nu} e^{-\beta E_{\nu}}{\mathrm{Im}}\bra{\nu}\hat{G}_1^+(E_{\nu})\ket{\nu}.
\end{align*}
This actually provides quite a practical way to calculate quantum mechanical Golden Rule rate constants for gas phase electronic predissociation problems.

\item
For the classical limit of the one-dimensional spin-boson model in Fig.~\ref{RateFig4} with $V_0(q)={1\over 2}kq^2$ and $V_1(q)={1\over 2}k(q-q_1)^2-\varepsilon$, we have
\begin{align*}
Q_r(T) &= {1\over 2\pi\hbar}\int {\mathrm{d}}p\int {\mathrm{d}}q\, e^{-\beta[p^2/2m+V_0(q)]}\\
&= {1\over\beta\hbar}\sqrt{m\over k},
\end{align*}
and
\begin{align*}
c_0(t) &= {1\over 2\pi\hbar}\int {\mathrm{d}}p\int {\mathrm{d}}q\,e^{-\beta[p^2/2m+V_0(q)]}e^{-i[V_0(q)-V_1(q)]t/\hbar}\\
&={1\over\hbar}\sqrt{m\over 2\pi\beta}\int {\mathrm{d}}q\,e^{-\beta V_0(q)}e^{-ikq_1(q-q^{\ddagger})t/\hbar},
\end{align*}
where $q^{\ddagger}=(\Lambda-\varepsilon)/kq_1$ with $\Lambda={1\over 2}kq_1^2$. Therefore
\begin{align*}
\int_{-\infty}^{\infty} c_0(t)\,{\mathrm{d}}t &={1\over\hbar}\sqrt{m\over 2\pi\beta}\int {\mathrm{d}}q\,e^{-\beta V_0(q)}{2\pi
\hbar\over kq_1}\delta(q-q^{\ddagger})\\
&=\sqrt{m\over k}\sqrt{2\pi \over\beta kq_1^2}\,e^{-\beta V_0(q^{\ddagger})}\\
&=\sqrt{m\over k}\sqrt{\pi\over\beta\Lambda}e^{-\beta(\Lambda-\varepsilon)^2/4\Lambda},
\end{align*}
which gives the standard Marcus theory result
\begin{align*}
k(T) &= {\Delta^2\over \hbar^2Q_r(T)}\int_{-\infty}^{\infty} c_0(t)\,{\mathrm{d}}t\\
&= {\Delta^2\over\hbar}\sqrt{\pi\beta\over\Lambda}e^{-\beta(\Lambda-\varepsilon)^2/4\Lambda}.
\end{align*}
\end{enumerate}

\end{Answer}

\begin{Answer}[ref=ex:rate-exercise5]

\begin{enumerate}
\item Here we are again considering the classical limit of a one-dimensional spin-boson model with $V_0(q)={1\over 2}kq^2$ and $V_1(q)={1\over 2}k(q-q_1)^2-\varepsilon$, so we can steal some of the intermediate results from Exercise~\ref{ex:rate-exercise4}3. In particular, the classical limit of $Q_r(T)$ is 
\begin{align*}
Q_r(T) = {1\over\beta\hbar}\sqrt{m\over k}
\end{align*}
and the classical limit of $e^{-\beta F(\lambda)}=c_0(i\lambda\hbar)$ is 
\begin{align*}
e^{-\beta F(\lambda)}&= {1\over\hbar}\sqrt{m\over 2\pi\beta}\int {\mathrm{d}}q\,e^{-\beta V_0(q)}e^{+\lambda[V_0(q)-V_1(q)]}\\
&={1\over\hbar}\sqrt{m\over 2\pi\beta}\int {\mathrm{d}}q\,e^{-\beta kq^2/2+\lambda[kq_1(q-q^{\ddagger})]}\\
&={1\over\beta\hbar}\sqrt{m\over k}e^{+\lambda^2kq_1^2/2\beta-\lambda kq_1q^{\ddagger}}\\
&=Q_r(T)\,e^{+\Lambda\lambda^2/\beta-[\Lambda-\varepsilon]\lambda},
\end{align*}
where $\Lambda={1\over 2}kq_1^2$ and $\Lambda-\varepsilon=kq_1q^{\ddagger}$. Hence
\begin{align*}
F(\lambda) = F(0)-{\Lambda\over\beta^2}\lambda^2+{[\Lambda-\varepsilon]\over\beta}\lambda,
\end{align*}
where $e^{-\beta F(0)}= Q_r(T)$. This is all we need to evaluate the Wolynes theory rate constant in Eq.~\ref{RateEq61}.

The saddle point condition $F'(\lambda_{\mathrm{sp}})=0$ gives
\begin{align*}
\lambda_{\mathrm{sp}} = {\beta\over 2}\left(1-{\varepsilon\over \Lambda}\right),
\end{align*} 
which decreases from $\lambda_{\mathrm{sp}}=\beta/2$ for symmetric electron transfer ($\epsilon=0$) to $\lambda_{\mathrm{sp}}=0$ for activationless electron transfer ($\epsilon=\Lambda$), and becomes negative in the Marcus inverted regime ($\epsilon>\Lambda$). The implications of this are discussed in Sec.~\ref{Non-adiabatic comments}.\\

The remainder of the calculation is simply to note that  $F(\lambda_{\mathrm{sp}}) = {[\Lambda-\varepsilon]/4\Lambda}$ and $F''(\lambda_{\mathrm{sp}}) = -{2\Lambda/\beta}$ give
\begin{align*}
k(T) &= {\Delta^2\over \hbar Q_r(T)}\sqrt{2\pi\over -\beta F''(\lambda_{\mathrm{sp}})}e^{-\beta F(\lambda_{\mathrm{sp}})}\\
&= {\Delta^2\over\hbar}\sqrt{\pi\beta\over\Lambda}e^{-\beta(\Lambda-\varepsilon)^2/4\Lambda},
\end{align*}
and thus that Wolynes theory does indeed give the correct (Marcus theory) result in the classical limit for the spin-boson model. 

\item
The Trotter product discretisation is based on the approximation
\begin{align*}
e^{-\beta_P\hat{H}} \simeq e^{-\beta_P\hat{V}/2}e^{-\beta_P\hat{T}}e^{-\beta_P\hat{V}/2},
\end{align*}
which has an error of $O(\beta_P^3)$ and gives
\begin{align*}
\bra{q} e^{-\beta_P\hat{H}} \ket{q'} &= e^{-\beta_P V(q)/2}\bra{q} e^{-\beta_P\hat{T}} \ket{q'} e^{-\beta_PV(q')/2}\\
&=\int {\mathrm{d}}p\,\left<q|p\right>e^{-\beta_Pp^2/2m}\left<p|q'\right>e^{-\beta_P[V(q)+V(q')]/2}\\
&= {1\over 2\pi\hbar} \int {\mathrm{d}}p\, e^{-\beta_P p^2/2m+ip(q-q')/\hbar}e^{-\beta_P[V(q)+V(q')]/2}\\
&= {1\over 2\pi\hbar}\sqrt{2\pi m\over\beta_P}\,e^{-\beta_P[m\omega_P^2(q-q')^2/2+V(q)/2+V(q')/2]}\\
&= {1\over 2\pi\hbar} \int {\mathrm{d}}p\,e^{-\beta_P[p^2/2m+m\omega_P^2(q-q')^2/2+V(q)/2+V(q')/2]},
\end{align*}
with $\omega_P=1/(\beta_P\hbar)$.\\

In order to derive Eq.~(1.64), we simply use this result $P$ times:
\begin{align*}
e^{-\beta_PF(\lambda_l)} &= {\text{tr}}\left[e^{-\lambda_l\hat{H}_1}e^{-(\beta-\lambda_l)\hat{H}_0}\right]\\
&= \int {\mathrm{d}}{\mathbf{q}} \bra{q_1} e^{-\beta_P\hat{H}_1} \ket{q_2}\cdots \bra{q_l}e^{-\beta_P\hat{H}_1}\ket{q_{l+1}}\\
&\phantom{xxx}\times \bra{q_{l+1}}e^{-\beta_P\hat{H}_0}\ket{q_{l+2}}\cdots\bra{q_P}e^{-\beta_P\hat{H}_0}\ket{q_1}\\
&= {1\over (2\pi\hbar)^P}\int {\mathrm{d}}{\mathbf{p}}\int {\mathrm{d}}{\mathbf{q}}\,e^{-\beta_PH_P^{(l)}({\mathbf{p}},{\mathbf{q}})},
\end{align*}
with $H_P^{(l)}({\mathbf{p}},{\mathbf{q}})$ defined as in Eq.~\ref{RateEq67}.

\item
From the definition of $F(\lambda)$ in Eq.~\ref{RateEq62},
\begin{align*}
-\beta F'(\lambda)e^{-\beta F(\lambda)} &= {{\mathrm{d}}\over {\mathrm{d}}\lambda} {\text{tr}}\left[e^{-(\beta-\lambda)\hat{H}_0}e^{-\lambda\hat{H}_1}\right]\\
&= {\text{tr}}\left[e^{-(\beta-\lambda)\hat{H}_0}[\hat{H}_0-\hat{H}_1]\,e^{-\lambda\hat{H}_1}\right]\\
&= {\text{tr}}\left[e^{-(\beta-\lambda)\hat{H}_0}[V_0(\hat{q})-V_1(\hat{q})]\,e^{-\lambda\hat{H}_1}\right]\\
&= {\text{tr}}\left[e^{-(\beta-\lambda)\hat{H}_0}\delta V(\hat{q})\,e^{-\lambda\hat{H}_1}\right]\\
&= {\text{tr}}\left[\delta V(\hat{q})\,e^{-\lambda\hat{H}_1}\,e^{-(\beta-\lambda)\hat{H}_0}\right],
\end{align*}
and therefore
\begin{align*}
-\beta F'(\lambda) &= {\displaystyle{{\text{tr}}\left[\delta V(\hat{q})\,e^{-\lambda\hat{H}_1}\,e^{-(\beta-\lambda)\hat{H}_0}\right]}\over\displaystyle{{\text{tr}}\left[e^{-\lambda\hat{H}_1}\,e^{-(\beta-\lambda)\hat{H}_0}\right]}}.
\end{align*}
Differentiating again and using the same argument gives
\begin{align*}
-\beta F''(\lambda) &= {\displaystyle{{\text{tr}}\left[\delta V(\hat{q})\,e^{-\lambda\hat{H}_1}\,\delta V(\hat{q})\,e^{-(\beta-\lambda)\hat{H}_0}\right]}\over\displaystyle{{\text{tr}}\left[e^{-\lambda\hat{H}_1}\,e^{-(\beta-\lambda)\hat{H}_0}\right]}}
-{\displaystyle{{\text{tr}}\left[\delta V(\hat{q})\,e^{-\lambda\hat{H}_1}\,e^{-(\beta-\lambda)\hat{H}_0}\right]^2}\over\displaystyle{{\text{tr}}\left[e^{-\lambda\hat{H}_1}\,e^{-(\beta-\lambda)\hat{H}_0}\right]^2}}.
\end{align*}
Equations~\ref{RateEq68} to \ref{RateEq70} in the text simply implement these expressions for $\lambda_l=l\beta_P$ with $l=0,\ldots,P$, where they can be written in terms of canonical averages of $\delta V(q_1)$ and $\delta V(q_1)\delta V(q_{l+1})$ in the system with the ring polymer Hamiltonian $H_P^{(l)}({\mathbf{p}},{\mathbf{q}})$.

\end{enumerate}
\end{Answer}

\backmatter

\setchapterstyle{plain}

\defbibnote{bibnote}{}
\printbibliography[heading=bibintoc, title=Bibliography, prenote=bibnote]

\printindex

\end{document}